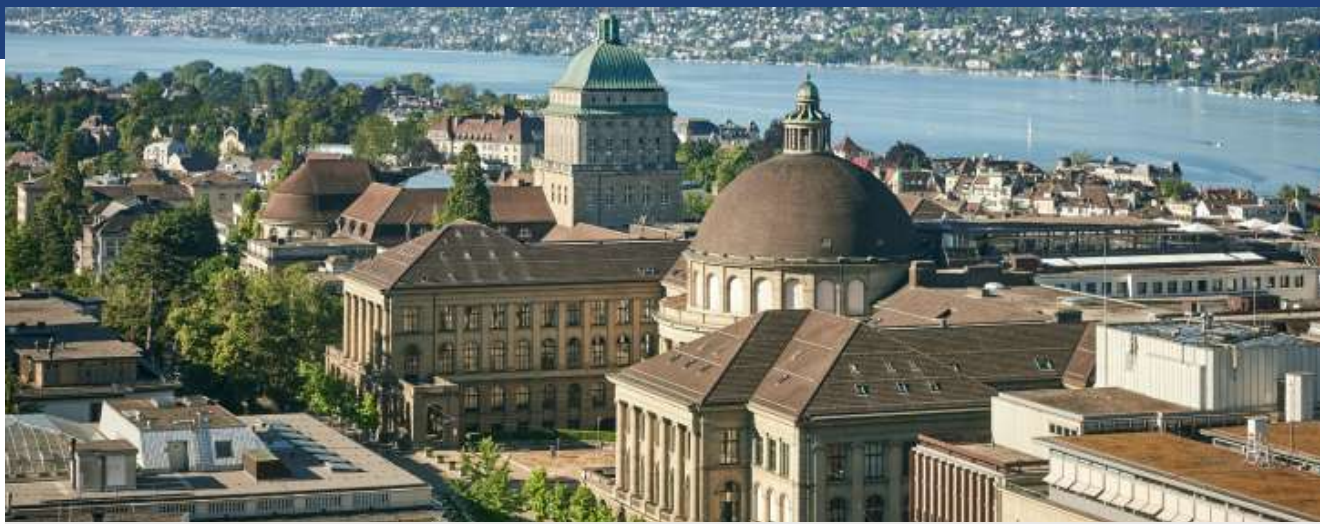

# ETH zürich

# Research Collection

Doctoral Thesis

# Lattice Boltzmann modeling of two-phase flow in macroporous media with application to porous asphalt

**Author(s):**
Son, Soyoun

ETH Library

# Lattice Boltzmann modeling of two-phase flow in macroporous media with application to porous asphalt

A thesis submitted to attain the degree of

**DOCTOR OF SCIENCES of ETH ZURICH**

(Dr. sc. ETH Zurich)

presented by

**SOYOUN SON**

MSc in Mechanical Engineering, Chung-Ang University

born on 29.01.1986

citizen of Republic of Korea

accepted on the recommendation of

Prof. Dr. Jan Carmeliet, examiner

Prof. Dr. Manfred Partl, co-examiner

Prof. Dr. Guenther Meschke, co-examiner

Dr. Qinjun Kang, co-examiner

Dr. Dominique Derome, co-examiner

2016

# Abstract

Porous asphalt (PA) is an open-graded porous material with a porosity reaching 20%, allowing fast drainage of rain and improving driving and acoustic conditions. However, the high porosity leads to significant contact with water resulting in a shorter life expectancy. To improve the durability and performance of PA, the distribution of water and its residence time have to be understood which entails capturing diverse multiphase phenomena such as gravity-driven drainage, capillary uptake and droplet wetting.

For these reasons, a numerical study is performed to analyze in detail the fluid transport mechanisms at play in PA, towards estimating the liquid distribution inside the nanometer- to millimeter-sized pore structure of PA. In this study, the lattice Boltzmann method (LBM), which is based on microscopic models and mesoscopic kinetic equations, is used for a detailed analysis of multiphase flow in complex porous domains. A main advantage of the multiphase LBM, here using the pseudopotential approach, is that the interface between different phases is automatically tracked by introducing an adequate equation of state. A multiphase single component LBM method, with parallel computing, has been developed which allows to study in three dimensions different phase separation phenomena on surfaces and in porous media successfully, including gravity effects and different boundary conditions.

The LBM is first validated with Laplace law and dynamic capillary intrusion test. Thereafter, 2D and 3D capillary uptake simulations are performed with LBM and validated with analytical solutions, varying contact angles, tube shapes and sizes. Pore meniscus and corner arc menisci are studied in both square and triangular tubes. In order to address the behavior of rain droplets on a PA surface, run-off, wetting, pinning and evaporation of single droplet are considered in terms of effects of variation of contact angle, surface wetting heterogeneity and structure. Finally, gravity-driven drainage in PA is studied with 2D and 3D LBM in accordance with temporal evolution of water distribution and the LB results are compared with experimental data, showing good agreement.

This study allows a better understanding of the diverse multiphase flow phenomena occurring in complex porous media, namely PA, at pore scale in saturated and unsaturated states, providing information towards improving the durability and performance of PA. The findings can be potentially extended to pavement design by considering real pore network documented with the complicated phenomena occurring inside PA as captured LBM.

# Résumé

Le revêtement bitumineux drainant (RBD) est un matériau poreux avec une porosité atteignant 20%, ce qui permet un drainage rapide de la pluie et une amélioration de la conduite et des conditions acoustiques. Cependant, cette porosité élevée résulte en un contact prolongé avec l'eau, réduisant ainsi l'espérance de vie du matériau. Pour améliorer la durabilité et la performance du RBD, les modes de distribution de l'eau et la durée de la présence d'eau doivent être connus, ce qui en retour demande une meilleure compréhension de phénomènes multiphasiques tels que le drainage par gravité, l'absorption capillaire et le mouillage par gouttes.

Pour ces raisons, une étude numérique est réalisée pour analyser en détail les mécanismes de transport de fluide dans le RBD, permettant l'estimation de la distribution du l'eau dans ce système poreux, allant du nanomètre au millimètre. Dans cette étude, la méthode Boltzmann sur réseau (MBR), qui est basée sur des modèles microscopiques et des équations cinétiques mésoscopiques, est utilisée pour une analyse détaillée de l'écoulement multiphasique dans des domaines complexes poreux. Un avantage principal de la MBR multiphasique, utilisant ici l'approche pseudopotentielle, est que l'interface entre les différentes phases est automatiquement suivie par l'introduction d'une équation d'état adéquate. Une méthode MBR multiphasique, en calcul parallèle, est développée permettant d'étudier en trois dimensions des phénomènes de séparation de phase sur les surfaces et dans les milieux poreux avec succès, comprenant les effets de la gravité et différentes conditions aux limites.

La MBR est d'abord validée avec la loi de Laplace et un test d'intrusion capillaire dynamique. Par la suite, des simulations d'absorption capillaire en deux et trois dimensions, validées avec des solutions analytiques, sont effectuées faisant varier l'angle de contact, la forme des tubes et leurs tailles. Les ménisques dans les pores et dans les arêtes sont étudiés pour des tubes carrés et triangulaires. Afin d'étudier le comportement des gouttes de pluie sur une surface RBD, le ruissellement, le mouillage et l'évaporation d'une goutte sont considérés faisant varier l'angle de contact, la mouillabilité hétérogène d'une surface et la structure de cette surface. Enfin, le drainage par gravité dans le RBD est étudié en 2D et 3D avec la MBR donnant l'évolution temporelle de la distribution de l'eau. Les résultats MBR sont comparés avec les données expérimentales, démontrant un bon accord.

Cette étude permet de mieux comprendre les divers phénomènes d'écoulement multiphasique qui se produisent dans les milieux poreux complexes, tel le RBD, à l'échelle des pores en états saturés et insaturés, fournissant des informations en vue d'améliorer la durabilité et la performance de revêtements. Les résultats peuvent être potentiellement utilisés pour la conception de chaussée en utlisant des simulations par réseaux de pores dont les propriétés seraient calculées par MBR.

# Acknowledgements

The research presented in this dissertation could become real with help and support from many people. I would like to express my gratitude to the people who helped and supported me.

First of all, the biggest thanks goes to my promoter Jan Carmeliet. He gave me the opportunity to work on this topic and provided constant support and guidance. I would like to thank Dominique Derome who shared her time, thought and enthusiasm for scientific discovery. She definitely helped me organize the whole research as well as the dissertation. Also, I would like to thank Qinjun Kang and Li Chen for being mentors to study lattice Boltzmann method (LBM) and providing fruitful comments. I would also like to thank Manfred Partl, Guenther Meschke and Qinjun Kang for being in the examination committees and for providing valuable comments. Furthermore, I would like to thank Lily Poulikakos for supporting the porous asphalt (PA) project and variable discussion.

For the fruitful discussion, I would like to thank Robert Guyer, Marc Prat, Manuel Marcoux and Benoit Coasne. Also, I would like to thanks Manuel Marcoux for providing experimental data of gravity-driven drainage.

I would like to thank Sreeyuth Lal, who started this SNF project and worked together, for providing some knowledge of PA and helping me with endless discussion.

I would like to thank Jan Carmeliet and Dominique Derome also for patiently reading the manuscript and for their critical and valuable comments, which definitely helped me improve this thesis. I thank Dominique Derome for French translate.

I would like to thank Martina Koch and Karin Schneider for their administrative support at Empa and ETH Zurich. Furthermore, I thank all my colleagues at Empa and ETH Zurich for variable discussion for work and study. From discussion, I had learned and understood various academic fields and obtained scientific knowledge. Furthermore, I would like to thank them for many enjoyable moments.

I would also like to acknowledge the Swiss National Science Foundation (SNF) for funding this project under the grant no. 200021-143651. The project was also supported by Swiss Federal Laboratories for Materials Science and Technology (EMPA). Their supports are greatly acknowledged.

I would like to thank my parents for their love, encouragement and endless support, which made it possible for me to come up to this point and to continue work in academic field. Finally, I would like to thank my sister, Sora and brother, Joohwan and my family to stand by my side and to believe me.

Soyoun Son
October 2016

# Nomenclature

This list of symbols is not exhaustive. Symbols that only appear locally in the text, or are self-explanatory, are not included.

**Roman symbols**

| | | |
|---|---|---|
| $a$ | pillar width/patch size | m or lattice |
| $A$ | area | $m^2$ or $lattice^2$ |
| $A_{corner}$ | corner fluid area | $m^2$ or $lattice^2$ |
| $A_{tot}$ | total wetted area | $m^2$ or $lattice^2$ |
| $Ai$ | wetted area on hydrophilic area (or hydrophobic area) | $m^2$ or $lattice^2$ |
| $b$ | pitch width | m or lattice |
| $Bo$ | Bond number | - |
| $c$ | lattice speed | lattice/time step |
| $c_s$ | sound speed | m/s or lattice/time step |
| $C$ | concentration of the vapor phase | $mg/m^3$ |
| $C_n$ | normalized curvature | m or lattice |
| $C_s$ | saturated vapor concentration | $mg/m^3$ |
| $C_\infty$ | vapor concentration far from the droplet surface | $mg/m^3$ |
| $d$ | distance between parallel plates | m or lattice |

| $D$ | groove depth | m or lattice (chap.4) |
|---|---|---|
| $D$ | contact droplet diameter | m or lattice (chap.5) |
| $D_d$ | diffusion coefficient | $m^2/s$ |
| $\mathbf{e}_i$ | discrete velocity | lattice/time step |
| $Fc$ | capillary force | N |
| $F_f$ | friction force | N |
| $F_g$ | gravity force | N |
| $\mathbf{F}_m$ | cohesive force | N |
| $\mathbf{F}_a$ | adhesive force | $kg \cdot m/s^2$ or lattice mass unit · lattice/time $step^2$ |
| $\mathbf{F}_b$ | body force | $kg \cdot m/s^2$ or lattice mass unit · lattice/time $step^2$ |
| $\mathbf{F}_{total}$ | total force | $kg \cdot m/s^2$ or lattice mass unit · lattice/time $step^2$ |
| $g$ | gravitational acceleration | $m/s^2$ or lattice/time $step^2$ |
| $G$ | Gibbs free energy | $kg \cdot m^2/s^2$ |
| $G_0$ | initial free energy | $kg \cdot m^2/s^2$ |
| $\mathbf{G}$ | interparticle interaction | - |
| $h$ | height of liquid column | m or lattice |
| $h_{max}$ | maximum height of liquid column | m or lattice |
| $H$ | groove height | m or lattice (chap. 4) |
| $H$ | droplet height | m or lattice (chap. 5) |
| $i$ | finite direction | - |
| $iter$ | iteration time step | time step |
| $I_m$ | evaporative flux | $m^3/s$ |
| *k* | permeability | Darcy |
| *l* | length | m or lattice |
| *lu* | lattice unit | - |
| *L* | side length of the tube | m or lattice |
| $L_{cap}$ | capillary length | m or lattice |
| $L_{contact}$ | side length of the corner arc meniscus wetting the side of the tube | m or lattice |
| *m* | mass | kg or lattice mass unit |
| *Ma* | Mach number | - |
| *n* | sides of n-sided polygonal tubes | |
| *n* | number of pillars | - (chap.5) |
| *p* | pressure | Pa |
| $p_c$ | capillary pressure | Pa |
| $p_G$ | gas pressure | Pa |
| $p_L$ | liquid pressure | Pa |
| *q* | mass flux | kg/s·m$^2$ |
| *r* | radius of capillary tube | m or lattice (chap. 4) |
| *r* | radius of droplet segment | m or lattice (chap. 5) |
| $r_{arc}$ | radius of curvature of the corner arc meniscus | m or lattice |
| $r_p$ | radius of capillary tube | m or lattice |
| *R* | contact droplet radius | m or lattice (chap. 5.3) |
| *R* | radius of spherical shaped droplet | m or lattice (chap. 5.4 and chap. 7) |
| **R** | ideal gas constant | - |
| *Re* | Reynold number | - |
| *S* | spreading parameter | - |
| $S_w$ | degree of saturation | - |
| *S* | perimeter | m or lattice |
| *t* | time | s |
| $t^*$ | non-dimensional time | - |
| $T$ | temperature | °C |
| $u$ | fluid velocity | m/s |
| **u** | macroscopic velocity | m/s or lattice/time step |
| $\mathbf{u}_r$ | real fluid velocity | m/s |
| $V$ | volume | m$^3$ |
| $w$ | solid-liquid interaction parameter | - |
| $w_o$ | solid-liquid interaction parameter for the surface | - |
| $w_i$ | solid-liquid interaction parameter for the groove | - |
| $W$ | excess energy | kg·m$^2$/s$^2$ |
| $x, y, z$ | Cartesian coordinates | m |

## Greek symbols

| | | |
|---|---|---|
| $\alpha$ | half corner angle | ° |
| $\alpha$ | anticlockwise angle of droplet | ° |
| $\beta$ | mass fraction | - |
| $\delta\theta$ | difference of contact angle | ° |
| $\delta F$ | depinning force | N |
| $\delta G$ | excess Gibbs free energy | kg·m$^2$/s$^2$ |
| $\delta x_{lb}$ | grid spacing | - |
| $\delta t_{lb}$ | time step | - |
| $\delta x$ | discrete space interval | - |
| $\delta t$ | discrete time interval | - |
| $\varepsilon$ | roughness | - |
| $\Delta x$ | spatial resolution | m |
| $\chi$ | parameter | - |
| $\theta$ | contact angle | ° |
| $\theta_{App}$ | apparent contact angle | ° |
| $\theta_l$ | local contact angle | ° |

| $\theta_{crit}$ | critical contact angle (for droplet depinning ) | ° |
| $\theta_0$ | equilibrium contact angle | ° |
| $\theta_c$ | critical contact angle ( for corner flow ) | ° |
| $\theta_{eff}$ | equivalent contact angle | ° |
| $\theta_o$ | contact angle of surface | ° |
| $\theta_i$ | contact angle of groove | ° |
| $\theta_p$ | contact angle of the meniscus inside the capillary | ° |
| $\xi$ | macroscopic velocity | lattice/time step |
| $\eta$ | dynamic viscosity | kg/m·s |
| $\nu$ | kinematic viscosity | $m^2/s$ |
| $\omega_i$ | weighting factor | - |
| $\psi$ | effective mass | - |
| $\rho$ | density | $kg/m^3$ or lattice mass unit/lattice$^3$ |
| $\gamma$ | surface tension | $kg/s^2$ |
| $\tau$ | relaxation time | - |
| $\varphi$ | level set function | - |
| $\Delta p$ | pressure difference | Pa |
| $\Omega$ | collision operator | $m^3$ |
| $\lambda$ | parameter as a function of contact angle of droplet $\theta$ | - |
| $\Gamma$ | closed curve | - |

**Subscripts**

| $APP$ | apparent |
| $bottom$ | bottom of capillary |
| $c$ | critical |
| $droplet$ | droplet |
| $g$ | gas phase |
| $in$ | inside of the droplet |
| $l$ | liquid phase |
| *lb* | lattice unit |
| *out* | outside of a droplet |
| *p* | physical unit |
| *pillar* | pillar |
| *SG* | solid-gas |
| *SL* | solid-liquid |
| *LG* | liquid-gas |
| *top* | top of capillary |
| *tot* | total |

## Acronyms

| | |
|---|---|
| *AC* | asphalt concrete |
| *BGK* | Bhantagar-Gross-Krock |
| *CAST* | coaxial shear test |
| *CBC* | convective boundary condition |
| *CFD* | computational fluid dynamics |
| *CFL* | Courant-Friedrichs-Lewy |
| *CHT* | constant head test |
| *EBC* | extrapolation boundary condition |
| *EDM* | exact-difference method |
| *EOS* | equation of state |
| *FC* | front capturing |
| *FHT* | falling head test |
| *FT* | front tracking |
| *IDT* | indirect Tensile tests |
| *LBGK* | lattice Bhantagar-Gross-Krock |
| *LBM* | lattice Boltzmann method |
| *LCPC* | rolling wheel compaction |
| *L-W* | Lucas-Washburn |
| *MIP* | mercury intrusion porosimetry |
| *MPI* | message passing interface |
| *MS-P* | Mayer and Stowe-Princen |

| | | |
|---|---|---|
| *μCT* | X-ray microcomputer tomography | |
| *NBC* | Neumann boundary condition | |
| *NR* | neutron radiography | |
| *OBC* | outflow boundary condition | |
| *PA* | porous asphalt | |
| *PG* | propylene glycol | |
| *PNM* | pore network model | |
| *SGC* | Superpave Gyratory Compactor | |
| *VOF* | volume of fluid | |

## Others

| | | |
|---|---|---|
| *f* | single-particle distribution function | - |
| $f_1, f_2$ | fraction of each surface (1 or 2) | - |
| $f^{eq}$ | equilibrium distribution function | - |
| $f_i$ | density distribution function | - |

# Contents

# 1. INTRODUCTION
## 1.1. Motivation

In various types of materials, such as fuel cells, sponge, wood, soil, and rocks, pores form a connected network, where fluids, such as air, $CO_2$, water and oil, can flow through. Fluid transport in porous media is relevant for diverse scientific, engineering and industrial fields such as petroleum engineering, biology, geophysics and building physics. Yet fluid transport in porous media is not fully understood due to its complexity.

Porous media can be of quite uniform pore size or can have pore sizes ranging over several orders of magnitude, from nanometer to millimeter scale. Fluid transport can be governed by different driving forces. In nanometer- or micrometer-sized pore system, surface tension and capillary forces are more dominant than gravity. In this case, capillary uptake and imbibition are key processes to understand liquid and gas transport in porous media. In contrast, in pores larger than a millimeter, gravity becomes a dominant driving force and gravity-driven drainage is an important process. In all cases, fluid transport can be characterized by the transport properties of the material, including porosity, vapor and liquid permeability and fluid capacity (retention).

Asphalt concrete (AC) is the material most commonly used for road pavement. Porous asphalt (PA) is a special type of AC with high porosity, of about 20%, applied as top layer of road pavement. Compared with AC, PA leads to improved driving conditions and reduced glare, aquaplaning risks and noise due to the macrotexture of its surface and its high porosity (Sandberg and Ejsmont 2002). However, its high porosity and large range of pores, from micrometer to millimeter size, expose PA to complex combination of mechanical, hygric, hydric and thermal loading conditions which play a significant role in its durability. Due to these loading conditions, PA has a much shorter life time, of approximately 10 years, compared to the one of dense asphalt, which reaches 20 to 30 years. To improve its performance and durability, the role of water in the oxidation and degradation process of PA, and especially the distribution and residence time of water within the PA should be understood. Fluid transport in PA has been documented in several studies, where the structure of PA was characterized with X-ray microcomputed tomography (μCT) and where fluid transport experiments including gravity-driven drainage and capillary uptake were measured by neutron radiography (Poulikakos, Saneinejad et al. 2013, Lal, Poulikakos et al. 2014, Jerjen, Poulikakos et al. 2015). However, estimating the exact time of residence of water/moisture, and its distribution within PA, after bulk water removal by gravity-driven drainage and redistribution by capillary action remains a challenge.

For these reasons, a numerical study is performed to analyze in detail the fluid transport mechanisms at play in PA and to estimate the liquid distribution inside the nanometer- to millimeter-sized pore structure of PA. In this study, the lattice Boltzmann method (LBM), which is based on microscopic models and mesoscopic kinetic equations, is used for a detailed analysis of multiphase flow in complex porous domains of different sizes ranging from nanometer to centimeter range. Multiphase fluid phenomena, such as capillary rise in polygonal tubes, sessile droplet evaporation and spreading on different types of surfaces and gravity-driven drainage within PA are studied in terms of effects of variation of contact angle, surface heterogeneity, pore geometry and pore size distribution.

This study leads to a better understanding of multiphase flow and fluid transport in complex porous media, and in particularly PA, in saturated and unsaturated states, providing information towards improving the durability and performance of PA. This study takes place within the framework of a Swiss National Science Foundation (SNF) project "Wetting and drying of porous asphalt pavement: a multiscale approach". The global goal of this project is to develop an integrated multiscale methodology, based on numerical models validated by experiments, to accurately capture the main physics of wetting and drying of PA at the material scale (micro scale). Such an integrated methodology is intended to be applicable, through a multiscale approach, in road engineering at the pavement scale (macro scale) and in the development and assessment of pavement solutions. Two doctoral students are funded by the project. While this doctoral student research focuses on modeling diverse two-phase flow mechanisms at the pore scale using lattice Boltzmann method (LBM), the other doctoral student, Mr. Sreeyuth Lal, focuses on the multiscale characterization of PA and on the experimental and numerical investigations of wetting and drying in PA.

## 1.2. Objective and methodology

The overall objective of the research in this thesis is to better understand the multiphase flow physics at pore sizes ranging from micrometer to centimeter scale that accompanies capillary rise, corner flow, sessile droplet evaporation and spreading and gravity-driven drainage in PA. The specific objectives of this work are:

- To further develop the three-dimensional Shan-Chen pseudopotential LBM with parallel computing, MPI, for the study of multiphase flow taking into account gravity;

- To develop a framework to validate and verify LBM by comparing with analytical solutions, for Laplace pressure, contact angle and capillary intrusion, and demonstrate the applicability of LBM to explore parametrically multiphase problems;

- To explore multiphase phenomena using LBM, through quantitative analysis and qualitative exploration. In particular,
  - o To document capillary uptake phenomena, with special attention to the phenomenon of corner flow, in the two- and three-dimensional geometries at pore scale.
  - o To model and understand the behavior of sessile droplets on different surfaces considering the influence of contact angle, surface geometry and surface structure.
  - o To explore and understand gravity-driven drainage in complex porous media.

This thesis addresses the following three points:

(1) Capillary uptake at pore scale, with the aim to understand capillary uptake phenomena in the two- and three-dimensional tubes, including corner flow, the effect of contact angle, polygonal geometry and tube size;

(2) Sessile droplet behavior of size ranging several order of magnitude from the nanometer to the micrometer scale under different driving forces, including run-off by gravity on grooved surfaces, stick-slip evaporation of a droplet on a set of micropillars, droplet spreading on heterogeneous surfaces, and droplet displacement on stepping stones for various contact angle and geometry;

(3) Fluid transport in PA: based on the knowledge of the different physics of multiphase flow investigated in points 2 and 3, the gravity-driven drainage in the complex geometry of PA is studied to characterize the temporal loss and remaining distributing of entrapped water. When available from literature or through collaboration, comparison with analytical expression or experimental data is performed.

## 1.3. Outline of dissertation

This thesis is composed of eight chapters. The thesis is structured as follows:

In chapter 2, multiphase flow in general, the Shan-Chen LBM and porous asphalt are explained and the state of art of multiphase flow research is reviewed. The chapter points out to gaps in scientific knowledge in terms of understanding the transport physics in PA.

In chapter 3, the lattice Boltzmann method (LBM), used in subsequent chapters, is described in detail. The basic concept of lattice Boltzmann equation (LBE) is explained with the kinetic theory and the Boltzmann equation. Then the Shan-Chen's single component multiphase LBM is introduced, including a description of the non-ideal equation of state (EOS), the forcing scheme and the different boundary conditions. The implementation and limitations of multiphase LBM are presented. Finally, the conversion between physical and lattice units, the parallel computing based on MPI and the post-processing procedure are described briefly.

In chapter 4, capillary uptake at pore scale is investigated. Two- and three-dimensional LBM simulations are conducted to analyze two liquid configurations: the pore meniscus and corner arc menisci. The effect of geometry, size and contact angles is studied and a comparison with analytical models is given.

In chapter 5, various phenomena related to sessile droplet behavior are studied with LBM and the simulation results highlight the role of gravity for droplet displacement, surface energy for the movement of droplets on heterogeneous surfaces, and pinning/depinning behavior of droplets on micropillar structures during evaporation. Furthermore, the influence of geometry, surface heterogeneity and contact angle on the droplet behavior is investigated.

In chapter 6, two- and three-dimensional gravity-driven drainage is simulated with LBM. From the LBM simulation results, the temporal mass fraction is determined taking into account the influence of different contact angles. The chapter includes a discussion of the LB results by a detailed comparison with experimental data, provided through collaboration, to verify and validate LBM.

In chapter 7, further explorations of multiphase phenomena, inspired by experimental studies, including droplet movement on a randomly heterogeneous surface, droplet "jumping" on hydrophobic structured channels, and drying in regular and hierarchical porous media are performed using two- and three-dimensional LBM. The qualitative description of each simulation highlights the potential of such type of investigation to study various complex multiphase phenomena.

Chapter 8 summarizes the main conclusions of this thesis and presents suggestions for further research.

# 2. STATE OF ART

## 2.1. Introduction

This chapter presents an overview of the state of the art of multiphase flow in macroporous media and of its computational modelling using lattice Boltzmann method (LBM). The macroporous material of choice, porous asphalt, is also described. Section 2.2 presents an overview on different aspects of multiphase flow including: 1) equilibrium of multiphase systems, 2) wetting on homogeneous or heterogeneous surfaces, 3) evaporation of a droplet on a surface, 4) capillary rise and 5) corner flow. In section 2.3, computational fluid dynamics (CFD) with interface tracking and lattice Boltzmann method (LBM), which will be described in more detail in chapter 3, are briefly introduced. Section 2.4 describes PA, its characteristics and moisture behavior, as investigated recently experimentally and numerically. Finally, in section 2.5, this chapter ends up highlighting the current gaps in knowledge towards understanding multiphase flow in porous media, and in PA in particular, that are targeted to be filled in this thesis.

## 2.2. Multiphase flow phenomena

In fluid mechanics, the motion of fluids and the forces acting on them are studied. Systems can consist of one or several components (e.g. water and oil) and of different phases of the component: the solid, liquid and gaseous phases. This study is limited to one component and two phases: the liquid and gaseous phases. The fluid mostly envisaged in this study is water, and thus as phases liquid water and water vapor, but since LBM uses non-dimensional parameters, the results may also apply to other fluids. Multiphase flow is the simultaneous flow of liquid and gas phases, a common phenomenon occurring in nature. Compared to single phase flow, the simulation of multiphase flow is complicated by the fact that the phase interface has to be tracked over time. To understand multiphase flow, the fundamental mechanisms at play and the force balances in both equilibrium and dynamic states must be addressed.

### 2.2.1. Equilibrium of a multiphase system

A liquid is a system of molecules in condensed state, which attract each other, the attraction referred to as cohesive force. In a liquid, a molecule is equally interacting with its neighbors resulting in a net zero force as shown in Fig.2.1 (a) for molecules situated in the bulk of the liquid. However, at the surface of the liquid, a molecule lacks about half of this interaction and the molecule experiences a net force towards the bulk of the liquid. The net force at its surface causes the liquid molecules to behave as a membrane, which exhibits a surface tension or surface energy. Surface tension has the dimension of force per unit length or of energy per unit area. The surface tension at the surface will tend to minimize the surface area of the liquid. That means that the liquid will pull into a spherical shape and an internal liquid pressure will be generated.

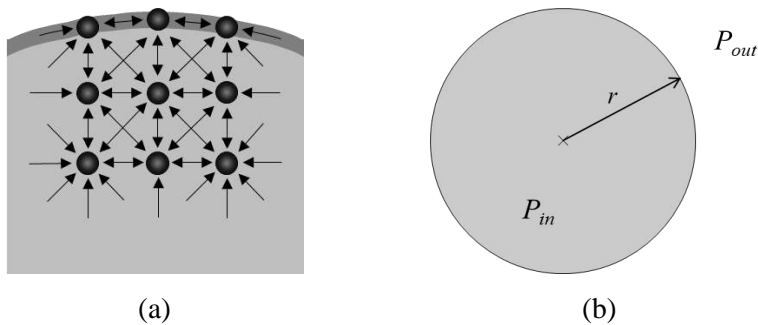

(a)                                (b)

Fig.2.1. (a) Forces on a liquid molecule within and at the surface of liquid and (b) difference in pressure between inside and outside of spherical drop with radius $r$.

The excess energy of the droplet in Fig.2.1 (b) can be expressed as:

$$\delta W = -p_{in}dV_{in} - p_{out}dV_{out} + \gamma dA \qquad (2.1)$$

where $p$ is pressure, $\gamma$ is the surface tension, $dV_{out}=dV_{in}=4\pi r^2\Delta r$ is the change of the droplet volume and $dA=8\pi r\Delta r$ is the change in surface area of the three-dimensional sphere. At equilibrium, $\delta W = 0$ and Eq. (2.1) reduces to:

$$\Delta p = p_{in} - p_{out} = \frac{2\gamma}{r} \qquad (2.2)$$

The pressure difference is referred to as Laplace pressure. Eq. (2.2) is known as the Young-Laplace equation, which is a fundamental equation to relate pressure difference (Laplace pressure) to surface tension. The pressure difference is inversely proportional to the droplet radius, meaning that smaller droplets show a larger Laplace pressure for the same surface tension.

### 2.2.2. Wetting on homogeneous surface

Wetting occurs when a liquid in contact with a solid and its vapor shows the tendency to maintain its contact with the solid surface. In the system solid, liquid and vapor, the degree of wetting, or wettability, is defined by the balance of adhesive and cohesive forces. Wetting occurs with the adhesive force is larger than the cohesive one and as seen in Fig.2.2, wetting can be divided into two regimes: total wetting (Fig.2.2. (a)) and partial wetting (Fig.2.2. (b)). Non-wetting occurs when the contact angle between liquid and solid equals 180° as shown in Fig.2.2 (c). The distinction between total and partial wetting is made using the spreading parameter S defined as:

$$S = \gamma_{SG} - \left(\gamma_{SL} + \gamma_{LG}\right) \qquad (2.3)$$

where $\gamma$ is the surface tension and the subscript $SG$, $SL$ and $LG$ refer to the surface tension of the solid-gas, solid-liquid and liquid-gas phases. When the spreading parameter is positive $S > 0$, the liquid completely wets the surface, which is called the total wetting regime (Fig.2.2 (a)). In contrast, when the spreading parameter is negative $S < 0$, the liquid keeps a spherical cap shape showing different contact angles, which is called the partial wetting regime (Fig.2.2 (b)).

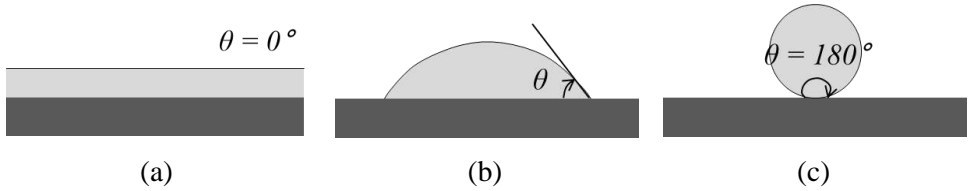

(a)                          (b)                          (c)

Fig.2.2. Schematic of three regimes of droplet on the surface: (a) total wetting at $\theta =$ 0°, (b) partial wetting at $0° < \theta < 180°$ and (c) non-wetting at $\theta = 180°$.

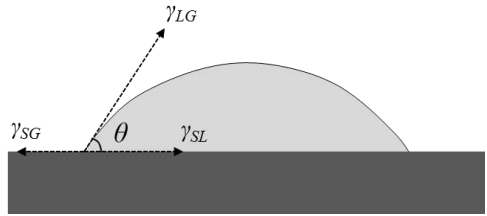

Fig.2.3. Schematic of liquid drop on the surface by the Young's equation

The thermodynamic equilibrium state between liquid, gas and solid phases in Fig.2.3 is described by Young's equation:

$$\gamma_{SG} = \gamma_{SL} + \gamma_{LG} \cos \theta \qquad (2.4)$$

with $\theta$ the contact angle. Young's equation is only valid for homogeneous smooth surfaces. For these surfaces, the contact angle can be used to define the degree of wetting or wettability. When the contact angle is below 90° ($0° < \theta < 90°$), the liquid spreads over a large area of the surface showing a high wettability and the surface is considered hydrophilic. Inversely, when the contact angle is over 90° ($90° \leq \theta < 180°$), the liquid spreads only over a small contact area on the surface showing a low wettability and the surface is considered hydrophobic.

In this section, the wetting phenomenon on homogeneous surface is considered by assuming a smooth surface with a well-defined contact angle value. In the following section, the wetting on heterogeneous surface will be discussed considering the chemical and topological characteristics of a solid surface.

### 2.2.3. Wetting on heterogeneous surface

Surfaces in nature are often chemically or topologically heterogeneous showing a different wettability on the surface or a different surface texture, respectively. This heterogeneous surface leads to more complicated wetting phenomena such as contact angle hysteresis and/or pinning-depinning of a droplet on the surface.

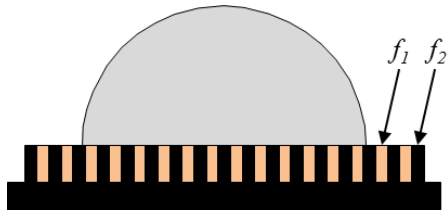

Fig.2.4. Wetting on chemically heterogeneous surface with Cassie model.

The Cassie equation describes the wetting of a droplet on a flat surface made of two materials (Fig.2.4) showing different contact angles $\theta_1$ and $\theta_2$ (Cassie and Baxter 1944). Cassie's equation gives the apparent contact angle $\theta_{App}$ on this heterogeneous surface as:

$$\cos \theta_{App} = f_1 \cos \theta_1 + f_2 \cos \theta_2 \tag{2.5}$$

where $f$ is a fraction of each surface and where the sum of $f_1$ and $f_2$ equals 1. The Cassie equation is only valid when the patches of each material are much smaller than the diameter of droplet.

The wetting on topologically heterogeneous surfaces, such as a structured surface consisting of micropillars, is described by the Wenzel and Cassie-Baxter models (Fig.2.5). The Wenzel state appears when the droplet penetrates the surface structure and wets the total surface (Fig.2.5 (a)). The Cassie-Baxter model describes the case where the droplet does not penetrate the structured surface, but sits on the surface showing pockets of air entrapped in the micro structure below the droplet.

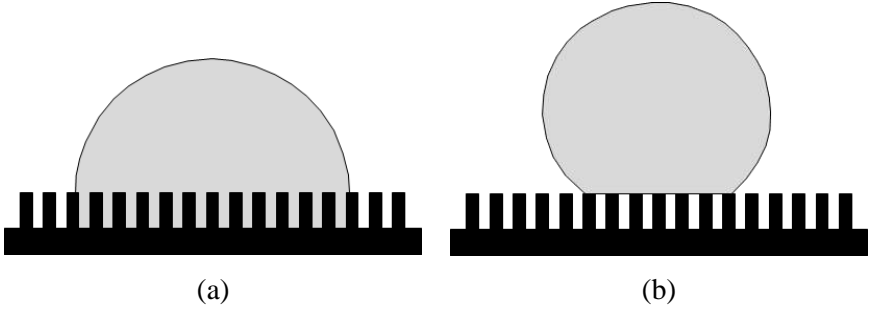

<div align="center">(a)        (b)</div>

Fig.2.5. Wetting on topologically heterogeneous surface with (a) Wenzel and (b) Cassie-Baxter models.

The Wenzel model (Wenzel 1936) is proposed to describe the wetting on a topologically heterogeneous surface in Fig.2.5 (a) by assuming that the roughness scale is much smaller than the drop size (De Gennes, Brochard-Wyart et al. 2013). Reaching equilibrium state after liquid penetration between pillars, the apparent contact angle $\theta_{App}$ can be calculated depending on the roughness ratio and the local contact angle given by Young's equation:

$$\cos \theta_{App} = \varepsilon \cos \theta_l \qquad (2.6)$$

where $\varepsilon$ is the roughness ratio which is the ratio of the actual to apparent surface areas. For a roughness ratio equal to 1 ($\varepsilon = 1$), the surface is smooth and the apparent contact angle equals the local contact angle. The Wenzel model shows two possible configurations: (1) on a hydrophilic surface ($\theta_l < 90°$), the apparent contact angle becomes smaller than the local contact angle and the surface becomes more hydrophilic ($\theta_{App} < \theta_l$); (2) on a hydrophobic surface ($\theta_l > 90°$), the apparent contact angle is bigger than the local contact angle and the surface becomes more hydrophobic ($\theta_{App} > \theta_l$). According to Wenzel model, the surface roughness magnifies either the hydrophilic or hydrophobic character of the surface.

The Cassie-Baxter model describes the case where the droplet stays on the structured surface without droplet penetration (Fig.2.5 (b)). This model is a simplified form of the Cassie model assuming that the second surface material is entrapped air. The apparent contact angle $\theta^*$ is given by (Cassie and Baxter 1944):

$$\cos\theta_{App} = f_1\left(\cos\theta_1 + 1\right) - 1 \tag{2.7}$$

It is important to mention again that these models are based on the assumption that the size of the surface structures is much smaller than the droplet radius.

### 2.2.4. Evaporation of a droplet

The evaporation of a droplet occurs by a phase change from liquid to vapor at the droplet interface with the surroundings. When evaporation occurs, the temperature at the evaporating surface lowers due to latent heat effects. When the droplet evaporation rate is sufficiently low, thermal equilibrium with the ambient atmosphere can be assumed (isothermal conditions). Then droplet evaporation occurs due to a concentration gradient from the interface to the surrounding environment; this process is called diffusion-controlled evaporation. Maxwell first described the diffusion-controlled evaporation of a spherical droplet in an infinite and uniform media at isothermal condition (Maxwell 1890, Seaver 1984, Erbil 2012). By assuming that the droplet radius is much larger than the mean free path of vapor molecules, the average mass loss rate is described as (Fuchs, Pratt et al. 1960):

$$-\frac{dm}{dt} = -4\pi R^2 D_d \frac{dc}{dR} \tag{2.8}$$

where $m$ is droplet mass, $R$ is the distance from the droplet, $D_d$ is diffusion coefficient and $c$ is concentration of the vapor phase. The concentration is equal to the saturated vapor concentration at the droplet surface $C = C_s$ $(R = R_s)$, while the vapor concentration far from the droplet surface is assumed to be constant and equal to $C = C_\infty$ $(R = \infty)$. Then, Eq. (2.8) can be written as:

$$I_m = -\frac{dm}{dt} = -4\pi R D_d\left(C_S - C_\infty\right). \tag{2.9}$$

A more detailed description of the evaporative flux at interface as a function of contact angle and location is given by (Stauber, Wilson et al. 2015). The flux at the liquid-vapor interface is given by (Deegan, Bakajin et al. 1997, Gelderblom, Bloemen et al. 2012) as:

$$I_m \sim A(\theta) \frac{D_d (C_S - C_\infty)}{R} \left( \frac{R - x}{R \cos \theta} \right)^{\lambda}$$

(2.10)

$$\text{where } \lambda = \frac{2\theta - \pi}{2\pi - 2\theta} \text{ as } x \to R$$

where $x$ is distance from the center of droplet, $A$ is droplet area and $\lambda$ is a parameter function of the contact angle $\theta$. In Eq. (2.10), the flux near the contact line is dependent on contact angle.

In Fig. 2.6, the evaporation flux over the droplet surface is illustrated. For $0° \le \theta < 90°$ or $-1/2 \le \lambda < 0$, the flux shows a larger value near the contact line, while at the apex of the droplet, smaller values are attained (Fig.2.6 (a)). At the contact line, a higher flux occurs and leads to a singularity. However, for $\theta = 90°$ or $\lambda = 0$, a uniform flux is observed (Fig.2.6 (b)). When $\pi/2 < \theta \le \pi$ or $\lambda > 0$, the flux is zero at the contact line as per Eq. (2.10).

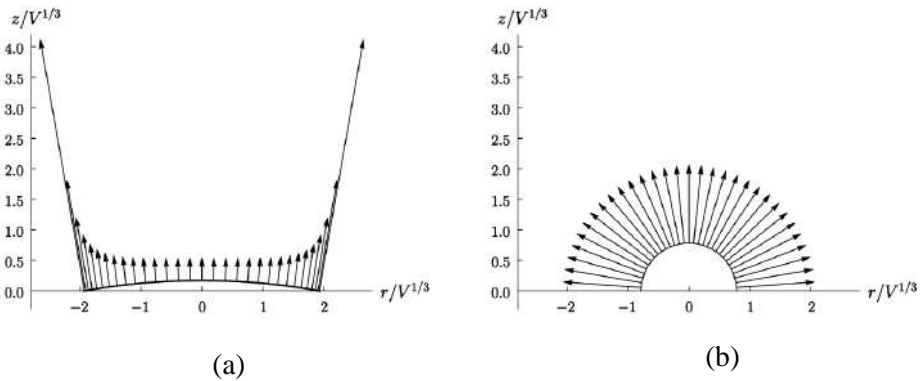

(a)                                        (b)

Fig.2.6. The evaporative flux from the free surface at different contact angles: (a) $\theta$ = 10° and (b) $\theta$ = 90° (Stauber, Wilson et al. 2015).

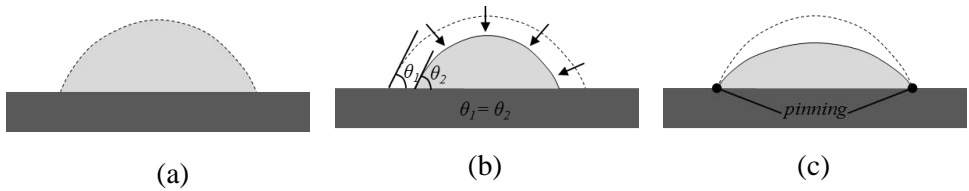

<div align="center">(a)   (b)   (c)</div>

Fig.2.7. Two different stages of evaporation of a sessile droplet: (a) liquid drop on the surface, (b) constant contact angle (CCA) mode and (c) constant contact radius (CCR) mode.

Droplet evaporation is characterized by two main modes: the constant contact angle (CCA) and the constant contact radius (CCR) modes as defined by Picknett and Bexton. In the CCA mode, the contact angle remains constant, whereas the contact line is decreasing as seen in Fig.2.7 (b). The CCA mode occurs when the surface is ideally homogeneous and perfectly smooth, and no contact angle hysteresis occurs. In contrast, in the CCR mode, the contact line stays at the same location due to its pinning on the surface. The contact angle decreases until reaching the receding contact angle, where it unpins as shown in Fig. 2.7 (c). This mode is usually observed when droplet evaporates on heterogeneous surface. A careful review of the evaporation of a sessile droplet on a solid surface is given in (Erbil 2012).

In nature, most surfaces are chemically or topologically heterogeneous resulting in contact line pinning and contact angle hysteresis during droplet evaporation (Extrand 2003, Forsberg, Priest et al. 2009, Zhang, Müller-Plathe et al. 2015).

Orejon et al. suggested an additional mode of droplet evaporation, referred to as the stick-slip evaporation mode (Orejon, Sefiane et al. 2011). The stick-slip mode shows alternatively CCR and CCA modes during droplet evaporation: 1) first, the droplet evaporates with decreasing contact angle and droplet height. The droplet remains pinned on the surface showing a constant contact area ('stick' phase); 2) when the contact angle reaches the receding contact angle, in this study called critical contact angle $\theta_{crit}$, the droplet suddenly unpins and slips towards the center of droplet. The contact radius is decreasing while the contact angle and contact height remain constant ('slip' phase). These two modes are repeated alternatingly until the droplet completely evaporates (Shanahan 1995, Orejon, Sefiane et al. 2011, Oksuz and Erbil 2014). In the stick-slip mode, the net force at the contact line is not zero (Varnik, Gross et al. 2011):

$$\delta F = \gamma \cos(\theta_0 - \delta\theta) - \gamma \cos\theta_0 \approx \gamma \sin\theta_0 \delta\theta \tag{2.11}$$

where $\delta\theta = \theta_0 - \theta_{crit}$ is the difference between initial equilibrium contact angle $\theta_0$ and the critical contact angle $\theta_{crit}$. The net force $\delta F$ increases with the difference between equilibrium and receding contact angle $\delta\theta$ as shown in Fig.2.8. When the contact angle of droplet reaches $\theta_{crit}$, the net force $\delta F$ equals to depinning force, when the droplet suddenly depins and starts to slip.

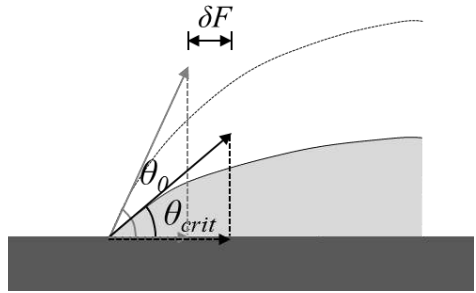

Fig.2.8. Schematic of liquid droplet on surface with net force $\delta F$, initial equilibrium contact angle $\theta_0$ and the receding contact angle $\theta_{crit}$ for stick phase.

Another approach to interpret depinning is to analyze the change of interfacial Gibbs free energy. The interfacial Gibbs free energy of the droplet showing a spherical cap geometry is given by (Shanahan 1995):

$$G = \gamma A + \pi R^2 (\gamma_{SL} - \gamma_{SV}) \text{ with } A = \frac{2\pi R^2}{1 + \cos\theta} \tag{2.12}$$

where $A$ is the liquid/vapor interfacial area of the droplet and $\pi R^2$ the contact area of the droplet with the surface. The excess free energy $\delta G = G(\theta) - G(\theta_0)$ per unit length of the contact line during the pinning phase can be calculated using a Taylor's expansion as given by (Shanahan 1995).

$$\delta G = \frac{\gamma R (\delta \theta)^2}{2(2 + \cos \theta_0)} \tag{2.13}$$

During the stick phase, the excess free energy increases due to the decrease of the contact angle $\theta$. When the excess free energy attains a critical value, sufficient energy is available to overcome the potential energy barrier and the contact line will slip to its next equilibrium position. During the slip phase the excess free energy decreases dramatically until it becomes pinned again. Fig.2.9 shows the evolution of the excess Gibbs free energy and contact angle versus time for stick and slip phases as obtained by LBM (see further in Chapter 5). The excess free energy is increasing, while the contact angle is decreasing during the stick phase. The droplet becomes unpinned when the excess free energy attains a maximal value when the contact angle reaches the critical contact angle.

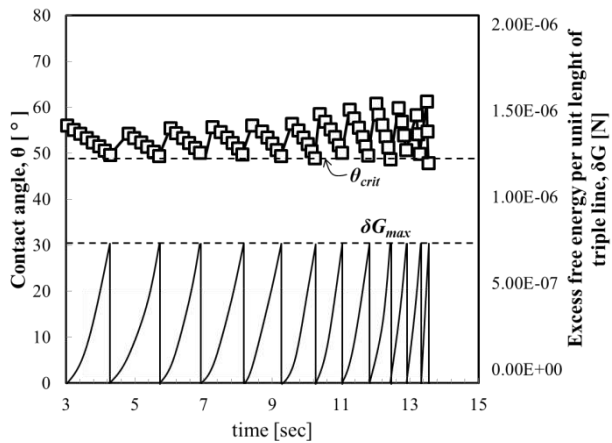

Fig.2.9. Temporal evolution of the excess free energy (square symbol) and contact angle (solid line) during stick and slip phases

In chapter 5, the behavior of droplets on structured surfaces will be studied motivated by the fact that the wetting of porous asphalt is mainly due to a wetting of rain droplets. The study includes: (1) the run-off of a droplet on a surface with groove; (2) the stick-slip behavior of an evaporating droplet on a structured surface with micropillars and; (3) the apparent contact angle of a droplet on a surface consisting of hydrophilic and hydrophobic patches. This study will help to better understand the wetting behavior of PA.

### 2.2.5. Capillary rise

Assume that the liquid in a capillary consisting of circular tube partially wets the surface with a contact angle $\theta$.

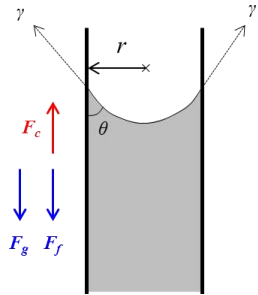

Fig.2.10. Force balance in the capillary tube with radius $r$ and contact angle $\theta$.

Capillary rise has been modelled by Washburn (1921) and Bosanquet (1923). In Lucas-Washburn (L-W) model (Washburn 1921), the flow in the tube is assumed to be a fully developed Poiseuille flow. The capillary, friction and gravitational forces are considered as the main driving forces at play as shown in Fig.2.10. The capillary force is given by:

$$F_c = 2\pi r \gamma \cos\theta \,. \tag{2.14}$$

The friction force can be written as:

$$F_f = \pi r^2 \Delta p , \qquad (2.15)$$

with the pressure difference $\Delta P$ calculated considering a Poiseuille flow in capillary tube:

$$\Delta p = \frac{8\eta}{r^2} uh \qquad (2.16)$$

and the gravity force

$$F_g = \pi r^2 \left( \rho g h \right), \qquad (2.17)$$

where $\gamma$ is interfacial tension, $R$ is radius of capillary tube, $\theta$ is a contact angle, $\eta$ is liquid viscosity, $\rho$ is liquid density, $u$ is velocity, $h$ is height of liquid column and $g$ is gravitational acceleration. At equilibrium state neglecting inertial effects, the sum of forces in Eqs. (2.14), (2.15) and (2.17) has to equal to 0:

$$\sum F = F_c - F_f - F_g = 0 . \qquad (2.18)$$

By substituting Eqs. (2.14), (2.15) and (2.17) in Eq. (2.18), we get:

$$8\pi\eta uh = 2\pi r \gamma \cos\theta - \pi r^2 h\rho g ,$$

$$\text{or } u = \frac{dh}{dt} = \frac{2r\gamma \cos\theta}{8\eta h} - \frac{r^2 \rho g}{8\eta} . \qquad (2.19)$$

After reaching equilibrium state, the velocity $u = dh/dt$ is equal to zero and the maximum height of liquid column can be determined:

$$u = \frac{dh}{dt} = \frac{2r\gamma \cos\theta}{8\eta h} - \frac{r^2 \rho g}{8\eta} = 0$$

$$\text{and } h_{max} = \frac{2\gamma \cos\theta}{r\rho g} . \qquad (2.20)$$

The capillary rise versus time as derived from Eq. (2.19) gives:

$$h(t) = \frac{2\gamma \cos \theta}{r\rho g}\left(1 - \frac{1}{e^{\left(tr^2\rho g/8\eta\right)}}\right).$$ (2.21)

When the capillary length, which is explained in the following section 2.2.6, is small, gravity can be neglected leading to:

$$u = \frac{dh}{dt} = \frac{2r\gamma \cos \theta}{8\eta h} .$$ (2.22)

The kinetics of capillary rise height in Eq. (2.19) follows a square root of time behavior:

$$h(t) = \sqrt{\frac{\gamma r \cos \theta}{2\eta}}\sqrt{t} .$$ (2.23)

As described in Eq. (2.20), the maximum capillary height is higher in tubes with small diameter and low contact angle. However, the capillary uptake rate is higher for larger pores (Eq. (2.23)).

### 2.2.6. Capillary length

To evaluate the importance of gravity effects in multiphase flow, the capillary length, based on the ratio between surface tension and gravity, is used (De Gennes 1985, Feng and Rothstein 2011):

$$L_{cap} = \sqrt{\frac{\gamma}{\rho g}}$$ (2.24)

For water at standard temperature and pressure, the capillary length is around 2 mm. When the characteristic length $l$ of the system is sufficiently smaller than the capillary length, $l < L_{cap}$, surface tension effects are dominant, while the effect of gravity can be neglected. This regime is called the capillary-dominant regime. On the contrary, when the characteristic length $l$ is larger than the capillary length, $l > L_{cap}$, the effect of gravity becomes important. This regime is called the gravity-dominant regime. In this study, both capillary- and gravity-dominant regimes are considered depending on the scale of the system in comparison with the capillary length. The detailed conditions for capillary- or gravity-dominant regime are described in each chapter.

### 2.2.7. Corner flow: pore and corner arc meniscus

Capillary rise in cylindrical tube was described in section 2.2.5. However, in polygonal tubes, also corner flow may arise. In this section, wetting by corner flow inside the polygonal tubes is described in equilibrium state.

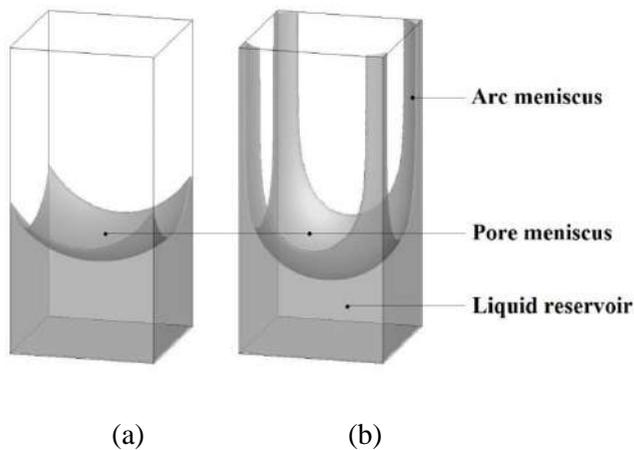

(a)                    (b)

Fig.2.11. Schematic representation of the two configurations of capillary flow in a square tube: (a) pore meniscus when the contact angle is larger than the critical contact angle, $\theta \geq \theta_c$ and (b) co-occurrence of arc and pore menisci when the contact angle is smaller than the critical contact angle, $\theta > \theta_c$.

In a partially filled polygonal tube, the liquid forms a pore meniscus and may additionally wets the corner forming a hemisphere which configuration depends on the contact angle between the liquid and the solid material. Concus and Finn (1974) identified the existence of a critical contact angle $\theta_c = \pi/n$ in n-sided polygonal tubes based on the Rayleigh-Taylor interface instability. When the contact angle $\theta$ is between $\pi/2$ and the critical contact angle, i.e. $\theta_c$ ($= \pi/n$) $\leq \theta < \pi/2$, the liquid wets the tube walls and the meniscus spans the total tube, resulting in a configuration named pore meniscus as presented in Fig.2.11 (a). In contrast, when the contact angle is smaller than the critical contact angle, i.e. $\theta < \theta_c$ ($= \pi/n$), the liquid additionally invades the edges or corners of the polygonal tube forming corner arc menisci as presented in Fig.2.11 (b) (Concus and Finn 1974, 1990). Corner arc menisci occur at each corner and move upward as a result of a capillary pressure gradient (Wong, Morris et al. 1992, Dong and Chatzis 1995). Princen proposed a model for capillary rise in triangular and square tubes for zero contact angle (Princen 1969 a, b). The Princen model predicts the height of the main meniscus from the balance between capillary force and gravity considering the liquid column at the center of the tube, while fingering of infinite height occurs in its corners. The capillary force can be written by considering a balance between the Laplace pressure:

$$F_{cap} = 4L\gamma \cos \theta \qquad (2.25)$$

and the hydrostatic pressure:

$$\frac{\gamma}{r} = \rho g z . \qquad (2.26)$$

The total mass of liquid $m$ in a square tube is given by:

$$m = \rho g h L^2 + 4 \rho g \int_h^\infty \left(1 - \pi/4\right) r_{arc}^2 (z) dz \qquad (2.27)$$

where $L$ is the size of the side of the rectangular polygon, $z$ is the coordinate along the height and $r_{arc}$ is the radius of curvature of the corner arc meniscus. Considering both presences of arc and main pore menisci, the Mayer and Stowe-Princen (MS-P) theory predicts the curvature radius of the arc meniscus as a function of the effective area and perimeter of the non-wetting phase (gas).

When corner arc menisci are formed at corner, the normalized curvature $C_n$ is given by (Ma, Mason et al. 1996):

$$C_n = \frac{(L/2)\cos(\alpha + \theta)}{L_{contact}\sin\alpha} \tag{2.28}$$

where $\alpha$ is the half corner angle dependent on the side parameter $n$ and $L_{contact}$ is the side length of the corner arc meniscus wetting the side of the tube. An analytical solution for the degree of saturation $S_w$ in function of curvature is given by (Ma, Mason et al. 1996):

$$S_w = \frac{\tan\alpha}{C_n^2}\left[\frac{\cos\theta}{\sin\alpha}\cos(\alpha + \theta) - \frac{\pi}{2}\left(1 - \frac{\alpha + \theta}{90}\right)\right] \tag{2.29}$$

Fig.2.12 shows curvature versus degree of saturation for the two contact angles in a log-log plot. When the curvature and contact angle decrease, the saturation degree at the corners increases and, at one point, the liquid covers the whole tube wall and the gas phase can become trapped by liquid.

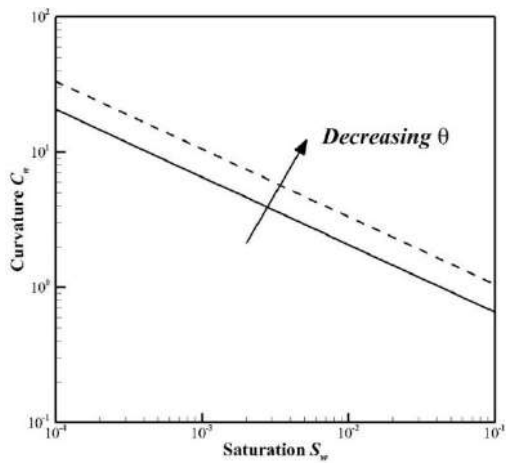

Fig.2.12. A relationship between curvature $C_n$ and fractional saturation $S_w$ for two different contact angles $\theta$.

This theory results in a better estimation of the interfacial area (Mayer and Stowe 1965, Princen 1969 a, b, Princen 1970) as shown by (Bico and Quéré 2002). Ma et al. (1996) investigated capillary flow in polygonal tubes during imbibition and drainage and suggested a relationship between liquid saturation and curvature of arc menisci in corners based on the Mayer and Stowe-Princen (MS-P) theory. They obtained a capillary pressure curve in terms of contact angle by taking into account contact angle hysteresis. In a recent work, Feng and Rothstein (2011) studied the pore meniscus height as a function of contact angles in a polygonal capillary tube when the contact angle is higher than the critical contact angle. Furthermore, they considered different geometries with either sharped or rounded corners showing the effect on meniscus height. They compared their simulation results with results obtained using Surface Evolver which is an open-source code for determining interphases shapes based on surface energy minimization. In polygonal tubes, the amount of liquid in the tube can be significantly different depending on taking into account corner flow when the contact angle is less than critical contact angle.

Capillary uptake in tubes will be first studied with LBM in Chapter 4 and used as a validation of the LBM. Special attention will be given to corner flow, which is believed to be also present in porous asphalt. In chapter 6, the LBM study will be further extended to two-phase flow in porous media aiming to analyze the influence of capillary forces on the liquid distribution in porous asphalt after drainage.

## 2.3. Numerical modeling for multiphase flow

Numerical modelling based on mathematical models has shown to be a powerful tool to analyze physical problems, especially when these are more difficult to analyze in detail by experiments due to limitations of temporal or spatial scale, complexity and cost. Numerical modelling is especially helpful for solving multiphase flow problems since it allows determining the movement of the interface between different phases. Multiphase flow problems in porous media have been described at the macroscale using continuum theory. However, in this thesis, the focus is on the modelling of multiphase flow at pore scale. Therefore, we limit the state of art to modelling of multiphase flow on the pore scale and describe the two most used methods: computational fluid dynamics and lattice Boltzmann method.

In this section, first computational fluid dynamics modelling (CFD) including interface tracking is presented. Then, the lattice Boltzmann method (LBM), especially the multiphase LBM, is briefly presented including previous researches in multiphase flow phenomena.

### 2.3.1. Computational fluid dynamics (CFD)

Fluid flow problems have been modeled by computational fluid dynamics (CFD), which solves the Navier-Stokes equations, introducing several assumptions such as incompressible or inviscid flow. CFD basically solves single phase flow problems. To solve multiphase flow, additional calculation steps to track the movement of the phase interface are included. Due to difficulty in calculating the location and dynamic movement of the interface between different phases, multiphase flow remains still a challenging problem compared to single phase flow. To solve multiphase flow, front tracking (FT) or front capturing (FC) methods are combined with Navier-Stokes equations to find the interface movement and its location at different times. The FT method is a Lagrangian approach which is straight forward to track the moving interface explicitly showing high accuracy (Tryggvason, Bunner et al. 2001). However, difficulties in the application of the Lagrangian approach arise when the interface becomes tortuous and breaks up. In these situations, the markers located on the interface, which allow distinguishing between the different phases, come so close to each other that singularities arise. As a result, the location of the interface becomes a not well-defined problem. The FC method is an Eulerian approach and defines the location of interface implicitly. The application of the FC method into numerical methods is quite easy and, as a result, this method is commonly used in solving multiphase flow. However, this method suffers from interface diffusion over several cells (Scardovelli and Zaleski 1999).

| | | | |
|---|---|---|---|
| *0* | *0* | *0* | *0* |
| *0.8* | *0.4* | *0.01* | *0* |
| *1* | *1* | *0.3* | *0* |
| *1* | *1* | *1* | *0* |

Fig.2.13. Schematic of distribution of fraction function in VOF method (as per Wikipedia).

Among the several FC methods, the Volume of Fluid (VOF) (Hirt and Nichols 1981) and the level set method (Sussman, Fatemi et al. 1998) are widely adopted in many studies. VOF is based on the application of marker and cell methods (Harlow and Welch 1965), where for each cell a fraction function is defined ranging between 0 to 1. The fraction function is 0 when the cell is totally occupied by gas or 1 when totally occupied by liquid. When the fraction function has a value between 0 and 1, the interface is located within the cell and the liquid density is calculated using the fraction function value (see Fig.2.13). However, it is well known that VOF may introduce some numerical diffusion requiring quite complex algorithms to solve it, making it less convenient for three-dimensional problems (Hirt and Nichols 1981). The level set method assumes a closed curve $\Gamma$ for the interface defining using a level set function $\varphi$ (Osher and Sethian 1988). Depending on the location, $\varphi$ is positive inside the curve, negative outside the curve and zero on curve. This method is simple to apply and can perform sharp interfaces, topological merging and breaking of interfaces in two- and three-dimensional problems (Osher and Fedkiw 2001). However, the level set method may show mass conservation problems (Osher and Sethian 1988). When level set method is applied on complex geometries or small scale problems, these limitations become particularly significant.

### 2.3.2. Lattice Boltzmann method (LBM)

The lattice Boltzmann method (LBM) is derived from the Boltzmann equation used in kinetic theory. Unlike CFD which solves macroscopic continuum equations (see Fig.2.14), LBM is based on microscopic models and mesoscopic kinetic equations using length and time scales as given in Fig.2.14 (Chen, Doolen et al. 1994, He and Doolen 2002).

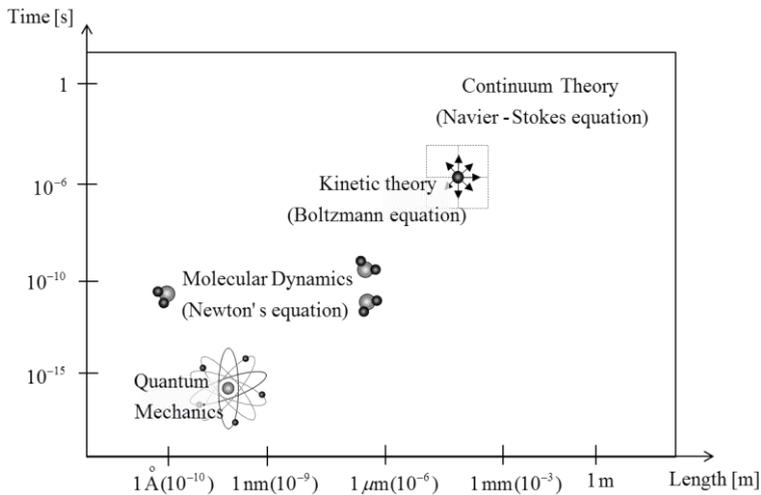

Fig.2.14. Length and time scale of different modelling methods.

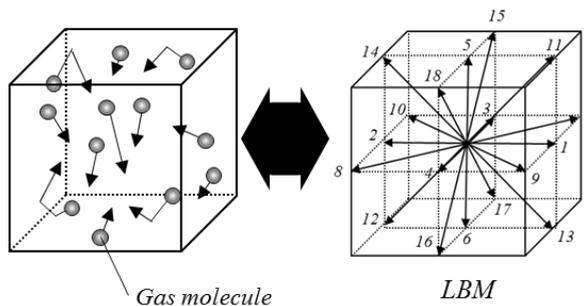

Fig.2.15. Schematic of comparison between Boltzmann model and LBM (Hecht and Harting 2010).

In LBM, the macroscopic dynamics of fluid particles are based on underlying microscopic processes such as a consecutive streaming and collision of the particles over the discrete lattice grid (see Fig.2.15).

In the last few decades, LBM has become a powerful numerical tool to solve single or multiphase flows, heat transfer, phase change and cavitation (Succi 2001, Sukop 2006, Aidun and Clausen 2010, Guo and Shu 2013). Furthermore, LBM has been successfully applied to study multiphase flow (Succi 2001), due to its automatic tracking of the fluid interfaces in a multiphase system (Shan and Chen 1993, 1994). The fluid-solid interactions can be implemented conveniently in the LBM without introducing additional complex kernels (Martys and Chen 1996). Owing to its constitutive versatility, LBM has developed into a powerful technique for fluid mechanics and is particularly successful in modelling spreading and wetting (Shan and Doolen 1995, Raiskinmäki, Koponen et al. 2000), gravity-driven droplet displacement (Kang, Zhang et al. 2002, 2005), bubble rising (Sankaranarayanan, Shan et al. 2002, Inamuro, Ogata et al. 2004) and fluid transport in porous media (Martys and Chen 1996, Pan, Hilpert et al. 2004). Several multiphase LB models have been developed including the color-gradient based LB method by Gunstensen et al. (1991) based on Rothman-Keller lattice gas model (Rothman and Keller 1988), the free-energy model by Swift et al. (1996), the mean-field model by He et al. (1999) and the pseudopotential model by Shan and Chen (1993, 1994).

### 2.3.3. The pseudopotential multiphase LBM

Comparing various multiphase LBM's, the pseudopotential multiphase LBM is often used due to its simplicity and versatility. This model represents microscopic molecular interactions at mesoscopic scale using a pseudopotential depending on the local density (Shan and Chen 1993, 1994). With such interactions, a single component fluid spontaneously segregates into high and low density phase (e.g., liquid and gas), when the interaction strength (or the temperature) is below the critical point (Shan and Chen 1993, 1994). The automatic phase separation is an attractive characteristic of the pseudopotential model, as the phase interface is no longer a mathematical boundary and no explicit interface tracking or interface capturing technique is needed. The location of the phase interface is characterized through monitoring the jump in fluid density from gas to liquid phase. The pseudopotential model captures the essential elements of fluid behavior, namely it follows a non-ideal equation of state (EOS) and incorporates a surface tension force. Due to its remarkable computational efficiency and clear representation of the underlying microscopic physics, this model has been used as an efficient technique for simulating and investigating multiphase flow problems, particularly for these flows with complex topological changes of the interface, such as deformation, coalescence and breakup of the fluid phase, or fluid flow in complex geometries (Chen, Luan et al. 2012).

Capillary flow has been studied effectively using the pseudopotential LBM. Sukop and Thorne (2006) performed a two-dimensional capillary rise simulation and compared their LB results with the theoretical capillary rise predictions. Raiskinmaki et al. (2002) and Thorne and Michael (2006) investigated capillary rise in a three-dimensional cylindrical tube using multiphase LBM. The effects of contact angle, tube radius and capillary number were studied with or without taking into account gravity. Their study provided a useful benchmark for other LBM studies of capillary rise by comparing it with the Washburn solution. Although previous studies showed interesting LBM works in capillary flow (Dos Santos, Wolf et al. 2005, Lu, Wang et al. 2013), these studies focused mainly on the position of the interface in the capillary column (cylindrical tubes or two parallel plates) and only limited studies exist on the influence of other phenomena such as corner flow.

For droplet displacement, Kang et al. (2002) investigated a two-dimensional droplet flowing down a channel with different Bond numbers. The effects of surface wettability, droplet size, density and viscosity ratio were studied. Mazloomi and Moosavi (2013) simulated the run-off of a gravity-driven liquid film over a vertical surface displaying U- and V- shaped grooves or mounds, showing the existence of a critical width for successful coating or covering with fluid. Azwadi and Witrib (2012) investigated the dynamic behavior of droplets with respect to contact angle, Bond number and tilting of the surface. Chen et al. (2014) studied the deformation and breakup of a droplet in a channel with a solid obstacle for different obstacle shapes, wettability, viscous ratio and Bond number. However, in the previous works, the displacement of an immiscible fluid with a low density ratio between liquid and gas, usually equal to 1, was investigated.

In porous media, Pan et al. (2004) investigated immiscible flow in porous structures including sphere packs and more complicated geometry. They studied the correlation between saturation and capillary pressure in terms of geometry, viscous ratio and capillary number and showed their LB results were in good agreement with experiments. Li et al. (2005) and Yiotis et al. (2007) both calculated relative permeability as a function of capillary number, wettability, viscous ratio and different geometries and validated their results comparing with experimental data and analytical solutions (Li, Pan et al. 2005, Yiotis, Psihogios et al. 2007). However, their porous media were modeled by packs of spheres or randomly generated geometries, which made it difficult to compare with experimental data of real porous media. Their studies are also limited to immiscible flow, not considering multiphase flow. Recently, Chen et al. (2014) thoroughly reviewed the theory and application of the pseudopotential LBM. The detailed concepts of the pseudopotential LBM will be discussed in chapter 3.

## 2.4. Porous Asphalt (PA)

Asphalt concrete (AC), composed of mineral aggregates and bituminous binder, is commonly used as a pavement material. PA achieves a mineral skeleton by densely packing showing relatively low void content as seen in Fig.2.16. (Young, Bentur et al. 1998). AC is characterized in terms of durability, consistency, temperature susceptibility, stiffness, stability, permeability, safety and aging depending on size, type and amount of components and void content (Poulikakos and Partl 2012).

Porous asphalt (PA) is an open-graded AC with a high porosity of 20 vol. % using less bitumen, i.e. approximately 6% (Figs. 2.16 and 2.17). It is used an environmental friendly road pavement. Due to its high porosity and permeability, PA fulfills the drainage function in terms of reducing aquaplaning risk and by improving driving conditions in wet state as shown in Fig.2.18.

Furthermore, PA helps traffic noise reduction, especially at high speeds on highways (Sandberg and Ejsmont 2002). The contact of tire with its macrotexture contributes to noise absorption and results in a noise reduction of 5-10dB (Sandberg 1999, Poulikakos, Gubler et al. 2006). Fig.2.19 shows that PA at the time of construction absorbs 7dB of noise in comparison to a reference dense graded AC mixture. Thus, PA is widely used in some countries for noise reduction as top layer of road pavements. In spite of these benefits, due to PA open structure, this material undergoes environmental and mechanical loads leading to a shorter life time of approximately 10 years, compared to the dense asphalt life expectancy of 20-30 years.

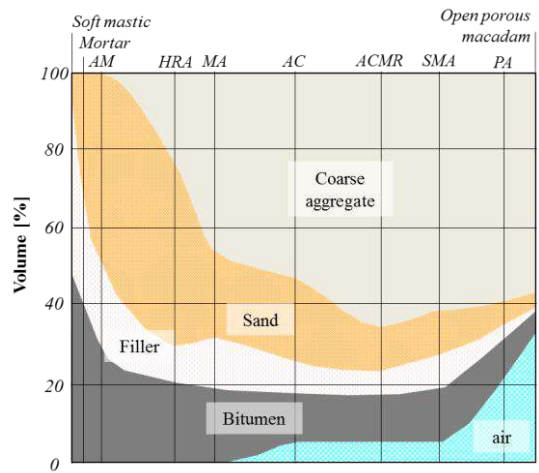

Fig.2.16. Volumetric distribution of constituents of various asphalt concrete road materials: HRA: Hot rolled asphalt, MA: Mastic asphalt, AC: Asphalt concrete, ACMR: Rauasphalt, SMA: Stone mastic asphalt and PA: Porous asphalt (Partl 2007, Poulikakos 2011).

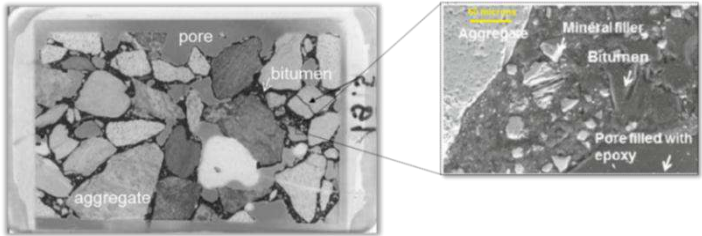

Fig.2.17. Cross section of porous asphalt and magnification (Poulikakos and Partl 2012)

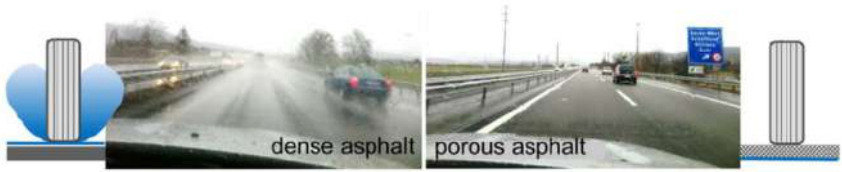

Fig.2.18. Comparison between dense asphalt and porous asphalt after rain event (Poulikakos, Gilani et al. 2013).

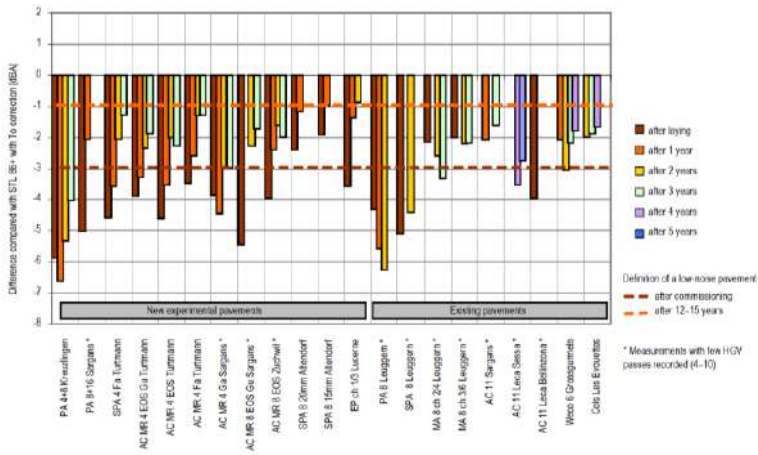

Fig.2.19. Absorption of noise of pavement under mixed traffic conditions. (Low-noise road surfaces in urban areas: Final report 2007)

Due to high void content (porosity), the internal structure of PA is exposed to air and water. This continuous exposition to moisture/air results in bitumen oxidation (hardening) and degradation, resulting in a reduction of service life of PA (Hoban 1985, Poulikakos, Gubler et al. 2006, Poulikakos and Partl 2012).

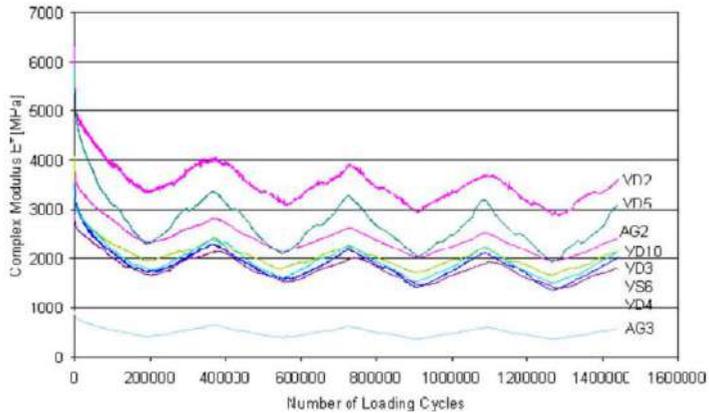

(a)

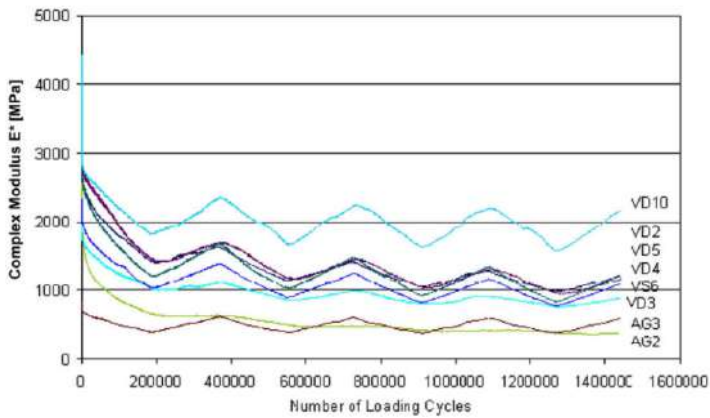

(b)

Fig.2.20. Evolution of complex modulus versus loading cycle at (a) dry and (b) wet state (Poulikakos and Partl 2009).

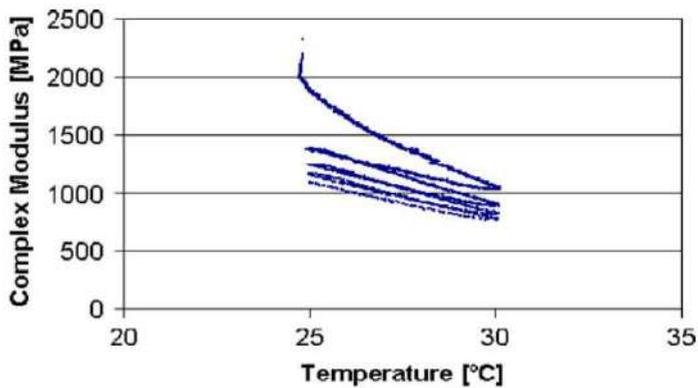

Fig.2.21. Evolution of complex modulus versus temperature cycle in wet state (Poulikakos and Partl 2009).

In saturated and unsaturated state, PA can become damaged and degraded depending on the area of interaction with water and the residence time of water within PA. At longer exposition to liquid water and moisture, water will penetrate into the bonded area between aggregate and bitumen. It will widely wet the hydrophilic aggregate and weaken the bond with the bitumen. Over time, bitumen is washed and stripped away from the aggregate surface due to broken bond. As a result, the durability of PA will significantly decrease and its lifetime will be shortened (Williams and Miknis 1998, Poulikakos and Partl 2010).

To evaluate the effect of water on PA, Poulikakos and Partl measured the complex modulus of PA by coaxial shear test (CAST) in both dry and wet states. In their results, the loss of modulus ranges between 8% to 15% in dry state, whereas it significant increases in wet state from 2% to 42% in Fig.2.20. Furthermore, they studied the effect on temperature on complex modulus in wet state and found greater temperature dependency, as shown in Fig.2.21 (Poulikakos and Partl 2009).

To understand the fundamental physics in PA and improve its performance, various experimental and numerical studies have been performed with PA.

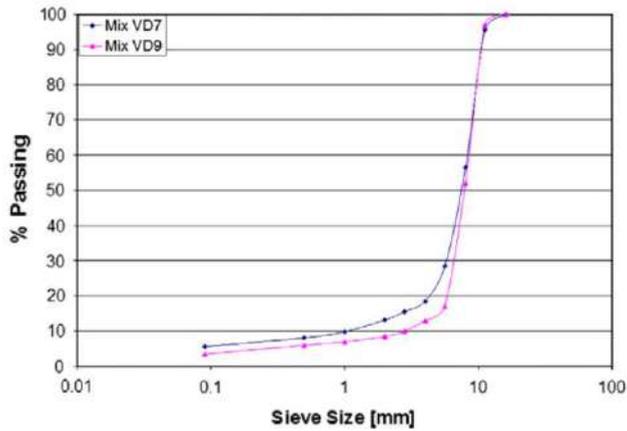

(a)

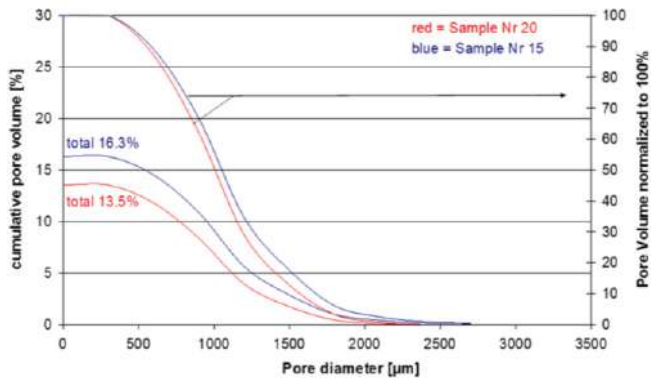

(b)

Fig.2.22. (a) Aggregate size distribution of porous asphalt and (b) continuous pore size distribution (c-PSD) of PA as a function of the cumulative pore volume and pore diameter for two different samples (Gruber, Zinovik et al. 2012).

Regarding the experimental approaches, Gruber et al. determined the continuous pore size distributions (c-PSD) of PA based on mercury intrusion porosimetry (MIP) (Fig.2.22) (Gruber, Zinovik et al. 2012). From the sieve and pore size distribution curves in their study, it can be observed that PA has a wide pore system ranging from 500 μm to 2 mm. Poulikakos et al. studied the effect of drying on water loss in PA using a micro-wind-tunnel and neutron radiography (NR) to determine the moisture content distribution (Poulikakos, Saneinejad et al. 2013). They evaluated the water loss and moisture distribution during drying and found that water loss is higher in open pores compared to closed pores and dead-end pores. The performance of PA was characterized by its complex modulus, water sensitivity, fatigue behavior through Indirect Tensile tests (IDT) and coaxial shear test (CAST) (Poulikakos and Partl 2009). Furthermore, the optimization of the void content distribution inside PA was investigated by compaction methods, Marshall tests, Superpave Gyratory Compactor (SGC) and Rolling Wheel compaction (LCPC) and evaluated using X-ray microcomputer tomography (µCT) (Partl, Flisch et al. 2007).

To determine the effect of different types of binders on PA, neat, SBS modified and high viscosity bitumen were investigated by rutting, IDT and Marshall tests (Liu and Cao 2009). The components of bitumen were also investigated in terms of polymer modifiers and fiber modifiers in terms of permeability and strength performed by falling head tests (FHT), constant head tests (CHT) and IDT tests. In their study, it was found that polymer modified PA showed a double strength and permeability (Faghri, Sadd et al. 2002). At microscale, the microstructure of PA was analyzed in terms of void content, binder film thickness, compaction degree and bitumen type. They showed that the microstructure is directly related with the presence of microcracks and voids which lead to failure and deterioration. Thus, by choosing proper compaction degree or bitumen mix type, the microstructure can be improved showing less microcracks (Poulikakos and Partl 2010). In saturated and unsaturated states, gravity-driven drainage and capillary uptake experiments were documented using neutron imaging. Neutron imaging provides the moisture content distribution with high moisture content, spatial and temporal resolution and the results were used for validation of computational models. In numerical approaches, the transport properties, including permeability and hydraulic conductivity of various porous media, have been calculated based on Darcy's law (Praticò and Moro 2008) and mixture theory (Krishnan and Rao 2001). Pore network model (PNM) based on networks of pores and throats has also been used to study Porous media (Blunt 2001).

PNM is suitable for the simulation of immiscible flow displacement at low capillary number (Dullien 2012) and invasion of wetting and non-wetting fluids in porous materials (Gostick, Ioannidis et al. 2007). Recently, Gruber et al. (2012) estimated permeability using computational fluid dynamic (CFD) and compared the results with the Darcy-Forchheimer model. In their study, the permeability and flow residence time is found to be related with the flow direction and compaction of PA. In chapter 6, gravity-driven drainage in PA will be investigated in more detail using LBM.

## 2.5.   Need for further study

Multiphase flow is a ubiquitous phenomenon with various industrial and academic applications. Thus, the study of multiphase flow using numerical and experimental techniques plays to a key role in order to understand this phenomenon. Many researches have studied multiphase flow numerically by applying continuum approaches. However, to find the motion and location of the interface between the different phases, additional methods including interface capturing and interface tracking have to be considered. Although such approach has been properly used for studying multiphase flow problems in many previous studies, it still remains a challenge to capture the interface in complicated problems due to limitations of capturing the sharp interface and due to the presence of diffusion at the interface. Thus, LBM is suggested as a powerful tool for solving multiphase flow over time since it can provide relatively sharp interfaces also in complex domains, such as porous media. From previous studies, it can be observed that LBM has been successfully applied to solve problems like droplets behavior on surfaces, capillary uptake, gravity-driven drainage and evaporation.

In this study, fundamental multiphase phenomena, such as droplet behavior on homogeneous and heterogeneous surfaces, capillary uptake in two- and three-dimensional polygonal tubes and droplet evaporation, are studied with LBM and compared with analytical solutions to verify and validate LBM. Then more complicated problems such as gravity-driven drainage in analogous PA are studied with LBM. PA's wide range of pore sizes and complicated network can be captured by μCT images and then used as computational domain for LBM. In dry and wet states, transport properties, residence time and distribution of liquid water will be characterized over time and compared with experimental data. At the end, for further exploration of LB works, diverse multiphase phenomena are studied using the two- and three-dimensional LBM.

# 3. MULTIPHASE LATTICE BOLTZMANN METHOD (LBM)

## 3.1. Introduction

This chapter describes the computational methodology used in this thesis presenting the derivation of the lattice Boltzmann equation (LBE) from the Boltzmann equation using the collision operator. Then, the basic concepts of the multiphase lattice Boltzmann model (LBM), especially the Shan-Chen single pseudopotential multiphase LBM, are described with the non-ideal Equation of State (EOS), the forcing scheme and the boundary conditions applied to the original pseudopotential LBM. Thereafter, the implementation of several tests including Laplace law, contact angle and capillary intrusion tests are explained in detail to obtain explicit relations between multiphase LBM and physical parameters. Furthermore, the constraints of the pseudopotential LBM for simulating multiphase flow are introduced using non-dimensional numbers. The parallel computing and types of cluster used in this thesis are briefly described at the end of this chapter.

For this thesis, the base code used was provided by Dr. Qinjun Kang, from Los Alamos National Laboratory. This single component multiphase LB code was further developed to allow taking into account gravity by adding a body force as described in section 3.3. Different boundary conditions were added such as velocity (pressure) and Neumann (outflow) boundary conditions (section 3.6). Finally a MPI version of the code was implemented (section 3.10). In addition to implementation in the code, several algorithms were developed, specifically for post-processing of two-phase flow with a sharp interface, for generating a computational domain from μCT images in section 6.3 or for analyzing the data such as finding the contact angle of non-spherical droplets in section 5.4.

## 3.2.   Basic concepts of LBE (lattice Boltzmann equation)

The lattice Boltzmann equation is derived from the Boltzmann kinetic equation (He and Luo 1997, Chen and Doolen 1998). The Boltzmann kinetic equation can be written as follows:

$$\frac{\partial f}{\partial t} + \xi \cdot \nabla f + F \cdot \frac{\partial f}{\partial \xi} = \Omega \,, \tag{3.1}$$

where $f$ is the single-particle distribution function, $\xi$ is the macroscopic velocity, $F$ is force field per unit mass acting on the particle and $\Omega$ is collision operator which counts the sum of all intermolecular interactions. In order to facilitate the numerical solution of the Boltzmann equation and to improve the computational efficiency, the Bhantagar-Gross-Krock (BGK) collision operator is applied to the Boltzmann equation (Bhatnagar, Gross et al. 1954, Succi 2001). The single relaxation time BGK collision operator is given by:

$$\Omega = \frac{1}{\tau}\left(f - f^{eq}\right), \tag{3.2}$$

where $f^{eq}$ is the equilibrium distribution function and $\tau$ is a relaxation time. By considering a finite direction $i$ for the macroscopic velocity and distribution functions and by applying the BGK collision operator, Eq. (3.1) can be discretized as the lattice BGK (LBGK) equation (He and Luo 1997):

$$f_i\left(\mathbf{x} + c\mathbf{e}_i\Delta t, t + \Delta t\right) - f_i\left(\mathbf{x}, t\right) = -\frac{1}{\tau}\left[f_i\left(\mathbf{x}, t\right) - f_i^{eq}\left(\mathbf{x}, t\right)\right], \tag{3.3}$$

where $f_i$ $(\mathbf{x},t)$ is the density distribution function and $f_i^{eq}$ $(\mathbf{x},t)$ is the equilibrium distribution function in the $i^{th}$ lattice velocity direction, $\mathbf{x}$ the position vector and $t$ the time. The relaxation time $\tau$ is related to the kinematic viscosity as $\nu = c_s^2(\tau\text{-}0.5)\Delta t$, where the lattice sound speed $c_s$ is equal to $c/\sqrt{3}$. The lattice speed $c$ is equal to $\delta x_{lb}/\delta t_{lb}$, with $\delta x_{lb}$ the grid spacing and $\delta t_{lb}$ the time step, which are both set equal to 1 leading to $c = 1$. Eq. (3.3) consists in a streaming and collision steps, respectively on the right and left sides of equation. The equilibrium distribution function is written as the discretized Maxwell-Boltzmann equilibrium distribution (Qian, d'Humières et al. 1992):

$$f_i^{eq} = \omega_i \rho \left[ 1 + \frac{1}{c_s^2}\left(\mathbf{e}_i \cdot \mathbf{u}\right) + \frac{1}{2}\left(\frac{\mathbf{e}_i \cdot \mathbf{u}}{c_s^2}\right)^2 - \frac{1}{2c_s^2}\mathbf{u}^2 \right], \tag{3.4}$$

where $\omega_i$ is the weighting factor, $\mathbf{e}_i$ is the discrete velocity, $\rho$ is the macroscopic density and $\mathbf{u}$ is the macroscopic velocity. In LBM, diverse discrete velocity models have been specified as $DnQm$ where $n$ is the space dimension and $m$ is the number of discrete velocities. In this thesis, the D2Q9 and D3Q19 models are used for the two- and three-dimensional LB simulations (see Fig.3.1).

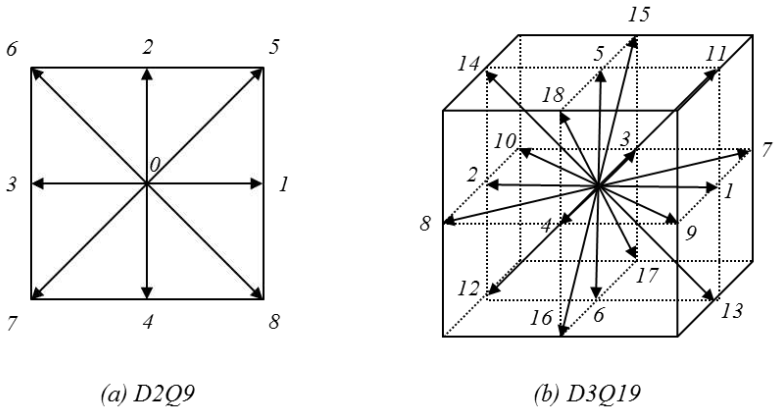

(a) D2Q9                    (b) D3Q19

Fig.3.1 Discrete velocity models (a) D2Q9 and (b) D3Q19.

The weighting factors for the D2Q9 model are:

$$\omega_i = \begin{cases} 16/36, & i = 0, \\ 4/36, & i = 1,...,4, \\ 1/36, & i = 5,...,8. \end{cases} \tag{3.5}$$

The weighting factors for the D3Q19 model are:

$$\omega_i = \begin{cases} 12/36, & i = 0, \\ 2/36, & i = 1,...,6, \\ 1/36, & i = 7,...,18. \end{cases} \tag{3.6}$$

The discrete velocities $\mathbf{e}_i$ for D2Q9 model are given by:

$$\mathbf{e}_i = \begin{cases} (0,0), & i = 0, \\ (\pm1,0),(0,\pm1), & i = 1,...,4, \\ (\pm1,\pm1,0), & i = 5,...,8, \end{cases} \tag{3.7}$$

and for the D3Q19 model by:

$$\mathbf{e}_i = \begin{cases} (0,0,0), & i = 0, \\ (\pm1,0,0),(0,\pm1,0),(0,0,\pm1), & i = 1,...,6, \\ (\pm1,\pm1,0),(\pm1,0,\pm1),(0,\pm1,\pm1), & i = 7,...,18. \end{cases} \tag{3.8}$$

The macroscopic density $\rho$ and velocity $\mathbf{u}$ are calculated as:

$$\rho = \sum_i f_i, \tag{3.9}$$

$$\rho\mathbf{u} = \sum_i f_i \mathbf{e}_i. \tag{3.10}$$

## 3.3.  Single component multiphase LBM

The Shan-Chen pseudopotential multiphase LBM is based on the concept of pairwise intermolecular interactions among fluid particles (Shan and Chen 1993, Shan and Chen 1994). Only the interactions between the nearest neighbors are considered:

$$\mathbf{G}(\mathbf{x},\mathbf{x}') = \begin{cases} \mathbf{G}, & |\mathbf{x}-\mathbf{x}'| \le c. \\ 0, & |\mathbf{x}-\mathbf{x}'| > c. \end{cases} \tag{3.11}$$

The parameter **G** reflects the strength of the interparticle interaction and its sign indicates attraction when **G** < *0* and repulsion when **G** > *0* (Shan and Chen 1993, Shan and Chen 1994, Raiskinmäki, Shakib-Manesh et al. 2002). If **G** surpasses a critical value, phase separation, which is one of the important attributions of the pseudopotential model, automatically appears (Dabbaghitehrani 2013). In single component multiphase LB model, the cohesive force $\mathbf{F}_m$ between nearest-neighbors fluid particles, which leads to phase separation (Thorne 2006), is defined as follows:

$$\mathbf{F}_m = -G\psi(\mathbf{x})\sum_{i}^{N}\omega\left(|\mathbf{e}_i|^2\right)\psi(\mathbf{x}+\mathbf{e}_i)\mathbf{e}_i,\tag{3.12}$$

where $|\mathbf{e}_i|^2 = 1$ at the four nearest neighbors and $|\mathbf{e}_i|^2 = 2$ at the next-nearest neighbors. The weight factors $\omega(|\mathbf{e}_i|^2)$ have the following values: $\omega(1) = 1/3$ and $\omega(2) = 1/12$. At $G < 0$, the attraction between particles increases leading to a large cohesive force. As a result, the cohesive force of the liquid phase is stronger than the cohesive force of the gas phase, leading to surface tension phenomenon (Thorne 2006). The effective mass $\psi(\mathbf{x})$ is defined by the non-ideal equation of state (EOS). In section 3.4 below, the choice of the EOS model used in this thesis is explained in detail. To consider fluid flow in porous media, the interactive force between the fluid and solid particles has to be incorporated. The adhesive force $\mathbf{F}_a$ between fluid and solid particles is described as follows (Martys and Chen 1996):

$$\mathbf{F}_a = -w\psi(\mathbf{x})\sum_{i}^{N}\omega\left(|\mathbf{e}_i|^2\right)s(\mathbf{x}+\mathbf{e}_i)\mathbf{e}_i,\tag{3.13}$$

where $w$ is a wettability factor, named in this thesis the solid-fluid interaction parameter, which reflects the strength of the interactive force between the fluid and solid phases. The LB model does not incorporate explicitly the contact angle $\theta$ (Lu, Wang et al. 2013) and, by varying the value of $w$, the envisaged contact angle can be obtained. The relationship between $w$ and $\theta$ is derived by comparing the LBM results for a droplet with the results of a modified empirical scheme based on Young's equation (see section 3.7.2). The wall density $s$ has a value equal to 0 and 1 for fluid nodes and solid nodes, respectively.

Gravitational effects are introduced using a body force $\mathbf{F}_b$ defined as:

$$\mathbf{F}_b = g\left(\rho(\mathbf{x}) - \rho_{gas}\right),$$ (3.14)

where $g$ is the body force per unit mass. In the LB method, the physical units are converted into lattice units as explained in section 3.9.

## 3.4. Equation of State (EOS)

The equation of state (EOS) describes the relation between the density of the gas and liquid phases for a given pressure and temperature. In single component multiphase LBM, the attractive force, which leads to phase separation, is characterized by a non-ideal EOS (Yuan and Schaefer 2006, Azwadi and Witrib 2012, Huang, Sukop et al. 2015). The choice of the EOS is directly related to the problem of numerical stability and is thus critical in LBM. The selection of a suitable EOS is based on different criteria (Yuan and Schaefer 2006, Chen, Kang et al. 2013). The first criterion is the determination of the density ratio between liquid and gas phases $\rho_l/\rho_g$. The second criterion is the reduction of spurious currents at the interface of different phases. Spurious currents are present in most multiphase models and higher density ratios promote larger spurious currents. The appearance of large spurious currents makes a numerical simulation unstable and leads to divergence. It is thus important in LBM with high density ratio to reduce as much as possible the appearance of these spurious currents. The third criterion relates to the selection of the temperature ratio $T_{min}/T_c$, where $T_c$ is the critical temperature. According to the Maxwell equal area construction rule, $T < T_c$ leads to the coexistence of two phases. At lower temperature ratios, spurious currents appear and the simulation becomes less stable. This temperature is actually a measure of the energy state of the particles. It is important to note that all LB simulations in this thesis are at isothermal condition. The last criterion relates to the agreement between a mechanical stability solution and thermodynamic theory. Choosing a proper EOS model reduces the appearance of spurious currents and leads to a thermodynamically consistent behavior (Yuan and Schaefer 2006). Recently, Yuan and Schaefer (2006) investigated the incorporation of various EOS models in a single component multiphase LB model and, based on the conclusion of their study, the Carnahan-Starling (C-S) EOS is selected and applied in this thesis. The C-S EOS generates lower spurious currents and applies to wider temperature ratio and density ratio ranges. The EOS is given as:

$$ p = \rho \mathbf{R} T \frac{1 + b\rho/4 + (b\rho/4)^2 - (b\rho/4)^3}{(1 - b\rho/4)^3} - a\rho^2 , \tag{3.15} $$

where $p$ is the pressure, $T$ is the temperature and $\mathbf{R}$ is the ideal gas constant equal to 1 in LBM. The attraction parameter $a = 0.4963(\mathbf{R}T_c)^2/p_c$ is chosen equal to 1 and the repulsion parameter $b = 0.1873\mathbf{R}T_c/p_c$ is chosen equal to 4, with $T_c = 0.094$ and $p_c = 0.0044$. The effective mass $\psi$ is calculated as:

$$ \psi(\rho) = \sqrt{\frac{2(p - c_s^2 \rho)}{Gc_o}} . \tag{3.16} $$

Substituting Eq. (3.15) into Eq. (3.16) yields

$$ \psi = \sqrt{\frac{2\left( \rho \mathbf{R} T \dfrac{1 + b\rho/4 + (b\rho/4)^2 - (b\rho/4)^3}{(1 - b\rho/4)^3} - a\rho^2 - \dfrac{\rho}{3} \right)}{\mathbf{G}c_0}} , \tag{3.17} $$

where $c_0$ equals 1 and $\mathbf{G}$ equals -1 to obtain a positive value inside the square root of Eqs. (3.16) and (3.17). Fig.3.2 shows the $p$-$V$ curve for C-S EOS at $T/T_c = 0.7$ with molecular volumes of liquid $V_l$ and gas $V_g$ phases determined from the Maxwell equal area construction rule.

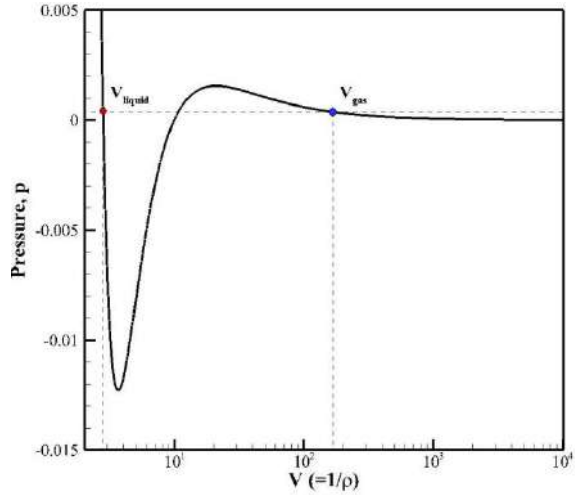

Fig.3.2. The $p$-$V$ curve for C-S EOS at $T/T_c = 0.7$ and $P_c = 0.044$.

## 3.5. Forcing scheme

In Shan-Chen pseudopotential LB model, the forcing scheme, incorporating the interactive forces, greatly affects the numerical accuracy and stability of the simulation. The original Shan-Chen LB model results in an inaccurate prediction of the surface tension, dependent on the chosen density ratio and relaxation time. When combining this model with a proper forcing scheme, the model can give an accurate surface tension prediction independent of relaxation time and density ratio. In recent studies, different forcing schemes for the Shan-Chen LB model have been compared by Li et al. (Li, Luo et al. 2012) and Huang et al. (Huang, Krafczyk et al. 2011). Based on these studies, the exact-difference method (EDM) developed by Kupershtokh et al. (Kupershtokh, Medvedev et al. 2009) is retained as the forcing scheme used in this thesis. For high density ratio with relaxation range of $0.5 < \tau \leq 1$, this method shows better numerical stability (Li, Luo et al. 2012). In EDM, a source term $\Delta f_i$ is added into the right term of the equilibrium distribution function Eq. (3.1) and is defined as:

$$\Delta f_i = f_i^{eq}\left(\rho, \mathbf{u} + \Delta\mathbf{u}\right) - f_i^{eq}\left(\rho, \mathbf{u}\right). \tag{3.18}$$

The increment of the velocity $\Delta\mathbf{u}$ is defined as:

$$\Delta\mathbf{u} = \frac{\mathbf{F}_{total}\Delta t}{\rho} \tag{3.19}$$

where $\mathbf{F}_{total}$ equals the sum of total forces in Eqs. (3.12), (3.13) and (3.14). By averaging the total force before and after a collision step, the real fluid velocity is calculated as:

$$\mathbf{u}_r = \mathbf{u} + \frac{\mathbf{F}_{total}\Delta t}{2\rho}. \tag{3.20}$$

## 3.6. Boundary conditions

In LBM, different boundary conditions can easily be taken into account in the model and incorporated within complex computational domains. To perform the different multiphase phenomena in complex porous media, the periodic, bounce back, velocity (pressure) and Neumann (outflow) boundary conditions are considered and introduced below.

### 3.6.1. Periodic boundary condition

The periodic boundary condition assumes that the opposite sides of a computational domain are connected. The fluid particles stream along their discrete direction and enter into the opposite boundary in domain. Fig.3.3 illustrates the periodic boundary condition of the D2Q9 model with the eight directions of distribution functions after the streaming step. The 3[rd], 6[th] and 7[th] distribution functions, which propagate from the red point on to the right side, move to the opposite boundary due to periodic boundary condition.

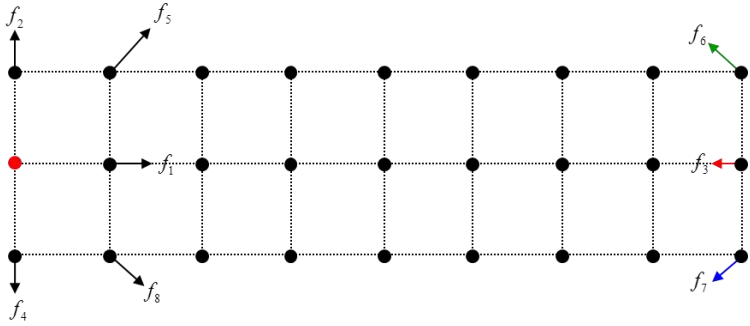

Fig.3.3 Illustration of the periodic boundary condition of the D2Q19 model with the eight directions of distribution functions after the streaming step.

When the periodic boundary condition is applied on the left and right sides, the distribution functions are determined as follows:

$$f_1(1, y) = f_1(nx, y)$$
$$f_5(1, y) = f_5(nx, y)$$
$$f_8(1, y) = f_8(nx, y)$$

$$f_3(nx, y) = f_3(1, y)$$
$$f_6(nx, y) = f_6(1, y)$$
$$f_7(nx, y) = f_7(1, y)$$

(3.21)

On top and bottom sides, the distribution functions are described as:

$$f_2(x,1) = f_2(x, ny)$$
$$f_5(x,1) = f_5(x, ny)$$
$$f_6(x,1) = f_6(x, ny)$$

$$f_4(x, ny) = f_4(x,1)$$
$$f_7(x, ny) = f_7(x,1)$$
$$f_8(x, ny) = f_8(x,1)$$

(3.22)

### 3.6.2. Bounce-back boundary condition

The bounce-back boundary condition is considered to simulate a no-slip boundary condition with zero velocity, as used in CFD models. However, in LBM, it is implemented in terms of streaming and collision steps. After the fluid particles encounter the bounce-back boundary conditions, the particles stream back in opposite direction. Due to its simplicity, the bounce-back boundary condition is easily applied into complex geometries, such as porous media. The main drawback is the dependency of this condition on the domain geometry. When the structure is curved or inclined and not of rectangular shape, the computational domain is approximated with a stair shape or zig-zag solid interface. As a consequence, a dense lattice resolution is required for curved or inclined geometries and high spatial and temporal computational costs follow (Bao and Meskas 2011, Zalzale 2014). To solve this problem, various bounce-back boundary condition models have been suggested by applying halfway bounce-back boundary with second-order accuracy and curved boundary. Nevertheless, the full bounce-back boundary condition which is the simplest bounce-back boundary condition is the only one considered in this thesis (Ladd 1994 a, Ladd 1994 b, Aidun and Lu 1995, Mei, Luo et al. 1999, Mei, Shyy et al. 2000). Fig.3.4 illustrates the bounce-back boundary condition with the direction of the distribution function of the fluid particle before and after streaming and collision. The $3^{rd}$, $6^{th}$ and $7^{th}$ distribution functions stream back along the opposite directions after a collision with the left wall.

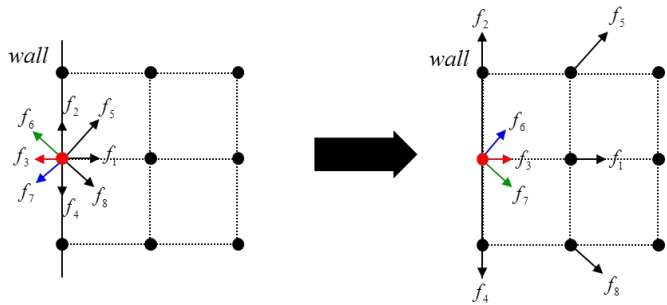

Fig.3.4 Illustration of the bounce-back boundary condition of D2Q19 model with the eight directions of distribution functions before and after the streaming step.

After collision and streaming, the distribution functions at the time step $t + \Delta t$ are described as follows:

$$f_1(x, y, t + \Delta t) = f_3(x, y)$$
$$f_3(x, y, t + \Delta t) = f_1(x, y)$$

$$f_2(x, y, t + \Delta t) = f_4(x, y)$$
$$f_4(x, y, t + \Delta t) = f_2(x, y)$$

(3.23)

$$f_5(x, y, t + \Delta t) = f_7(x, y)$$
$$f_7(x, y, t + \Delta t) = f_5(x, y)$$

$$f_6(x, y, t + \Delta t) = f_8(x, y)$$
$$f_8(x, y, t + \Delta t) = f_6(x, y)$$

### 3.6.3. Velocity (pressure) boundary condition

Velocity or pressure boundary conditions are used to describe inlet and outlet boundaries in LBM. In order to introduce a diffusion-controlled evaporation or condensation in the computational domain, the Zou-He velocity and pressure boundary conditions are used due to their simplicity as they are based on the idea of bounce-back of the non-equilibrium part (Zou and He 1997, Huang, Sukop et al. 2015). In the study of Huang et al (2015), the accuracy and stability of the boundary conditions were verified by performing two- and three-dimensional Poiseuille flow simulations by applying velocity and pressure boundary conditions and the simulation results showed reasonable agreement with the analytical solution. Velocity boundary conditions along the y-direction with an inlet at the left side are first presented. After streaming, the distribution functions $f_0$, $f_2$, $f_3$, $f_4$, $f_6$ and $f_7$ are known, but the distribution functions $f_1$, $f_5$ and $f_8$ are unknown, since these distribution functions stream from outside into the domain. Assuming the velocity known at the boundary, relations can be derived for these unknown distribution functions. The velocity $\mathbf{u} = (u_x, u_y)$ is assumed to be specified at the left wall, where the velocity in y-direction $u_y$ is assumed to be 0. Introducing the known distribution functions and velocities into Eqs. (3.9) and (3.10), the following equations are obtained:

$$f_1 + f_5 + f_8 = \rho - \left(f_0 + f_2 + f_3 + f_4 + f_6 + f_7\right)$$
$$f_1 + f_5 + f_8 = \rho u_x + \left(f_3 + f_6 + f_7\right)$$
$$f_5 - f_8 = \rho u_y - \left(f_2 - f_4 + f_6 - f_7\right) = -\left(f_2 - f_4 + f_6 - f_7\right)$$

(3.24)

However, to solve these equations, more constraints are required since the density value remains an undetermined value. Therefore, the bounce-back rule for the non-equilibrium part in the direction normal to the boundary is used:

$$f_1 - f_1^{(eq)} = f_3 - f_3^{(eq)}.$$

(3.25)

Substituting Eqs. (3.4), (3.5) and (3.7) into Eq. (3.25), the unknown distribution functions $f_1$, $f_5$ and $f_8$ are found to be:

$$f_1 = f_3 + \frac{2}{3}\rho u_x,$$
$$f_5 = f_7 - \frac{1}{2}\left(f_2 - f_4\right) + \frac{1}{6}\rho u_x,$$
$$f_8 = f_6 + \frac{1}{2}\left(f_2 - f_4\right) + \frac{1}{6}\rho u_x.$$

(3.26)

In velocity boundary condition, the inlet velocity, $u_x$, is specified and Eq. (3.24) allows to determine the density as:

$$\rho = \frac{\left[f_0 + f_2 + f_4 + 2\left(f_3 + f_6 + f_7\right)\right]}{1 - u_x}.$$

(3.27)

In pressure boundary condition, the density, $\rho$, is specified and Eq. (3.24) allows to determine the velocity in x-direction as:

$$u_x = 1 - \frac{\left[f_0 + f_2 + f_4 + 2(f_3 + f_6 + f_7)\right]}{\rho}.$$

(3.28)

### 3.6.4. Neumann (outflow) boundary condition

The outflow boundary condition (OBC) describes an open boundary from which fluid can flow out of the domain. Among different OBC models, a proper OBC is chosen based on the fact (a) that fluid is smoothly moving out from the domain without showing distortion and (b) that the numerical procedure is accurate and stable. Recently, Lou et al. (2013) studied the application of three types of OBC's to a two-phase LBM: the Neumann boundary condition (NBC), the convective boundary condition (CBC) and the extrapolation boundary condition (EBC) (Lou, Guo et al. 2013). They compared these three types of OBCs by simulating a moving droplet in an infinite long channel and a droplet passing through a channel with an obstacle. Their results show that the CBC is the most stable and accurate one, while the EBC showed the poorest stability and accuracy. The NBC is found to be quite stable and simple to implement, but the droplet shape near the outlet boundary is less smooth and becomes distorted. In this thesis, NBC is used as the OBC due to its simplicity and flexible implementation. In the NBC, the derivative of the variables is set to zero, or:

$$\frac{\partial \chi}{\partial x} = 0 \,, \tag{3.29}$$

where $\chi$ represents the following variables: distribution function $f_i$, density $\rho$ and velocity $\mathbf{u}$. The distribution function on the right side where $x$ equals $nx$ then becomes:

$$f_i\left(nx, y\right) = f_i\left(nx - 1, y\right), \tag{3.30}$$

The macroscopic variables are then given as:

$$\begin{aligned} \rho(nx, y) &= \rho(nx - 1, y) \\ \mathbf{u}(nx, y) &= \mathbf{u}(nx - 1, y) \end{aligned} \,. \tag{3.31}$$

## 3.7. Implementation of multiphase LBM

The multiphase LBM is a bottom-up approach, meaning that no explicit relations for surface tension or contact angle are specified (Lu, Wang et al. 2013). By choosing LB parameters, such as relaxation time $\tau$, liquid-gas interaction value $\mathbf{G}$ and solid-fluid interaction parameter $w$, the surface tension and contact angles can be found by simulating adequate physical experiments. In this section, known analytical relationships are compared to LB results to determine the physical properties as function of the LBM parameters and also as a mean to validate and verify the LBM.

First, Laplace law is used to study the surface tension. Then, the equilibrium contact angle as a function of the solid-fluid interaction parameter $w$ is studied (Ghassemi and Pak 2011, Mahmoudi, Hashemi et al. 2013, Huang, Sukop et al. 2015). Finally, the dynamic intrusion in a capillary is studied and the LB results are compared with Washburn's equation.

### 3.7.1. Laplace law: surface tension

In this test, two-dimensional liquid droplets with different radii are simulated and located at the center of a gas domain, as shown in Fig.3.5 (a). The domain shows periodic boundary conditions on all sides. Gravity is not taken into account. According to Laplace law, the pressure difference between the liquid and gas phases $\Delta p$ is equal to:

$$\Delta p = \frac{\gamma}{r}. \tag{3.32}$$

The pressure difference is calculated for 5 different droplet radii: 20, 25, 30, 40 and 50 lattices. The liquid droplet is located in the center of a computational domain of $201 \times 201$ lattice$^2$. Fig.3.5 (a) shows the obtained density contour for the droplet with a radius of 30 lattices at equilibrium state. The pressure along the central axis of the computational domain is plotted in Fig.3.5 (b). A pressure fluctuation occurs near the phase interface due to the sharp density change between the liquid and gas phases. However, this fluctuation does not affect the resulting pressure difference, as shown in the study of Huang et al (Huang, Krafczyk et al. 2011). Fig.3.5 (c) plots this pressure difference as a function of the inverse of the droplet radius and compares this result with Eq. (3.32). A linear relation, as expected by Laplace law, is obtained. The slope represents the surface tension $\gamma$ and is equal to 0.0152 lattice units. LB simulations are performed for temperature ratios ranging between 0.75 and 0.95. To validate the LB results, the obtained surface tension values are compared with values obtained by Huang et al (Huang, Krafczyk et al. 2011) in similar simulations. A good agreement can be observed with their results showing differences of less than 3%. The density of liquid and gas phases, as well as the density ratio as function of the temperature ratio obtained by LBM, are given in Fig.3.5 (d).

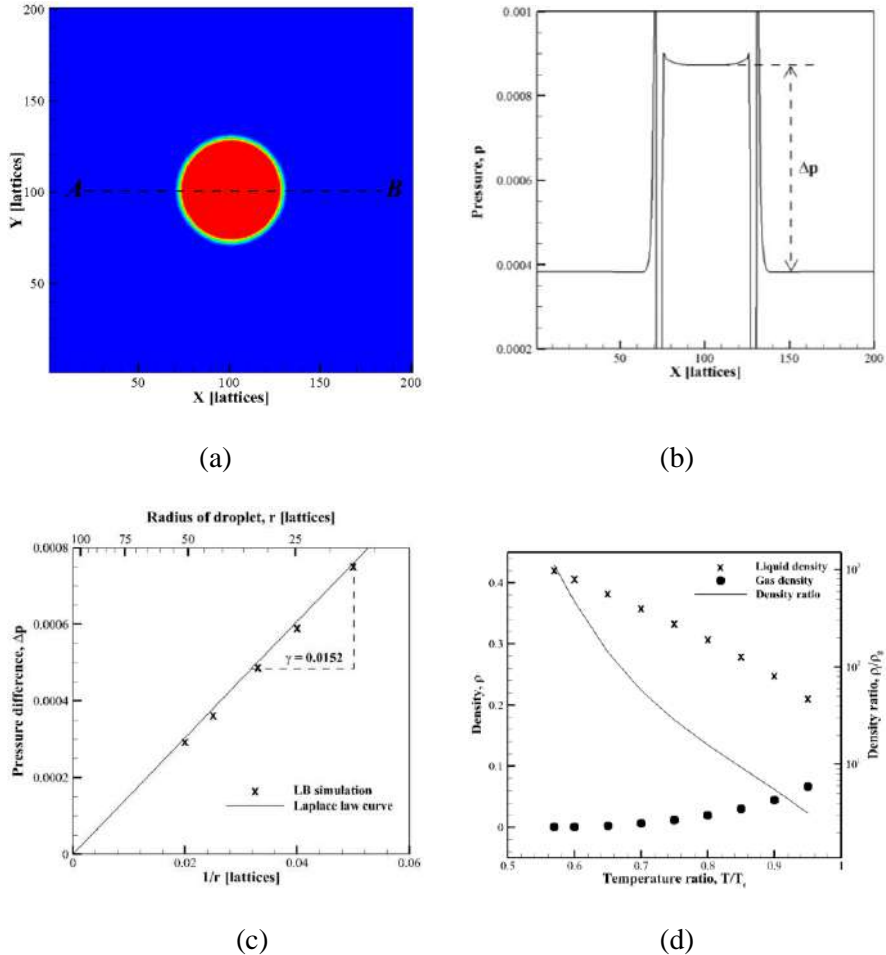

(a)         (b)

(c)         (d)

Fig.3.5. Comparison of LB results with Laplace law: (a) density contour at equilibrium state for droplet radius $r = 30$, (b) pressure along the central axis of the computational domain (line A-B), (c) LBM pressure difference versus droplet radius $r$ at $T/T_c = 0.7$ compared with Laplace law and (d) LBM gas and liquid densities and density ratio $\rho_l/\rho_g$ versus temperature ratio $T/T_c$ with critical density $\rho_c = 0.1304$.

In the LB simulation, the initial densities are set by the Maxwell equal area construction method. The original pseudopotential LBM can only achieve the maximum density ratio of the order of O (10). However, to perform real multiphase flow where density ratio of the order of O (1 000), higher density ratio is required and it can be performed by using the EOS and forcing scheme presented in sections 3.4 and 3.5 (Chen, Kang et al. 2014). In this thesis, the highest density ratio which gives stable results equals 1 365 corresponding to a temperature ratio $T/T_c = 0.57$. Since further increasing the density ratio (or decreasing the temperature ratio) leads to high spurious currents and numerical instability (Yuan and Schaefer 2006), for the following LBM studies, the temperature ratio is limited to a range from 0.7 to 0.85 corresponding to a density ratio ranging 59.1 to 9.4.

### 3.7.2. Young's equation: contact angle/wettability

The equilibrium contact angle of a liquid droplet on a flat horizontal solid plate is determined by LBM for different values of the solid-fluid interaction parameters $w$. The surface is partially wetting or hydrophilic when the contact angle $\theta < 90^\circ$ and $w$ is negative, and the liquid tends to spread as a film on the solid surface. In contrast, the surface is non-wetting or hydrophobic when $\theta > 90^\circ$ and $w$ is positive, and the liquid tends to form a spherical droplet resting on the solid surface. Huang et al. (Huang, Thorne Jr et al. 2007) proposed an empirical scheme to determine the contact angle directly from the solid-fluid interaction parameter $w$ in LBM. This scheme is an application of Young's equation and shows good agreement with values obtained from multiphase LBM (Huang, Thorne Jr et al. 2007). However, the developed scheme applies only for a multicomponent multiphase LBM, which is not the case in this study. Joshi and Sun (Joshi and Sun 2009) recently suggested an empirical scheme for single component multiphase LBM, but they use a EOS different from the one used in this study. Therefore, a modified empirical scheme based on previous studies and Young's equation is suggested. Young's equation can be written as:

$$\cos(\theta) = \frac{\gamma_{SG} - \gamma_{SL}}{\gamma_{LG}} = \frac{\mathbf{F}_a}{\mathbf{F}_m} \ . \tag{3.33}$$

The ratio of the interfacial tensions is described as the ratio of adhesion to cohesive forces as given respectively by Eqs. (3.12) and (3.13). A series of simulations are carried out where an initially semicircular droplet is placed on a horizontal solid surface. The solid-fluid interaction parameter $w$ is changed to obtain different contact angles. The simulations are performed in a domain of $201 \times 201$ lattice$^2$ with the top and bottom boundaries modelled as bounce-back boundary conditions and the left and right boundaries as periodic boundaries. The initial radius of the semicircular droplet is chosen to be 30 lattices at $T/T_c = 0.7$. After reaching equilibrium, the contact angles are measured using Image J with LB-ADSA (Stalder, Melchior et al. 2010). The obtained contact angles are plotted as a function of $w$ in Fig.3.6. In the considered range of $w = -0.15$ to $w = 0.05$, or a contact angle range between 47.8° and 105.2°, an excellent agreement with Eq. (3.33) is obtained.

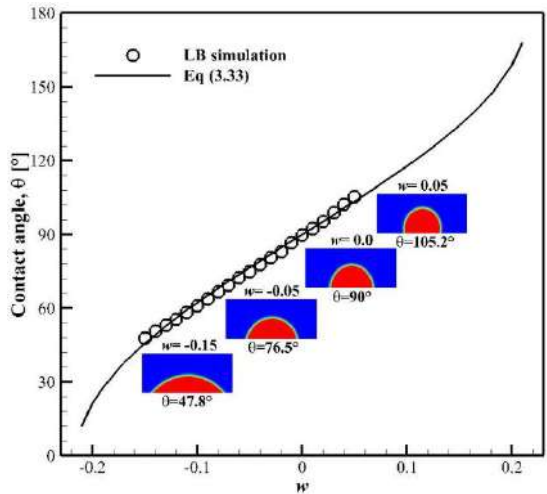

Fig.3.6. Contact angle test as a function of $w$ at $T/Tc = 0.7$ with respective inserted snapshots: comparison of equilibrium contact angles $\theta$ obtained by LBM with Eq. (3.33).

### 3.7.3. Washburn equation: dynamic capillary intrusion

The capillary intrusion test is chosen to assess the capacity of the pseudopotential LBM to simulate a moving contact line problem (Liu, Valocchi et al. 2013). For simplicity, the intrusion test uses a two-dimensional problem, i.e. two parallel plates (Fig.3.7 (a)). The contact line moves as a result of the balance between the pressure difference across the phase interface and the viscous force experienced by the intruding liquid. Neglecting the influence of gas viscosity, gravity and inertial force, this force balance results in (Diotallevi, Biferale et al. 2009, Pooley, Kusumaatmaja et al. 2009):

$$\gamma \cos(\theta) = \frac{6}{d} \eta x \frac{dx}{dt}, \tag{3.34}$$

where $\theta$ is the equilibrium contact angle between liquid and solid, $d$ is the width between plates, $\eta$ is the dynamic viscosity of the liquid and $x$ is the position of the interface.

The dynamic viscosity is obtained by multiplying the kinematic viscosity $v$ with the liquid density. Fig.3.7 (a) illustrates the two-dimensional computational domain of the capillary intrusion test of $400 \times 20$ lattice$^2$ following the geometry used in a previous study (Liu, Valocchi et al. 2013). Periodic boundary conditions are imposed on all boundaries of the computational domain. The parallel plates of the capillary are positioned at lattices 100 to 300 of the domain. The boundaries of the plates are treated as solid walls with bounce-back boundary conditions, represented by thick black lines in Fig.3.7 (a). The interface between the liquid and gas phases is tracked versus time. Fig.3.7 (b) shows the position of the interface versus time (iteration steps) as obtained by LBM. A good agreement with the analytical solution (Eq. (3.34)) is obtained.

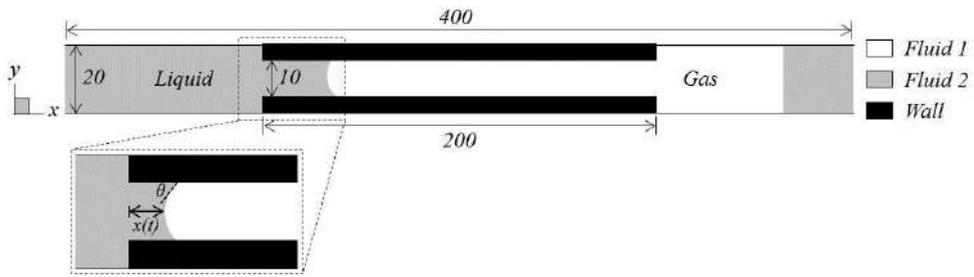

(a)

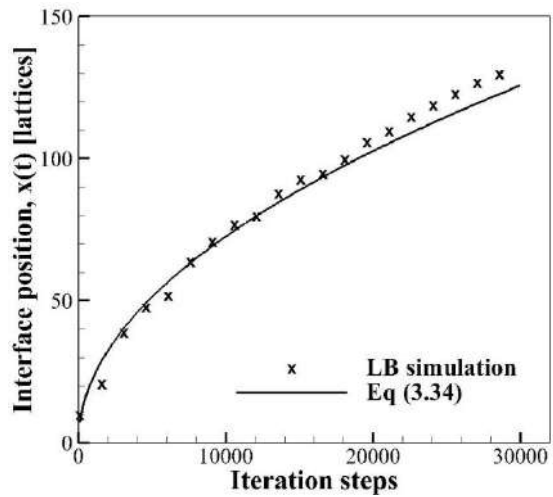

(b)

Fig.3.7. Dynamic capillary intrusion test: (a) computational domain and (b) comparison between LB results and analytical solution of the position of phase interface as a function of time.

## 3.8.  Limitations of the multiphase LBM

Unlike other numerical methods, in LBM, physical properties use LBM units, by converting physical units into lattice units [lu] and time is considered in time steps [ts]. Therefore, parameters in LBM are defined using non-dimensional numbers such as relaxation time, sound speed, velocity and body force. Therefore, in this section, three non-dimensional numbers used in the thesis are presented: the Mach number

*Ma*, the Reynold number *Re* and the Bond number *Bo*. The range of these non-dimensional numbers where LBM shows stability and convergence is given.

In each numerical method, stability is one of the most necessary conditions to obtain accurate solutions. LBM shows some constraints for stability and convergence due to its discrete system with finite propagations. LBM is derived from the discrete Boltzmann equation with a finite propagation speed, also called discrete velocity. Accordingly, the Courant-Friedrichs-Lewy (CFL) condition has to be satisfied for convergence in solving partial differential equation numerically. The CFL $= c\Delta t/\Delta x$ in LBM equals 1, since the lattice speed, grid spacing and time steps are assumed to be equal to 1 in the LBM (Yuan 2005). By considering the CFL condition, the Mach number, the ratio of fluid velocity to sound speed, has to be much lower than 1 (Dabbaghitehrani 2013) :

$$Ma = \frac{u}{c_s} << 1 \,. \tag{3.35}$$

Joshi and Sun (Joshi and Sun 2009) showed that an accurate solution in the incompressible flow regime can be obtained when the Mach number is less than 0.1. To ensure that flow is incompressible in LBM, the velocity should then be sufficiently smaller than the speed of sound, or $u << c_s$, where $c_s, = 1/\sqrt{3} = 0.577$. The fluid velocity can be defined using the Reynold number which reflects the balance between viscous and inertial forces:

$$\text{Re} = \frac{\rho u l}{\eta} = \frac{u l}{v} \,, \tag{3.36}$$

where *L* is a characteristic length. For a stable simulation, the maximum velocity is recommended to remain below 0.1 [lu/ts]. Smaller velocities lead to better and more stable simulation in LBM (Dellar 2003, Thorne 2006). Thus, by considering this velocity constraint, the velocity, which is converted from the physical value, is limited to 0.1 [lu/ts] for all subsequent simulations.

For the body force, the Bond number, which gives the ratio between the gravitational force and surface tension, is used:

$$Bo = \frac{\rho g l^2}{\gamma}. \tag{3.37}$$

In single component multiphase LBM, the Bond number must be less than 10 ($Bo <$ 10) to ensure stability (Orr, Powers et al. 2015).

## 3.9. Conversion between physical and lattice units

In LBM, variables are represented in terms of lattice units. To solve physical phenomena efficiently, a conversion between physical and lattice units should be supposed. This conversion can be performed through non-dimensional numbers and a non-dimensional parameter, the relaxation time, $\tau$ as described in section 3.2. In the following the previous study, the conversion between physical units ($p$) and lattice Boltzmann units ($lb$) is briefly described (Latt 2008). The discrete space interval is defined as the ratio between physical length $l_p$ divided by lattice number $N$:

$$\delta x = \frac{l_p}{N} [-], \tag{3.38}$$

where lattice number $N$ equals the numbers of grid points minus one. In LBM, the kinematic viscosity $v_{lb}$, is decided by relaxation time $\tau$ and described as

$$v_{lb} = c_{s,lb}^2 \left( \tau - \frac{1}{2} \right) \delta t_{lb} \left[ \frac{lu^2}{ts} \right], \tag{3.39}$$

where $c_{s,lb}$ is the speed of sound, equal to $c/\sqrt{3} = 1/\sqrt{3}$, where the velocity $c$ equals $c = \delta x_{lb}/\delta t_{lb} = 1/1 = 1$, with $lu$ for lattice unit and $ts$ for lattice time step. In this thesis, which uses single component LBM, only one relaxation time is considered. To guarantee a good stability of the LB simulation, the relaxation time is chosen equal to 1 and the kinematic viscosity $v_{lb}$ equals to

$$v_{lb} = \frac{1}{3}\left(1 - \frac{1}{2}\right) = \frac{1}{6}\left[\frac{lu^2}{ts}\right] . \tag{3.40}$$

The relation between the kinematic viscosity in physical unit $v_p$ and lattice unit $v_{lb}$ is expressed as

$$v_p = v_{lb} \frac{\delta x^2}{\delta t}\left[\frac{m^2}{s}\right]. \tag{3.41}$$

Following Eq. (3.41), the discrete time interval $\delta t$ can be calculated as

$$\delta t = v_{lb} \frac{\delta x^2}{v_p}[s]. \tag{3.42}$$

Using Eqs. (3.38) and (3.42) respectively for $\delta x$ and $\delta t$, the conversion from physical to lattice units, or vice versa, can be performed effectively. For example, the velocity in lattice units can be calculated from the velocity in physical units as

$$u_{lb} = u_p \frac{\delta t}{\delta x}\left[\frac{lu}{ts}\right]. \tag{3.43}$$

The other type of conversion uses non-dimensional numbers. In physical and LB systems, the Reynold number which is widely used non-dimensional number is as follows

$$\mathrm{Re} = \frac{u_p l_p}{v_p} = \frac{u_{lb} l_{lb}}{v_{lb}}[-]. \tag{3.44}$$

By applying Eqs. (3.39) and (3.40) and physical fluid properties in Eq. (3.44), a velocity in physical unit $u_p$ is translated into a velocity in lattice unit $u_{lb}$ which is simply applied into the LB simulation.

## 3.10. Numerical procedure and parallel computing

In this thesis, all two-dimensional simulations are done on a single CPU, while all three-dimensional simulations are performed on a high performance computing cluster. In the three-dimensional studies, larger size of the computational domain and longer simulation time are required to solve the problem. Thus, for three-dimensional problems, parallel computing is used to reduce the computational cost and increase computational efficiency. Furthermore, as mentioned before, an advantage of the LBM is that it can be easily parallelized comparing with other numerical methods. The three-dimensional LBM code further developed in this thesis is written in FORTRAN 90 using Message Passing Interface (MPI). All numerical simulations are run by MPI on the high performance computing cluster MUSTANG at Los Alamos National Laboratory (LANL). The processor of the cluster is a AMD Opteron 6 176 and the operating system is Linux. This cluster aggregate performance peaks at 352 TFlop/s with 102.4 TB of memory for 38 400 cores (37 080 total CPUs) or 1 545 nodes with 24 CPU cores per node. Each simulation requires different numbers of processor cores depending on the grid size and iteration time steps needed, while the wall time is fixed to a maximum of 16 hrs. The details on the processor cores used in each different simulation are given for each simulation in the next chapters.

## 3.11. Post-processing procedure for multiphase phenomena

In numerical modeling of multiphase phenomena, the determination of the liquid and gas phase interface is key. LBM studies successfully multiphase flow, due to its capacity in tracking the fluid interfaces. However, the phase interface in LBM is not sharp as density gradually decreases from liquid to gas over two to three lattices as shown in Fig.3.8 (a). In this thesis, some of the simulation results are post processed to remove the density information and provide clear gas and liquid interface. In such post processing, the position of the phase interface is located where the density equals the average density of the liquid and gas phases after reaching equilibrium state. This is illustrated on Fig. 3.8 (b).

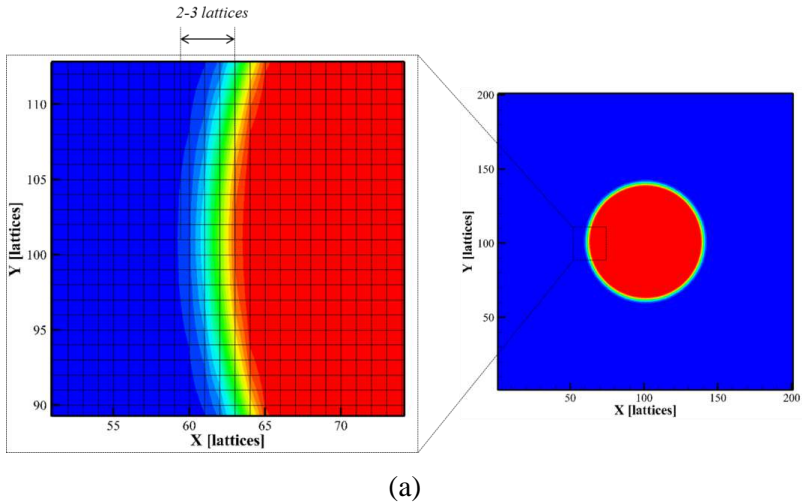

(a)

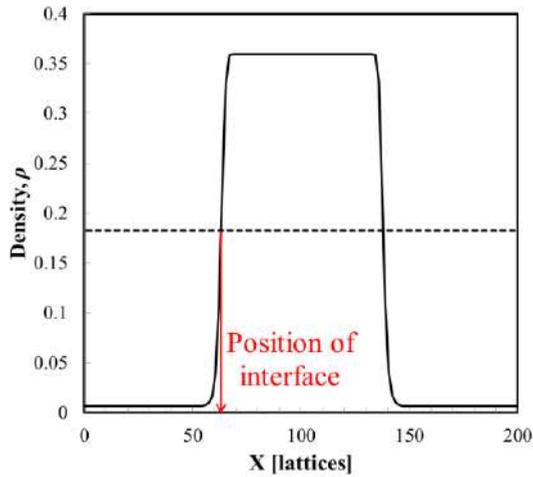

(b)

Fig.3.8. At equilibrium state, (a) density contour at equilibrium state of the interface between liquid and gas phases showing that the change of density occurs over approximately 3 lattices and (b) density profile across the center line. The location where the density profile equals the averaged density of the liquid and gas phases is considered the interface location.

## **3.12. Summary and link to next chapter**

In this chapter, the pseudopotential multiphase LBM with the non-ideal EOS and forcing scheme was introduced. Special attention was given to the adequate formulation the boundary conditions. Parametric studies including Laplace law, the contact angle of a droplet on a surface and the dynamic capillary intrusion were performed. LBM has been validated showing a good agreement with analytical solutions. Constraints of the pseudopotential multiphase LBM were explained in detail using non-dimensional numbers and ranges of application were specified. Finally, unit conversion, model limitations and interface post-processing were explained. In the next chapters, the pseudopotential LBM will be used to study different two-phase phenomena related to the study of fluid transport in capillary, on surfaces and in a porous medium analogous to porous asphalt.

# 4. CAPILLARY UPTAKE AT PORE SCALE
## 4.1. Introduction

One of the main driving forces for liquid transport in porous materials is capillarity arising from the pressure difference between the liquid and gas phases at the interface between both phases. When porous asphalt is newly produced, it is hydrophobic under normal conditions, since all aggregates are covered with bitumen, resulting in a full hydrophobic pore surface. However, the pore surface inside PA may be partly hydrophilic when the binder is not covering all aggregates due to incomplete mixing. Once the porous asphalt is installed as pavement, it will be exposed to the environment: oxygen, temperature, water, and mechanical loads. Under these loadings, the bitumen will age and the aggregates can become partly stripped of the binder, making the PA partly hydrophilic. As a result, porous asphalt may become wet also due to capillary action, as was seen in a 7-years-old PA (Lal, Poulikakos et al. 2014).

In this chapter, capillary action is studied using LBM for 2D and 3D tubes showing different geometrical cross-sections, size and contact angles. Capillary rise is simulated in tubes (2D parallel plates and 3D polygonal tubes) and the LB results are compared with analytical solutions. In polygonal tubes, depending on the wetting contact angle, the liquid can also fill the corners, which is known as corner flow. Corner flow is studied using LBM in square and triangular tubes relating the meniscus curvature with the degree of saturation. This chapter ends up with LB simulation of capillary rise in circular tubes and in pore structures inspired by PA. All systems in this chapter are characterized with length conditions below the capillary length, meaning that the influence of gravity can be neglected.

## 4.2.    Capillary rise between parallel plates

In this section, capillary rise between parallel plates is studied and compared with analytical solutions to verify and validate the LB results. The influence of contact angle and pore size is studied.

### 4.2.1.   Simulation set-up and boundary conditions

The computational domain is discretized in $400 \times 400$ lattice$^2$ as illustrated in Fig.4.1. A grid sensitivity analysis was performed discretizing the same domain for four different grid sizes: $100 \times 100$, $200 \times 200$, $300 \times 300$ and $400 \times 400$ lattice$^2$. Fig.4.2 shows that no significant differences can be observed for grid sizes larger than $300 \times 300$ lattice$^2$. A grid of $400 \times 400$ lattice$^2$ is chosen for the LBM simulations. Two parallel plates are located 50 lattices above the bottom. The plates have a height of 300 lattices and the width $d$ between the plates is 40 lattices. The bottom quarter of the domain is initially filled with liquid to act as a liquid reservoir as shown in Fig.4.1. The plates are partially submerged in this reservoir with a length of 50 lattices.

By applying a density ratio $\rho/\rho_c = 59.1$ at $T/T_c = 0.7$, the liquid and gas density is 0.359 and $6.07 \times 10^{-3}$ lattice units, respectively. For the contact angle inside the plates, three different contact angles of 48°, 62° and 75° are considered, corresponding to solid-fluid interaction parameters $w$ of -0.15, -0.1 and -0.05 as determined in the contact angle test in chapter 3. The surface on the outside of the plates has a contact angle of 90° or the neutral solid-liquid interaction parameter $w$ of 0 to prevent wetting. Periodic boundary conditions are imposed on all sides with exception to the top and bottom sides, which are treated as bounce back boundary conditions.

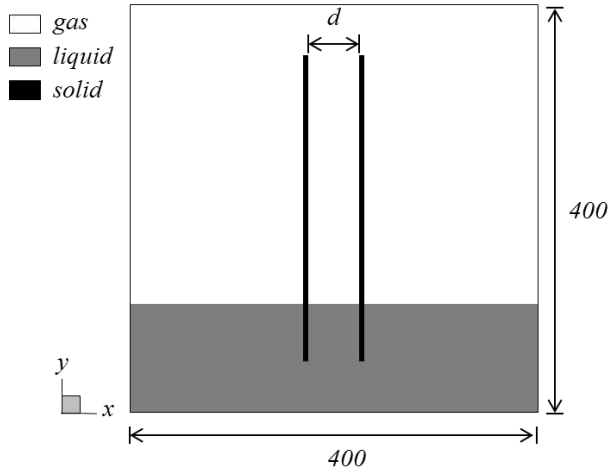

Fig.4.1. Schematic representation of the 2D computational domain for capillary rise between immersed parallel plates.

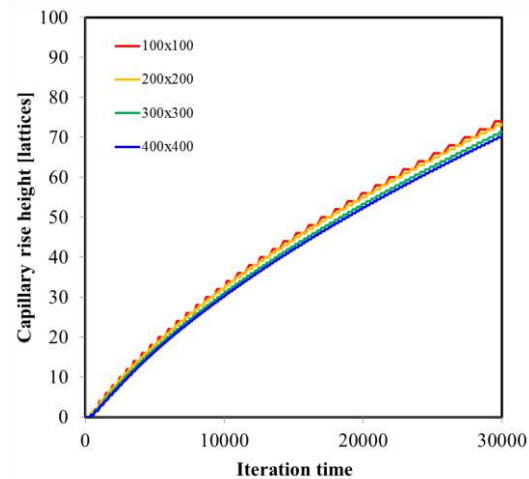

Fig.4.2. Grid sensitivity test of capillary rise between the 2D parallel plates.

### 4.2.2. Results and comparison with analytical solutions

In Fig.4.3, the capillary rise height is shown versus time for the three widths between the parallel plates. The capillary rise height is defined as the difference between the liquid level inside and outside the tube. The widths are 10, 20 and 40 lattices. The contact angle equals 75º.

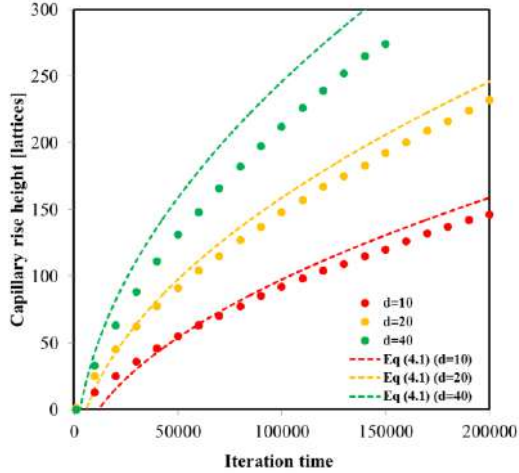

Fig.4.3. Capillary rise height versus time for three widths between the parallel plates: 10, 20 and 40 lattices. The contact angle is 75° corresponding to a solid-liquid interaction parameter of -0.05.

The capillary rise is slowest for the small width, and fastest for the large width. The faster uptake for the tube with larger width between the plates can be explained by the fact that the capillary forces overrule the friction forces (see Eqs. (2.14) and (2.15)). The LB results show in general a good agreement with the analytical solution (Eq. (2.23)) for parallel plates:

$$h(t) = \sqrt{\frac{4d\gamma\cos\theta}{12\eta}}\sqrt{t} \ . \tag{4.1}$$

However, for higher width between the plates, the capillary rise predicted by LBM starts to deviate from the analytical solution as iteration time increases. Two explanations for the difference can be formulated. First the analytical solution is based on the assumption that the reservoir is infinite, while, in LBM, the reservoir is limited, which leads to lower suction heights in LBM. Second, in LBM, the plates are immersed in the reservoir, leading to an entry resistance. In Fig.4.4, liquid velocity streamlines are shown for two different widths at different iteration times. The color represents the velocity magnitude. As the liquid rises between plates, vortices develop at the base of the plates giving rise to the entrance resistance. The vortices and their velocities increase with width between the plates as shown in Fig.4.4 (b) leading to a larger entry resistance.

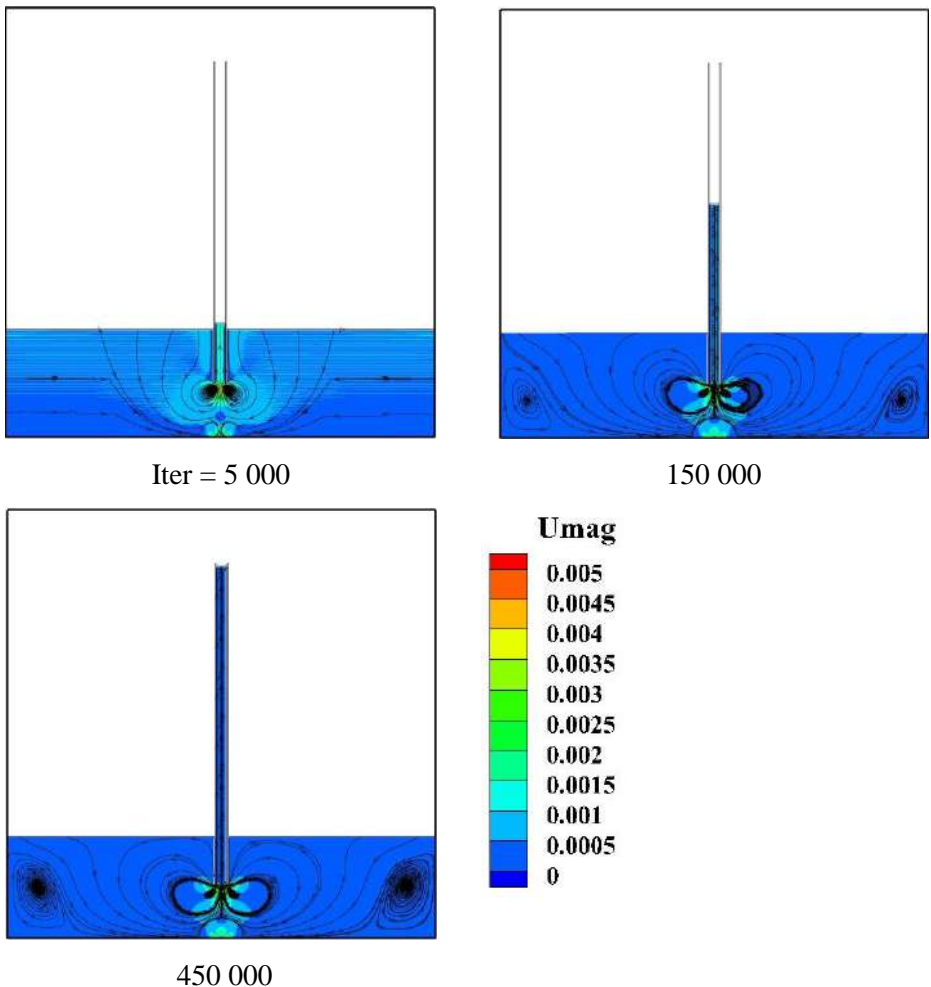

Iter = 5 000                    150 000

450 000

(a)

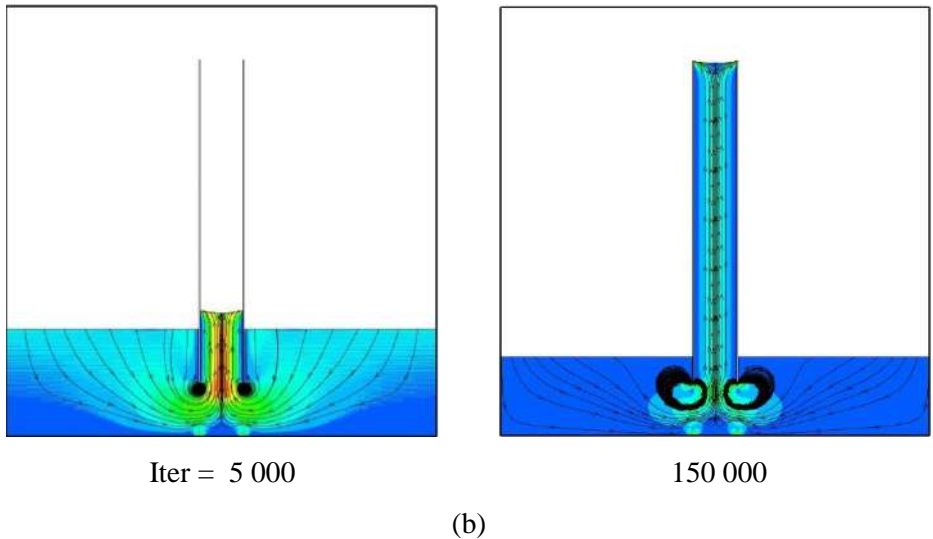

<div align="center">Iter = 5 000        150 000</div>

<div align="center">(b)</div>

Fig.4.4. Liquid velocity magnitude profiles and stream lines for widths between the plates of (a) 10 and (b) 40 lattices at different iteration times.

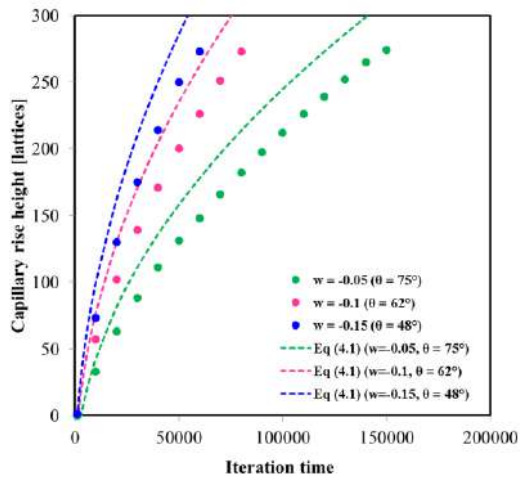

Fig.4.5. Capillary rise height versus iteration time for different contact angles of 48°, 62° and 75° or solid-fluid interaction parameter *w* of -0.15, -0.1 and -0.05. The width between the two parallel plates is 40 lattices.

In Fig.4.5, the capillary rise height is given versus time for three different contact angles: 75º, 62º and 48º, for a constant width of 40 lattices. Capillary rise is faster for more hydrophilic surfaces, because in this case the capillary force (see Eq. (2.14)) becomes relatively larger compared with the friction forces. Again the deviations between LBM and the analytical solutions are attributed to the limited reservoir condition and the entry resistance.

## 4.3.  3D square tube

In this section, capillary rise in a square tube is studied using 3D LBM. The results from LB are compared with analytical solutions for different tube sizes and contact angles.

### 4.3.1.  Simulation set-up and boundary conditions

The domain size is $300 \times 300 \times 300$ lattice$^3$ with spatial resolution $\varDelta x = 1$ μm per lattice chosen based on a grid sensitivity analysis. The tube has a height of 235 lattices, a width of 30 or 40 lattices, and is located in the middle of domain, 38 lattices above the bottom. The bottom quarter of the domain is filled with liquid, representing a limited reservoir as shown in Fig.4.6. Liquid and gas densities are 0.359 and $6.07 \times 10^{-3}$ lattice units respectively corresponding to a density ratio $\rho/\rho_c = 59.1$ at $T/T_c = 0.7$. The contact angle is 75° corresponding to a solid-fluid interaction parameter $w$ of -0.05. This contact angle is applied to all solid surfaces. Periodic boundary conditions are imposed on all sides except the top and bottom sides, which are treated as bounce back boundaries.

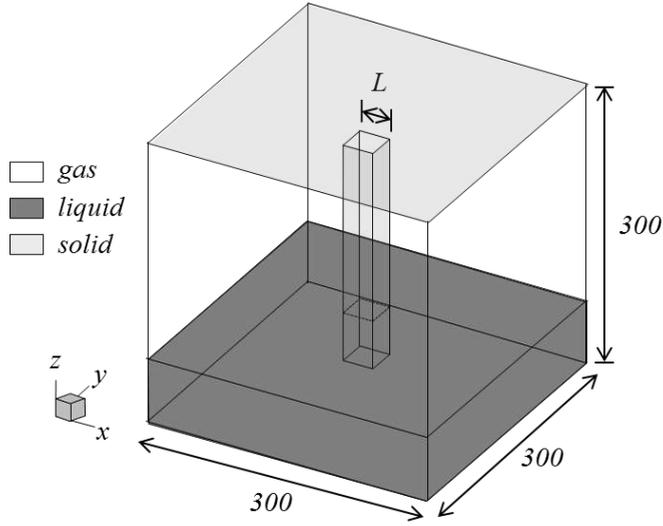

Fig. 4.6. Schematic of capillary rise in the 3D square tube.

In this section, all 3D simulations are run by parallel computing based on MPI (Message Passing Interface) on the high performance computing cluster of Los Alamos National Laboratory (LANL). The cluster aggregate performance is 352 TFlop/s with 102.4 TB of memory for 38 400 cores. Each simulation is run on 216 processor cores ($6 \times 6 \times 6$) and requires 16 hrs to run 50 000 time steps.

### 4.3.2. Results and comparison with analytical solutions

Capillary rise in a 3D square tube is given by (Ichikawa, Hosokawa et al. 2004):

$$h(t) = \sqrt{\frac{64 L \gamma \cos \theta (\pi - 2)}{\pi^5 \eta}} \sqrt{t} \ . \tag{4.2}$$

In LB results, the height is the difference between levels inside and outside the tube to take into account changes in the reservoir level due to its finite size.

Fig.4.7 shows the capillary rise height versus time for two different tube sizes of 30 and 40 lattices. In the larger tube, the uptake is faster as explained in the previous section.

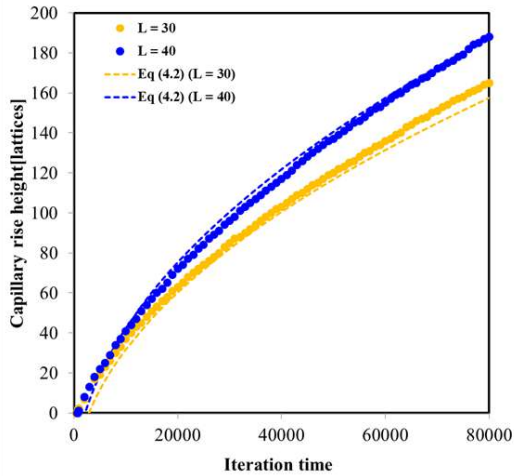

Fig.4.7. Capillary rise height versus time, for 3D square tube with widths of 30 and 40 lattices, with contact angle of 75° or a solid-fluid interaction parameter *w* of -0.05.

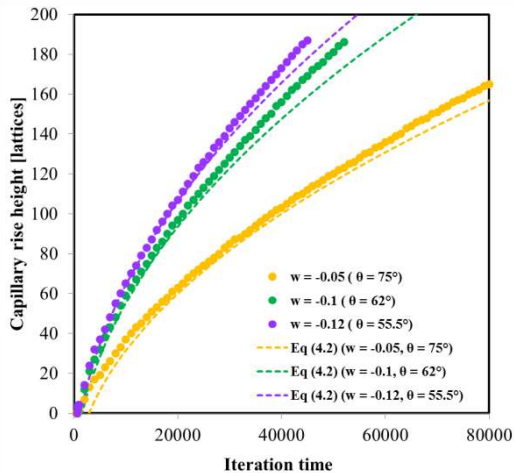

Fig.4.8. Capillary rise height versus time for three different contact angles of 55.5°, 62° and 75° corresponding to solid-fluid interaction parameters *w* of -0.12, -0.1 and -0.05 respectively. The square tube has a width of 30 lattices.

The 3D LB results show a good agreement with the analytical solution, better than the one seen in the 2D case of the previous section. In 2D, the difference was attributed to the change in level of the limited reservoir and the resulting entrance resistance. In 3D, the level of the reservoir decreases less, since the ratio of liquid volume taken up in the tube versus the volume of the reservoir is much smaller in 3D compared to 2D. This means that the level in the reservoir will change relatively less in 3D than in 2D leading also to a smaller entrance resistance in the former case. As a consequence, the 3D LB results show an overall better agreement with the analytical solution than in 2D does.

In Fig.4.8, the effect of the contact angle on capillary rise is studied. Three different contact angles are considered: 75º, 62º and 55.5º corresponding to a solid-fluid interaction parameter $w$ of -0.05, -0.1 and -0.12 respectively. The tube has a width of 30 lattices. An overall good agreement is observed with the analytical solutions. The differences at the end of the uptake process are limited to 4.1, 6.4 and 4.5 % for the contact angles of 75º, 62º and 55.5º respectively.

In conclusion, the 3D LB results for capillary rise in a square tube show an overall good agreement with the analytical results.

## 4.4. 3D polygonal tube: steady state meniscus and occurrence of corner flow

*This section is based on the journal paper: Son, S., L, Chen., Q, Kang., D, Derome. and J, Carmeliet. (2016). Contact Angle Effects on Pore and Corner Arc Menisci in Polygonal Capillary Tubes Studied with the Pseudopotential Multiphase Lattice Boltzmann Model. Computation; **4**(1): 12.*

The capillary behavior in a polygonal tube with $n$ sides depends on the critical contact angle defined as $\theta_c = \pi/n$ (Concus and Finn 1974). When the equilibrium contact angle $\theta$ is in between $\pi/2$ and the critical contact angle, the liquid meniscus spans the total tube, referred to as pore meniscus configuration (Fig.4.9 (a)).

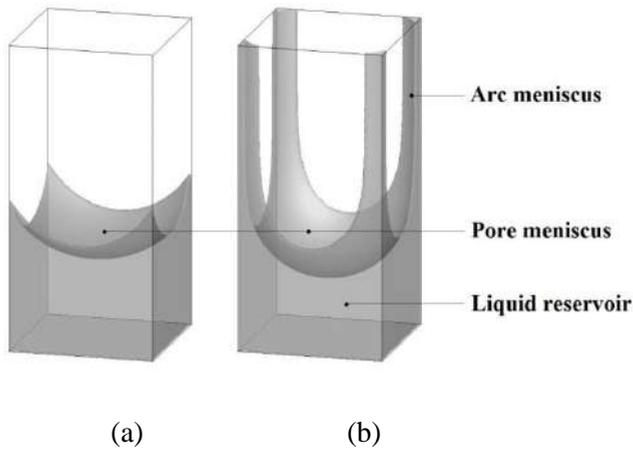

(a)                    (b)

Fig.4.9. Schematic representation of the two liquid configurations in a square tube: (a) pore meniscus when the contact angle is larger than the critical contact angle, $\theta \geq \theta_c$; and (b) co-occurrence of pore and corner arc menisci when the contact angle is smaller than the critical contact angle, $\theta < \theta_c$.

In contrast, if the contact angle is smaller than the critical contact angle $\theta < \theta_c$, the liquid will, in addition to forming a pore meniscus as shown in Fig.4.9 (b), also invade the edges or corners of the polygonal tube, forming corner arc menisci (Concus and Finn 1974, Concus and Finn 1990).

### 4.4.1. Simulation set-up and boundary conditions

In this study, capillary rise in two different polygonal tubes is simulated using LBM: a square ($n = 4$) and triangular ($n = 3$) tube. The cross-sections are circumscribed by a circle with a radius $r = 100$ lattices for the square tube and a radius of $r = 200$ lattices for the triangular tube, as shown in Fig.4.10 (a) and (b). By changing the values of the contact angle, cases with contact angles larger and smaller than the critical contact angle $\theta_c = \pi/n$ (45° for square, 60° for triangular tube) are considered. For the square tube, the domain size is $142 \times 142 \times 300$ lattice$^3$ for the pore meniscus case and $142 \times 142 \times 500$ lattice$^3$ for the corner arc menisci case. The spatial resolution $\Delta x$ of 1 μm per lattice is chosen based on a mesh grid sensitivity analysis for the corner arc menisci case, as presented below at the end of section 4.4.3. For the triangular tube, the regular lattice grid results in a zigzag boundary, at least for two boundaries when the mesh is aligned to one of the triangle sides. This zigzag boundary introduces an artificial roughness, which in combination with a full bounce-back boundary condition produces mesh-dependent results, as will be shown below. The bounce-back boundary condition represents a no-slip boundary condition with zero velocity at the wall. To improve the quality of the results, two measures are taken. First, the spatial resolution is increased: $\Delta x$ equals 0.5 µm per lattice. As a result, the domain consists of $292 \times 290 \times 600$ lattice$^3$ for the pore meniscus case and $292 \times 290 \times 1\,000$ lattice$^3$ for the corner arc meniscus case. Second, the mesh is turned with an angle of 15° to decrease the effect of side roughness (see Fig.4.10 (c)). However, even when applying these measures, the corners show some roughness, especially in corner 1.

An alternative would be to apply a different boundary condition, such as the curved, the half bounce-back or the moving boundary condition, as these methods allow tracking the interface independently from the mesh (Mei, Luo et al. 1999, Mei, Shyy et al. 2000). However, the implementation of these boundary conditions, and their analysis, was considered out of the scope of this study.

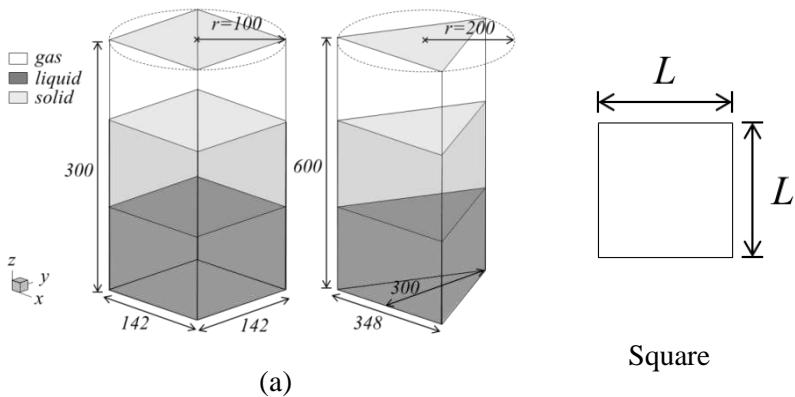

(a)

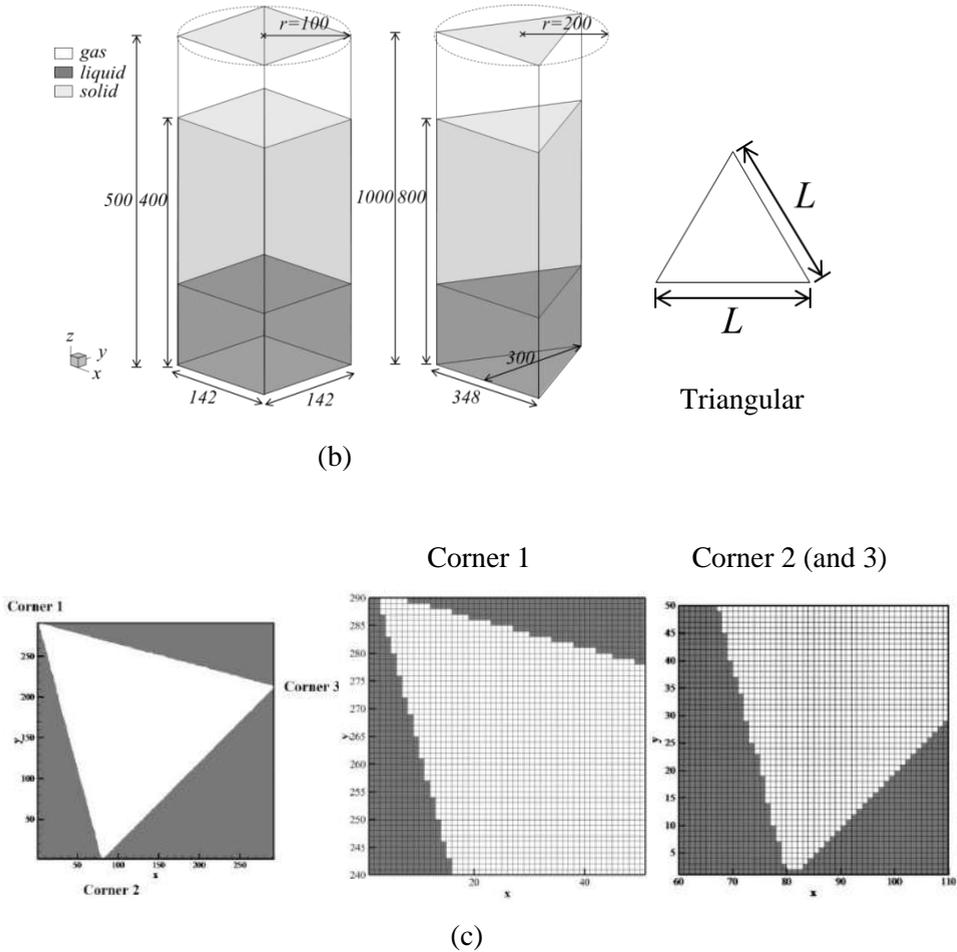

Corner 1        Corner 2 (and 3)

(c)

Fig.4.10. Schematic geometry of polygonal tubes: (a) computational domain for pore meniscus case; (b) computational domain for corner arc menisci case; and (c) computational mesh details for the triangular tube at the three corners.

The polygonal tube is initially filled by liquid to a height of 100 lattices from the bottom, as shown in Fig.4.10 (a) and (b). The redistribution of the liquid is then calculated by LBM. The liquid and gas densities are 0.28 and 0.0299 lattice units, respectively corresponding to a density ratio $\rho/\rho_c = 9.4$ at $T/T_c = 0.85$. Different contact angle ranges are applied. For the square tube, the contact angle ranges from 42.6° to 136.5° as related to a solid-fluid interaction parameter $w$ ranging from −0.08 to 0.06. For the triangular tube, the contact angle ranges from 59.8° to 125.6° as related to a solid-fluid interaction parameter $w$ ranging from −0.05 to 0.05. As shown in Fig.4.10 (a) and (b), bounce-back boundary conditions are imposed on all sides, except for the top 100 lattices on the three or four vertical sides where periodic boundary conditions are imposed to simulate an open capillary tube.

All numerical simulations are run by parallel computing based on Message Passing Interface (MPI) on the high performance computing cluster of Los Alamos National Laboratory (LANL). The cluster aggregate performance is 352 TFlop/s with 102.4 TB of memory for 38 400 cores. Each simulation is run on 120 or 200 processor cores for pore meniscus or corner arc meniscus simulations in the square tubes and on 400 or 800 processor cores for pore meniscus or corner arc meniscus simulations in the triangular tubes and requires 16 hours to run 20 000 or 40 000 time steps, respectively.

### 4.4.2. Results of pore meniscus

When the contact angle is larger than the critical contact angle, $\theta \geq \theta_c$, the liquid wets the tube walls and a pore meniscus is formed in the tube. Fig.4.11 shows, as example, snapshots of pore menisci for square and triangular tubes with hydrophilic and hydrophobic surfaces after reaching steady state. For the square configuration, the meniscus is regular (Fig.4.11 (a) − (d)), while for the triangular configuration (Fig.4.11 (e) and (f)) the pore menisci show different heights at each corner, especially at a small contact angle (hydrophilic cases). This observation is explained by the artificially introduced wall roughness for the triangular tube, as also observed by other authors such as Dos Santos *et al.* (Dos Santos, Wolf et al. 2005). It is found that corner 1 in Fig.4.10 (c), which has the highest roughness, shows the lowest height, while corners 2 and 3 show the same height.

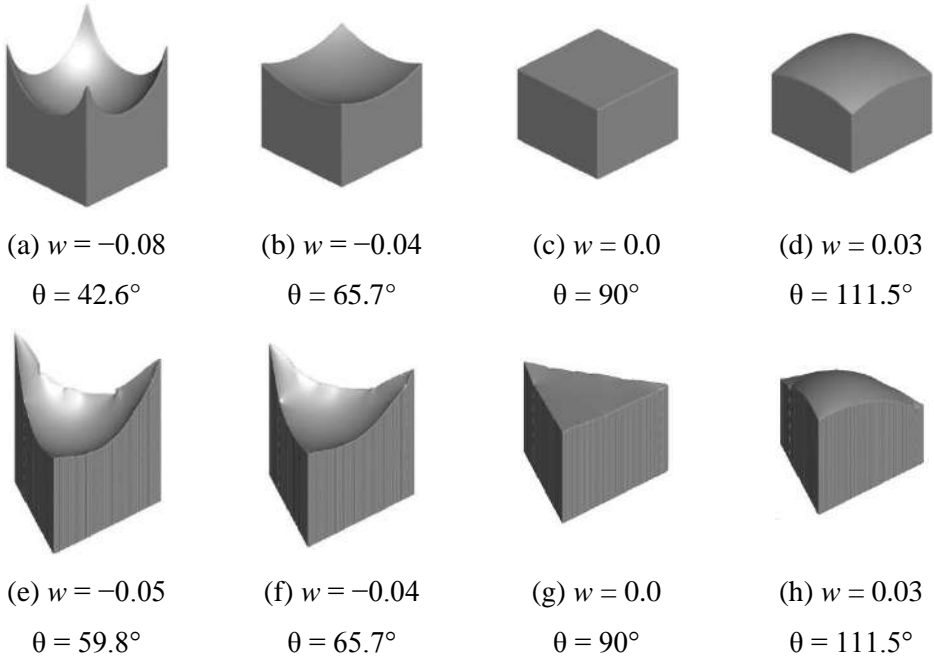

(a) $w = -0.08$    (b) $w = -0.04$    (c) $w = 0.0$    (d) $w = 0.03$

$\theta = 42.6°$     $\theta = 65.7°$     $\theta = 90°$     $\theta = 111.5°$

(e) $w = -0.05$    (f) $w = -0.04$    (g) $w = 0.0$    (h) $w = 0.03$

$\theta = 59.8°$     $\theta = 65.7°$     $\theta = 90°$     $\theta = 111.5°$

Fig.4.11. Liquid configurations in square and triangular tubes for different contact angles after reaching steady state.

Results are presented in terms of height of pore meniscus versus cosine of the contact angle after reaching equilibrium. The height is defined as the difference between the bottom and the top of the meniscus (see insets of Fig.4.12 (a) and (c)). Since the height for the triangular tube is not equal in all corners, the average of the heights in the three corners of the tube is determined. The curves show an S-shape, meaning that at very high (low) contact angles, the height decreases (increases) even more. Fig.4.12 (b) shows, for the square tube, profiles of the pore meniscus along the diagonal for different contact angles. With decreasing contact angle (more hydrophilic), the absolute value of height increases. This can be explained by the fact that, with decreasing contact angle, the adhesive force $\mathbf{F}_a$ between solid and fluid increases resulting in an increase of the height.

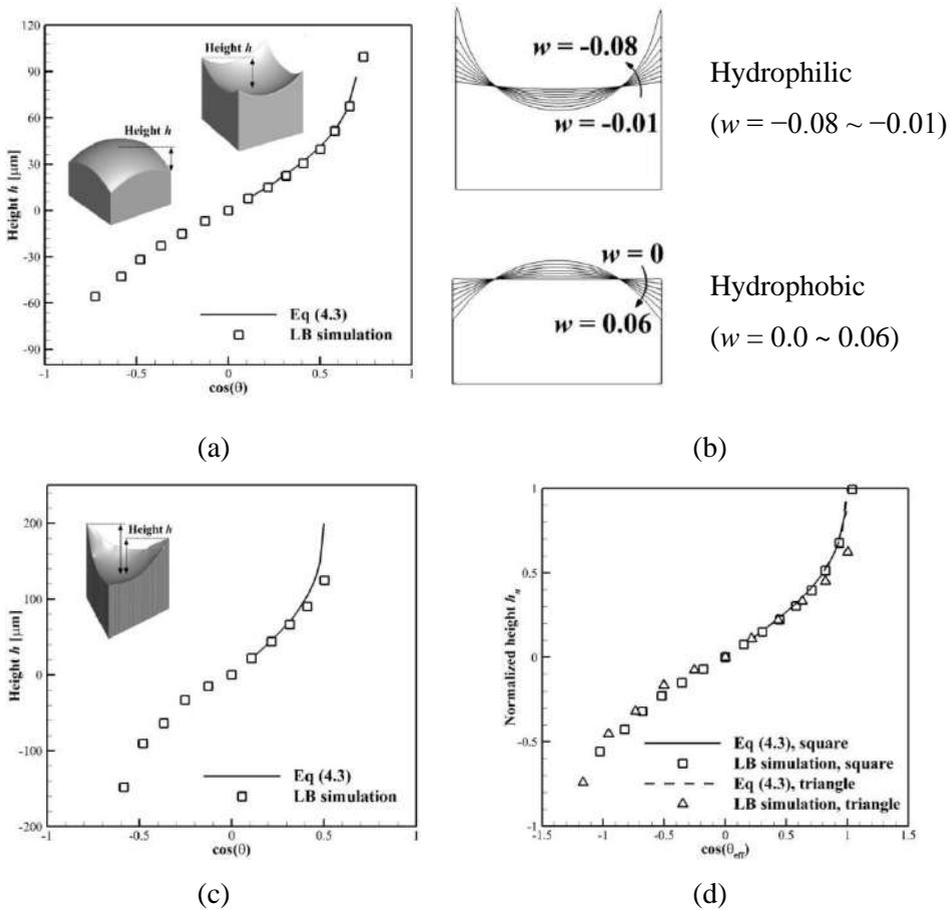

(a)                                            (b)

(c)                                            (d)

Fig.4.12. Height of the pore meniscus $h$ as a function of cosine of contact angle $\theta$. Comparison between simulation results and analytical solution: (a) square tube; (b) diagonal profiles for square tube, for different solid-fluid interaction parameters $w$; (c) triangular tube. (d) Normalized height of the pore meniscus as a function of cosine of the effective contact angle $\theta_{eff}$ for square and triangular tubes and comparison with analytical solutions.

An analytical solution for the height $h$ for a $n$-sided polygonal tube in the hydrophilic case ($\theta < \pi/2$) is given by (Feng and Rothstein 2011):

$$h = r \sin \alpha \left[ 1 - \sqrt{1 - \left( \cos \theta / \sin \alpha \right)^2} \right] \Big/ \cos \theta \,, \tag{4.3}$$

where $\alpha$ is the half of the corner angle:

$$\alpha = \left( n - 2 \right) \pi / \left( 2n \right). \tag{4.4}$$

For the square tube, the LBM heights are in good agreement with the analytical solution. For the triangular tube, the simulated average height is a little lower than the analytical solution for higher values of $\cos \theta$ (more hydrophilic). This difference is explained by the zigzag boundary and the artificially introduced roughness. Since the height is under-predicted in one corner, the average value is also too low. This is in agreement with the observations of Quéré (Bico and Quéré 2002) showing that the hydrophilicity of a hydrophilic surface increases with roughness.

Eq. (4.3) can be rewritten in a normalized form as:

$$\frac{h}{r} = \frac{1}{\cos \theta_{eff}} \left[ 1 - \sqrt{1 - \left( \cos \theta_{eff} \right)^2} \right], \tag{4.5}$$

with $\cos \theta_{eff} = \cos \theta / \sin \alpha$ and $\theta_{eff}$ defined as the equivalent contact angle. Fig.4.12 (d) shows the normalized height $h/r$ versus cosine of the equivalent contact angle. The analytical solutions and LB results for square and triangular tubes collapse onto a single curve. This shows that the LB results for the triangular tube, although suffering from the artificial roughness introduced, agree well over the total hydrophobic and hydrophilic ranges with the results of the square tube, which does not suffer from an artificial roughness.

### 4.4.3.  Co-occurrence of pore and corner arc meniscus

When the contact angle is smaller than the critical contact angle, $\theta < \theta_c$, the liquid invades the corners forming corner arc menisci. We consider two contact angles of 22° and 32° ($w = -0.12$ and $-0.10$), both lower than the critical contact angle for the square and triangular tubes respectively. Fig.4.13 (a) and (b) show snapshots of the pore and corner arc menisci as a function of time (iteration step) for the square and triangular tubes. Fig.4.13 (c) shows diagonal profiles of the menisci and horizontal cross-sections of the meniscus at one corner as a function of time for the square tube. For both tubes, at the early stage, the liquid invades the corners at a small thickness and reaches the top of the tube in a short time. With increasing time, the corner arc menisci thicken while their curvature decreases. At the same time, the pore menisci at the bottom evolve from a more flat shape to a concave shape. For the triangular tube, corner arc menisci develop only at two corners, while one corner does not show the presence of a corner arc meniscus, or it does so only at a late time. As mentioned before, this observation is attributed to the artificial roughness introduced by the zigzag surfaces, which is higher in corner 1 than in corners 2 and 3 (see Fig.4.10 (c)), where the former corner is not invaded by liquid. The profiles in Fig.4.10 (c) show that the thickness of the corner arc menisci is not constant over the height, since at the bottom its thickness is influenced by the pore meniscus and at the top by the edge of the tube. It is also remarked that the thickness of the corner arc meniscus at equilibrium depends on the initial liquid volume present in the tube. In the case of an infinite reservoir, the corner arc menisci of two adjacent corners will join at the end of the thickening process. The cross-sections show that the thickness and curvature for the more hydrophilic surface ($\theta = 22°$) are higher compared to the less hydrophilic case ($\theta = 32°$) at the same time step.

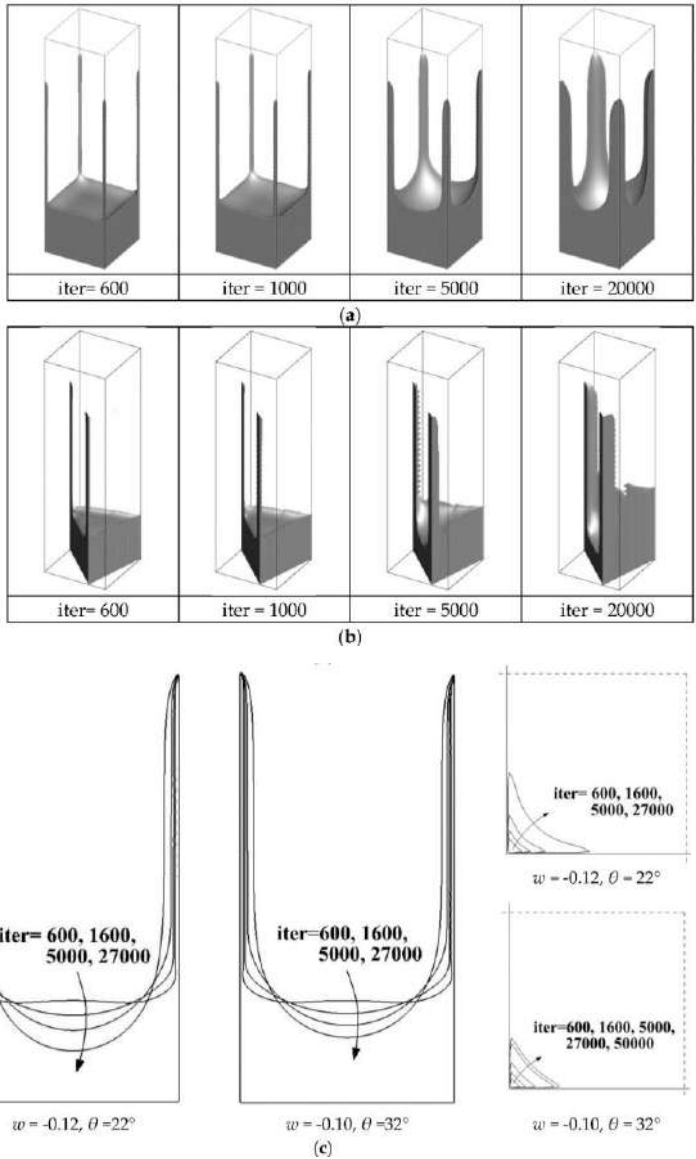

Fig.4.13. Liquid configuration versus time (iteration count) for $\theta = 22°$: (a) square tube; (b) triangular tube; and (c) diagonal profiles and horizontal cross-sections of a corner arc menisci for square tube at different iteration steps for $\theta$ of 22° and 32°.

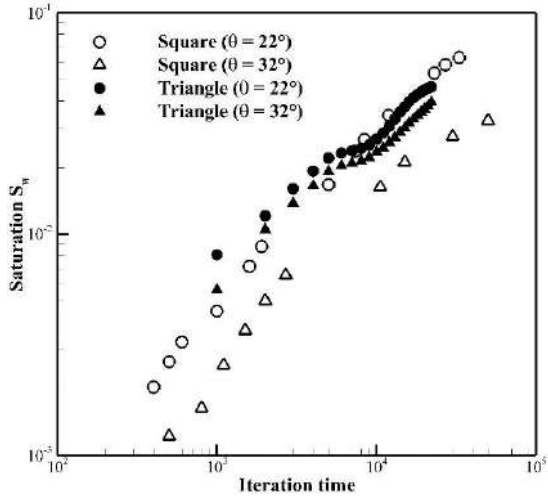

Fig.4.14. Log-log plot of degree of saturation $S_w$ versus time for contact angles $\theta$ of 22° and 32° for square and triangular tubes.

Fig.4.14 shows the time evolution of the degree of saturation for the square and triangular tubes for the contact angles of 22° and 32° in a log-log plot. The degree of saturation is defined as the ratio of the cross-area occupied by liquid at corners to the area of the full cross-section of the tube and is calculated at the mid-height of the corner arc menisci. The curves for the two geometries and two contact angles show similar shapes. The results show that the corner filling process is faster at an early time and then slows down somewhat. As expected, the degree of saturation at a lower contact angle (more hydrophilic) is higher compared to the degree of saturation at a higher contact angle. The influence of contact angle is smaller when the corner angle is smaller, as seen for the triangular tube.

Further, the normalized curvature of the corner arc menisci $C_n$ is given by (Ma, Mason et al. 1996):

$$C_n = \frac{(L/2)\cos(\alpha+\theta)}{L_{contact}\sin\alpha},$$
(4.6)

where $L$ is the side length of the tube, $\alpha$ is the half corner angle dependent on the side parameter $n$, $\theta$ is the contact angle and $L_{contact}$ is the side length of the corner arc meniscus wetting the side of the tube. The contact length $L_{contact}$ is determined from the LBM results at mid-height of the corner arc menisci. The phase interface in LBM is not sharp but gradually decreases from liquid to gas density over three to five lattices. The position of a phase interface is evaluated at the average density between liquid and gas. Therefore, there is an uncertainty on the contact length $L_{contact}$ of around two lattices (Thorne and Michael 2006).

Fig.4.15 (a) shows the normalized curvature versus degree of saturation for the two contact angles 22° and 32° for the square tube. The results for the triangular tube are not represented, since the contact length could not be determined unambiguously due to the artificial roughness problem of the wall, as mentioned above. An analytical solution for the degree of saturation $S_w$ in function of the curvature is given by (Ma, Mason et al. 1996):

$$S_w = \frac{\tan\alpha}{C_n^2}\left[\frac{\cos\theta}{\sin\alpha}\cos(\alpha+\theta) - \frac{\pi}{2}\left(1 - \frac{\alpha+\theta}{90}\right)\right]. \tag{4.7}$$

In Fig.4.15 (a), the LB simulation results are compared with the analytical solution in a log-log plot and an overall good agreement is observed. At a low degree of saturation, the LB results overpredict the curvature slightly, which is attributed to the uncertainty (error) in determining the contact length $L_{contact}$. At a small contact length, an error of two lattices can have a non-negligible effect, as the length $L_{contact}$ appears in Eq. (4.6) in the denominator. Based on these, a slight underestimation of the contact length is expected.

Finally, the mesh sensitivity study as mentioned before is presented. Three meshes were selected for the square tube with a contact angle $\theta = 22°$, i.e. below the critical angle, to study the most critical case of corner arc meniscus formation. Each mesh differs in resolution with a factor of 2: a coarser mesh of $72 \times 72 \times 250$ lattice$^3$ ($\Delta x$

= 2 μm / lattice), a reference mesh of 142 × 142 × 500 lattice³ ($\Delta x$ = 1 μm/lattice) and a finer mesh of 284 × 284 × 1 000 lattice³ ($\Delta x$ = 0.5 μm/lattice). Fig.4.15 (b) gives the curvature versus degree of saturation for the three meshes and compares these LBM results with the analytical solution. An overall good agreement is obtained, which shows that the reference mesh is fine enough to produce mesh-insensitive results for the square case. For the triangular tube, not shown here, the results are more mesh-sensitive since the resolution also determines the artificial roughness introduced. This is the reason why, for the triangular tube, the finest mesh, which is a compromise between calculation time and accuracy is chosen.

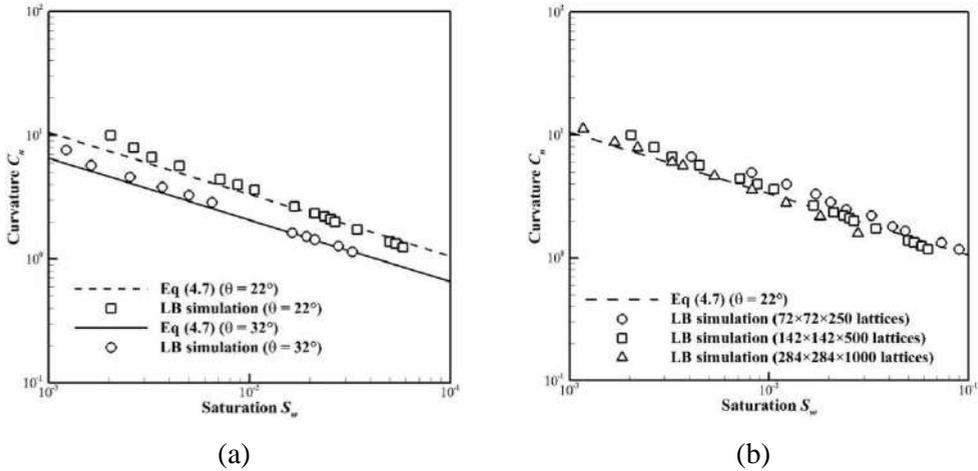

(a)                                    (b)

Fig.4.15. Log-log plot of curvature $C_n$ versus degree of saturation $S_w$ for square tube and comparison with analytical solution: (a) for different contact angles $\theta$ of 22° and 32°; and (b) grid sensitivity analysis for coarse, reference and fine mesh.

## 4.5.    Exploration of corner flow in more complex geometries

In this section, corner flow is further explored for more complex geometries. In the previous section, it was found that for straight tubes liquid first invades the corners of the tube forming corner arc menisci. The corner menisci further thicken with time.

In this section, the liquid configuration of corner arc menisci and their thickening process is studied for more complex geometries.

### 4.5.1. Simulation set-up and boundary conditions

In this study, flow in a single open corner, which is formed at the crossing of two planes, is analyzed. Three configurations are considered: a straight path, a straight path with a small U-bent (referred to as U-bent path) and a staircase path as shown in Fig.4.16. To allow corner flow, the contact angle is chosen to be 22°, thus smaller than critical contact angle $\theta_c = \pi/4 = 45°$, which corresponds to a solid-fluid interaction parameter $w$ of -0.12. The domain size is $210 \times 50 \times 1\,000$ lattice³ for the straight and U-bent paths and $320 \times 50 \times 1\,100$ lattice³ for the staircase path. All lattices have a spatial resolution $\Delta x$ of 1 μm per lattice.

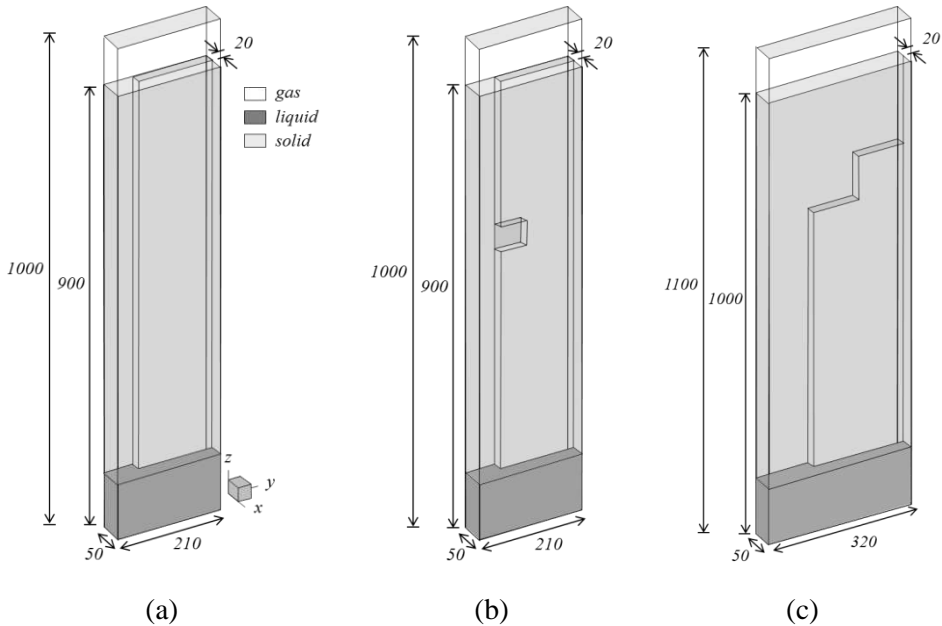

<p style="text-align:center">(a)        (b)        (c)</p>

Fig.4.16. Computational domains for three different geometries: (a) straight; (b) U-bent; and (c) staircase paths.

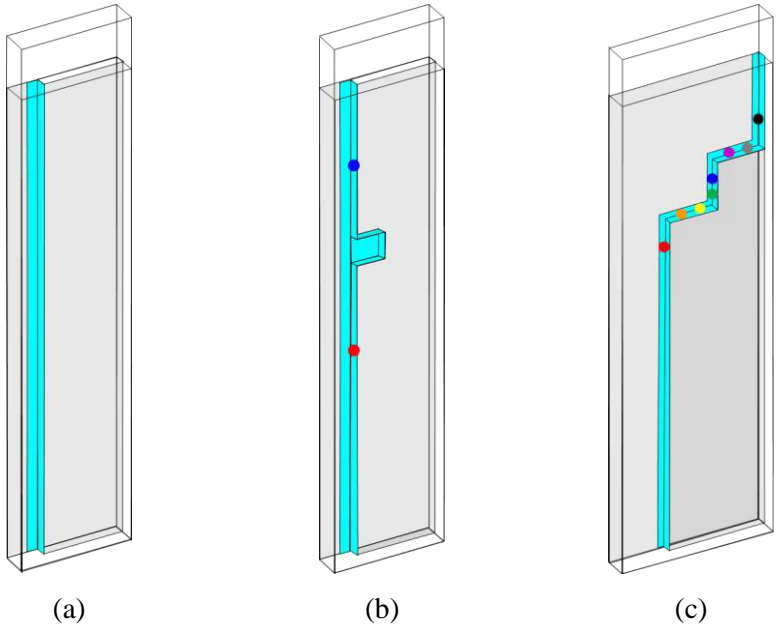

<div align="center">(a)           (b)           (c)</div>

Fig.4.17. Surfaces with hydrophilic contact angle of 22° are indicated in light blue for: (a) straight; (b) U-bent; and (c) staircase path computational domains. Locations along the paths for which results are provided are indicated with color dots.

The corner has a depth of 20 lattices. The surfaces intended for corner flow have a contact angle of 22° and are highlighted with light blue in Fig.4.17. All other surfaces have a contact angle of 90°, or a neutral solid-fluid interaction parameter $w$ of 0, to prevent wetting. As shown in Fig.4.16, bounce-back boundary conditions are imposed on all sides, except for the top 100 lattices on the three or four vertical sides where periodic boundary conditions are imposed to simulate an open capillary tube. The domain is initially filled by liquid to a height of 100 lattices from the bottom. Liquid and gas densities are 0.28 and 0.0299 lattice units, respectively corresponding to a density ratio $\rho/\rho_c = 9.4$ at $T/T_c = 0.85$.

All numerical simulations are run by parallel computing based on Message Passing Interface (MPI) on the high performance computing cluster of Los Alamos National Laboratory (LANL). The cluster aggregate performance is 352 TFlop/s with 102.4 TB of memory for 38 400 cores. Each simulation is run on 120 for straight and U-bent path and 250 processor cores for the staircase path and requires 16 hrs to run 50 000 time steps for all cases.

### 4.5.2. Results

First, the curvature of liquid-gas interface of the invading liquid in the corner is studied at different stages of the wetting process.

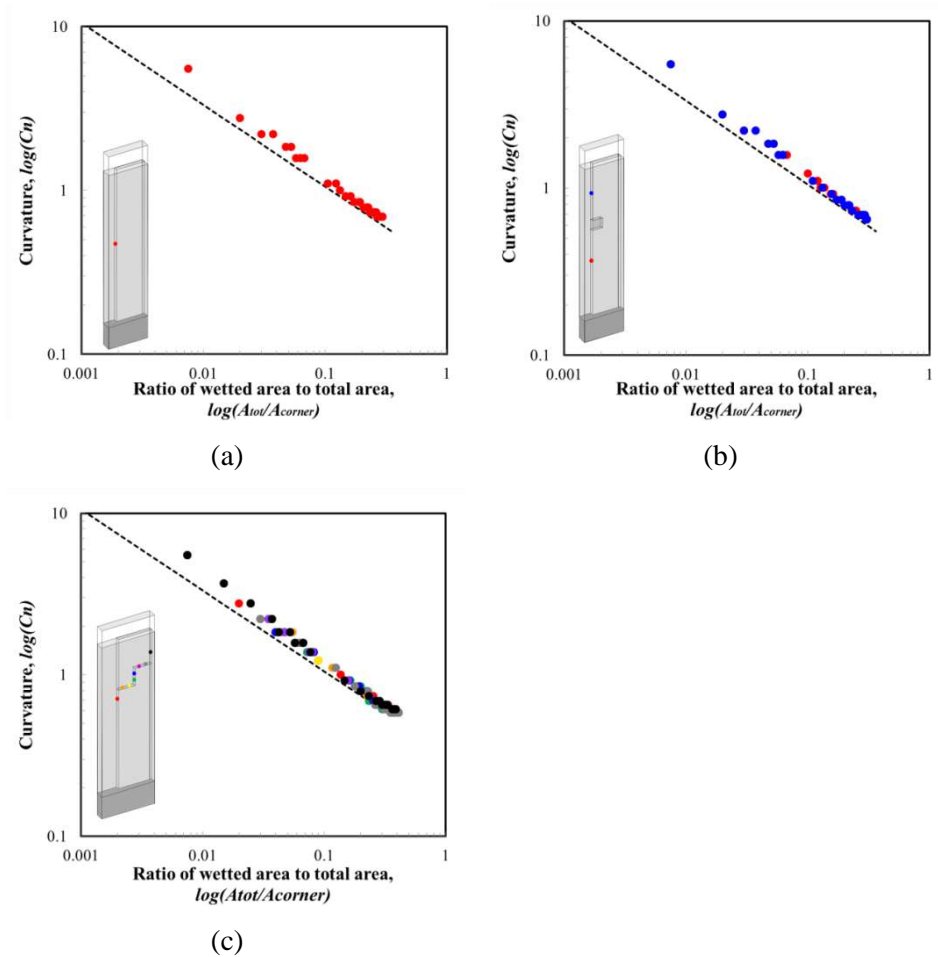

(a)  (b)

(c)

Fig.4.18. Log-log plot of curvature $C_n$ versus ratio of wetted area to total area for (a) straight; (b) U-bent; and (c) staircase paths and comparison with analytical solution.

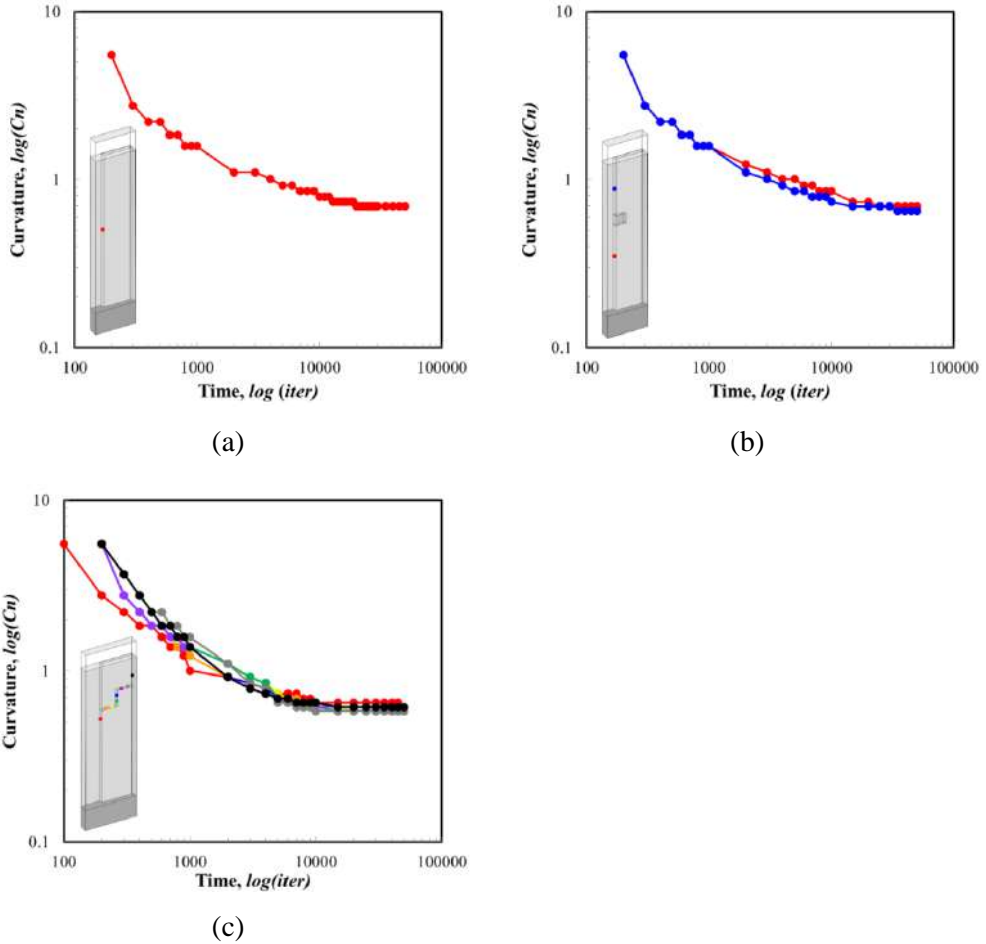

(a)

(b)

(c)

Fig.4.19. Log-log plot of curvature $C_n$ versus time for (a) straight; (b) U-bent; and (c) staircase paths.

For the open corner, saturation is defined as the ratio of the cross section occupied by liquid at the corner to the area of the full cross section of the corner (i.e. 20 x 20 lattice$^2$). The curvature and saturation degree is determined at mid-height of the straight path, below and above the U-bent and at eight points along the staircase path, as shown in the insets of Figs 4.18-20. The LBM results are compared with the analytical solution described by Mayer and Stowe-Princen (MS-P) theory, as per Eq.

(4.7), in a log-log plot and an overall good agreement is observed. Fig.4.18 (b) and (c) show that the results for the U-bent and staircase path are similar to the straight path ones. The LB results slightly over predict the curvature at low ratio of wetted area to total area, as also seen in section 4.4.3, probably due to the uncertainty (error) in determining the contact length $L_{contact}$. As mentioned above, at small contact length, an error of two lattices can have a non-negligible effect, as the length $L_{contact}$ appears in Eq. (4.6) in the denominator. Based on these, the contact length is expected to be slightly underestimated at low ratio of wetted area to total area.

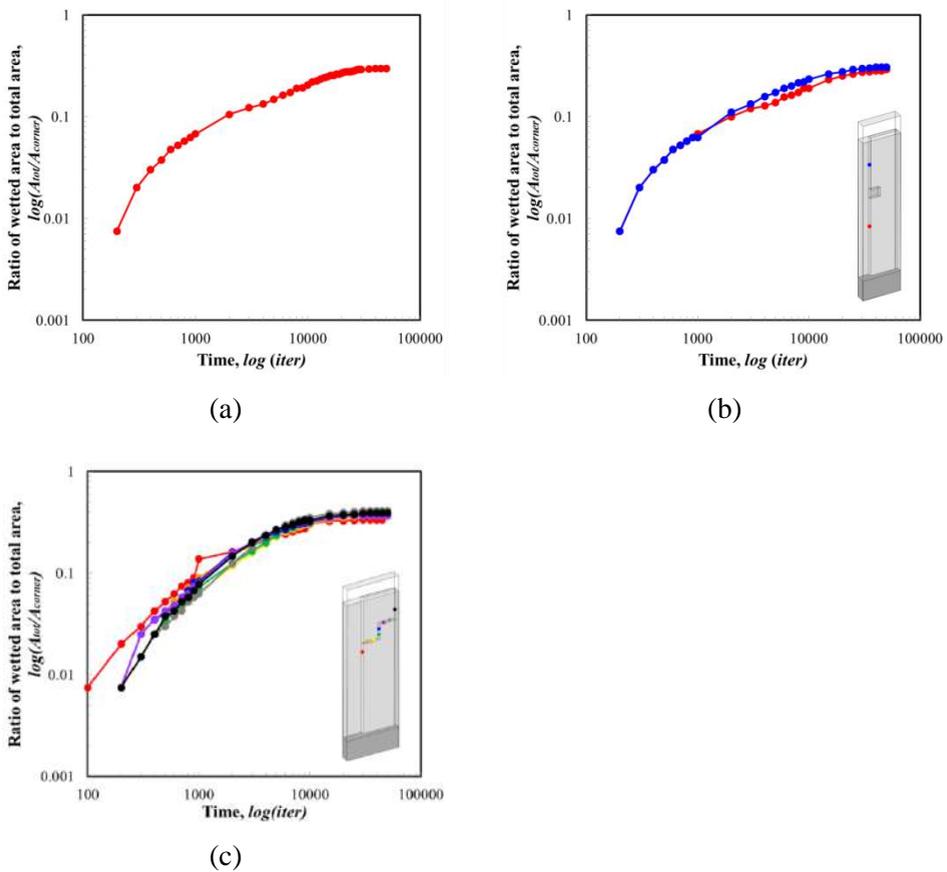

(a)　　　　　　　　　　　　　　(b)

(c)

Fig.4.20. Log-log plot of ratio of wetted area to total area R versus time for (a) straight; (b) U-bent; and (c) staircase paths.

The triangular liquid prisms in the corner thicken with time due to the on-going corner flow. The curvature and the wetted area ratio are plotted versus time in Figs. 4.19 and 4.20, respectively. Fig.4.19 shows the log-log plot of curvature versus time for the three path configurations at the points of observation. In all points, the curvature decreases over time, as liquid is wetting more and more the corners. During the first 1 000 iterations, the curvature is decreasing more rapidly, as an equivalent change in added liquid volume has more impact on corner meniscus curvature when it is added early on in the process. At the end, after iteration time 10 000, the corners in all sections are saturated by liquid and the curvature no longer evolves. For corner flow before and after the U-bent, only a little delay in the curvature decreases above compared to below the U-bent is observed. For the corner flow along the staircase path, only small variations between the different points of observations are observed, especially between the first point of observation and the higher ones. Overall, one may conclude that the thickening process along the path is similar for all three configurations.

Fig.4.20 presents the ratio of wetted area to total area versus time in a log-log plot for the same points of observation along the three path configurations. It is remarked that the change in curvature observed in Fig. 4.19 is inversely related to the change in ratio of wetted area to total area in Fig. 4.20. Therefore, it is logical that the graphs of wetted area versus time show an initial rapid increase, which slows down further in the wetting process. Again all curves show a similar behavior with small differences for the wetting process at different locations.

Finally, the mass flux is determined for the straight path in the middle of the section. The mass flux is determined as the product of liquid density, wetted area at the corner and average fluid velocity. The mass flux is initially very small ranging between 0.001 and 0.01 lattice units. Some scatter is observed in the results at these very low flux values, which is due to some uncertainties in the determination the flux close to the wall, where the density and the velocity drop. After iteration time 10 000, a steep increase in mass flux is observed. The same curves were obtained for all other complex paths and points of observations, but are not reported here for brevity. The steep increase in mass flux can be explained by the fact that, when the wetting films in the corners have sufficiently thickened, the frictional losses for fluid flow will decrease and, once the frictional forces are overruled by the capillary forces, a steep increase in mass flux will result.

To understand the effect of complicated geometry on corner flow, the staircase path is considered to analyze liquid configuration and curvature, ratio of wetted area to total area and mass flux for different iteration times in Fig. 4.21. The different locations are counted from 1 to 8 as shown in Fig. 4.21 (a). In Figs.4.21 (b) and (c), the larger curvature and lower ratio of wetted area to total area can be observed at points of 3, 4, 7 and 8 until iteration 1 000. This can be explained by the junction points between horizontal and riser sides which make it difficult for the liquid to evolve due to the smaller area of corner. However, with increasing iteration times, the curvature and ratio show no significant difference for different locations since the corners are fully saturated with liquid. In terms of mass flux, as shown in Fig.4.21 (c), the larger curvature and lower ratio of wetted area to total area at points of 3, 4, 7 and 8 result in smaller mass flux at iterations 200, 500 and 1 000. Thereafter, the mass flux is quite constant for all points in the staircase paths. Thus, it can be concluded that due to the geometry of corners, especially at the junction point between the horizontal and riser sides, the evolution of liquid is difficult at initial stage. However, with increasing iteration time and saturation of liquid, the effect of geometry on the liquid configuration is not significant in the LB results.

Overall, one may conclude that the corner flow process versus time is quite similar for all three configurations, showing that the corner flow process is quite independent of the configuration studied. This means that the complexity of the path the corner flow follows has only a negligible effect on how the liquid configuration evolves versus time. This finding could be of significant impact in understanding unsaturated flow in complex porous media like porous asphalt, if hydrophilic conditions emerge.

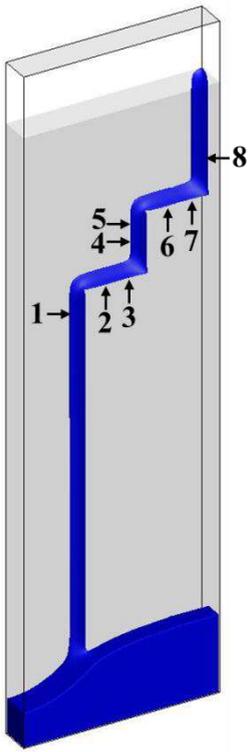

(a)

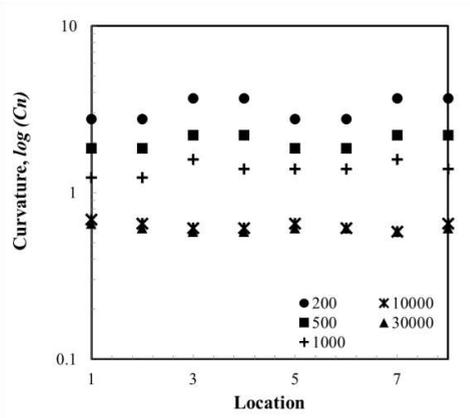

(b)

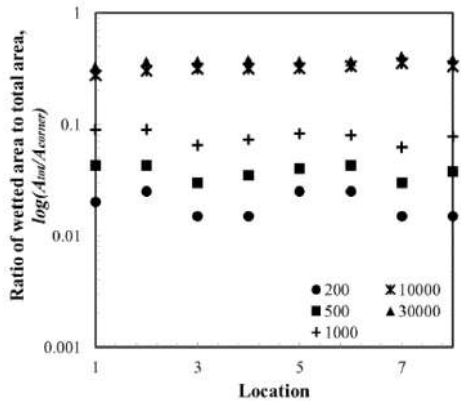

(c)

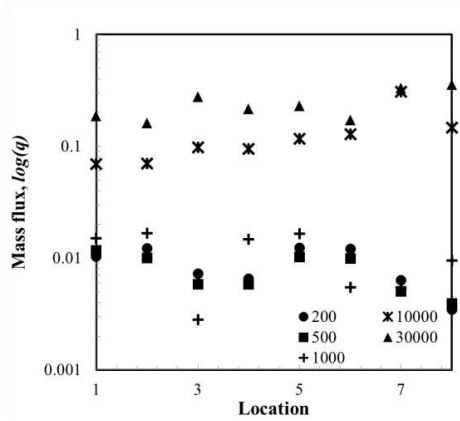

(d)

Fig.4.21. (a) Schematic of liquid distribution at equilibrium state and log-log plot of temporal evolution of (b) curvature; (c) ratio of wetted area to total area; and (d) mass flux at different points in the staircase path.

## 4.6. Capillary rise in a cylindrical tube

In this section, the capillary uptake in a cylindrical tube is simulated. The 3D LB results for different tube radii and contact angles are compared with analytical solutions mentioned in section 2.2.5.

### 4.6.1. Simulation set-up and boundary conditions

For the circular tube, the 3D domain size is $300 \times 300 \times 300$ lattice$^3$ with a spatial resolution $\Delta x$ of 1 μm per lattice. This discretization is chosen based on a grid sensitivity analysis comparing the normalized capillary rise height in tube as a function of time for different discretization (Fig.4.22). It is observed that no significant difference exists between the three domain sizes of $100 \times 100 \times 100$, $200 \times 200 \times 200$ and $300 \times 300 \times 300$ lattice$^3$. The domain size of $300 \times 300 \times 300$ lattice$^3$ is chosen further in this study. The circular tube has a height of 235 lattices and is located in the middle of the domain, 38 lattices above the bottom surface. The reference radius of tube is 15 lattices. Radii of 5, 10 and 20 lattices are considered to study the effect of tube size on capillary rise. The bottom quarter of the domain is filled with liquid, thus forming a limited reservoir of 75 lattices high. The tube is initially filled with liquid to a height of 37 lattices as shown in Fig.4.23. Liquid and gas densities are 0.359 and $6.07 \times 10^{-3}$ lattice units respectively, corresponding to the density ratio $\rho/\rho_c = 59.1$ at $T/T_c = 0.7$. The contact angle is 75°, corresponding to a solid-fluid interaction parameter $w$ of -0.05, which is applied to the entire solid surface and also inside and outside tube. Boundary conditions are as follow: bounce back boundary conditions on top and bottom sides and periodic boundary conditions on the other sides.

The 3D simulations are run by parallel computing based on Message Passing Interface (MPI) on the high performance computing cluster of Los Alamos National Laboratory (LANL). The cluster aggregate performance is 352 TFlop/s with 102.4 TB of memory for 38 400 cores. Each simulation is run on 216 processor cores (6 × 6 × 6) and requires 16 hrs to run 50 000 time steps.

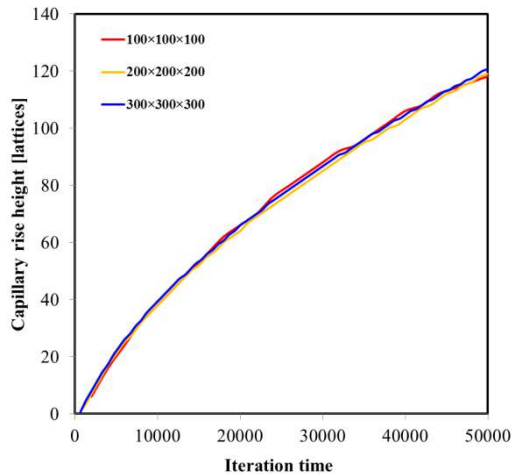

Fig.4.22. Grid sensitivity test of capillary rise in a circular tube with different grid resolutions.

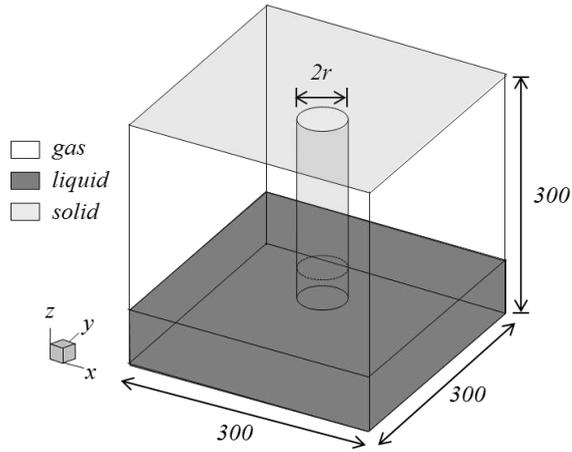

(a)

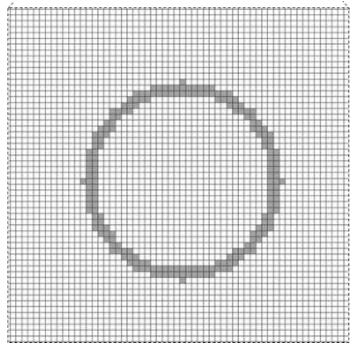

(b)

Fig.4.23. (a) Schematic representation of the computational domain of capillary rise in a 3D cylindrical tube. (b) Computational mesh detail for a tube cross-section in x-y direction.

### 4.6.2. Results and comparison with analytical solutions

In this section, the capillary rise in a cylindrical tube is modelled in 3D and the LB results are compared to analytical solutions for different contact angles and tube sizes.

Fig.4.24 shows the capillary rise height versus time for three different tube radii: 10, 15 and 20 lattices. As expected, the rate of capillary uptake increases with the diameter of tube as shown in Fig.4.24. In Fig.4.25, the effect of contact angle on capillary uptake height is plotted for two different contact angles: 62° and 75° for a tube radius of 15 lattices. As expected, the LB results show a faster capillary rise for smaller contact angle $\theta$ or a wettability representing a more hydrophilic behavior. The higher capillary rise can be observed in both LB results and analytical solution for lower contact angle.

The LB results are compared with the analytical solution of Eq. (2.23). The LB results show a higher uptake rate in comparison with the analytical solution. The difference between LB results and the analytical solution is attributed to the artificial wall roughness introduced by zigzag surface as discussed in section 4.5. The detailed zigzag discretization of cylindrical tube is plotted in Fig. 4.23 (b). The cross section of cylindrical tube shows that the mesh is aligned to the regular lattice grid along the circular tube shape. The application of full bounce-back boundary conditions on the zigzag surface results in a more hydrophilic behavior, which is in agreement with the observations of Quéré (Bico and Quéré 2002) showing that roughness increases the hydrophilicity.

In conclusion, the simulation of the cylindrical tube shows that the version of LBM used in this thesis suffers from the misalignment of the discretization to the real boundary, referred to as the zig-zag problem and introducing artificial roughness speeding up the uptake process. As mentioned above, this problem can be solved using different implementations of the boundary condition, such as the curved, the half bounce-back or the moving boundary condition (Mei, Luo et al. 1999, Mei, Shyy et al. 2000). However, the implementation of these boundary conditions and their analysis was considered out of the scope of this study.

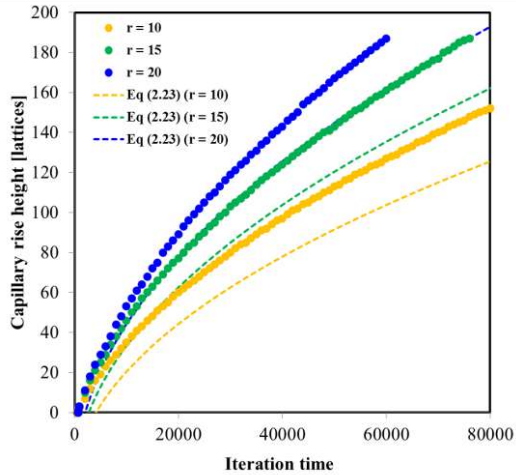

Fig.4.24. Capillary rise height versus iteration time for different radii of cylindrical tube: 10, 15 and 20 lattices. The contact angle is 75° or a solid-fluid interaction parameter of -0.05.

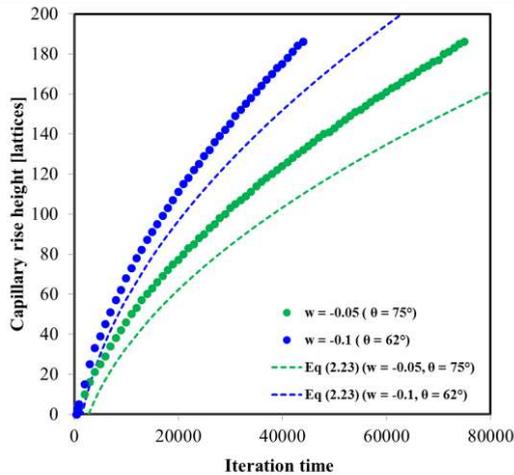

Fig.4.25. Capillary rise height versus iteration time for different contact angles of 62° and 75° or solid-fluid interaction parameter *w* of -0.1 and -0.05 in the cylindrical tube with radius of 15 lattices.

# 4.7. Capillary uptake in porous asphalt-inspired geometry

In this section, LBM is used to explore more complex capillary uptake processes in porous media. The 2D and 3D LBM studies are inspired by flow in PA and the LB results are illustrated with snapshots of liquid configurations overtime and discussed only qualitatively.

## 4.7.1. Simulation set-up and boundary conditions of capillary uptake in porous media

In the previous sections, capillary rise in singular capillary systems was investigated for the following systems: parallel plates, square and triangular tubes, cylindrical pores and corners. However, the focus of this project is on multiphase flow in PA, i.e. a macroporous material with complex pore structure. As an exploration step, in this section, capillary uptake in a 2D academic porous medium is simulated.

The porous domain is a 2D representation of porous domain inspired by the pore structure of PA. The 2D computational domain spans $701 \times 474$ lattice$^2$ with a spatial resolution $\Delta x$ of 12.24 µm per lattice. The size of porous medium is $501 \times 274$ lattice$^2$ or $6\ 120 \times 3\ 340$ µm$^2$ and is located 105 lattices from the bottom as shown in Fig.4.26. The bottom part of the domain acts as a liquid reservoir. The porous media is partially immersed in a reservoir with a height of 131 lattices. Note that the reservoir is thus limited. By applying a density ratio $\rho/\rho_c = 59.1$ at $T/T_c = 0.7$, liquid and gas densities are 0.359 and $6.07 \times 10^{-3}$ lattice units respectively. The contact angle equals 63° corresponding to a solid-fluid interaction parameter $w = -0.1$ as determined in the contact angle test in chapter 3. Periodic boundary conditions are imposed on the left and right sides, and the top and bottom sides are treated as bounce back boundaries.

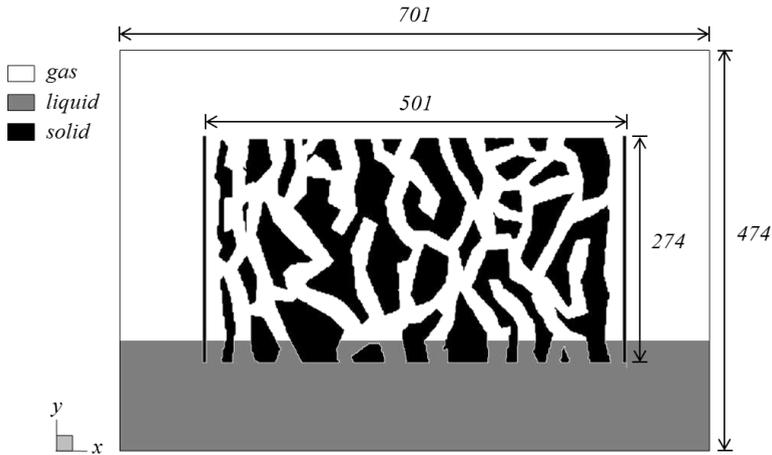

Fig.4.26. Schematic representation of 2D computational domain for capillary uptake in porous media.

### 4.7.2. Results of 2D capillary uptake in porous media

In Fig.4.27, snapshots of the capillary uptake process are shown for different iteration times. As capillary uptake takes place, capillary menisci are formed in the pores. However, it is observed that the uptake process comes to a halt in different pores. For instance, the capillary uptake stops in the two middle pores at iteration time 6 000, when a meniscus with larger curvature is formed. This larger curvature results in a lower capillary suction, which seems to be insufficient to prolong the uptake process in the larger pore above. The stopping of uptake process is also observed in other pores as time passes. This halting of the uptake process is discussed in more detail below.

It is noted that the level of the finite reservoir lowers with time as more and more liquid is taken up by the pores. Moreover, a meniscus at the outside of the porous medium is formed, since the outer side of the porous medium has also a contact angle of 63° and will be wetted. Due to the finite size of the reservoir and the periodic boundary conditions, the reservoir acts as a large pore with a contact angle of 63°at the sides. A Laplace pressure over the meniscus will arise and the water surface will not act anymore as a free water surface with zero Laplace pressure. As a consequence, at a certain time, the Laplace pressure over the reservoir surface equals the capillary pressure in the pores and the capillary uptake process in the pores comes to a halt.

<table>
<tr><td>(a) iter = 1 000</td><td>(b) 4 000</td></tr>
<tr><td>(c) 6 000</td><td>(d) 9 000</td></tr>
<tr><td>(e) 16 000</td><td>(f) 30 000</td></tr>
</table>

Fig.4.27. Dynamic behaviors of capillary uptake in porous domain at different iteration times.

### 4.7.3. Simulation set-up and boundary conditions of 3D capillary uptake in PA

In this section, capillary uptake in a 3D porous domain as obtained from 3D X-ray tomography of a PA sample is simulated. The simulation is based on experimental work of Lal et al. (2014) (Lal, Poulikakos et al. 2014). Water uptake in PA was tracked over time using NEUtron Transmission RAdiography (NEUTRA). Their study showed that naturally aged PA specimen becomes hydrophilic and takes up water by capillary action. This 3D exploratory LBM study of capillary uptake in PA is based on their work.

The PA sample studied in LBM has the same geometry as a part of the specimen of Lal et al. (2014) but measures $200 \times 109 \times 40$ lattice$^3$ ( $4\ 000 \times 2\ 180 \times 800\ \mu m^3$). So the sample simulated is actually scaled down from the real geometry ($45\ 000 \times 25\ 000 \times 10\ 000\ \mu m^3$) by a factor 12 approximately. The spatial resolution $\Delta x$ is 20 $\mu m$ per lattice. The PA sample is located 53 lattices above the bottom as shown in Fig.4.28. The total domain size measures $271 \times 150 \times 61$ lattice$^3$.

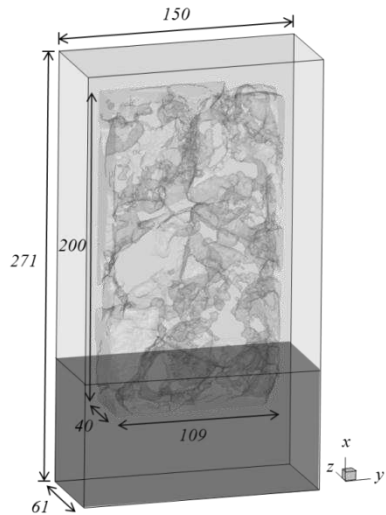

Fig.4.28. Schematic representation of 3D computational domain for 3D capillary uptake in PA.

The bottom part of the domain is filled with water to act as a limited liquid reservoir. The PA domain is partially immersed into the reservoir with a height of 77 lattices. By applying a density ratio $\rho/\rho_c = 59.1$ at $T/T_c = 0.7$, liquid and gas densities are 0.359 and $6.07 \times 10^{-3}$ lattice units respectively. The contact angle is 63° corresponding to a solid-fluid interaction parameter $w$ of -0.1 as determined in the contact angle test in chapter 3. Bounce back boundary conditions are imposed on all sides.

### 4.7.4. Results of capillary uptake in PA

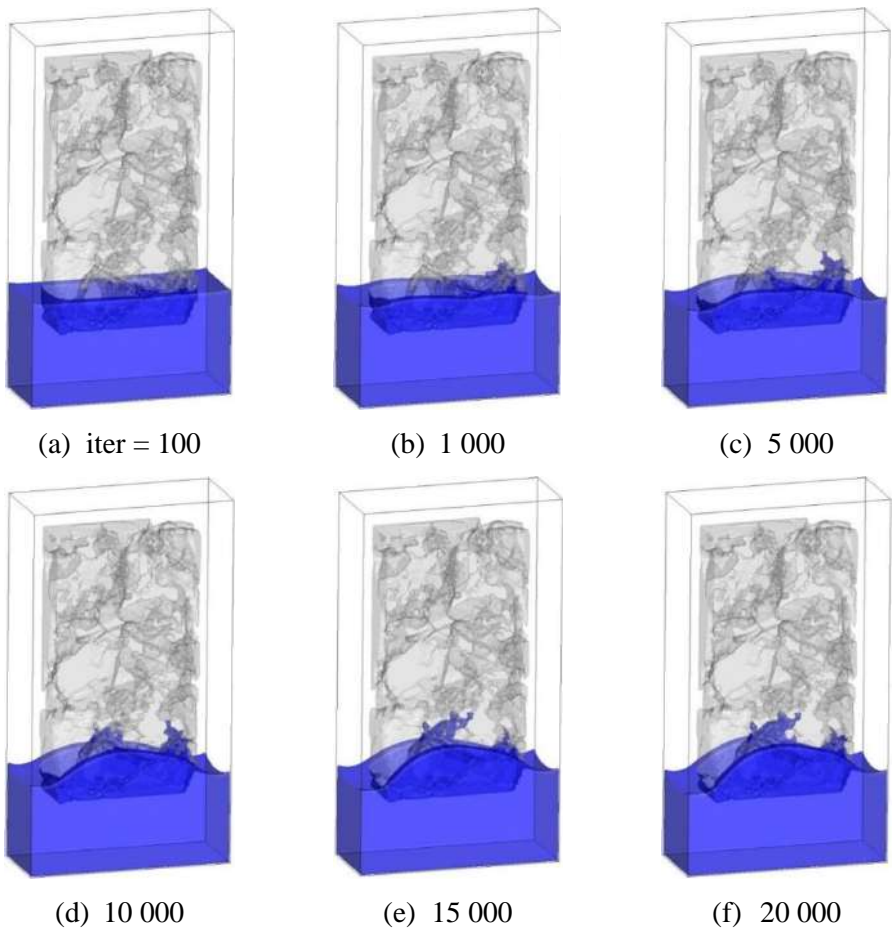

(a) iter = 100          (b) 1 000          (c) 5 000

(d) 10 000          (e) 15 000          (f) 20 000

Fig.4.29. Dynamic behaviors of capillary uptake in 3D PA at different iteration times.

Fig.4.29 shows snapshots of the capillary uptake process in 3D at different iteration times until an equilibrium liquid configuration is reached. Detailed liquid configurations are documented in Fig.4.30 with snapshots of isosurfaces (middle) and cross sections in x-y direction (right) at different iteration times.

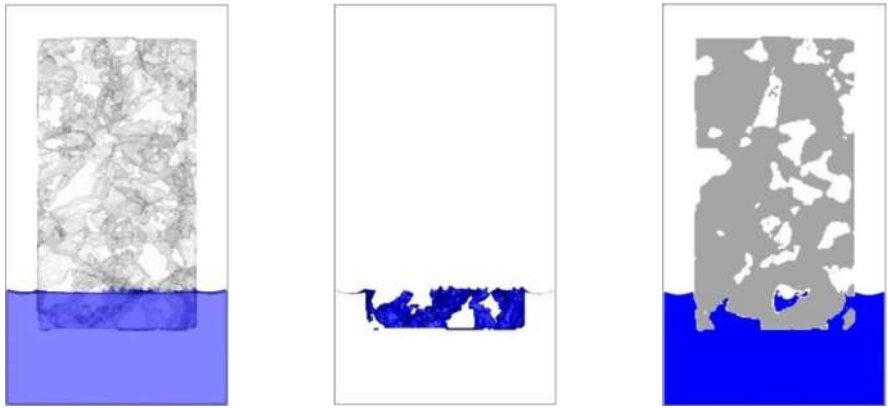

(a)  iter = 100

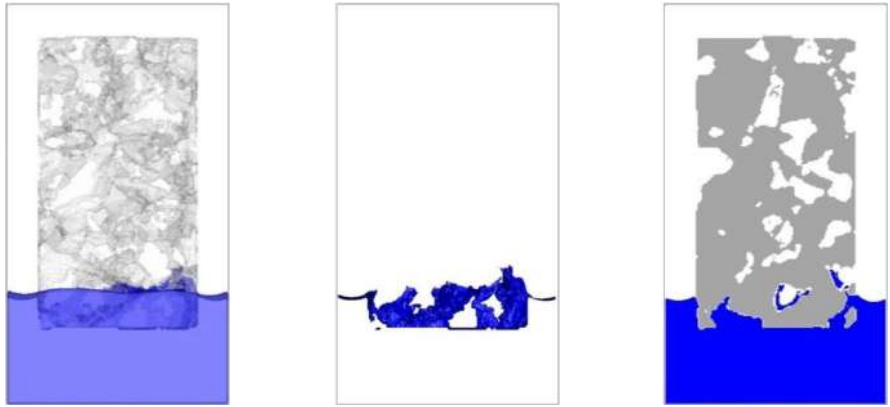

(b)  1 000

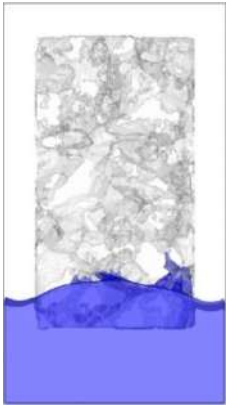 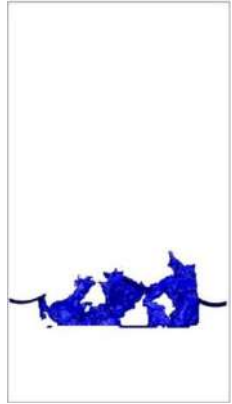 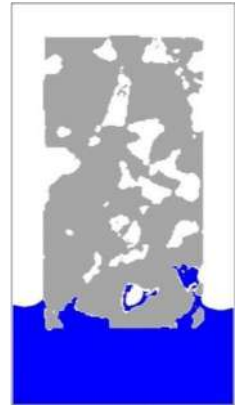

(c)  5 000

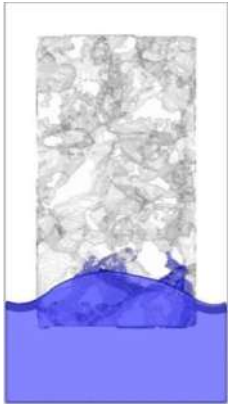 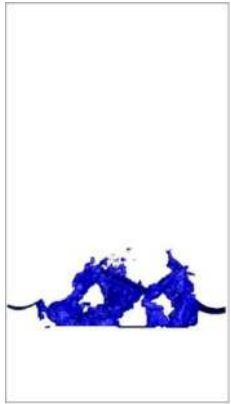 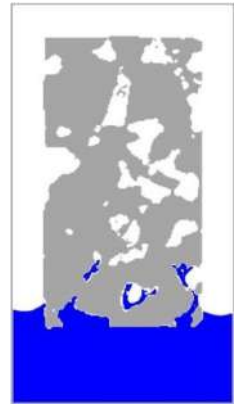

(d)  10 000

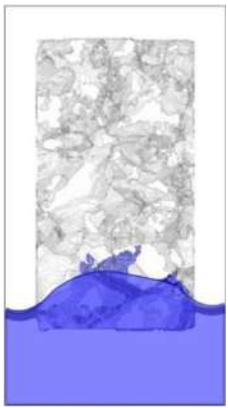 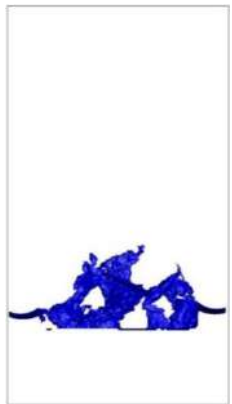 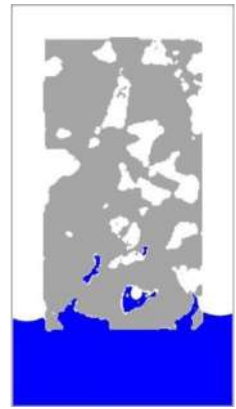

(e)  15 000

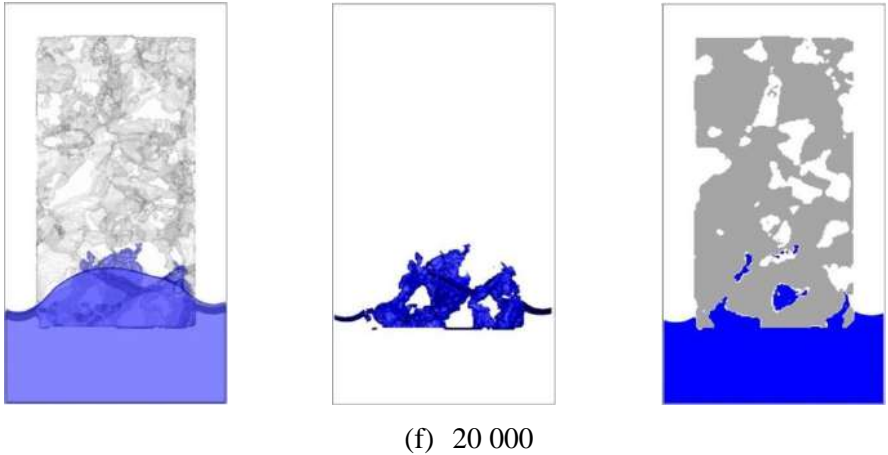

(f)  20 000

Fig.4.30. Liquid configurations of capillary uptake in x-y direction: (left) 3D view from front, (middle) isosurfaces of liquid phase viewed from front, and (right) cross section in the middle plane at different iteration times.

At iteration time 100, the bottom part of the specimen is immerged in the liquid reservoir and no capillary uptake has taken place yet. By iteration time 1 000, liquid uptake starts on the right side of PA. It is observed that the walls surrounding the specimen become also wetted showing a contact angle of 63°. With increasing iteration time, the middle and right sides of PA start to fill up, while the level of the finite liquid reservoir further decreases. Again it is observed that the capillary uptake process in the pores comes to a halt, when the Laplace pressure over the reservoir surface will equal the capillary pressure in the pores.

In conclusion, the two last simulations show that the capillary uptake process is highly influenced by the limited volume of liquid available in the reservoir and by the Laplace pressure over the meniscus appearing in the reservoir. These limitations can be removed by implementing two measures. First, the contact angle of the outside surface of the porous medium can be set to neutral wetting conditions (contact angle of 90°) to prevent a meniscus to be built up in the reservoir. This mitigation measure was already applied in the simulation of the capillary suction in between two parallel plates (see 4.2.2). Another mitigation measure is to supply liquid from the bottom side in order to keep the liquid level constant. In the frame of this thesis, only the former mitigation measure was implemented.

## 4.8.    Conclusion

Capillary rise and corner flow processes at pore scale have been studied with 2D and 3D LBM and extended to further exploration work by considering uptake in porous media inspired by PA. The LB results are validated with analytical solutions in terms of capillary rise height versus time, saturation and curvature at corners and mass flux. In the polygonal tube, two liquid configurations can occur depending on the critical contact angle: 1) the formation of pore meniscus at contact angle larger than critical contact angle and 2) the occurrence of a pore meniscus and corner arc menisci in the corners at contact angle smaller than critical contact angle. Also in this case, LBM results were found to agree well with analytical predictions. For capillary rise between parallel plates (2D) and in square tubes (3D), an overall good agreement with analytical solutions is observed. However, triangular and circular pores show some disagreement, which is attributed to the misalignment of the discretization of the lattice to the boundary, introducing artificial roughness, leading to a speed up of the uptake process. Further, corner flow in three different corner configurations was explored: straight path, straight path with U-bent and staircase path. The corner flow process in these different configurations is found to be in all cases similar showing the negligible effect of specific corner path on how the liquid configuration evolves versus time, as long the corner size and contact angle remain equal.

Finally, this section ends up with the exploration of capillary uptake in more complex porous media inspired by PA. The importance of properly modelling the boundary conditions is mentioned. Overall, these studies highlight the potential of simulating capillary uptake in PA by LBM.

In the next chapters, LBM will be used to study in detail diverse droplet phenomena (chapter 5), as well as drainage of liquid from PA (chapter 6), and other multiphase phenomena related to PA (chapter 7).

# 5. DROPLETS
## 5.1. Introduction

In this chapter, droplet-related phenomena occurring in PA are studied using 2D and 3D LBM. Wetting of PA is caused mainly by rain impingement. Since PA is initially hydrophobic, droplets can sit on the porous surface and start to evaporate from the surface. When the bitumen covering the aggregates peel off due to aging, PA may become partly hydrophilic, showing a pore surface with different wettability ranging from hydrophobic to hydrophilic. In this case, droplets may run down by gravity or be taken up by capillarity, wetting the inside of PA.

In this study, the phenomena are however simplified and the study is an academic abstraction of reality. The following phenomena are studied. (1) The droplet displacement into PA is mimicked by simulating the run-off of droplets on a surface with grooves, which can be thought of as an abstraction of dead-end pores. The influence of groove size, wettability of the surface and groove, and tilt angle of the surface is investigated. (2) When a droplet sits on the surface of PA, the droplet may be taken up by capillary suction when PA is hydrophilic. When the surface is hydrophobic, the droplet may remain pinned on the PA surface, eventually showing stick-slip behavior during evaporation. These phenomena are studied by depositing a droplet on the set of capillary pillars. The evaporation of the droplet is studied using 2D LBM. (3) When PA ages, the PA pore surface may change from total hydrophobic to partially hydrophilic and hydrophilic. This phenomenon is mimicked by analyzing the behavior of a droplet deposited on a surface with heterogeneous wettability, showing a regular checkboard pattern of hydrophilic and hydrophobic patches using 3D LBM. The effect of patch size and droplet radius on droplet shape and local contact angle is studied.

First, in section 5.2, the gravity-driven run-off of a droplet on a vertical surface with or without groove is studied for different contact angles, Bond number and tilting angle. In section 5.3, the evaporation of a droplet deposited on a 2D set of micro pillars is investigated in terms of evolution of contact radius, contact angle and excess free energy for different pillar and pitch widths. In section 5.4, the spreading of a droplet deposited on a checkerboard heterogeneous surface with regular hydrophilic and hydrophobic patches is studied. The LB results are compared with analytical solutions when possible to validate and verify LBM.

## 5.2. Run-off on vertical surface

*This section is based on the journal paper: Son, S., L, Chen., D, Derome. and J, Carmeliet. (2015). Numerical study of gravity-driven droplet displacement on a surface using the pseudopotential multiphase lattice Boltzmann model with high density ratio. <u>Computers & Fluids;</u> 117: 42-53.*

Droplet and film movement on surfaces has been widely investigated using LBM. Kang et al. (2002) investigated a 2D droplet flowing down a channel with different Bond numbers, droplet size and viscosity ratio. Mazloomi and Moosavi (2013) simulated the run-off of a gravity-driven liquid film over a vertical surface displaying U- and V- shaped grooves or mounds, defining the critical width for successful coating or covering the surface with fluid. The dynamic behavior of droplets was investigated by Azwadi and Witrib (2012) for different contact angles, Bond numbers and tilting of the surface. Recently, Li et al. (2014) studied the deformation and breakup of a droplet in a channel with a solid obstacle, considering different obstacle shapes, wettability, viscous ratio and Bond number. In this study, the gravity-driven droplet displacement on surface with and without grooves is investigated. The effect of Bond numbers, groove geometry, contact and tilt angles of the surface on dynamic droplet behavior is studied.

### 5.2.1. Simulation set-up and boundary conditions

In this section, droplet displacement due to gravity on a vertical solid surface with and without groove is studied. Particularly the study focuses on the influence of surface conditions such as contact angle and geometry, height and width of the groove and tilt angle of the surface. Contact angle hysteresis is not accounted, although it could be accounted using a dynamic contact angle function of capillary number and velocity (Raiskinmäki, Shakib-Manesh et al. 2002). The dimensionless Bond number $Bo$ is defined as the ratio between the gravitational force and surface tension $\gamma$:

$$Bo = \frac{\rho_l g l^2}{\gamma}, \tag{5.1}$$

where $\rho_l$ is the liquid density, $l$ is the characteristic length and $g$ is the gravity acceleration. In this study, the Bond number $Bo$ equals 3.78 and 7.55, corresponding respectively to a gravitational acceleration of $1 \times 10^{-4}$ or $2 \times 10^{-4}$ lattice units.

### 5.2.2. Run-off on a flat vertical surface

First, the dynamic behavior of a droplet moving down on a vertical flat surface is investigated with 2D LBM. The contact angles considered are 74.3º, 88.7º and 101.6º in accordance with a solid-liquid interaction parameter $w$ of respectively -0.05, 0.0 and 0.035, as determined from the contact angle test in chapter 3. The domain size is $900 \times 10\ 800\ \mu m^2$, or $101 \times 1\ 201\ lattice^2$, with a spatial resolution $\Delta x$ of 9 µm per lattice. The droplet radius is equal to 360 µm or 40 lattices. Fig.5.1 shows the position of the droplet versus time for different contact angles or solid-liquid interaction parameters at two gravitational accelerations (a) $1 \times 10^{-4}$ and (b) $2 \times 10^{-4}$ lattice units. The initial shape of the droplet is different corresponding to its contact angle. As time proceeds, the droplet deforms due to gravity, viscous and adhesion forces. The shape of droplet changes from a semicircle to a half-tear drop. For the cases with different wettability but same gravitational force, the droplet shows a similar shape until the 2 000th iteration step. After that, the droplet becomes more and more elongated. Fig.5.2 shows the droplet length as a function of time for the two gravity values. The droplet length is defined as the length between the current front position and the initial position of the droplet. The droplet length for different contact angles shows similar values until the 3 000th iteration step. In this initial period, the motion of the droplet is mainly dominated by gravity. The viscous force is small as the droplet velocity is still small. The adhesion force is also weak as the contact line between the droplet and the solid surface is still short. As the droplet becomes longer and moves down faster, viscous and adhesion forces become more dominant. The droplet length increases as the contact angle increases. This is caused by the lower adhesion force for higher contact angles. As shown in Fig.5.2, after about 5 000th iteration steps, the relation of droplet length versus time is approximately linear, indicating a constant velocity of the droplet. At higher contact angles, the velocity increases due to the smaller adhesion force between liquid and solid phases. From Fig.5.2, it can also be seen that higher gravity values lead to a higher velocity of the droplet.

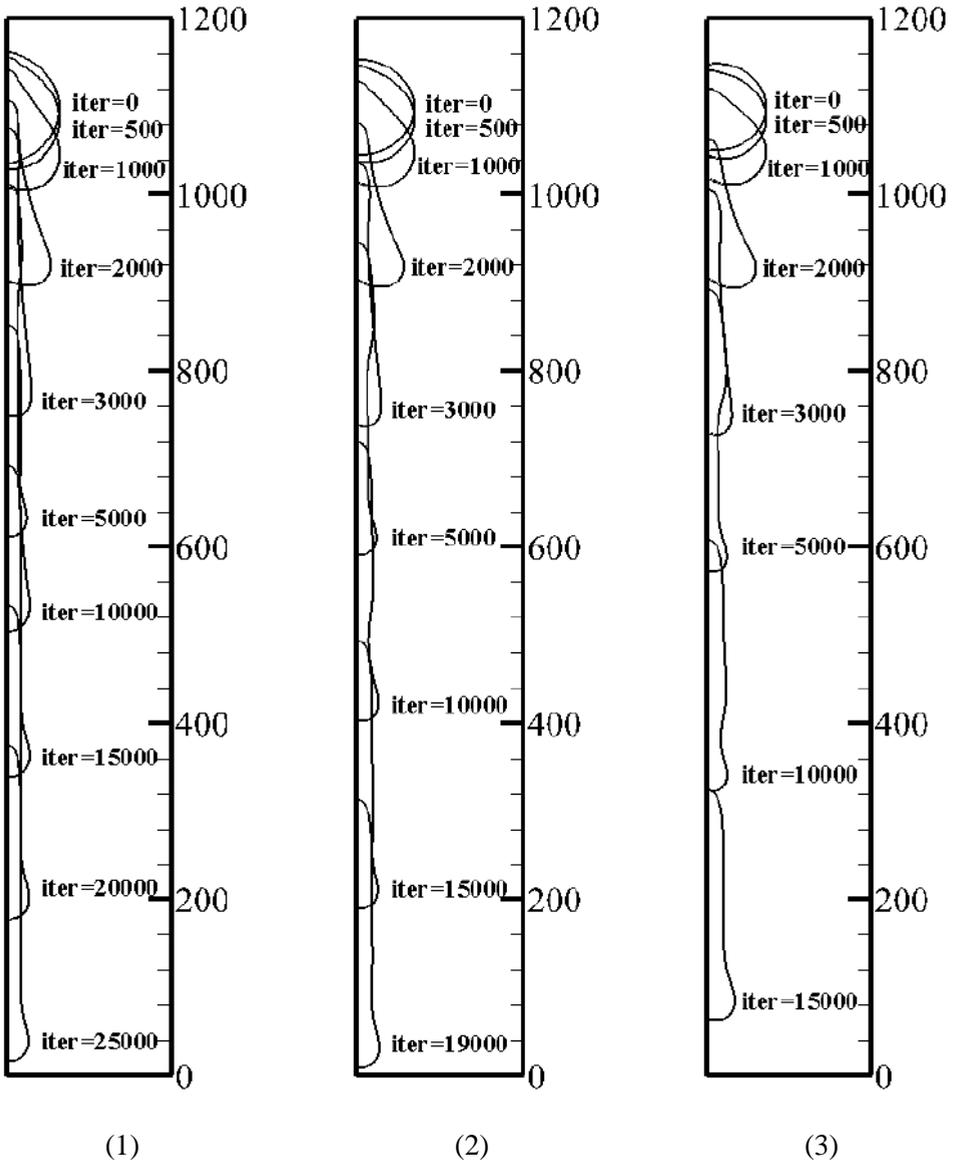

(1)        (2)        (3)

(a)  gravitational acceleration $g = 1 \times 10^{-4}$ lattice units.

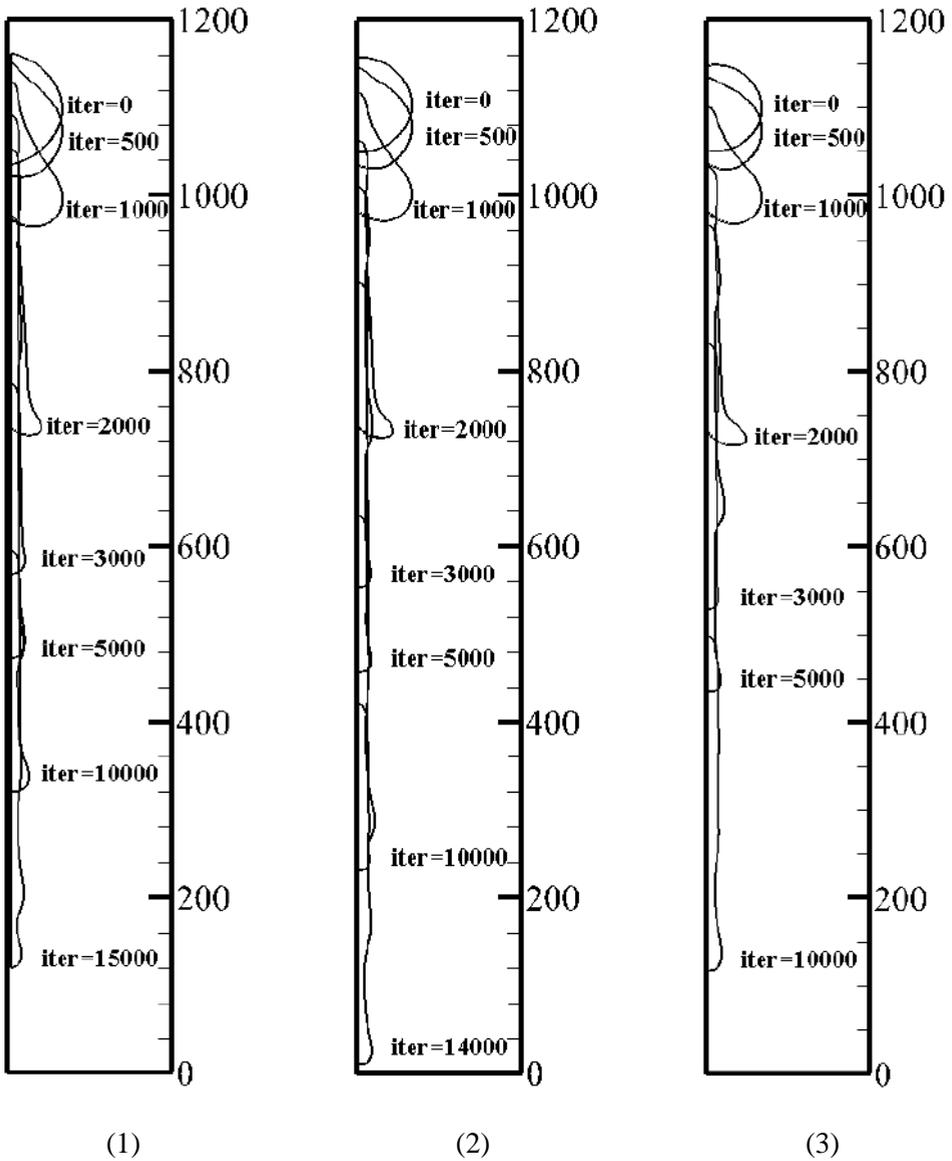

(1)        (2)        (3)

(b) gravitational acceleration $g = 2 \times 10^{-4}$ lattice units.

Fig.5.1. Shape and position of a gravity-driven droplet on flat vertical surface, for different contact angles (or solid-liquid interaction parameters): (1) $\theta = 74.3^{\circ}$ ($w = -0.05$), (2) $\theta = 88.7^{\circ}$ ($w = 0$) and (3) $\theta = 101.6^{\circ}$ ($w = 0.035$) and different gravitational acceleration: (a) $1 \times 10^{-4}$ and (b) $2 \times 10^{-4}$ lattice units at different iteration times.

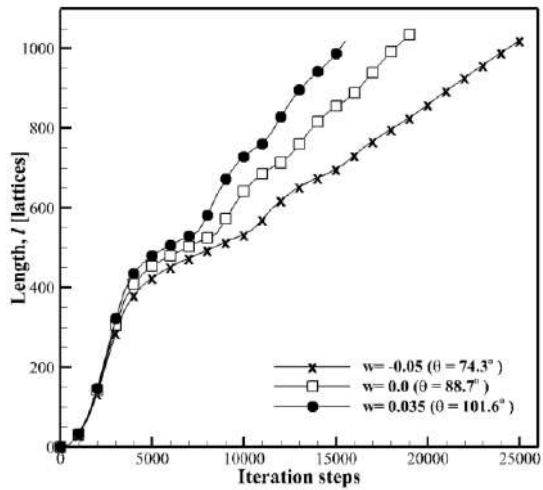

(a) gravitational acceleration $g = 1 \times 10^{-4}$ lattice units.

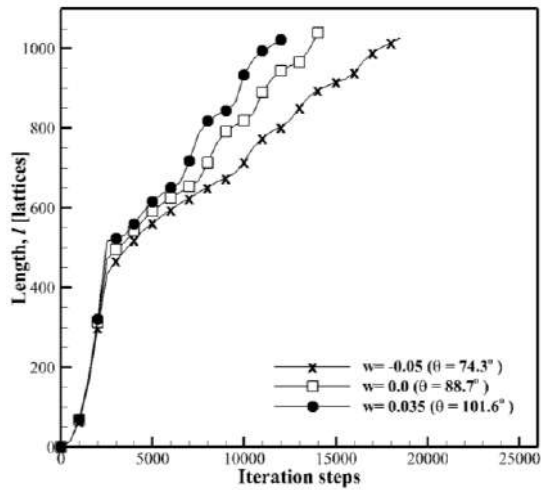

(b) gravitational acceleration $g = 2 \times 10^{-4}$ lattice units.

Fig.5.2. Droplet length versus time (iteration steps) for a gravity-driven droplet on a flat vertical surface. Different contact angles (or solid-liquid interaction parameters) of $\theta = 74.3^\circ$ ($w = -0.05$), $\theta = 88.7^\circ$ ($w = 0$) and $\theta = 101.6^\circ$ ($w = 0.035$) are considered for gravitational accelerations of (a) $1 \times 10^{-4}$ and (b) $2 \times 10^{-4}$ lattice units.

## 5.2.3. Run-off on a grooved surface

### 5.2.3.1. Effect of groove height and depth

In this section, the effect of groove height and depth on the dynamic behavior of a droplet moving down a vertical grooved surface is analyzed. Fig.5.3 shows the computational domain with $1\,080 \times 10\,800$ µm$^2$, discretized by $121 \times 1\,201$ lattice$^2$. The discretization is chosen based on a grid sensitivity analysis. The grid sensitivity analysis is conducted by comparing the remaining liquid fraction in a groove of $360 \times 360$ µm$^2$ at a gravity acceleration of $g = 2 \times 10^{-4}$ lattice units. The remaining liquid fraction for different lattice numbers $N$ is shown in Fig.5.4. No significant difference is observed when the lattice number is higher than 40 or accordingly a spatial resolution $\Delta x$ higher than 9 µm. For further simulations therefore a minimal spatial resolution $\Delta x$ of 9 µm is used. The groove is located at $y = 401$ lattices. This position is chosen as that is the location where the droplet attains a constant velocity, as shown in Section 5.2. For the base case, the size of the groove is $H \times D = 40 \times 40$ lattice$^2$. The center of the liquid droplet with radius of 40 lattices is located on the surface at $y = 1\,101$ lattices. Liquid and gas densities are 0.359 and $6.07 \times 10^{-3}$ lattice units respectively in accordance with the density ratio $\rho/\rho_c = 59.1$ at $T/T_c = 0.7$. The contact angle of 74.3° or a solid-liquid interaction parameter of -0.05 is imposed to the entire solid surface. Boundary conditions are: bounce back boundary conditions on left and right sides and periodic boundary conditions on top and bottom sides. As a reminder, the bounce back boundary condition is a no-slip boundary condition with zero velocity at the wall. The analysis focusses on the amount of liquid entering and remaining in the groove. The remaining liquid fraction is defined as the ratio of the volume of the liquid remaining in the groove at the end of the dynamic process to the initial volume of the droplet. Fig.5.5 plots the remaining liquid fraction with inserted snapshots of liquid distribution for different groove heights and gravitational accelerations. Fig.5.6 shows snapshots of the droplet moving down near/in the groove for different groove heights.

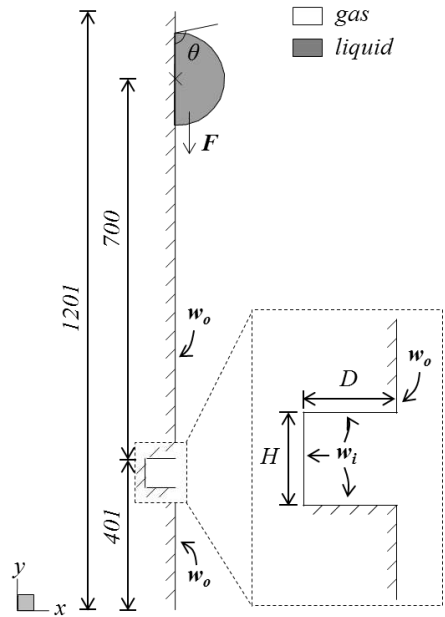

Fig.5.3. Schematic of the computational domain of a gravity-driven droplet running down on a vertical grooved surface.

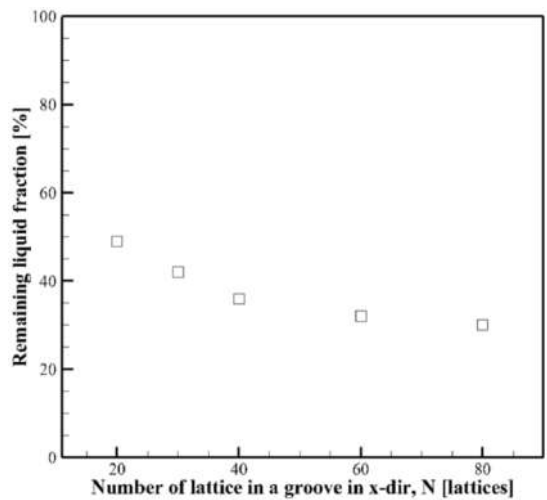

Fig.5.4. Grid sensitivity analysis of the remaining liquid fraction in a groove of 360 × 360 μm$^2$ for different lattice numbers $N$.

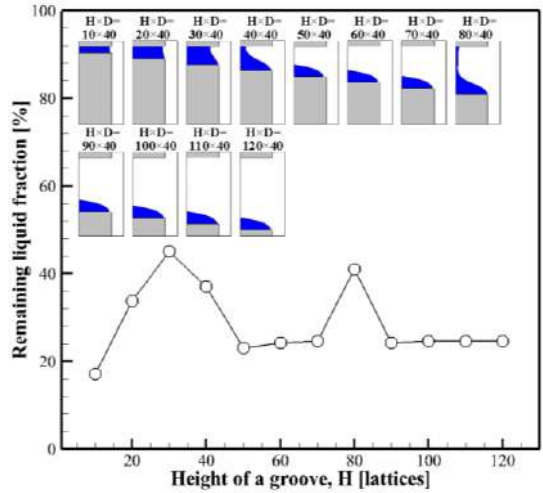

(a) gravitational acceleration $g = 2 \times 10^{-4}$ lattice units.

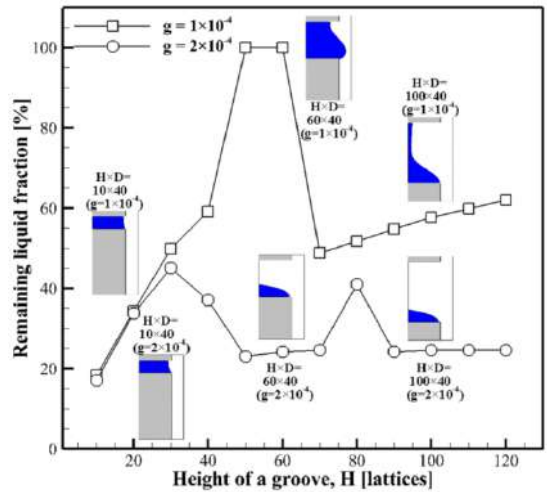

(b) gravitational acceleration $g = 1 \times 10^{-4}$ and $2 \times 10^{-4}$ lattice units.

Fig.5.5. Remaining liquid fraction with inserted snapshots of liquid distribution in the groove for different groove heights $H$ and two gravitational accelerations of (a) $2 \times 10^{-4}$ and (b) $1 \times 10^{-4}$ and $2 \times 10^{-4}$ lattice units.

In Fig.5.5 (a), the remaining liquid fraction shows a complex trend, indicating a complex dynamic behavior of the droplet in/near the groove. For a relatively narrow groove, the groove is totally filled ($H \leq 20$ lattices) or largely filled ($H \leq 30$ lattices). Since for small groove heights, the initial droplet volume is larger than the volume of the groove, the remaining liquid fraction is lower than 1 although the groove is almost totally filled with liquid (Fig.5.6 (a)). With increasing groove height, more and more liquid is entrapped in the groove and the remaining liquid ratio increases with groove height. This regime is called the 'height controlled regime'.

When the groove becomes higher ($40 < H \leq 70$ lattices), the filling pattern changes and only the bottom surface of the groove is finally covered by liquid water, as shown in the inserts of Fig.5.5. As represented in Fig.5.6 (b) at $H = 60$ lattices, the liquid first moves due to adhesion forces into the groove and then drips down under gravity at iteration 16 000. When the liquid droplet reaches the bottom surface of the groove, a liquid bridge is formed between the top and the bottom of the groove at iteration 19 000. The bridge does not break and drags most of the liquid water down from the top surface of the groove at iteration 19 900. A negligible amount of liquid water remains adhered at the top surface of the groove. For the groove heights from 50 to 70 lattices, the remaining liquid fraction is almost constant. This regime is called the 'bottom surface controlled regime'.

When the groove height increases to 80 lattices (Fig.5.6 (c)), the dynamic behavior of the droplet changes. A liquid bridge is again formed, but now the bridge breaks at iteration 17 000. As a result, a larger portion of the liquid remains adhered on the top surface of the groove. Afterwards, the amount of liquid on the top surface grows as it is still fed by the upstream film flow. A second droplet forms and contacts with the liquid on the bottom surface of the groove, forming a second liquid bridge, which breaks later at iteration 18 300. Finally, a portion of the liquid remains adhered at the top surface of the groove, which then moves due to adhesion forces into the groove, running down on the vertical side of the groove at iteration 26 100. At iteration 30 000, a continuous liquid film on the back side of the groove is formed, which feeds the remaining liquid on the bottom surface. This results in an increase of the remaining liquid fraction, as shown in Fig.5.5 (a). This regime is called the 'top-bottom surface controlled regime'.

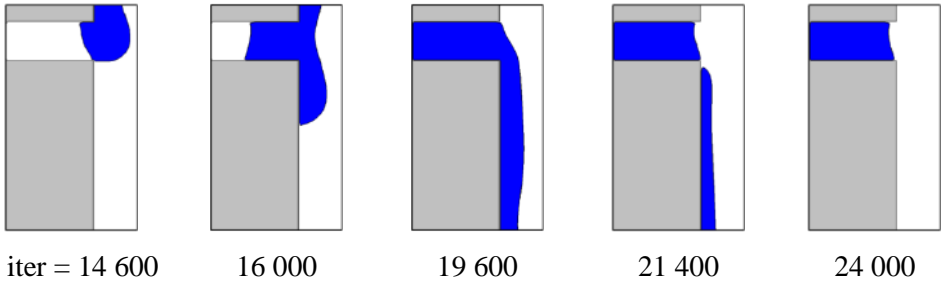

iter = 14 600     16 000     19 600     21 400     24 000

(a) $H \times D = 20 \times 40$ lattice$^2$

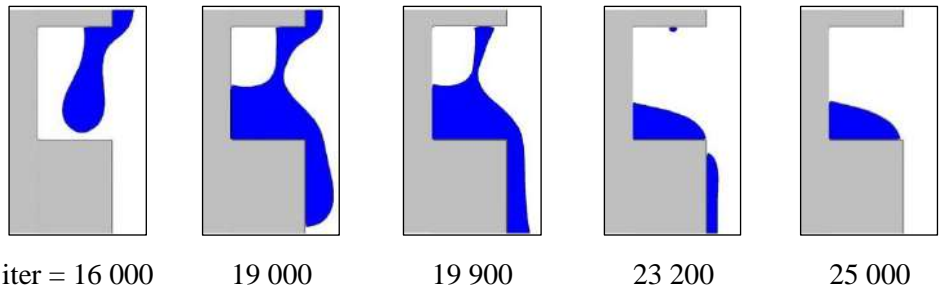

iter = 16 000     19 000     19 900     23 200     25 000

(b) $H \times D = 60 \times 40$ lattice$^2$

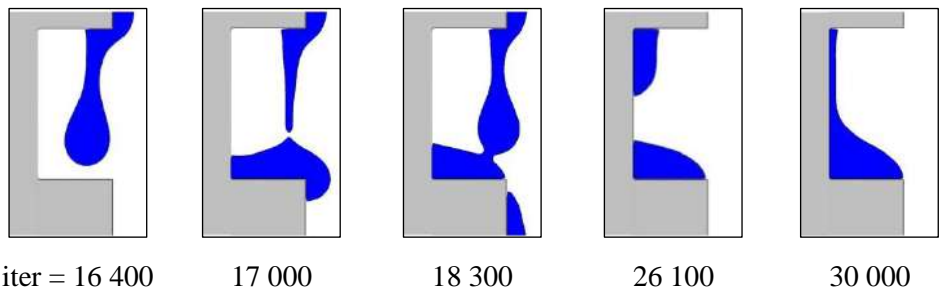

iter = 16 400     17 000     18 300     26 100     30 000

(c) $H \times D = 80 \times 40$ lattice$^2$

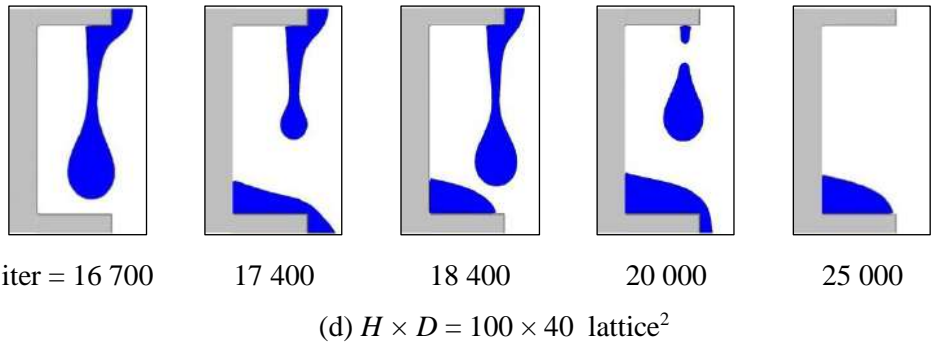

iter = 16 700      17 400      18 400      20 000      25 000

(d) $H \times D = 100 \times 40$  lattice$^2$

Fig.5.6. Dynamic behavior of a droplet in a groove for different heights $H$ of (a) 20; (b) 60; (c) 80; and (d) 100 lattices at different iteration times.

When the height of the groove further increases ($H \geq 90$ lattices), three droplets are formed and drip down on the bottom surface respectively at iterations 16 700, 17 400 and 20 000, as shown in Fig.5.6 (d). No liquid bridge is formed. This regime is called 'top surface controlled regime'. Compared with the 'top-bottom surface controlled regime', the remaining liquid fraction decreases, as there is no liquid film forming on the backside of the groove.

Fig.5.5 (b) compares the remaining liquid fraction for two different gravitational accelerations. We observe that the behavior differs quite substantially depending on gravitational acceleration. At a gravitational acceleration of $1 \times 10^{-4}$ lattice units, only two regimes occur, which are identified as 'height controlled regime' and 'top-bottom surface controlled regime'. The groove is almost totally filled for $H \leq 60$ lattices, showing that the droplet is completely trapped in the groove at $H = 50$ and 60 lattices. When the height of groove increases ($H \geq 70$ lattices), a continuous liquid film is formed in the groove. We conclude that, at low gravitational acceleration, the droplet displacement in/near the groove shows a more simple behavior compared to the behavior at higher gravitational acceleration.

The effect of the groove depth $D$ on the remaining liquid fraction is illustrated in Fig.5.7 with insets of remaining liquid fraction in the groove for two different gravitational accelerations. The height of the groove is 40 lattices, leading to a 'bottom surface controlled regime' as discussed above.

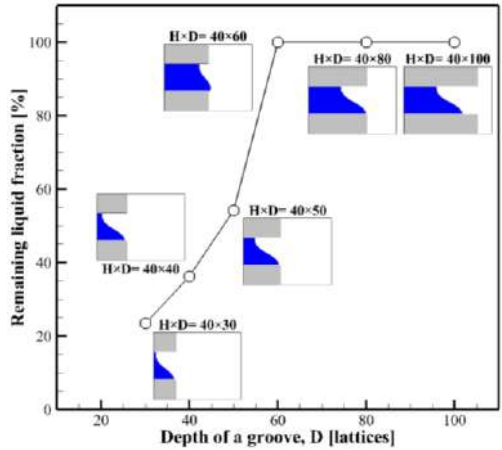

(a) gravitational acceleration $g = 2 \times 10^{-4}$ lattice units.

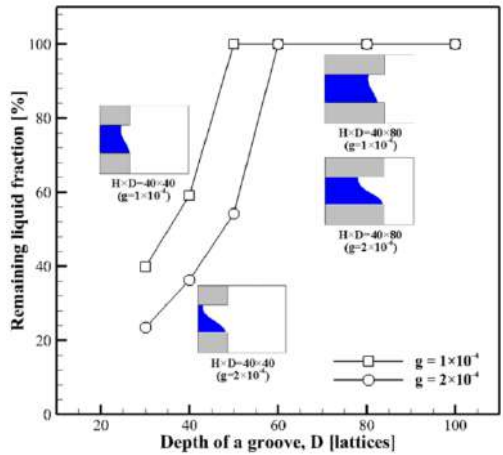

(b) gravitational acceleration $g = 1 \times 10^{-4}$ and $2 \times 10^{-4}$ lattice units.

Fig.5.7. Remaining liquid fraction and liquid distribution in the groove for different depths of the groove $D$ for two gravitational accelerations of (a) $2 \times 10^{-4}$ and (b) $1 \times 10^{-4}$ and $2 \times 10^{-4}$ lattice units with respective inset images.

In all cases, part of the droplet is trapped in the groove. The remaining liquid fraction increases with depth for a gravitational acceleration of $1 \times 10^{-4}$ lattice units until a depth of 50 lattices, while for a gravitational acceleration of $2 \times 10^{-4}$ lattice units, the remaining liquid fraction increases until a depth of 60 lattices. When the depth of the groove further increases, the droplet is completely trapped in the groove and the remaining liquid fraction reaches a value of one. For other regimes for different heights, analogous results are obtained. The results are for brevity not discussed.

### 5.2.3.2. Effect of the wettability

In this section we focus on the effect of the contact angle or solid-liquid interaction parameter of the surface and groove on the droplet dynamics. The term "surface" refers here to the wall excluding the groove. Different combinations of solid-liquid interaction parameters of the surface and groove are investigated as listed in Table. 5.1. In Table.5.1, $w_o$ and $w_i$ indicate the solid-liquid interaction parameters for the surface and the groove, respectively. For all the cases, the groove has a height and depth of $40 \times 40$ lattice$^2$.

Fig.5.8 shows the remaining liquid fraction and liquid distribution in the groove for different combinations of solid-liquid interaction parameters. For a hydrophilic groove ($w_i$ = -0.1), the liquid remains on the top and bottom surfaces of the groove. The adhesion force due to the hydrophilic nature of the groove is sufficiently strong to keep the liquid captured in the groove. For a neutral ($w_i$ = 0) and hydrophobic groove ($w_i$ = 0.035), the remaining liquid fraction in the groove is controlled by the wettability of the surface. When the surface is hydrophobic ($w_o$ = 0.02), 22.6% of the liquid remains in a neutral groove ($w_i$ = 0) and 20.8% in a hydrophobic groove ($w_i$ = 0.035). When the surface is hydrophilic ($w_o$ = -0.05), no liquid is captured in the neutral and hydrophobic groove, and the droplet passes over the groove resulting in a zero remaining liquid fraction. This different behavior can be explained by the difference in adhesion force between the surface and groove. On a hydrophilic surface, the liquid enters the hydrophobic/neutral groove, but the adhesion force of the groove is too low to keep the liquid captured in the groove. In contrast, on a hydrophobic surface, the adhesion force of the hydrophobic/neutral groove is higher and the groove captures some liquid. Liquid enters the groove and breaks into two parts, one part captured in the groove and a part moving down on the surface by gravitational force.

Table.5.1. Cases of different solid-liquid interaction parameters (or contact angles) of the surface $w_o$ and the groove $w_i$ .

| | $w_o$ | Contact angle of surface $\theta_o$ [deg] | $w_i$ | Contact angle of groove $\theta_i$ [deg] | Wettability |
|---|---|---|---|---|---|
| Case 1 | | | -0.1 | 63.1 | both hydrophilic |
| Case 2 | -0.05 | 74.3 | 0.0 | 88.7 | hydrophilic/ neutral |
| Case 3 | | | 0.035 | 101.6 | hydrophilic/ hydrophobic |
| Case 4 | | | -0.1 | 63.1 | hydrophobic/ hydrophilic |
| Case 5 | 0.02 | 97.4 | 0.0 | 88.7 | hydrophobic/ neutral |
| Case 6 | | | 0.035 | 101.6 | both hydrophobic |

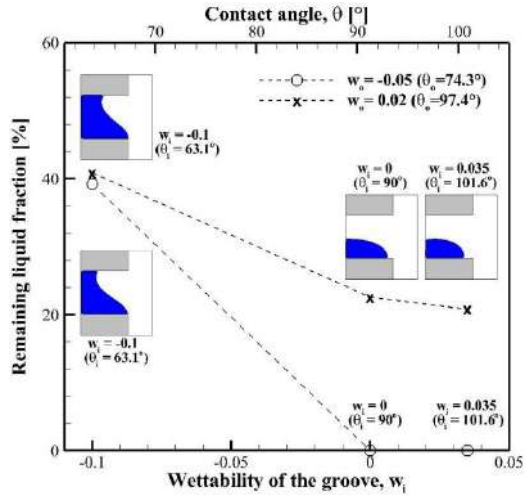

Fig.5.8. Remaining liquid fraction and liquid distribution in the groove with different solid-liquid interaction parameters of the surface $w_o$ and groove $w_i$.

### 5.2.3.3. Tilt angles

Finally, the effect of different tilt angles of the surface with groove on the droplet dynamics is studied. The force on a tilted surface, which drives the movement of the droplet, equals the projection of the gravitational force on this surface. This force component is maximal for a surface of $90\,^o$ and decreases with decreasing tilt angle. Inversely, the force component along x-direction equals zero for a surface of $90\,^o$ and increases with decreasing tilt angle. This force in x-direction leads to a trapping of liquid inside the groove. In the study, three tilt angles of 45°, 60° and 75° are considered. Fig.5.9 plots the remaining liquid fraction for the different tilt angles. For all cases, the remaining liquid fraction decreases when tilt angle increases, due to a decrease of the force in x-direction promoting the capture of liquid in the groove. The effect of wettability (solid-liquid interaction parameter $w$) of the groove on the remaining liquid fraction is not significant for smaller tilt angle, showing a similar remaining liquid fraction of 60 % in the groove. However, as the tilt angle increases, the effect of the wettability of the groove on the remaining liquid fraction becomes more significant and leads to differences of 40 to 50 % at a tilt angle of 75 º.

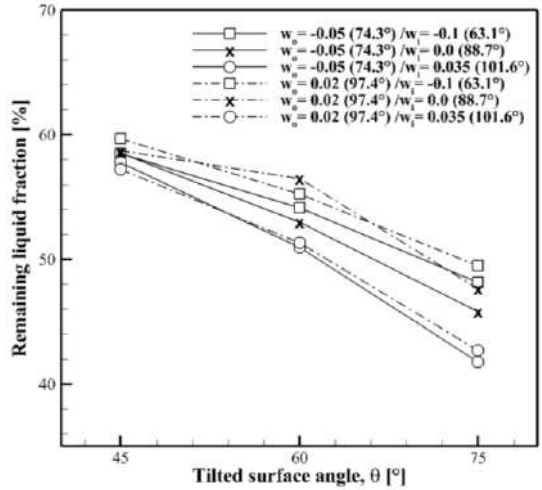

Fig.5.9. Remaining liquid fraction in the groove with different tilt angles of the surface and the solid-liquid interaction parameter of surface, $w_o$ and groove, $w_i$.

## 5.3.  Evaporation of droplet on micropillars

In this section, the dynamic behavior of a droplet on a set of micropillars during evaporation is investigated using 2D LBM for different sizes and distances between pillars. The LB results are analyzed in terms of temporal evolution of droplet radius, critical contact angle and excess Gibbs free energy. Furthermore, the time evolution of contact angle, capillary pressure and velocity streamlines inside the droplet and in the capillaries in between the micropillars is analyzed during stick-slip mode.

As introduction to the study of the evaporation of a droplet on a set of micropillars, some qualitative LBM results are first discussed. Fig.5.10 shows the evaporation of a droplet on a perfectly smooth surface with single surface wettability and no contact angle hysteresis. A CCA (constant contact angle) mode is observed, which means that the contact diameter and the height of droplet continuously decrease while the contact angle with the surface remains constant. A detailed explanation of CCA model studied with the pseudopotential LBM is described in (Joshi and Sun 2010).

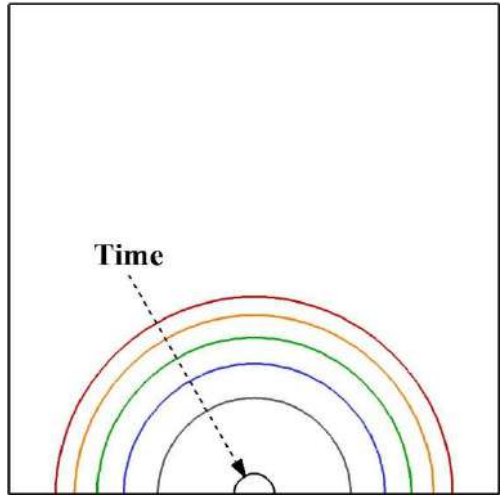

Fig.5.10. Evolution of droplet evaporation in CCA mode.

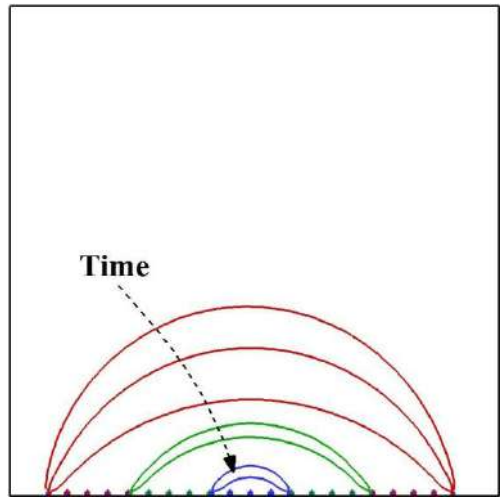

Fig.5.11. Evolution of droplet evaporation in CCR mode with a distance between the needles of 5 lattices.

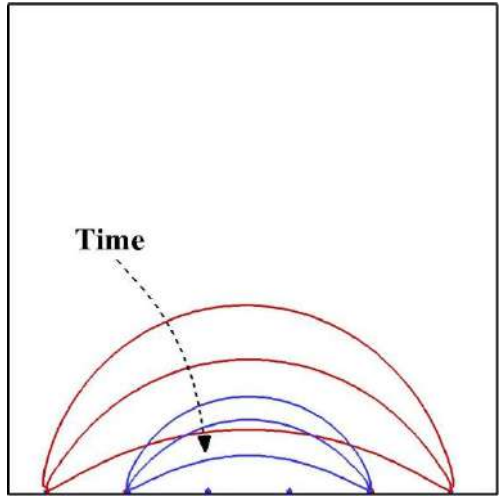

Fig.5.12. Evolution of droplet evaporation in CCR mode with a distance between the needles of 20 lattices.

In Figs. 5.11 and 5.12, the droplet evaporation on a surface with artificial roughness is simulated by including small needles on the surface. The needles have a width of 1 lattice and a height of 2 lattices. The distance between the needles is 5 or 20 lattices in Figs.5.11 and 5.12, respectively. The droplets in both cases remain pinned on the needles. When evaporation continues, the contact angle and the droplet height decrease while the contact diameter remains constant. This is referred to as CCR (constant contact radius) mode. Depinning of droplet occurs when the contact angle reaches its critical contact angle. The critical contact angle is lower for higher distance between the needles (Fig.5.12) compared to short distance (Fig.5.11), and thus depends on the artificial roughness introduced (here the distance between the needles). The droplet becomes again pinned after sliding over a certain distance. This pinning/depinning process is repeated until the droplet is totally evaporated. It is found that the critical contact angle is constant for all depinning cycles for a surface with same roughness. A detailed explanation of pinning/depinning process of a droplet on a set of micropillars will be given in the next section.

### 5.3.1. Simulation set-up and boundary conditions

The LBM 2D domain has a size of $3\,000 \times 3\,000$ µm$^2$ or $601 \times 601$ lattice$^2$ with the spatial resolution $\Delta x$ of 5 µm as illustrated in Fig.5.13. The series of micropillars is located at $y = 50$ lattices from the bottom surface. The pillars have a height $l$ of 50 lattices, a pillar width $a$ and pitch width $b$, which is the distance between two pillars. A droplet with radius of 490 µm or 98 lattices is deposited on the middle pillar. The density ratio $\rho/\rho_c = 59.1$ at $T/T_c = 0.7$ results in liquid and gas densities of 0.359 and $6.07 \times 10^{-3}$ lattice units respectively. The contact angle is 69.3º in accordance with a solid-liquid interaction parameter of -0.07 as indicated in the contact angle test in chapter. 3. The boundary conditions imposed are periodic boundary conditions at the lateral sides (left and right) and a bounce back boundary condition at the bottom side. To generate a diffusive evaporation of the droplet, the Zou-He velocity boundary condition is imposed at the top side as a constant vapor phase velocity of 0.0107 m/s or $8.9 \times 10^{-3}$ lattice units, resulting in a constant evaporation flux. Remark that vapor can flow from the bottom side of the pillars around the set of pillars to the velocity boundary condition at the top side.

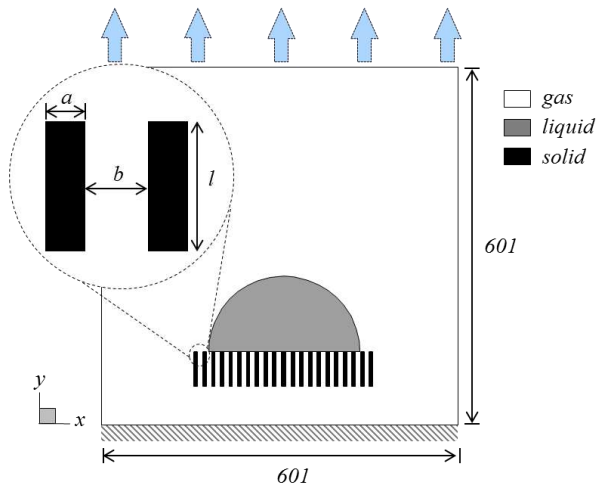

Fig.5.13. Schematic representation of computational domain of evaporating droplet on a series of micropillars with width $a$, height $l$ and pitch $b$ between pillars.

### 5.3.2. Results

Fig.5.14 is a schematic representation the evaporation process of a droplet deposited on a set of micropillars, as simulated with LBM. Once the droplet contacts the surface, the droplet invades the micropillars due to capillary suction (see Fig.5.14 (a)). Then, the droplet gets pinned on the surface, which is called the stick mode (see Fig.5.14. (b)). As evaporation evolves, the droplet suddenly depins and moves over the surface to the next pillar. This process is called the slip mode (see Fig.5.14. (c)). It is noted that the liquid in between the pillars forms a capillary meniscus both at the bottom side and at the top side after the depinning of the droplet.

In Fig.5.15, the time evolution of the total liquid volume, the fluid volume of the droplet sitting on the pillars and the liquid volume inside the capillaries in between the pillars are plotted. As mentioned before, capillary suction occurs first, resulting in a fast decrease of the droplet volume with increasing liquid volume in the space in between the pillars. The liquid volume in between the pillars increases with the pitch width $b$. After the space between the pillars is totally filled with liquid, the evaporation results in a linear decrease of the droplet volume versus time due to the constant evaporation flux imposed on the top boundary. The slope of the different droplet evaporation curves is similar for all cases, indicating the constant evaporation rate. After some time, the droplet is totally evaporated and liquid remains only in between the pillars showing a capillary meniscus. The capillary liquid will also evaporate, but this process is not documented. Due to the different volumes uptaken in the space in between the pillars, the evaporation time of droplet is different for the different cases: the evaporation goes on the longest for the smaller pitch width. The little jumps seen in the graphs are attributed to the depinning and fast sliding of the droplet over the surface.

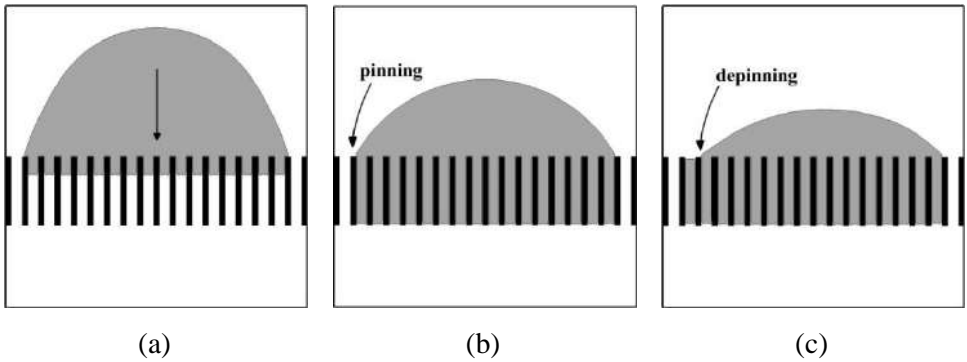

Fig.5.14. Schematic of the evaporation process of a droplet deposited on a set of micropillars: (a) capillary suction; (b) stick (pinning) mode on pillars; and (c) slip (depinning) mode on pillars.

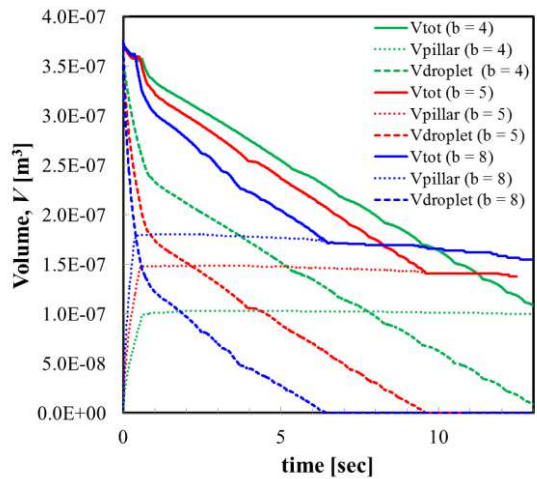

Fig.5.15. Total liquid volume, fluid volume of the droplet sitting on the pillars and liquid volume in the space in between the pillars versus time for three different pitch widths of 4, 5 and 8 lattices.

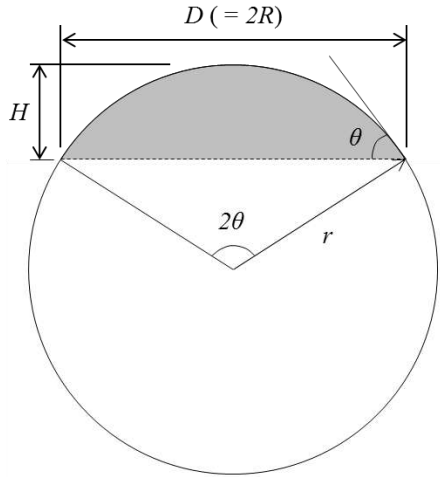

Fig.5.16. Schematic of spherical cap with radius *r*, contact radius *R*, contact diameter *D* and height *H*.

The contact angle of the droplet during evaporation is calculated using the cap method, which uses the nomenclature illustrated in Fig.5.16 (Berthier and Beebe 2007). It is noted that the LB results are 2D simulations and that the equations in this chapter are given for the 2D case. The contact radius *R* and the height *H* of the cap are measured using the LB results. The circle radius *r* can be determined from the droplet contact radius *R* and height *H* as follows:

$$r = \frac{\left(R^2 + H^2\right)}{2H}.$$ (5.2)

The contact angle $\theta$ is then given by:

$$\theta = \cos^{-1}\left(1 - \frac{H}{r}\right).$$ (5.3)

Figs.5.17 and 5.18 show the contact radius and contact angle versus time for nine cases with different pillar and pitch widths. Two phases are distinguished: an initial phase where the contact radius remains constant, while the contact angle reduces (CCR mode) until the occurrence of the first sliding, followed by a second phase where a stick-slip behavior is observed. The stick-slip evaporation mode shows alternatively CCR and CCA modes (Shanahan 1995, Orejon, Sefiane et al. 2011, Oksuz and Erbil 2014): longer periods with constant contact radius and decreasing contact angle alternate with short periods with constant contact angle and decreasing contact radius. In the following, focus will be given only on the stick-slip phase, and the initial phase before the first slip is not considered. In the stick-slip phase, the droplet is first pinned on the surface, while the contact angle decreases continuously until a critical contact angle is reached (CCR mode). At this contact angle, the droplet starts to move reducing its contact radius, until the droplet is pinned again. As a result, the temporal evolution of the contact radius shows a staircase shape with repetitions of CCR and CCA modes until the droplet is totally evaporated.

Figs.5.17 (a) - (c) show the contact radius versus time for different pillar widths *a* and same pitch width *b*. With increasing pillar width, the droplet remains pinned during longer time and then slides over a larger distance. Figs.5.17 (h), (b) and (i) show the contact radius versus time for different pitch widths *b* and same pillar width *a*. With increasing pitch width, the droplet remains pinned during shorter time and then slides over a larger distance. These dependencies on pillar and pitch widths will be further explored below.

In Figs. 5.18, it is observed that the contact angle decreases until a critical contact angle is reached. At the critical contact angle, the droplet gets depinned and slips over the surface. After the droplet becomes pinned again (mostly at the next pillar), a higher contact angle is attained again. It is observed that the critical contact angle is constant over all stick-slip cycles for a given pillar geometry. This critical contact angle is indicated in the figures by a dashed line.

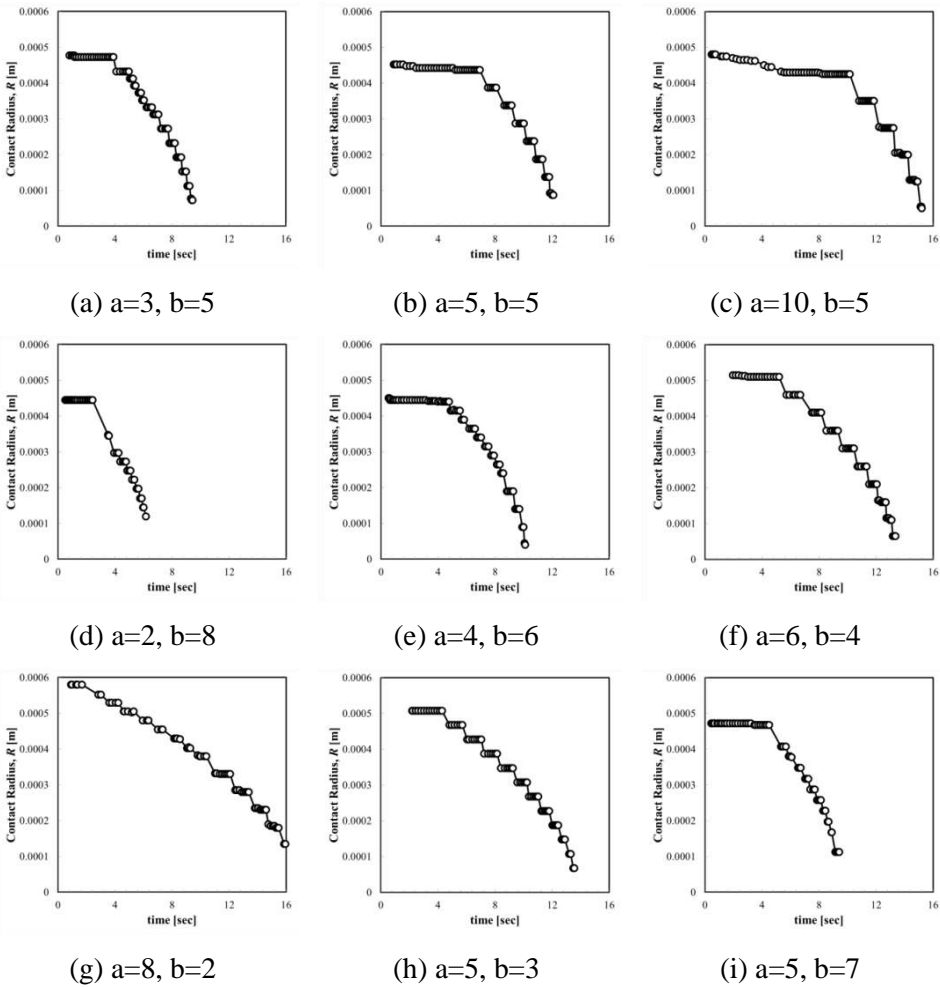

(a) a=3, b=5      (b) a=5, b=5      (c) a=10, b=5

(d) a=2, b=8      (e) a=4, b=6      (f) a=6, b=4

(g) a=8, b=2      (h) a=5, b=3      (i) a=5, b=7

Fig.5.17. Evolution of droplet contact radius versus time on pillars with different width *a* and pitch width *b*.

Figs.5.18 (a) - (c) show that, for the same pitch width, the critical contact angle is almost constant, which leads to the observation that the critical contact angle only depends on the pitch width *b*. In Fig.5.19, the critical contact angle is plotted versus pitch width *b*, showing that the critical contact angle decreases with increasing pitch width. To analyze these observations in more detail, the results are further interpreted in view of excess Gibbs free energy.

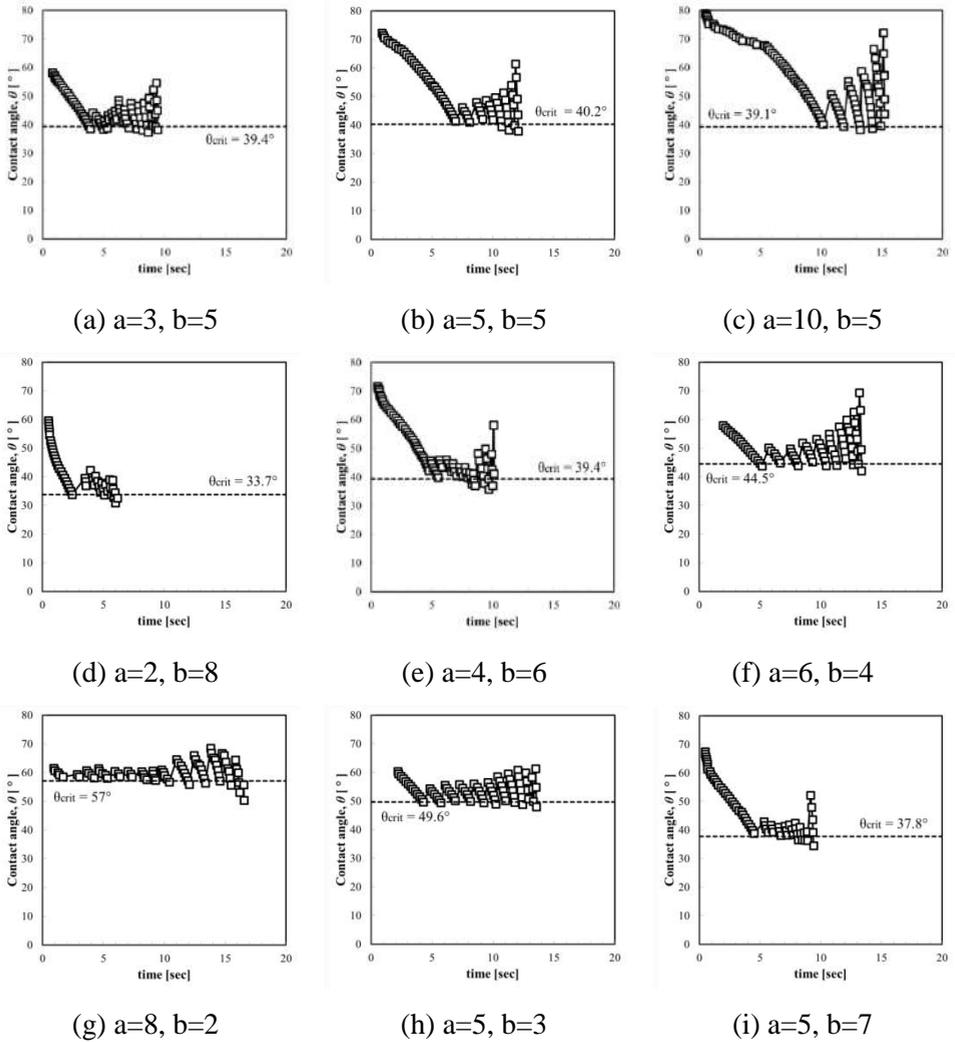

(a) a=3, b=5          (b) a=5, b=5          (c) a=10, b=5

(d) a=2, b=8          (e) a=4, b=6          (f) a=6, b=4

(g) a=8, b=2          (h) a=5, b=3          (i) a=5, b=7

Fig.5.18. Evolution of droplet contact angle versus time on pillars with different width *a* and pitch *b*. The critical contact angle is indicated by a dashed line.

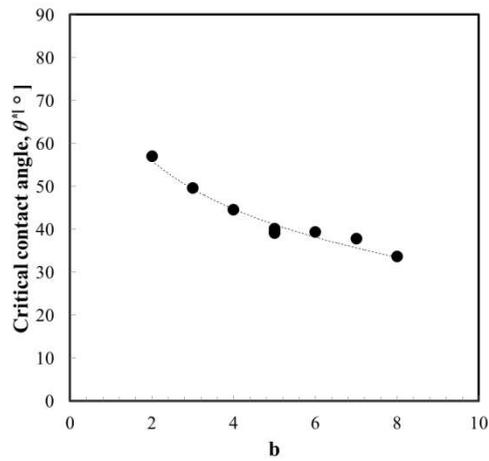

Fig.5.19 Critical contact angle at droplet depinning versus pitch width *b*.

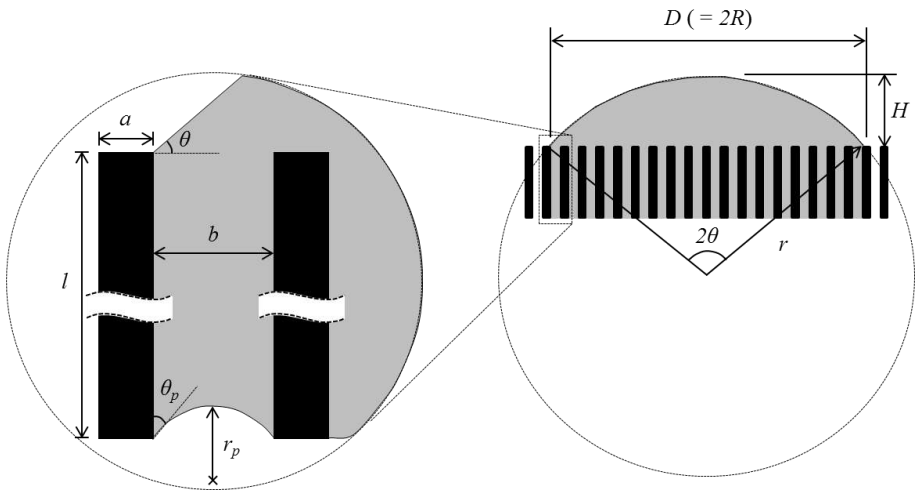

Fig.5.20. Schematic of the evaporating droplet on a set of capillary pillars with pillar width *a*, height *l*, pitch width *b* and contact angle *θ*. The droplet has contact radius *R*, height *H* and the segment radius *r*. The meniscus in the capillary shows a radius $r_p$.

The derivation of the Gibbs free energy for a spherical droplet evaporating on a smooth surface is given in chapter 2. This model is adapted to take into account a 2D droplet sitting on a set of micropillars where the space in between the pillars is also filled as shown in Fig.5.20. This means that not only the interfacial Gibbs free energy of the droplet on top of micropillars has to be considered, but also the interfacial Gibbs free energy of the liquid contacting the solid inside the pillars, and the surface energy of the menisci in the capillaries.

The total Gibbs free energy is given as the sum of: 1) the interfacial energy of the droplet on the pillars, 2) the liquid/solid interfacial energy of the liquid in between the pillars and 3) the interfacial energy of the capillary meniscus inside each capillary between the pillars. The interfacial energy of the droplet on the pillars can be written as:

$$G_1 = \gamma \, S_1 = \gamma \left(2r\theta\right), \tag{5.4}$$

where $S_1$ is the surface of liquid-vapor interface of the droplet, $\gamma$ is the liquid/vapor interfacial tension, $r$ is radius of the droplet segment and $\theta$ the contact angle. The liquid/solid interfacial energy of the liquid in between the pillars is given by:

$$G_2 = \left(\gamma_{SL} - \gamma_{SV}\right) \times \left((n-1)a + 2nl\right) \tag{5.5}$$

where $\gamma_{SL}$ is the solid/liquid interfacial tension, $\gamma_{SV}$ is the solid/vapor interfacial tension, $a$ is the width of pillar, $l$ is height of the pillars, $n$ is the number of pillars filled with liquid. Using Young's equation, Eq. (5.5) becomes:

$$G_2 = \gamma \cos\theta_0 \times \left((n-1)a + 2nl\right) \tag{5.6}$$

with $\theta_0$ the equilibrium contact angle. The interfacial energy of the capillary meniscus for all capillaries between the pillars (see small schematic in Fig.5.20) is given by:

$$G_3 = \gamma \, S_3 \, n = \gamma \left(\left(\pi - 2\theta_p\right)r_p\right)n \tag{5.7}$$

where $S_3$ is the liquid-vapor surface of the capillary meniscus, $r_p$ is the radius of segment in one capillary and $\theta_p$ is the contact angle of the meniscus inside the capillary, which as shown in Fig.5.20, will be considered as a constant for each case. The excess Gibbs free energy $\delta G$ is defined as

$$\delta G = G_o - G(\theta), \tag{5.8}$$

where $G(\theta)$ is the total Gibbs free energy dependent on the contact angle of the droplet, and $G_0$ is the initial Gibbs free energy where $\theta$ equals the equilibrium contact angle $\theta_0$. When depinning occurs as shown in Fig.5.21, the Gibbs free energy is updated at each new stick phase as follows:

$$G_0 = G\left(\theta = \theta_{0,i}\right), \tag{5.9}$$

where $\theta_{0,i}$ is the equilibrium contact angle attained at the moment the droplet becomes pinned again, and where $i$ is the order of the pinning cycle (see Fig.5.21).

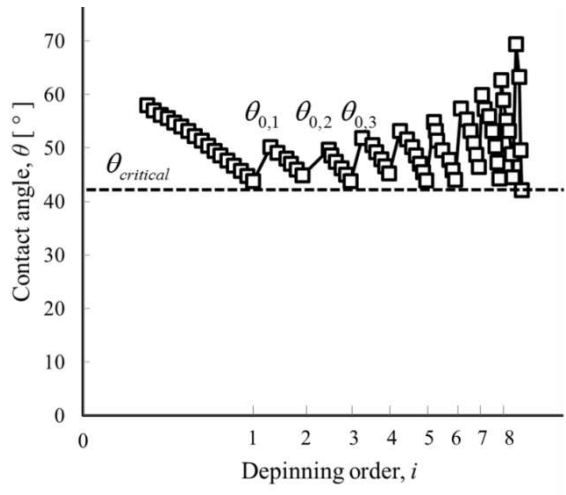

Fig.5.21. Schematic graph of evolution of contact angle of evaporating droplet versus time, with equilibrium contact angle $\theta_{0,i}$ at the beginning of each pinning phase and critical contact angle $\theta_{crit}$ i.e. the angle when the droplet becomes depinned.

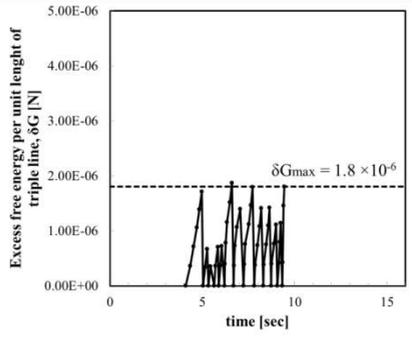

(a) a=3, b=5

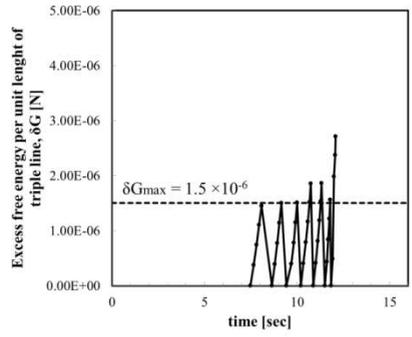

(b) a=5, b=5

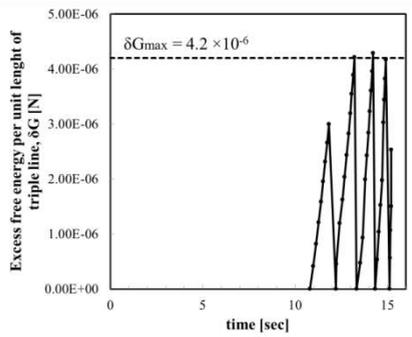

(c) a=10, b=5

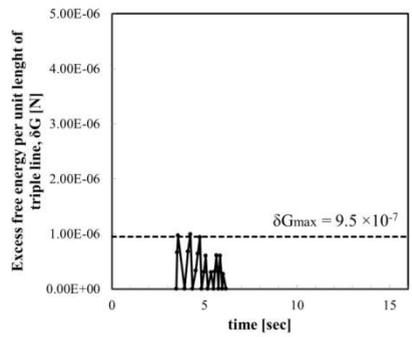

(d) a=2, b=8

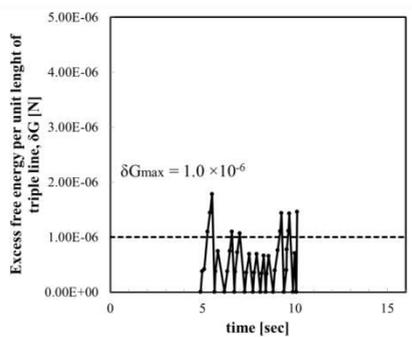

(e) a=4, b=6

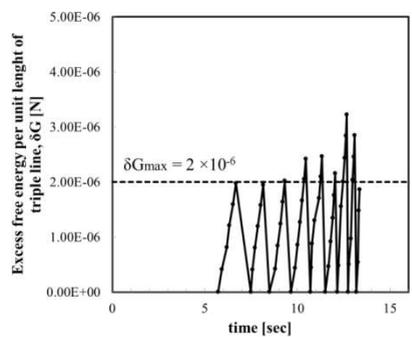

(f) a=6, b=4

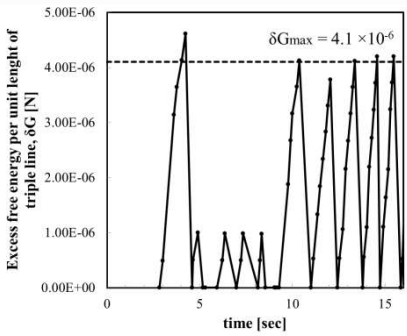

(g) a=8, b=2

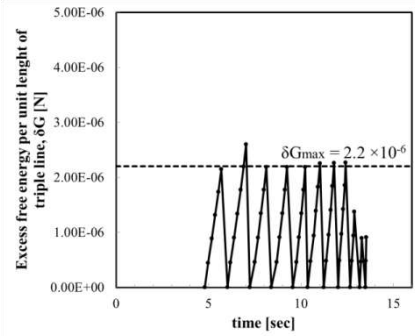

(h) a=5, b=3

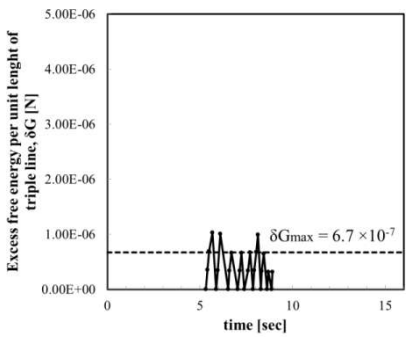

(i) a=5, b=7

Fig.5.22. Evolution of excess Gibbs free energy versus time of a droplet on pillars with different widths *a* and pitch widths *b*.

In Fig.5.22, the excess Gibbs free energy is plotted versus time for different pillar and pitch widths. The excess Gibbs free energy increases during the stick phase. When the excess Gibbs free energy attains its maximum, there is sufficient energy available to overcome the energy barrier, the droplet depins and the triple line moves to its new equilibrium position. At the new equilibrium condition, the excess Gibbs free energy equals zero again and the process is repeated. Some scatter is observed in the excess Gibbs free energy, which is explained by some artefacts. Normally, the droplet depins symmetrically and the triple line slides over one pillar distance at both sides. However, sometimes the depinning process becomes asymmetric, and only one triple line slides over one or more pillars, while the other side remains pinned.

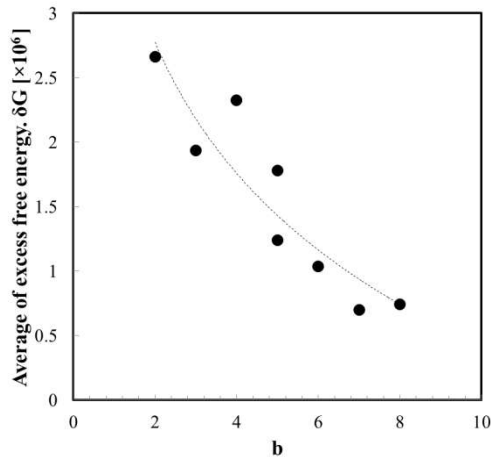

Fig.5.23. Relation between maximum excess free energy and pitch width *b* between pillars.

Fig.5.23 gives the relation between the maximum excess Gibbs free energy and the pitch width. It is noted that the maximum excess Gibbs free energy does not depend on the pillar width. The curve shows a clear trend with exception with one outliner, where the depinning process was highly asymmetrical. The maximum excess Gibbs free energy decreases with increasing pitch width. It is concluded that the pitch width is the main influencing parameter that controls the depinning of a droplet on a set of micropillars, its critical contact angle and maximum excess Gibbs free energy. The dependency of the depinning on pitch width will be further analyzed below.

To understand the process of evaporation, pinning and depinning of a droplet sitting on a set of micro pillars, the fluid transport processes in the droplet and in between the micropillars are studied in more detail. Three cases with different pitch widths of 4, 7 and 10 lattices are selected. The pillar width in these cases is $a = 6$ lattices for pitch width of 4 lattices and $a$ = 5 lattices for pitch widths of 7 and 10 lattices. The results are plotted versus normalized time $t^*$ which is the ratio of time to the total time of a pinning/depinning cycle.

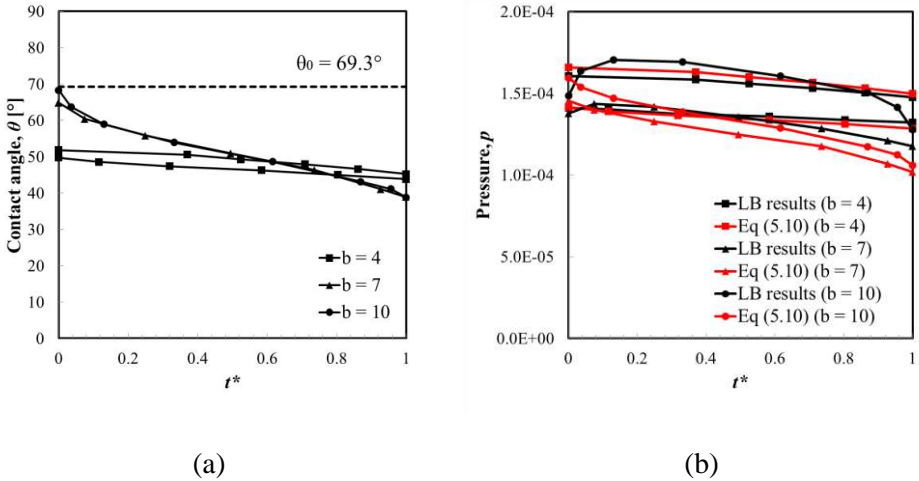

(a)                                      (b)

Fig.5.24. (a) Temporal evolution of contact angle for three pitch widths of 4, 7 and 10 lattices for a characteristic depinning event and (b) temporal evolution of Laplace pressure of droplet for three pitch widths of 4, 7 and 10 lattices for a characteristic depinning event. Comparison with Laplace Eq. (5.10).

The contact angle of the droplet is plotted versus normalized time for different pitch widths in Fig.5.24 (a). The contact angle decreases with time due to evaporation in CCR mode until depinning occurs at the critical contact angle. It is remarked that the contact angle at $t^*$ = 0 does not equal the equilibrium contact angle as imposed in the LBM for a pitch width $b$ = 4. For pitch widths of $b$ = 7 and $b$ =10, the initial contact angle at $t^*$ = 0 equals the equilibrium contact angle. This difference can be explained by the fact that the remaining volume of the droplet sitting on the pillars is not the same for different pitch widths, since different liquid volumes are taken up from the droplet to the capillaries (see e.g. Fig. 5.15). It is also noted that when the droplet slides to its new position, the sliding distance is smaller for smaller pitch width. It was found that the contact angle for smaller pitch widths will be lower considering the smaller sliding distance of the droplet to its new equilibrium position, while it will not differ a lot from the equilibrium contact angle for larger pitch widths. As already pointed out in Fig.5.19, the critical contact angle decreases with increasing pitch widths. More complete results compared with critical contact angles for the other pitch widths are given in the Addendum A1 'Contact angle of droplet'.

The Laplace pressure of the droplet has also been determined from the LB results, defined as the pressure difference between the liquid pressure in the bulk of the droplet and surrounding gas pressure. According to the Laplace equation, the Laplace pressure for the droplet is given as:

$$\Delta p_{drop} = p_L - p_G = \frac{\gamma}{r} = \frac{\gamma \sin\theta}{R} \, , \tag{5.10}$$

with $r$ the droplet radius, $R$ the contact droplet radius and $\theta$ the contact angle. In Fig.5.24 (b) the Laplace pressure as obtained directly from the LB results is compared to the Laplace pressure determined using Eq. (5.10), where the contact angle determined from the LB results is used. A relatively good agreement is obtained. The differences may be attributed to the difficulty in determining reference values for liquid and gas pressure. First the liquid pressure inside the droplet does not show a constant value. Moreover, the pressure profile at the interface shows some instability, as explained in section 3.7, which makes the determination of the reference pressures for gas and liquid phases more difficult. Considering these uncertainties, one may conclude the agreement is satisfactory to validate the LBM.

The time evolution of the contact angle $\theta_p$ (subscript $p$ standing for pore) and capillary pressure $p_c$ at the meniscus inside the capillaries is determined from the LB results for the three pitch widths. Three capillaries are considered: the capillary at the edge where the droplet is pinned, one capillary next to the edge capillary (referred to as 'next edge') and the middle capillary.

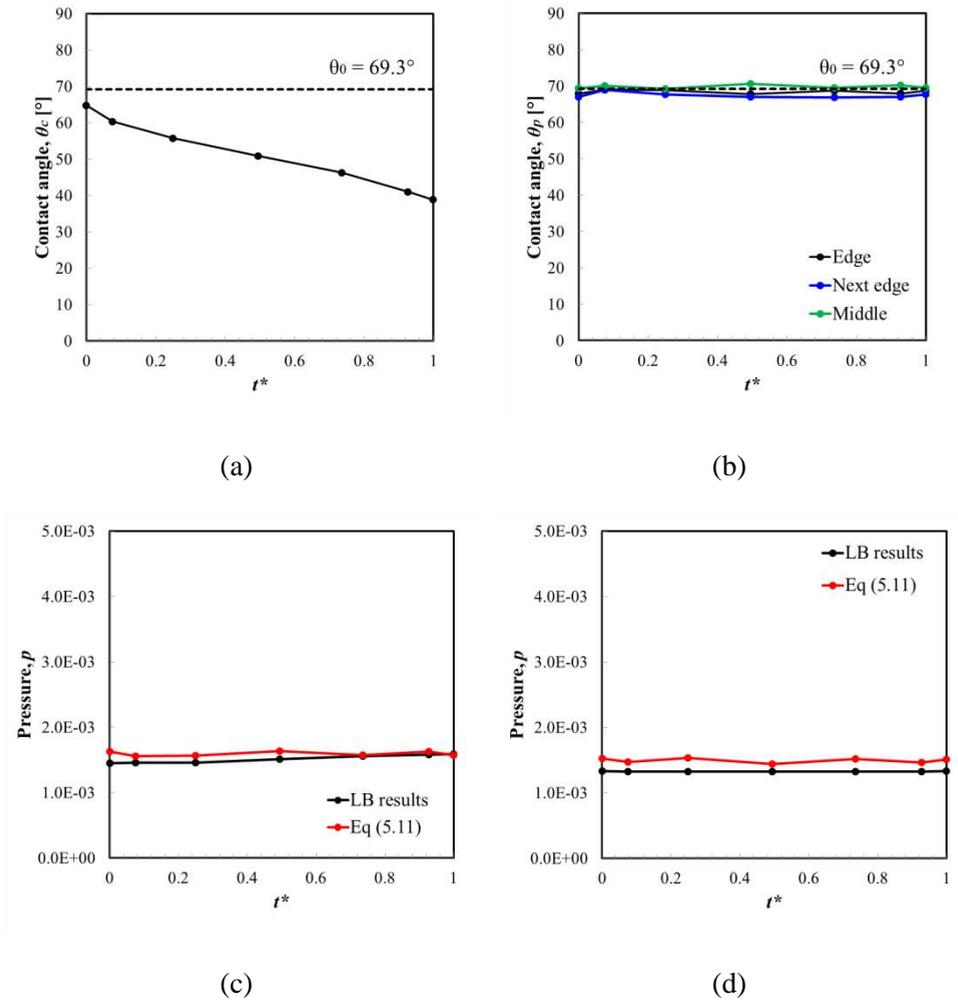

(a)                                                    (b)

(c)                                                    (d)

Fig.5.25. Temporal evolution of (a) contact angle of the droplet; (b) contact angle of meniscus in edge, next to edge and middle capillary; and (c - d) capillary pressure in edge and middle capillaries, respectively. Pitch width is 7 lattices.

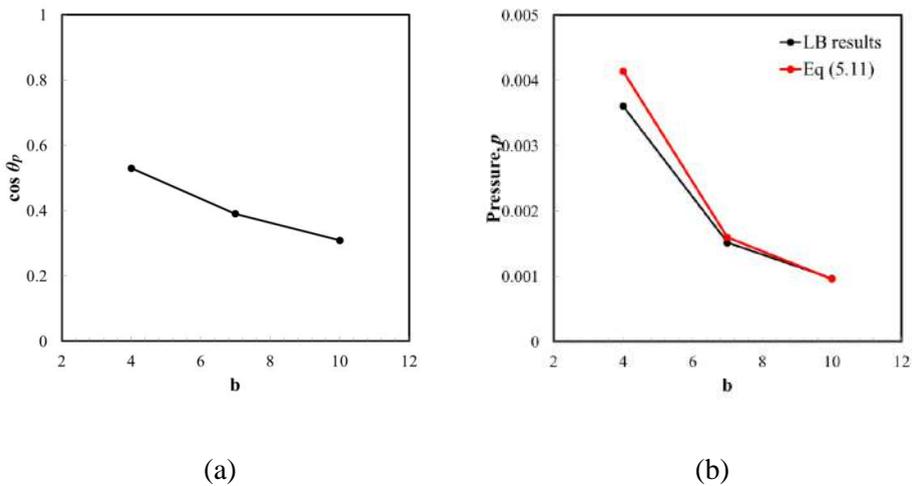

(a)                                          (b)

Fig.5.26. (a) Average contact angle in the capillary versus pitch width and (b) capillary pressure versus pitch width and comparison of LB results with Laplace law (Eq. (5.11)).

The capillary pressure has been determined as the pressure difference between the gas and liquid phases $p_c = p_G - p_L$ at opposite sides of the meniscus. The contact angle $\theta_p$ is the determined using the method LB-ADSA in Image J (Stalder, Melchior et al. 2010). The temporal evolution of the capillary contact angle and the capillary pressure is given in (Fig.5.25 (b) and Figs.5.25 (c - d) respectively, for a pitch width of 7 lattices. More complete results for the other pitch widths are given in the Addendum A2 'Contact angle in the capillaries between the micropillars' and A3 'Capillary pressures at meniscus inside capillaries'. For comparison, Fig.5.25 (a) gives the temporal evolution of the contact angle of the droplet. It is observed that in contrast to the contact angle of the droplet, the contact angle $\theta_p$ in the capillary remains almost constant over time. The contact angle is similar for all investigated capillaries from edge to middle, for a given micropillar geometry. The average contact angle in the capillaries as function of the pitch width is given in Fig.5.26 (a). It is found that the contact angle in the capillary increases with increasing pitch width, or equivalently $\cos\theta_p$ decreases with increasing pitch width. The contact angle in the capillaries is found also to be slightly from the equilibrium contact angle of 69.3°, as imposed in LBM. The dependence of the contact angle $\theta_p$ in the capillary on the pitch width will be further explained below.

The capillary pressure at the meniscus in the capillary is given by Laplace law:

$$p_c = p_G - p_L = \frac{\gamma \cos \theta_p}{b/2},$$
(5.11)

The capillary pressure as determined from the LB results is plotted in Fig.5.26 (b) and compared with Eq. (5.11). The capillary pressure as expected decreases with increasing pitch width, and a good agreement with Eq. (5.11) is obtained, validating LBM. Fig.5.27 gives snapshots of the fluid flow inside the droplet and of the vapor flow at the droplet surface, representing velocity streamlines at three different points of time for pitch widths of 7 and 10 lattices. An internal fluid flow in the droplet is observed moving from the evaporating droplet surface towards the capillaries below. The evaporative vapor flux at the droplet surface is also shown and indicates a singularity at the triple point of the droplet. Due to this high evaporative flux, the fluid flow shows a vortex close to the triple point as shown in more detail in the enlarged snapshots in Figs.5.28 (a - b). The high vapor flux at the triple point increases with time, due to the decrease in droplet contact angle. This evaporative flux has to remain compensated by the internal fluid flow in the droplet and capillaries. When the evaporative flux at the triple point becomes very large, the liquid from the capillary at the edge will also start to flow upwards to the triple point to maintain mass conservation. This means that the flow in the edge capillary, showing initially a downwards flow, may change flow direction from downwards to upwards. This is clearly demonstrated when comparing the snapshots of Figs.5.27 (a) to (b) and 5.28 (a) to (b). Fig.5.27 (d) shows the average velocity in the edge capillary versus time. Negative velocities point to a downwards flow and positive values to an upwards flow. The velocity is initially downwards (negative) for some time, but then starts to decrease and finally becomes positive at the end of the pining process. When the internal flow in the droplet and the edge capillary cannot compensate anymore for the high evaporative flux at the triple point, the droplet will depin, trying to find another equilibrium state.

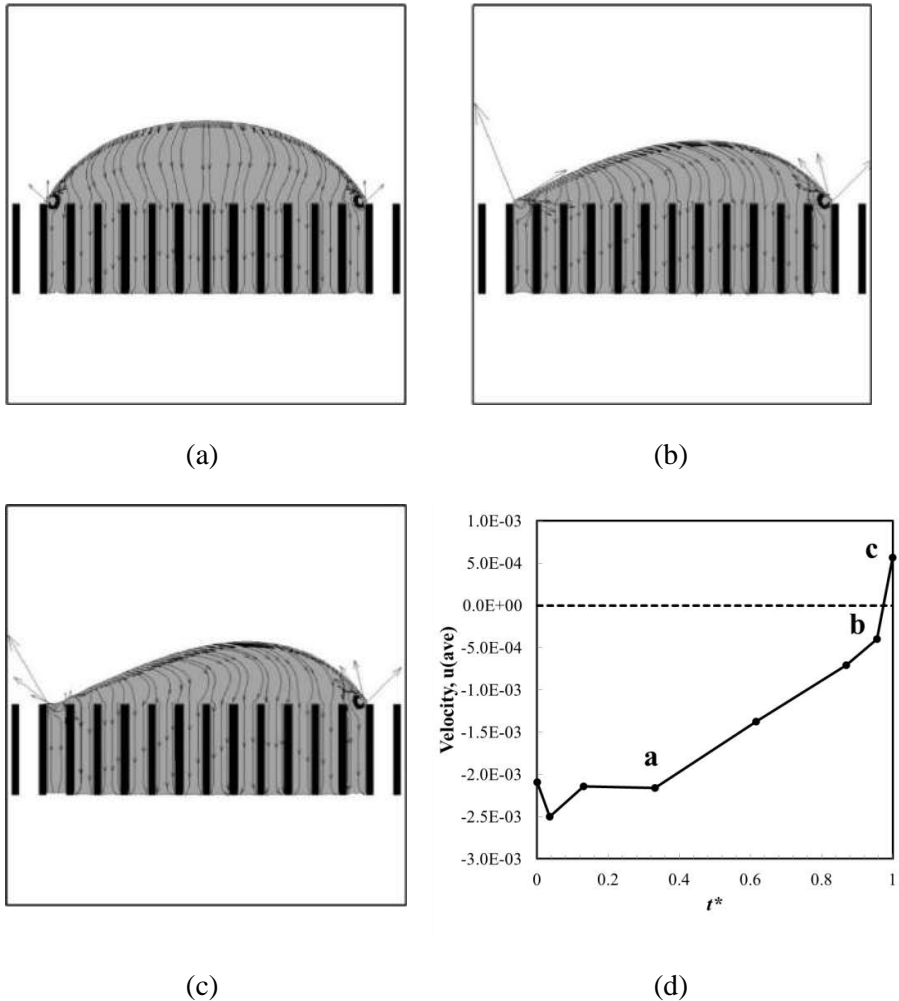

(a)                                        (b)

(c)                                        (d)

Fig.5.27.A (10 lattices). (a - c) Snapshots of velocity streamlines at different times and (d) average velocities inside the capillary at the edge for a pitch width of 10 lattices.

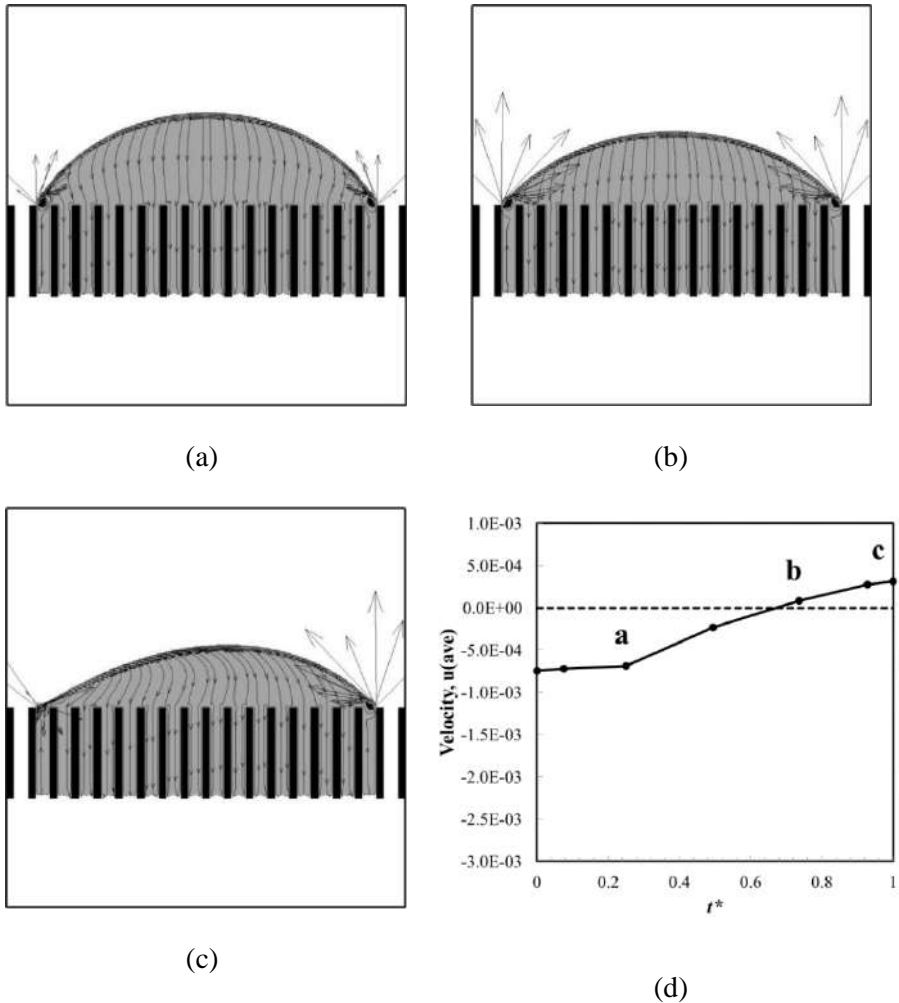

(a)                                              (b)

(c)

(d)

Fig.5.27.B (7 lattices). (a - c) Snapshots of velocity streamlines at different times and (d) average velocities inside the capillary at the edge for a pitch width of 7 lattices.

Fig.5.28 (left) gives the temporal evolution of the fluid pressure in the top (solid line) and bottom (dashed line) part of the edge capillary. The liquid pressure at the bottom of the capillary remains quite constant over time, while the pressure at the top decreases continuously, finally leading to a change in flow direction. More complete results of liquid pressures at different positions in the droplet and capillaries for the other pitch widths are given in Addendum A4'<u>Liquid pressure in the droplet</u>'.

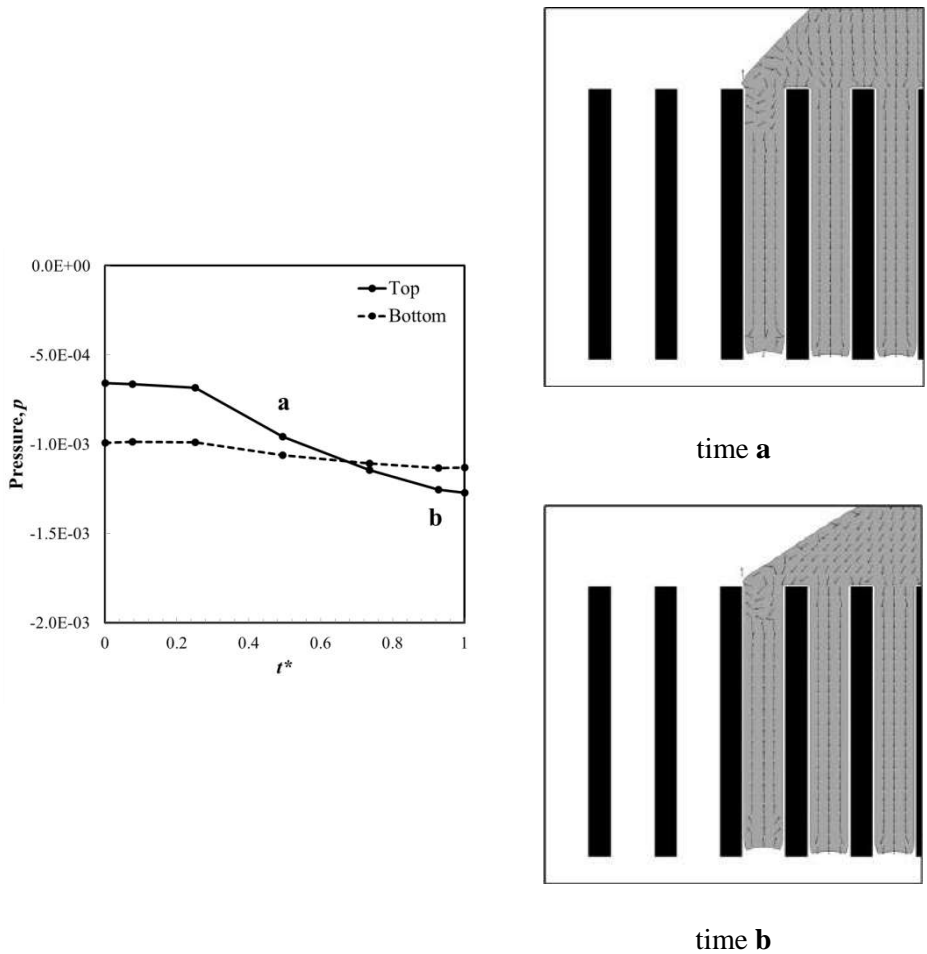

time **a**

time **b**

Fig.5.28. (Left) Evolution of liquid pressure in edge capillary with pitch width of 7 lattices. (Right) Velocities streamlines in the fluid at the points of time **a** and **b** as indicated in the left figure.

The average velocity in the capillary can be determined assuming fully developed Poiseuille flow:

$$u = k\, \frac{p_{L,bottom} - p_{L,top}}{b}\,, \tag{5.12}$$

with $p_{L,bottom}$ and $p_{L,top}$ the liquid pressure at the bottom and top of the capillary, respectively. The permeability $k$ follows the cubic law:

$$k = \frac{b^3}{12\,\eta\,l}\,, \tag{5.13}$$

with $\eta$ the viscosity. The pressures $p_{L,bottom}$ and $p_{L,top}$ are obtained from LB results. Then the velocity is determined using Eq. (5.12).

Fig.5.29 gives the average velocity versus time inside the capillary at the edge, the one next to the edge and in the middle capillaries for pitch widths of 4, 7 and 10 lattices. In general, a very good agreement is obtained between the velocity obtained directly from LB results and the velocity determined by the cubic law (Eq. (5.12)), validating again LBM.

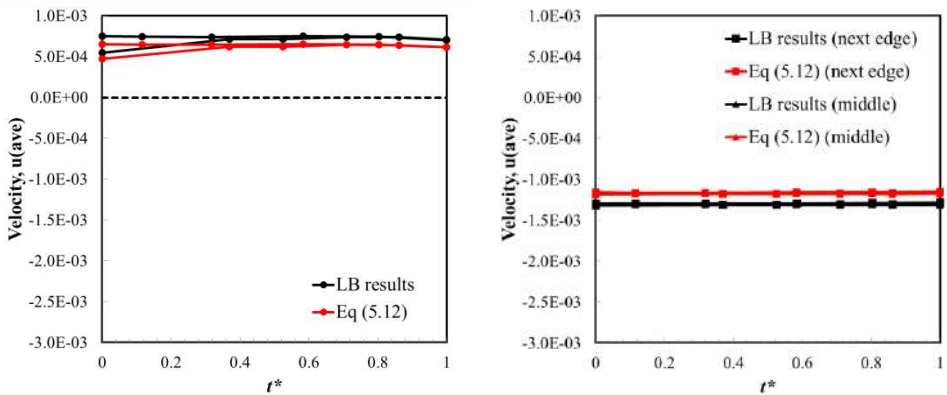

(a) b = 4 lattices

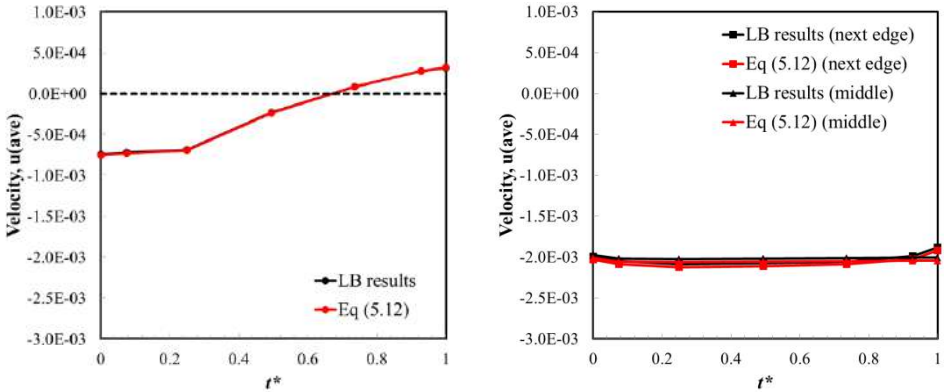

(b) b = 7 lattices

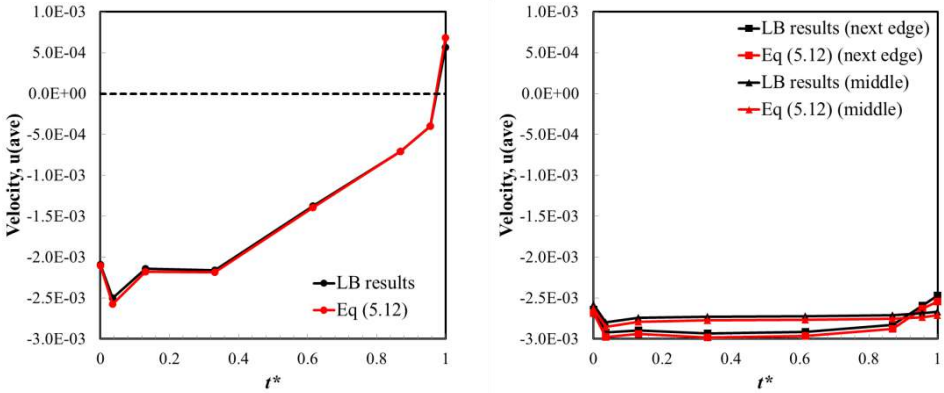

(c) b = 10 lattices

Fig.5.29. Average velocities versus time inside the capillaries at the edge (left column), and in the one next to the edge and in the middle (right column) for pitch widths of (a) 4; (b) 7; and (c) 10 lattices.

It is observed that the velocity in the middle capillary and capillary next to the edge remains almost constant over time. The velocity is negative showing that the flow is downwards from the droplet through the capillaries to the meniscus, where evaporation occurs.

The time evolution of the flow in the capillary at the edge is more complex. For a small pitch width, the velocity is positive all time, meaning it is upwards towards the triple point. For higher pitch widths, the average velocity is initially downwards (negative) and then reduces until it changes direction before the droplet depins. As noted above, the change in flow direction in the edge capillary is explained by the large evaporation flux at the triple point due to a singularity (as explained in section 2.2.4) forcing an upwards flux.

The difference in fluid velocities in the capillaries with different pitch widths may help to explain the dependence of the contact angle $\theta_p$ in the capillaries on pitch width as observed in Fig.5.26 (a). As shown in the schematic in Fig.5.30, due to contact angle hysteresis, the contact angle will be lower when the flow is from the gas to the liquid phase (top figure in Fig.5.30). The contact angle will be larger when the flow is from the liquid to the gas phase (bottom figure). This is exactly what is observed in Fig. 5.26. When the flow is upwards, i.e. when the pitch width is smaller, the flow is from gas to liquid phase and the contact angle is smaller. The contact angle is larger when the flow is from the liquid to the gas phase, or in case of micropillars when the liquid flow is downwards, which occurs preferentially for large pitch widths.

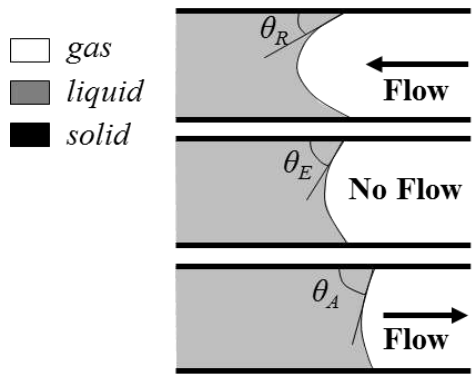

Fig.5.30. Schematic of contact angle hysteresis in a capillary dependent on the flow direction from gas to liquid phase.

It is concluded that the singularity of the evaporative flow at the triple point of the droplet interface highly determines the depinning of the droplet, since it drives the internal liquid flow in the droplet and capillaries. With decreasing contact angle of the droplet as droplet evaporation goes on, the evaporative flux at the triple point increases very fast, creating a vortex at the triple point which is fed by flow from the surrounding droplet and the capillary at the edge. The liquid flux in the edge capillary may change direction due to this high evaporative flux from downwards, to pointing towards the triple point. Due to this internal fluid flow in the capillary, also the contact angle at the meniscus of the capillary changes depending on the change of flow direction. Once the internal liquid flow cannot compensate for the high evaporative flux in the triple point, the droplet will depin. Since the compensating liquid flow in a large capillary will be larger compared to a small capillary, the droplet on a set of micropillars with large pitch width will depin later leading to a smaller critical contact angle. This observation explains why the critical contact decreases with the pitch width between the pillars.

## 5.4. Droplet on heterogeneous surface

In this section, the local contact angle of a droplet deposited gently on a surface with heterogeneous wettability is investigated using 3D LBM. Commonly, the apparent contact angle of a heterogeneous surface is described by Cassie equation (Eq. (2.5)). The assumption of this equation is that the patches with different wettability or contact angle are sufficient small compared to the droplet diameter. The apparent contact angle can then be obtained considering a small displacement of the contact line crossing several patches in relation to the surface fraction of both patches. However, this assumption breaks down when the size of the patches increases. In this section, the apparent contact is analyzed for a checkerboard surface with hydrophilic ($\theta = 23°$) and hydrophobic ($\theta = 121°$) patches for different droplet and patch sizes.

### 5.4.1. Simulation set-up and boundary conditions

The surface has a heterogeneous regular checkerboard pattern with patch size $a$ (Fig.5.31). The patches are alternating hydrophilic or hydrophobic with contact angles of 23° and 121° or solid-liquid interaction parameter $w$ of −0.1 to 0.05. Two domain sizes are used: $300 \times 300 \times 300$ lattice$^3$ for a droplet radius of 50 or 100 lattices and a domain of $400 \times 400 \times 400$ lattice$^3$ for a droplet radius of 150 lattices. The droplet radius refers to the initial radius of a hemisphere droplet before deposition. Bounce-back boundary conditions are imposed on top and bottom sides, while the other sides are treated as periodic boundary conditions. A hemisphere droplet with radius $R$ is initially located in the middle of the bottom surface.

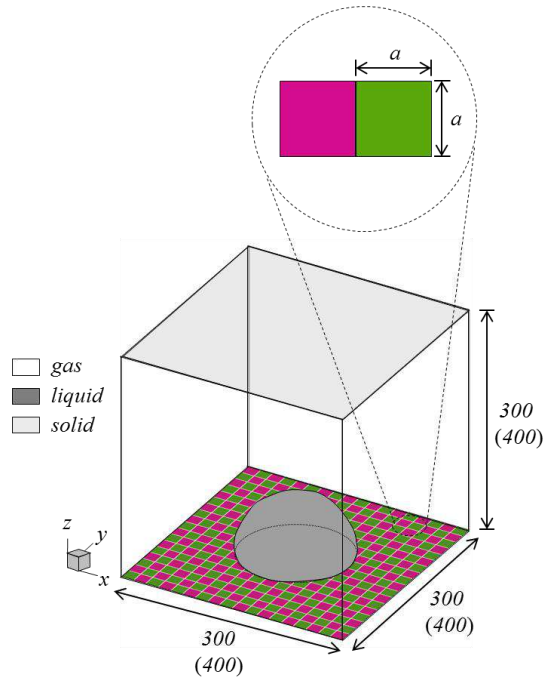

Fig.5.31. Schematic geometry of droplet on heterogeneous surface with patch size $a$. The patches are alternating hydrophilic (magenta) and hydrophobic (green) with contact angles of 32° and 125° or solid-liquid interaction parameter $w$ of −0.1 to 0.05.

The droplets in this study have a radius $R$ of 50, 100 or 150 lattices. The densities of liquid and gas are 0.28 and 0.0299 lattice units, respectively corresponding to a density ratio $\rho/\rho_c = 9.4$ at $T/T_c = 0.85$. To reach equilibrium state, all cases are run for 50 000 time steps. All numerical simulations are run by parallel computing based on Message Passing Interface (MPI) on the high performance computing cluster of Los Alamos National Laboratory (LANL). The cluster aggregate performance is 352 TFlop/s with 102.4 TB of memory for 38 400 cores. Each simulation is run on 1 000 (10 × 10 × 10) or 4 000 (20 × 20 × 10) processor cores for two different domain sizes and requires 16 hrs to run 50 000 time steps.

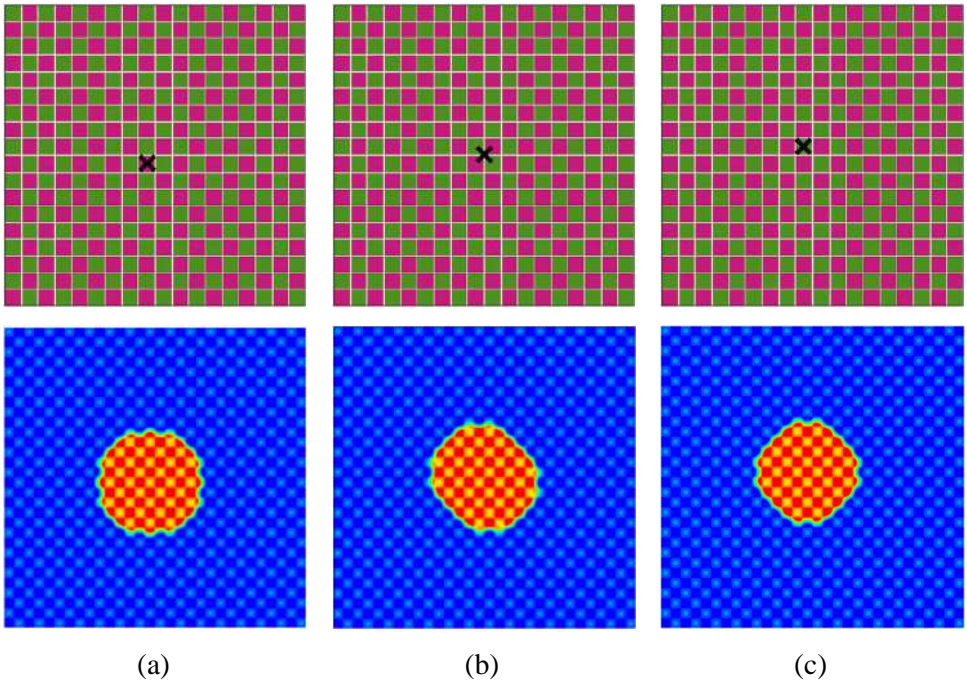

|          (a)          |          (b)          |          (c)          |

Fig.5.32. Illustration of the effect of the deposition of the center of droplet at different locations: (a) on center of hydrophilic patch (magenta), (b) on the crossing between hydrophilic and hydrophobic patches and (c) on the center of the hydrophobic patch (green) for a droplet radius of 50 lattices and patch size of 10 lattices. The bottom row shows the different attained shapes of the droplet contact area.

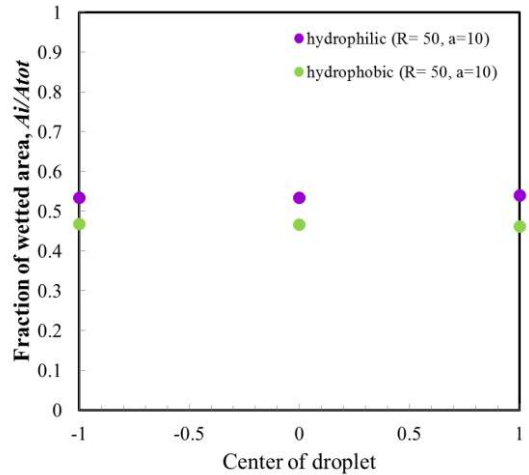

Fig.5.33. Fraction of the wetted area on hydrophilic or hydrophobic patches for different initial locations of the deposition of the center of the droplet: on center of hydrophilic patch (-1), on crossing of hydrophobic and hydrophilic patches (0) and on the center of the hydrophobic patch (1).

It was found that the droplet attained a different final shape whether it was deposited with its center aligned with  (a) the center of a hydrophilic patch, (b) the crossing between hydrophilic and hydrophobic patches or (c) the center of a hydrophobic patch (as illustrated in Fig.5.32 top row). Depending on the case of initial position, the droplet will attain different shapes of the contact area at equilibrium. (Fig.5.32). However, the fraction of wetted area, defined as the ratio of wetted area on hydrophilic area (or hydrophobic area) to total wetted area $A_i/A_{tot}$, was determined and the results in Fig.5.33 show that no significant difference exists among the three different cases. In the further analysis, the droplet is deposited on the crossing between hydrophilic and hydrophobic patches.

### 5.4.2. Results

Six patch sizes and three droplet radii are considered, resulting in 18 cases. Fig.5.34 shows the droplet shape at equilibrium and the wetted surface for the six different patch sizes of 3, 5, 10, 25, 50 and 150 lattices. The droplet radius equals 50 lattices, resulting in ratios of patch size versus droplet radius $a/R$ ranging from 0.02 to 3. In the figure, a hydrophilic patch covered with gas is shown in light blue and with liquid in red while a hydrophobic patch covered with gas is shown in dark blue and with liquid in yellow color.

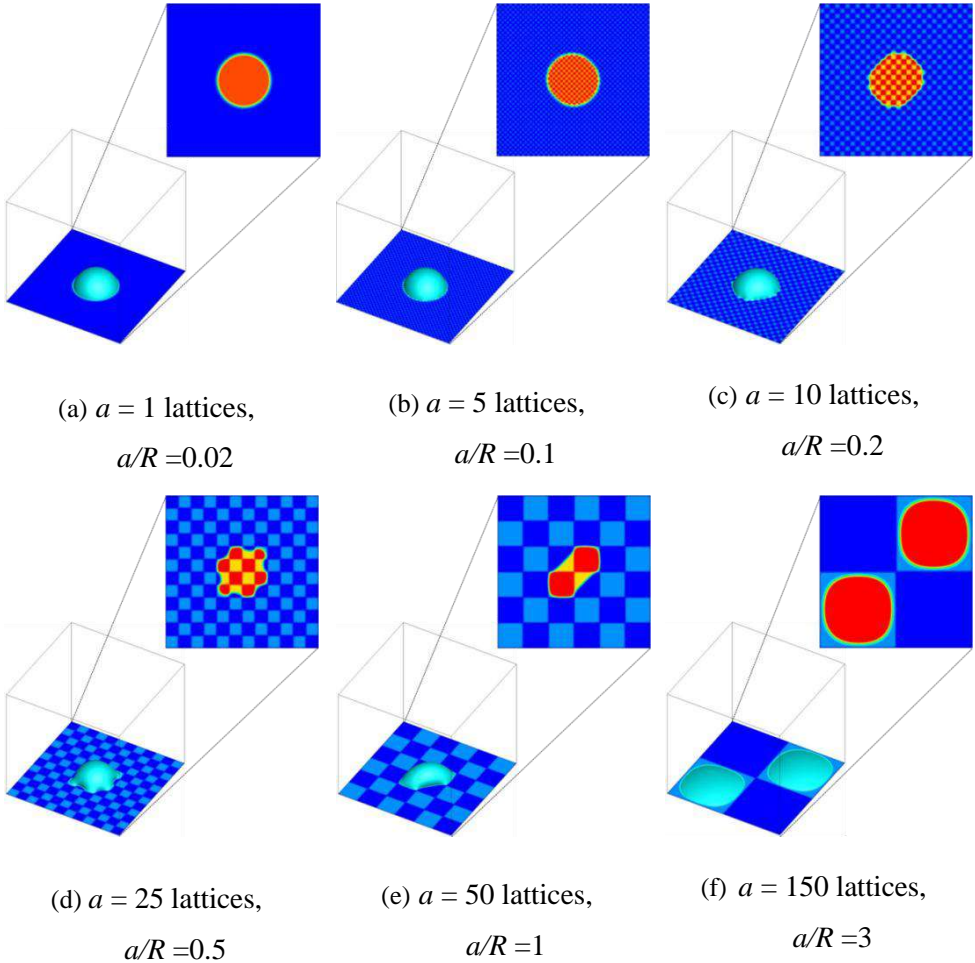

(a) $a = 1$ lattices, $a/R = 0.02$

(b) $a = 5$ lattices, $a/R = 0.1$

(c) $a = 10$ lattices, $a/R = 0.2$

(d) $a = 25$ lattices, $a/R = 0.5$

(e) $a = 50$ lattices, $a/R = 1$

(f) $a = 150$ lattices, $a/R = 3$

Fig.5.34. Droplet shape and wetted pattern (contact area) with droplet radius $R$ of 50 lattices on checkboard surface with different patch sizes $a$, resulting in different ratios $a/R$. A hydrophilic patch covered with gas is light blue and with liquid red while a hydrophobic patch covered with gas is dark blue and with liquid yellow.

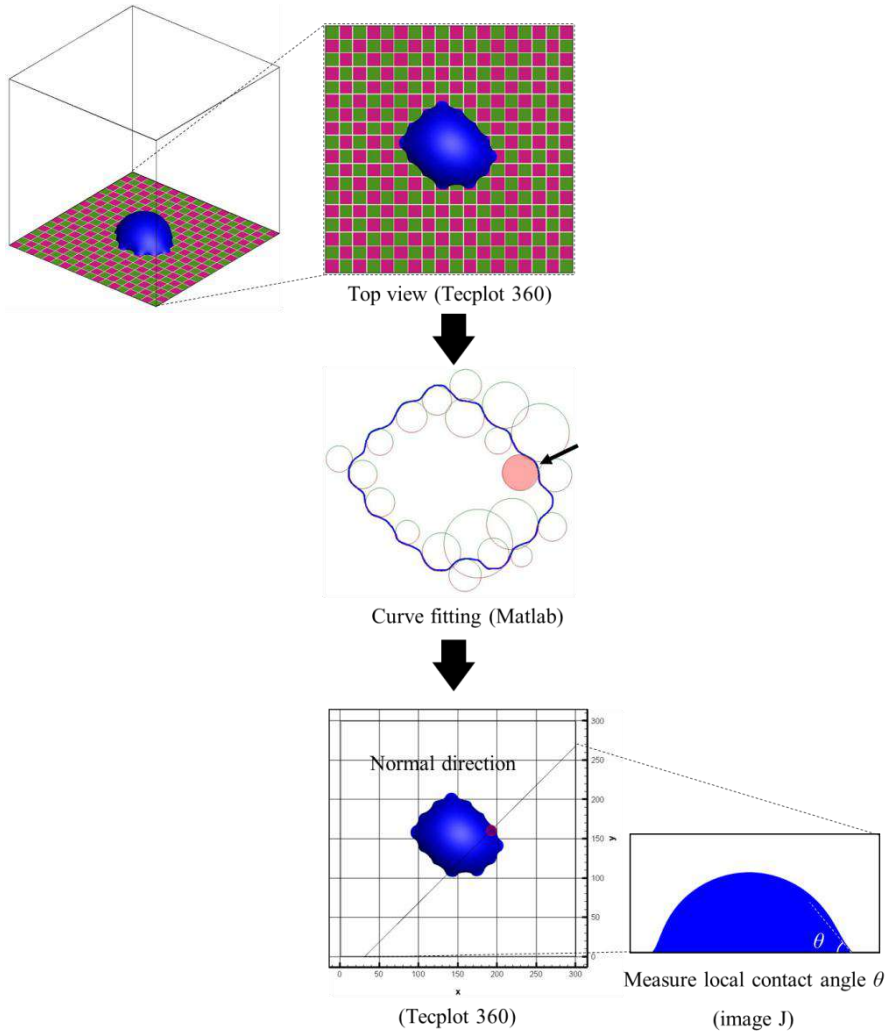

Top view (Tecplot 360)

Curve fitting (Matlab)

Normal direction

(Tecplot 360)

Measure local contact angle $\theta$

(image J)

Fig.5.35. Schematic representation of the procedure to determine the local contact angle of the distorted droplet at equilibrium state on a heterogeneous surface. The contact line is extracted using own FORTRAN code. Then circles are fitted to the contact line to determine its curvature using a own Matlab procedure. Based on these circles, vertical planes lying normal to the contact line are determined using Tecplot 360. Finally, the local contact angle is determined on the cross section of the droplet at the said vertical plane using the method LB-ADSA in Image J (Stalder, Melchior et al. 2010).

The droplet shows a tendency to wet the hydrophilic patches preferentially (red) resulting in a distortion of the hemispheric droplet. The degree of distortion increases with increasing patch size as shown in Fig.5.34 (a) - (e). When the patch size is three times larger than the droplet radius, $a/R = 3$ (case (f) in Fig.5.34), the droplet breaks into two parts and wets only the hydrophilic patches. In contrast, at significantly smaller patch size than droplet radius, ratio $a/R = 0.02$, the droplet shows a hemispherical shape, as described by the Cassie state (case (a) in Fig.5.34).

The local contact angle of the distorted droplet is determined in three steps as shown in Fig.5.35. At equilibrium state, the contact line of bottom droplet is determined by Tecplot 360 and an in-house code written in FORTRAN 90. Knowing the contact line, the distorted curve for each patch is fitted by a circle using MATLAB (see the middle schematic of Fig.5.35) and the radii and center of these circles are determined. From this information, a vertical plane lying normal to each circle can be determined, which then is used to determine the cross section of the droplet using Tecplot 360 (see the bottom of schematic of Fig.5.35). Finally, the local contact angle is determined using the method LB-ADSA in Image J (Stalder, Melchior et al. 2010). The location of the local contact angles of along the curved contact area line is given an angular coordinate in the anticlockwise direction, as shown in Fig.5.36.

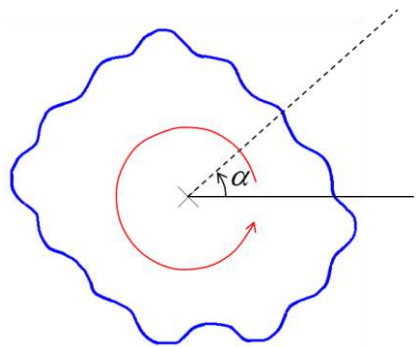

Fig.5.36. Schematic of the anticlockwise angle $\alpha$ used to provide an angular coordinate for the location of each determined local contact.

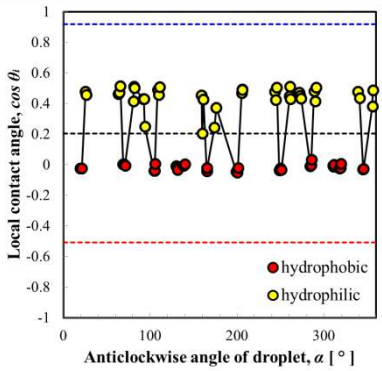

(a)  *a* = 5 lattices, *a/R* =0.1

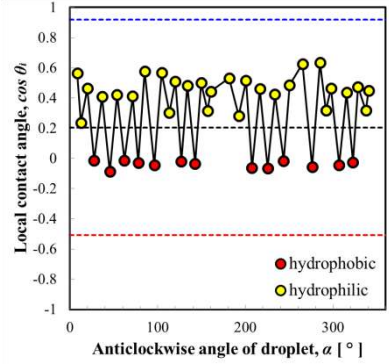

(b)  *a* = 10 lattices, *a/R* =0.2

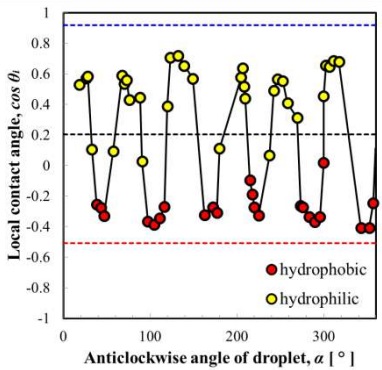

(c)  *a* = 25 lattices, *a/R* =0.5

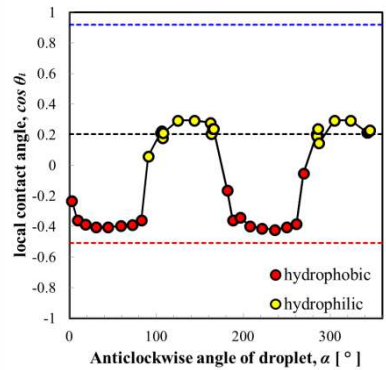

(d)  *a* = 50 lattices, *a/R* =1

Fig.5.37. Cosine of contact angle cos $\theta_i$ plotted in function of the anticlockwise angle $\alpha$ for a droplet of radius of 50 lattices. Also represented values are the cosines of the contact angles of a hydrophilic patch of 23° (blue dashed line), a hydrophobic patch of 121° (red dashed line) and  the apparent contact angle of 78° (black dashed line) according to Cassie's equation.

Fig.5.37 shows the cosine of the local contact angles on hydrophilic (in yellow) and hydrophobic (in red) patches at regular interval along the contact area line for four ratios of *a/R*, as no curvature is observed for the lowest patch size of 3 lattices. The cosines of the contact angle for hydrophilic ($\theta = 23°$) and hydrophobic ($\theta = 121°$) surfaces are also shown, as well as the cosine of the contact angle predicted by Cassie's equation. Remark that the surface fraction of hydrophilic and hydrophobic surfaces is equal: $f_1 = f_2 = 0.5$. The apparent contact angle as predicted by Cassie's equation equals then 78°. It is observed that the local contact angles vary between two limits. The maximum limit of the local contact angle attained on hydrophobic patches is higher than Cassie's contact angle, while the minimum limit attained on hydrophilic patches is lower than Cassie's apparent contact angle. The maximum and minimum contact angle limits depend on the patch size. The maximum limit of the local contact angle decreases with patch size on hydrophobic patches and inversely, the minimum limit of the local value on hydrophilic patches increases with path size, resulting in smaller interval between the limits of the local angles with decreasing patch size. From Figs.5.37 (a) and (b), it is noted that the local contact angle limits on hydrophilic and hydrophobic patches are not symmetrically spread around Cassie's contact angle: i.e. the maximum limit is closer to Cassie's contact angle, while the minimum limit is further away from Cassie's angle. This asymmetry will be discussed in more detail below. In Fig.5.37 (d), it is shown that the contact angles for $a/R = 1$ follow a strange behavior. The reason is that in this case the local contact angles are difficult to determine accurately for all positions around the contact line. As seen in Fig.5.34 (e) for $a/R = 1$, the liquid droplet, especially at the contact line, shows a quite particular shape. The droplet shape on the hydrophilic patch is mainly determined by the fact that the droplet is forced to remain in between the borders of the hydrophilic square patch and does not display much curvature. As a consequence, the analysis in the following will remain limited to ratios of patch size versus droplet diameter smaller than 0.6.

Fig.5.38 shows the maximum and minimum limits of the local contact angle as function of the ratio of patch size versus droplet diameter $a/R$. All data for the different droplet radii of 50, 100 and 150 lattices are included. It is interesting to note that the droplets with different radii show a similar behavior when the maximum (minimum) limits of the local contact angle are plotted versus the ratio *a/R*. With decreasing ratio *a/R*, the limits of the local contact angle converges to the Cassie's contact angle. With increasing ratio *a/R*, the local contact angle limits diverge from the Cassie value. Again it is observed that the curves for hydrophilic and hydrophilic patches are asymmetrically located around the Cassie's contact angle.

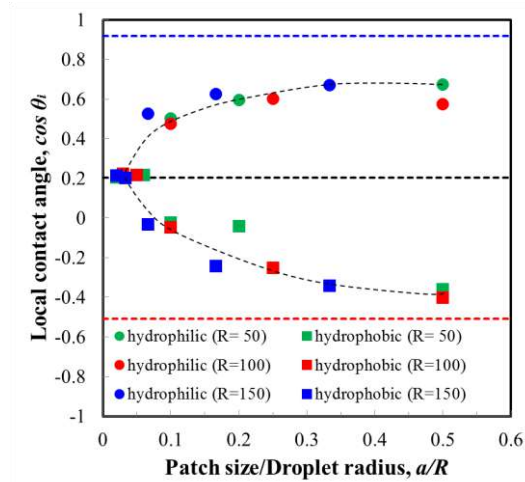

Fig.5.38. Cosine of the local contact angle limit versus ratio *a/R* between patch size *a* and droplet radius *R* of 50, 100 and 150 lattices. The cosines of the contact angles of hydrophilic patch of 23° (blue dashed line), hydrophobic patch of 121° (red dashed line) and of the apparent contact angle of 78° (black dashed line) according to Cassie' equation are also shown.

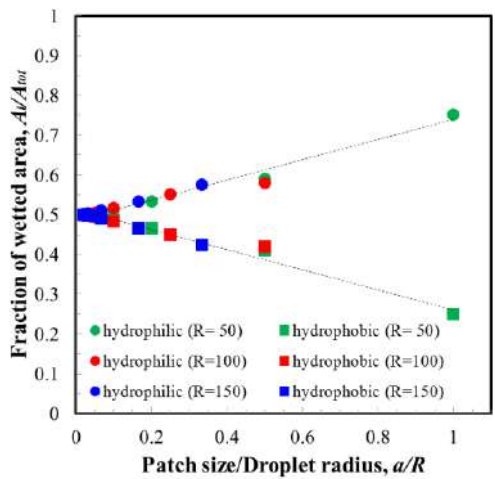

Fig.5.39. Fraction of wetted area on hydrophilic and hydrophobic patches versus ratio between patch size and droplet radius of 50, 100 and 150 lattices.

Fig.5.39 shows the fraction of wetted area as a function of the ratio of patch size to droplet radius $a/R$. The fraction of wetted area is defined as the ratio of wetted area on hydrophilic area (or hydrophobic area) to total wetted area $A_i/A_{tot}$. It is observed that all results for the different droplet radii coincide into a single linear curve versus $a/R$. The results show that when the patch size increases, the droplet will preferentially wet the hydrophilic patches, resulting in a larger wetted area fraction for the hydrophilic patches. When the ratio $a/R$ is very small, the fraction attains a value of 0.5, meaning that both surfaces are wetted according to their fraction $f_1 = f_2 = 0.5$. This observation is in agreement with the assumption of Cassie's equation, where it is assumed that a surface of patches is wetted equal to their fraction, when the droplet is sufficient large compared to the patch size.

Fig.5.40 gives the fraction of wetted area on hydrophilic or hydrophobic patches versus cosine of the local contact angle limit for droplet radii of 50, 100 and 150 lattices. This figure is obtained combining Figs. 5.38 and 5.39. The results for different droplet radii coincide into a single curve, which is asymmetric with respect to the hydrophobic and hydrophilic regions. In the hydrophobic region, the cosine of the local contact angle increases gradually attaining Cassie's contact angle at a fraction of wetted area at 0.5. In the hydrophilic region, the cosine of the local contact angle remains quite long almost constant and only decreases rapidly when attaining the fraction of wetted area of 0.5 at the Cassie's contact angle. This asymmetric curve may be explained by the fact that wetting on hydrophilic patches is from an energetic point a much more favorable process than the wetting of hydrophobic patches driven by the wetting of the hydrophilic patches next to them. In conclusion, it is clearly shown that the ratio between patch size and droplet radius $a/R$ is a key parameter to define droplet spreading on checkerboard heterogeneous surface. It is also shown that the Cassie's equation is a limit only attained when the ratio $a/R$ is very low, meaning the patch size has to be much smaller than the droplet radius.

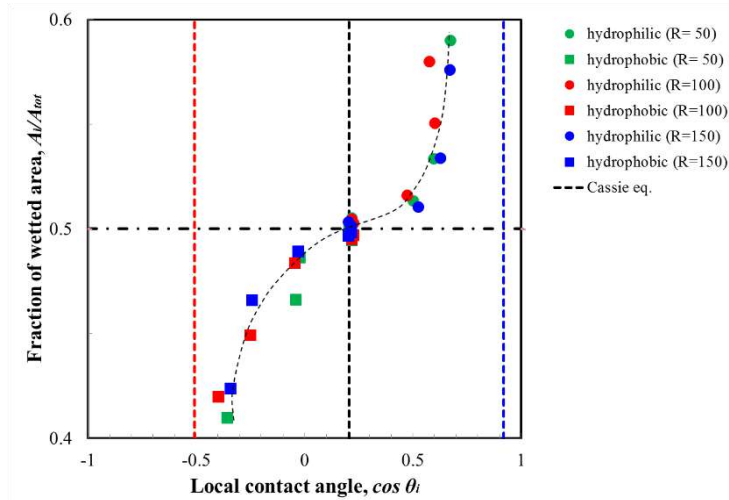

Fig.5.40. Fraction of wetted area on hydrophilic or hydrophobic patches versus the limit of the local contact angle for droplet radii of 50, 100 and 150 lattices with represented contact angle from Cassie's equation (black dashed line) and fraction of wetted area of 0.5 (black dashed dotted line).

# 5.5. Conclusion

Diverse droplet phenomena, similar to these that may occur in PA have been studied with 2D and 3D LBM. For the study of droplet run-off on the surface with groove, four different regimes, namely (1) height controlled regime; (2) bottom surface controlled regime; (3) top-bottom surface controlled regime; and (4) top surface controlled regime, are identified depending on the size of the groove. Furthermore, the effects of solid-liquid interaction parameters of the surface and groove on the run-off of the liquid droplet as well as on the remaining liquid fraction in the groove are analyzed.

The evaporating droplet on the set of micropillars is studied with 2D LBM varying pillar and pitch widths. First, part of the droplet is taken up by capillarity in the space between the micropillars, followed by a stick-slip behavior, showing alternatingly constant contact radius (CCR) and constant contact angle (CCA) modes. The droplet is found to depin at a critical contact angle or at a maximum excess Gibbs free energy, when there is sufficient energy available to overcome an energy barrier, resulting in a move of the triple line to a new equilibrium position. Depinning is found to occur when the internal liquid flow from the bulk of the droplet and from the capillaries cannot compensate anymore for the high evaporative flux at the triple point. The width of the pitch is found to play a significant role in the control of the pinning/depinning cycles, the maximum excess Gibbs free energy and critical contact angle.

Finally, the wetting of a droplet on a checkerboard heterogeneous surface with regular hydrophilic and hydrophobic patches is studied. Different patch sizes and droplet radii are considered to investigate the effect of ratio between patch size and droplet radius on local contact angle at equilibrium state. The local contact angle on hydrophilic (or hydrophobic) patch increases (or decreases) with decreasing patch/droplet ratio. Furthermore, more hydrophilic patches (and less hydrophobic patches) are wetted with increasing ratio since the droplet sits preferentially on the hydrophilic patches. Furthermore, the results for different droplet radii for the relationship between local contact angle limit and wetted area coincide into single S-shape curve, which is asymmetric with respect to hydrophobic and hydrophilic regions.

By considering different droplet wetting and movement on structured surfaces, this study displays the varied behavior of droplets that could all be found when considering a complex material like PA. The LBM study continues by considering cases of liquid uptake by capillaries.

# 6. SIMULATION OF GRAVITY-DRIVEN DRAINAGE IN QUASI-2D POROUS ASPHALT (PA) AND COMPARISON WITH EXPERIMENT

## 6.1. Introduction

In this chapter, gravity-driven drainage in PA is studied with 2D and 3D LBM and compared with experimental data using a model PA sample. After adding a body force in the LBM as shown in section 5.2 and looking at droplet flow over grooves, the aim here is to study gravity-driven drainage in the complex pore system of PA using the improved LBM. To this aim, an experimental dataset documenting gravity-driven drainage in a micro-channel device is described in section 6.2. The device is manufactured by 3D printing based on a slice from a micro tomographic dataset of a PA sample. In the following sections 6.3 and 6.4, 2D and 3D gravity-driven drainage is simulated and both LB results are compared with experimental data qualitatively in terms of water configuration and quantitatively in terms of distribution of fraction of mass inside PA versus time. Finally, whether LB results can be verified and validated by comparing with these experimental data is discussed.

## **6.2. Gravity-driven experiment**

Given the need for a gravity-driven drainage dataset documenting water configuration during drainage, different imaging options were considered to capture the water configuration. As neither X-ray nor neutron tomography options could not be fast enough to capture water flow in real PA, the option to use normal photography in a quasi-2D transparent macro porous medium in which water could drain was retained.

The gravity-driven drainage experiments were performed by Dr. Manuel Marcoux at the Institut de Mécanique des Fluides de Toulouse (IMFT, France) on a micro-channel specimen also prepared by Dr. Marcoux. The image and data analysis was performed by the author of this thesis.

### **6.2.1. Sample preparation**

The gravity-driven drainage microfluidic device is built using an image of a real PA geometry with PA11 which has a maximum aggregate size of 11 mm and a porosity of 20%. A sample cut out of a main slab to dimensions of $180 \times 10 \times 30$ mm$^3$ is imaged with the X-ray microcomputed tomography (X-ray μCT) setup of Empa. The setup consists of an X-ray source (X-ray tube ''XT9225-TEP'', Viscom), an XYZ linear stage (composed of three linear stages ''LS-270'', Micos) for positioning the specimen, a rotation table (''UPR-160 F air'', Micos) and an X-ray detector (''XRD 1621 CN3ES'', Perkin-Elmer). The specimen is mounted at a distance of 503.2 mm from the X-ray source and a distance of 467 mm from the detector. From a pixel size of 200 μm and a geometrical magnification of ~1.9, a spatial resolution of 103.7 μm is obtained in the final 3D dataset. The chosen tube parameters are an acceleration voltage of 200 kV and a nominal current of 100 μA. In order to reduce artefacts, the X-ray spectrum is hardened by means of a 1 mm Cu filter. For each scan, a region of interest of 1 000 $\times$ 2 000 pixels is chosen given the elongated geometry of the objects. 720 radiographic images are recorded from different viewing angles distributed over 360° in 0.5° steps. Each image takes 10 seconds and the total scan time is 120 minutes. The X-ray detector is calibrated before the start of the measurement and no additional dark and flat field corrections are necessary. After ring and beam hardening artefact corrections, the 3D spatial distribution of the attenuation coefficient is calculated with an in-house Feldkamp code (Feldkamp, Davis et al. 1984). The preparation of the PA11 specimen and the acquisition of μCT images are performed by the other PhD student of this SNF project, Sreeyuth Lal.

The middle slice of the μCT dataset is selected to generate the center layer of the micro fluidic device. The slice is converted into a binary image with aggregate in black color as shown in Fig.6.1. This information is converted into a STereoLithography (STL) file format readable for printing by additive manufacturing the gravity-driven drainage microfluidic device. HD 3 500 3DSystems printer has a resolution of $33 \times 33 \times 29$ microns. The microfluidic device is composed of two different materials: polydimethylsiloxane (PDMS) in sheets of about 2 mm and 3DSystems Visijet resin which is polymerized by UV during additive manufacturing. The parallel sides of the device are of PDMS to get transparency and allow photographic recording of the water configuration inside the sample over time. The aggregate part in grey color in Fig.6.2 is made of resin and printed by the 3D printer to generate the complex geometry of PA. The two covers and aggregate parts are assembled together, what is the distance between the plates of 1 cm. The two materials are both hydrophobic with contact angles of 102.5° for PDMS and 115° for the resin. These contact angles are measured using sessile water droplets on flat surfaces of each material. To allow liquid draining out the left side of the device, the sample is opened on the top and left sides but the bottom and right sides are closed up with resin.

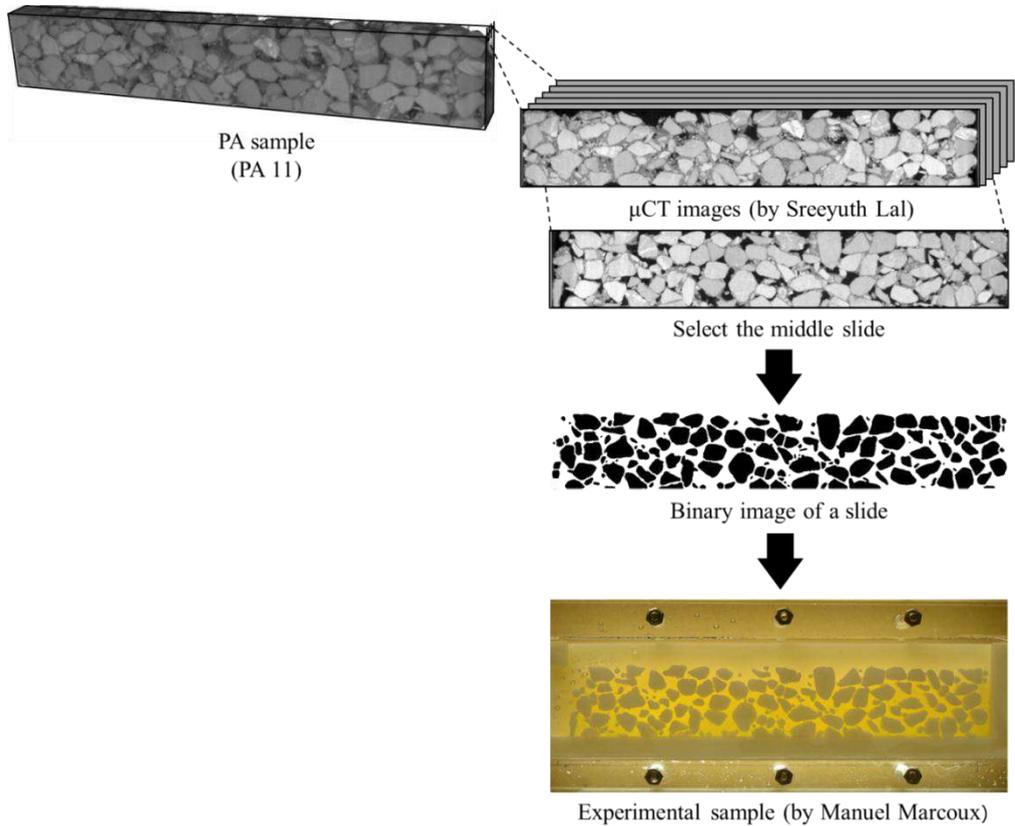

PA sample
(PA 11)

μCT images (by Sreeyuth Lal)

Select the middle slide

Binary image of a slide

Experimental sample (by Manuel Marcoux)

Fig.6.1. Sequence of imaging for microfluidic device: PA sample (PA 11), acquisition of μCT images (Sreeyuth Lal), selection of image, segmentation to binary image, microfluidic pattern (in device prepared by Dr. Manuel Marcoux).

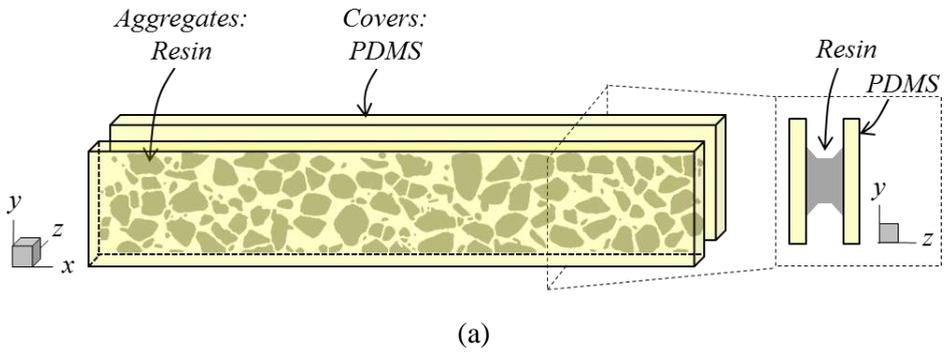

(a)

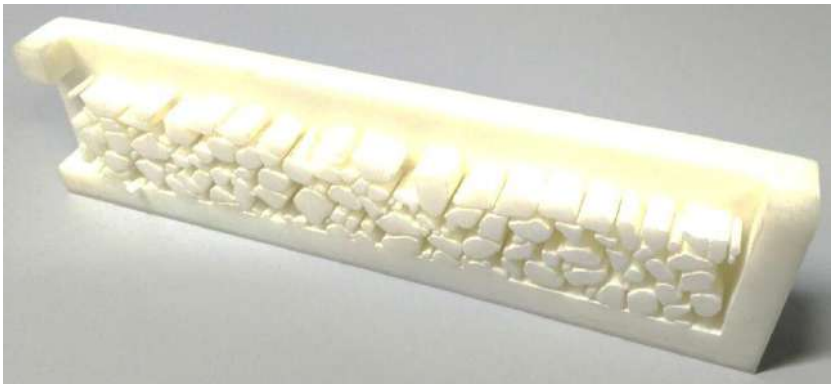

(b)

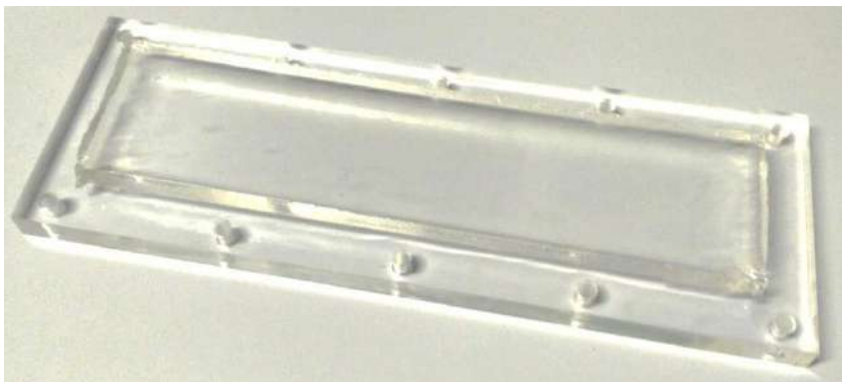

(c)

Fig.6.2. (a) Schematic representation of microfluidic device made of (b) PDMS (aggregates and sides) and (c) resin (transparent side cover).

### 6.2.2. Experimental procedure

The microfluidic device is kept dry until it is mounted on the sample holder. The micro-channel is filled with distilled water dyed with methylene blue injected in the top-right corner until the patterned section is overfilled by liquid. The drainage procedure is documented with digital camera ($3\,680 \times 2\,456$ pixels) taking one image per second. The region of interest of $3\,402 \times 567$ pixels provides a sample size of $180\,000 \times 30\,000\ \mu m^2$. The experiment is finished when water stops draining out from the sample. Drainage is completed in 28 to 33 seconds approximately. The experiment was repeated 2 times, so two drainage sequences are analyzed below.

### 6.2.3. Image processing

The mass of water remaining inside the microfluidic device is directly measured from the images at different times by Image J, as presented next. First, for an image at any time during drainage, the water is segmented by subtracting the dry image from the current image, the difference being the pixels of the image that are water. Multiplying the number of pixels filled with water with the pixel size, the depth of the sample and water density, the total mass of water is obtained:

$$
\begin{aligned}
m[kg] &= total\ \ number\ of\ \ pixels \times \Delta x^2 \times depth \times \rho_l \\
&= total\ \ number\ of\ \ pixels \times \left(52.9 \times 10^{-6}\,m\right)^2 \\
&\quad \times \left(10000 \times 10^{-6}\,m\right) \times 998\left[kg/m^3\right]
\end{aligned}
\tag{6.1}
$$

with a spatial resolution $\Delta x$ of 52.9 μm, a depth of 10 mm and liquid density $\rho_l$ of 998 kg/m$^3$ for room temperature conditions at 20°C approximately.

### 6.2.4. Results

To compare the experimental data with the LB results, the fraction of outflow mass is plotted versus non-dimensional time. The mass fraction $\beta$ is given by the ratio of water mass at any time to the initial mass inside the microfluidic device as follows:

$$
\beta = \frac{m(t)}{m_{initial}} .
\tag{6.2}
$$

The non-dimensional time is defined as the ratio of time to the time when the fraction of outflow is equal to 0.65, which means 65% of liquid remains in the domain as correspond to when the gravity-dominant drainage is finished. The non-dimensional time is expressed as follows:

$$t^* = \frac{t}{t(\beta = 0.65)}$$ 

(6.3)

Fig.6.3 plots the temporal evolution of mass fraction inside the microfluidic device versus non-dimensional time from two series of experimental data. After drainage starts, the liquid flows fast out from the device since gravity effects are dominant. When $t^*$ is between 1 and 3, the drainage continues at a reduced rate and the mass fraction decreases now more gradually. After $t^*$ larger than 3, gravity-driven drainage stops and the mass fraction remains at around 0.4. In Fig.6.4, detailed liquid configurations at three different times from two repeated experiments are given. In all these images, water is seen to leave the patterned area at the bottom left opening. Over time, as the liquid level decreases, the bulk of the liquid remains connected, although some liquid gets trapped on the top surface and in between aggregates. At $t^* = 4$, some liquid still remains inside PA even though the liquid level is higher than the position of the left bottom exit. This can be explained by the fact that the surface tension force of water islands sitting on the hydrophobic surface is so high, that the islands cannot be driven through narrows by gravity force. In the bottom and left side of the experimental domain, the distance between aggregates is so narrow that it blocks liquid overflow after $t^* = 4$. The detailed configuration of liquid will be discussed in following sections by comparing with LB results.

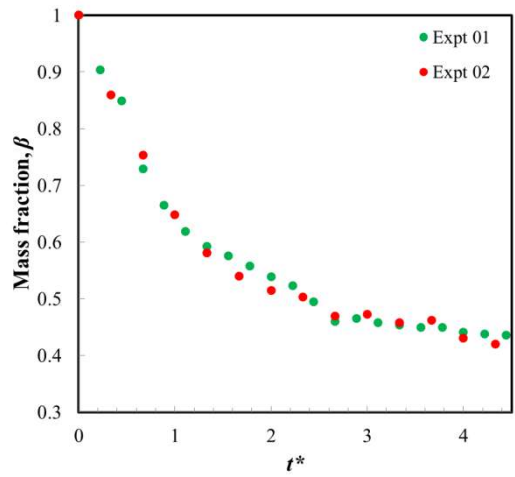

Fig.6.3. Two series of temporal evolution of mass fraction in microfluidic device used for the gravity-driven drainage experiments.

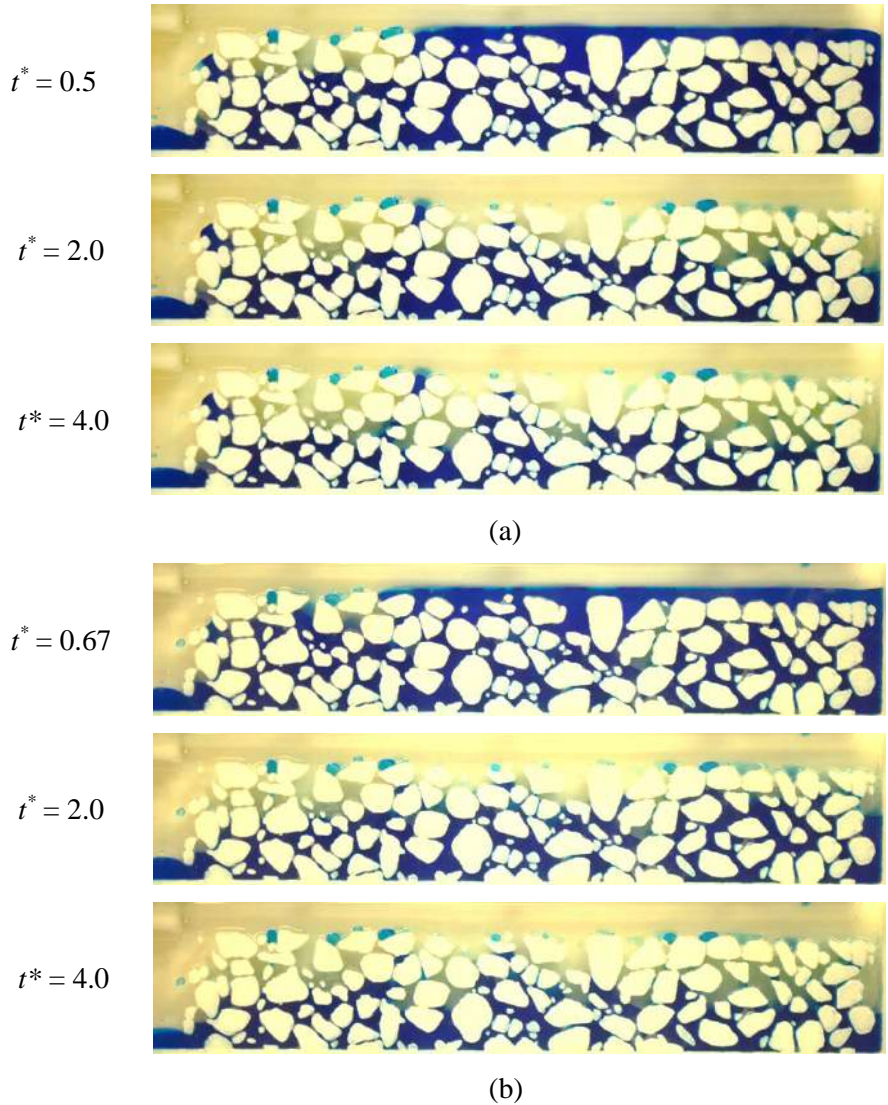

$t^* = 0.5$

$t^* = 2.0$

$t^* = 4.0$

(a)

$t^* = 0.67$

$t^* = 2.0$

$t^* = 4.0$

(b)

Fig.6.4. Water liquid configurations in PA versus (a) time of $t = 2$ seconds or $t^* = 0.5$, $t = 9$ seconds or $t^* = 2$ and $t = 18$ seconds or $t^* = 4$ from the first experiment and (b) time of $t = 2$ seconds or $t^* = 0.67$, $t = 6$ seconds or $t^* = 2$ and $t = 12$ seconds or $t^* = 4$ from the second experiment.

## 6.3. LBM results of 2D gravity-driven drainage

Given that the flow experiment is a quasi-2D configuration of drainage in PA, drainage is first simulated in 2D.

### 6.3.1. Simulation set-up and boundary conditions

The 2D computational domain is 1 300 × 300 lattice$^2$ with a spatial resolution $\Delta x$ of 150 µm per lattice. The size of the porous medium is 1 187 × 200 lattice$^2$ or 178 050 × 30 000 µm$^2$, so the sample is reproduced at true scale. The porous domain is flush with the right of the domain and is situated at 113 lattices from the left side and at the bottom of the domain as shown in Fig.6.5.

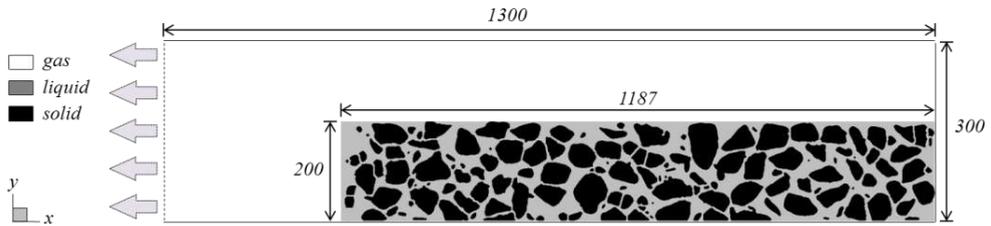

Fig.6.5. Schematic representation of the 2D computational domain for gravity-driven drainage in PA.

To study gravity-driven drainage with this 2D LB domain, two initial configurations of liquid phase are considered. First, the porous domain is initially filled using the experimental initial liquid configuration, as shown in Fig.6.6. Second, the domain is fully filled as represented by the grey zone in Fig.6.5. The free domain at the left of the material allows the liquid to drain. The densities of liquid and gas are 0.28 and 0.0299 lattice units, corresponding to a density ratio $\rho/\rho_c = 9.4$ at $T/T_c = 0.85$. Two different contact angles of 100° and 126° corresponding to solid-fluid interaction parameters $w$ of 0.02 and 0.05 are considered. These contact angles were chosen to analyze the sensitivity of the results to the wettability of the material as the original experiment has two different materials. Bounce back boundary conditions are imposed on all sides except the left side which is treated as an outlet boundary, or a Neumann boundary condition, applying a zero gradient as explained in section 3.6.4.

The gravity acceleration is set equal to $9.38 \times 10^{-4}$ lattice units, using the unit conversion explained in section 3.6. The gravity force, defined as gravity acceleration multiplied with density, is applied as a body force term as mentioned above in section 3.3 (Eq. (3.13)).

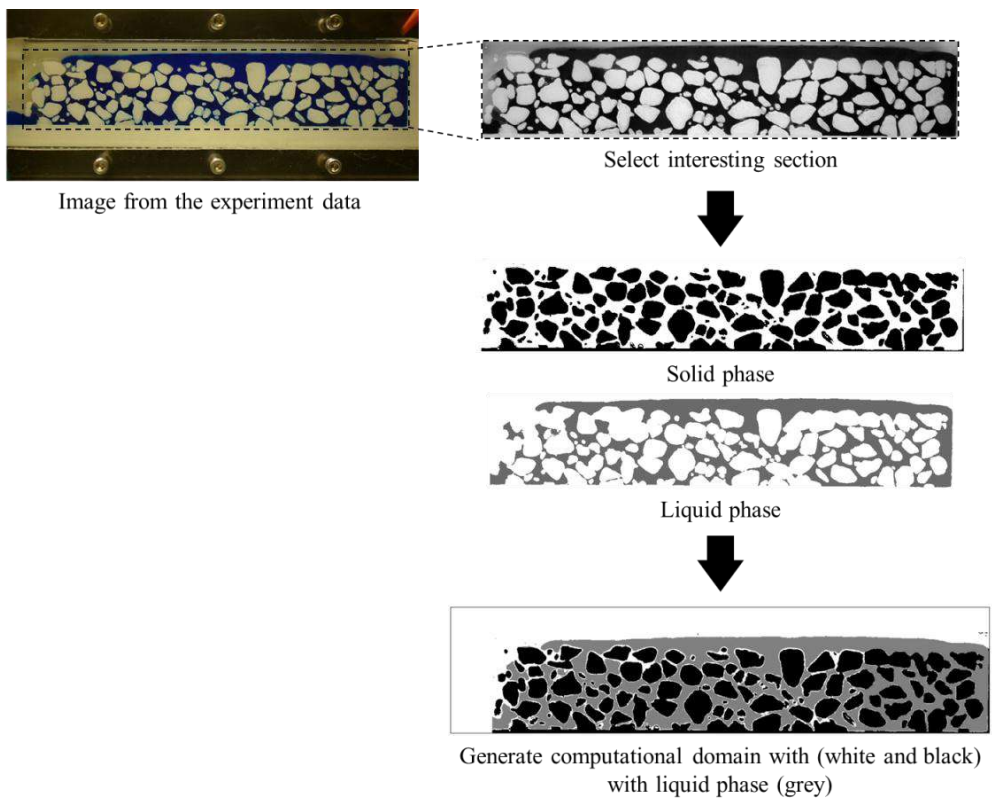

Image from the experiment data

Select interesting section

Solid phase

Liquid phase

Generate computational domain with (white and black) with liquid phase (grey)

Fig.6.6. Procedure to generate of 2D computational domain and to initialize the liquid phase inside the computational domain from the experimental image.

### 6.3.2. Computational results

Fig. 6.7 shows the mass fraction from LB results for two different contact angles of 100° and 126° and two different liquid phase initializations. The LB results are compared with the experimental data.

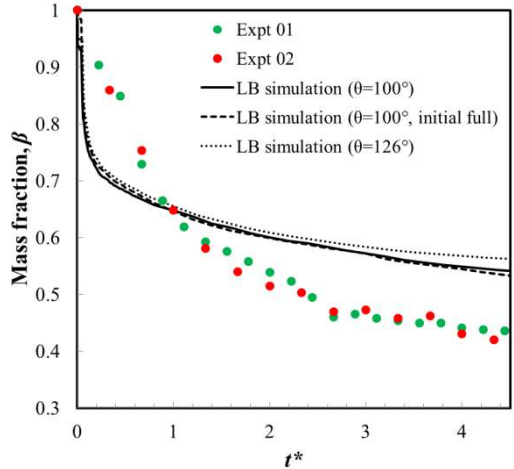

(a)

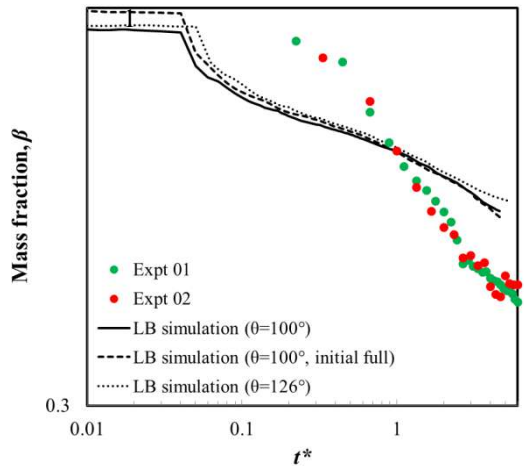

(b) logarithmic scale

Fig.6.7. Evolution of mass fraction in porous medium from LB simulations and gravity-driven drainage experiments versus non-dimensional time.

The data is presented versus time in Fig.6.7 (a) and versus log of time in Fig.6.7 (b). An important discrepancy between the LB results and the experimental data can be observed. The LB results show an initial too fast liquid outflow until $t^* \sim 0.3$. Thereafter, the mass fraction continues to reduce but is too slow. Comparing all LB results, only small differences can be observed all along the drainage process for the two different contact angles of 100° and 126°, the simulation with 126° being slightly slower. No significant change can be seen between the LB results for the two different liquid initializations (same contact angle of 100°), but the drainage is slightly faster for the total saturated initial configuration, which starts from $\beta = 1$. It is concluded that the mass fraction evolution versus time is only little affected by the contact angle and initial configuration of the liquid phase.

To understand the source of discrepancy between the LB results and the experimental data, measured and simulated water configurations at different iteration times are plotted in Fig.6.8. In the experimental series, the water level is at almost the same height along the porous medium after $t^* = 1.0$, although water leaves the porous medium to the left. In the LB results, the top level of the water domain forms a gradual slope towards the left, which remains throughout the total drainage process. Additionally, in the experimental series, the liquid phase becomes more distributed forming water island as some liquid remains trapped between the aggregates. In comparison, the water remains connected into one zone in the LB results. It is observed that more liquid remains inside the porous medium in the LB results at equilibrium state compared to the experimental data. A possible explanation for the difference could be the specific configuration of the aggregates at the bottom left of the system. There seems to be more restriction to flow in the computational domain than in the experimental device, which is essentially 3D. As a result of this obstruction in the computational domain, a higher mass fraction of liquid remains inside the PA over time, as observed in the LB results.

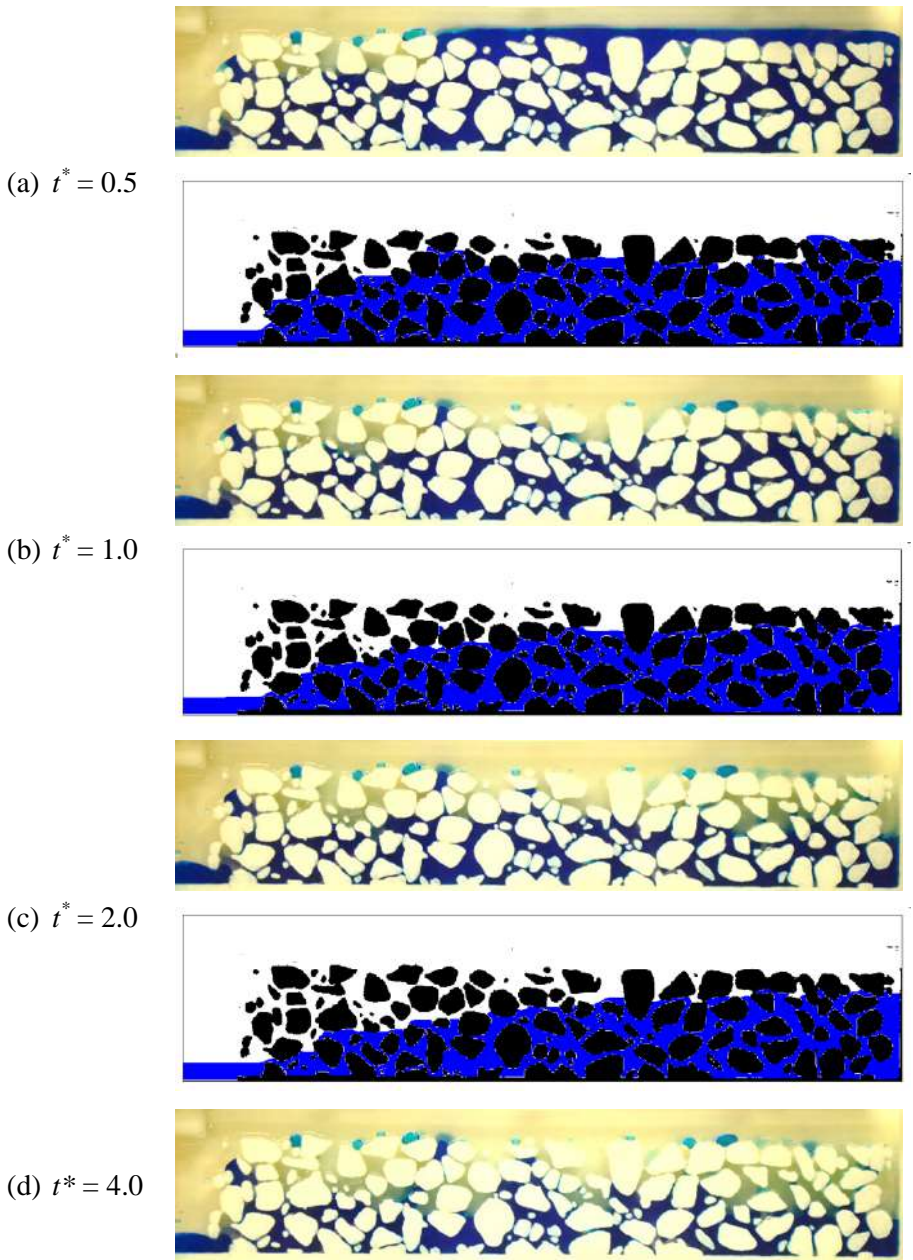

(a) $t^* = 0.5$

(b) $t^* = 1.0$

(c) $t^* = 2.0$

(d) $t^* = 4.0$

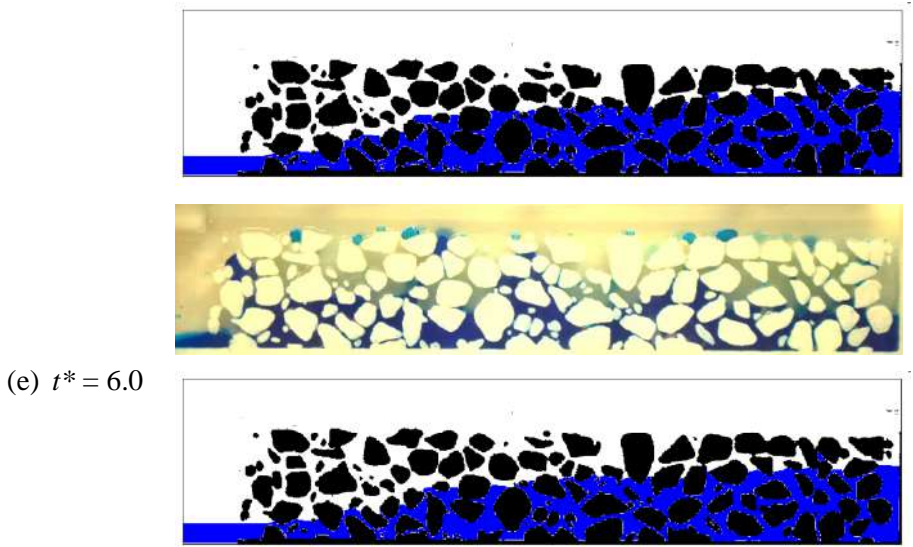

(e) $t^* = 6.0$

Fig.6.8. Comparison of water liquid configurations in porous medium obtained in the first experiment and from LB simulations at time (a) $t^* = 0.5$; (b) $t^* = 1$; (c) $t^* = 2$; (d) $t^* = 4$; and (e) $t^* = 6$.

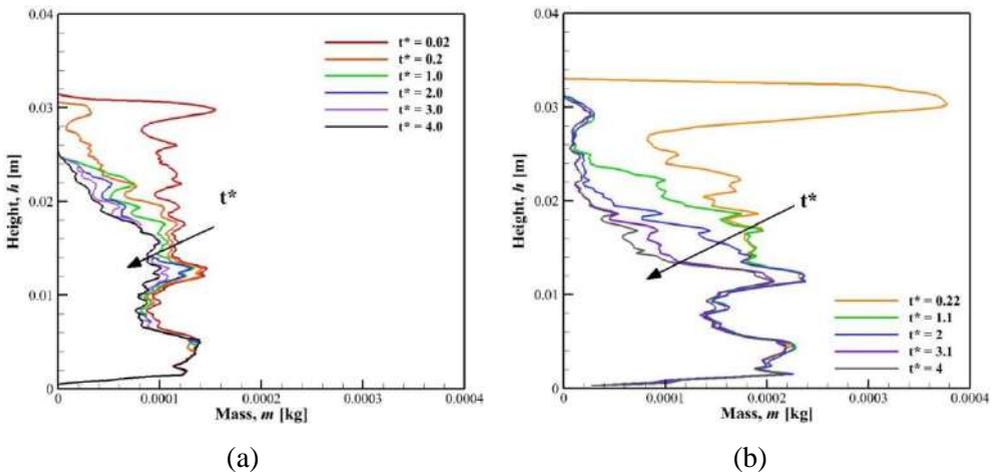

(a)                 (b)

Fig.6.9. Temporal evolution of total remaining water mass along height from (a) 2D LB results and (b) the first experimental data.

Finally, two additional analyses are carried out to better capture the overall trends of the drainage process. First, the mass distributions along the height averaged over the total length of the specimen are plotted at different times in Fig.6.9. The data selected for this comparison are the first experiment and the LB results for a contact angle of 100° and an initial liquid configuration as obtained from the experimental image (see Fig. 6.6). It is noted that the experimental result shows a high peak in the top part of the specimen at $t^* = 0.22$ due to water sitting on top of the specimen resulting from water overflow during pouring. This peak is not obtained in the LB results, since the overflow of water on the top surface is not modelled. Also at later stages, some small droplets remain on the top surface contributing to a small peak at height of 0.03 m in the experimental data (see Fig.6.8 (b) – (d)). Qualitatively the two graphs indicate a similar behavior, where the top half of the domain shows a constant reduction in mass, while the mass in the bottom half being remains quite constant. The quite constant mass profiles at the bottom of the system are attributed to the fact that liquid moves to the bottom by gravity for both cases, replenishing any departed liquid. The large drops in the experimental mass profiles between the four first profiles indicate large amount of liquid leaving the systems, while these drops are much less evident in the LB results. Furthermore, the absence of trapped water islands in the LB results explains the more uniform and regular decrease in mass profiles in the top half of the system.

In a second analysis, the liquid transport from right to left side is studied in more detail by dividing the domain into six sections along the length of the specimen as shown in Fig.6.10 (a). The mass fraction in each section is plotted versus time for both LB results (Fig.6.10 (b)) and experimental data (Fig.6.10 (c)). The overall inclined liquid configuration seen in the LB results yields a clear progression of fluid transport from right to left seen by a clear delineation of the curves. The more horizontal fluid configurations seen in the experiment result in curves that are quite grouped together, not allowing to identify a logical sequence in the flow out of the different sections. The LB results in contrast show significant different mass fraction, with exception to the initial stage in Fig.6.10 (b). The difference between experimental data and LB results indicates that the simulation is not capturing all the details of the drainage process and the assumption of 2D simulation could be an oversimplification. Therefore, a 3D simulation is considered in the next section.

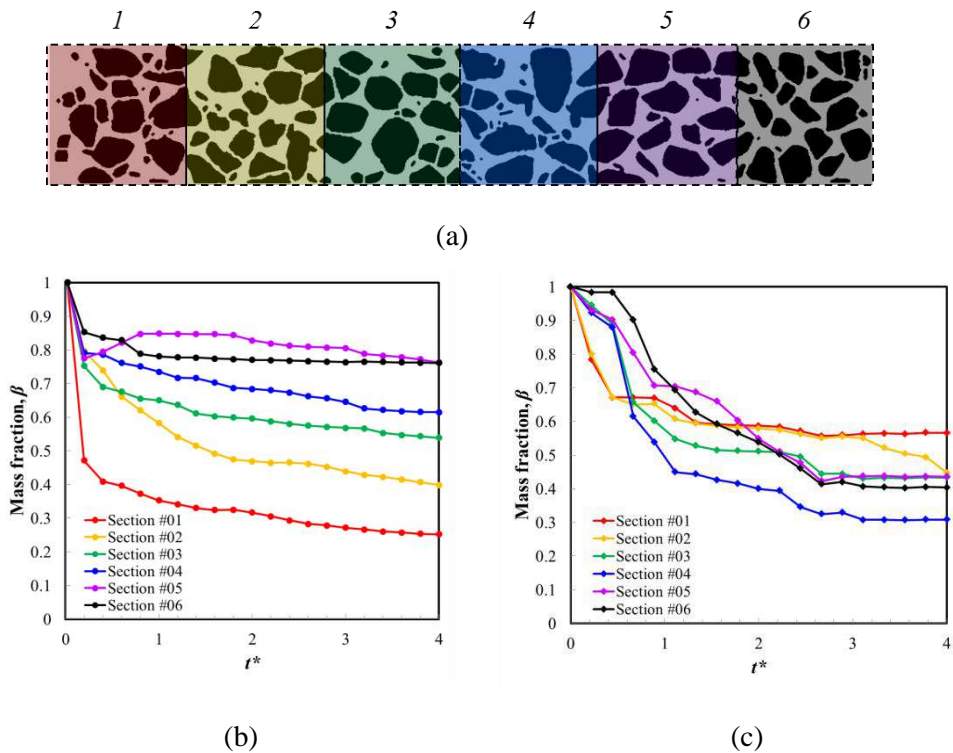

(a)

(b)                                           (c)

Fig.6.10. Schematic representation of the six sections (a) and profiles of remaining mass fraction in each section as a function of time from (b) 2D LB results and (c) the first experimental data.

## 6.4.    3D gravity-driven drainage and drying in PA

In this section, gravity-driven drainage in the quasi-2D system is studied with 3D LBM. The detailed liquid configurations at different times are compared with experimental data, in the aim of verifying and validating the LBM.

### 6.4.1. Simulation set-up and boundary conditions

For the simulation of the 3D gravity-driven drainage, the domain size is $700 \times 300 \times 30$ lattice$^3$ with spatial resolution $\Delta x$ of 300 μm per lattice. The size of the porous medium is $600 \times 100 \times 30$ lattice$^3$ or $180\ 000 \times 30\ 000 \times 10\ 000$ μm$^3$ as illustrated in Fig.6.11. For gravity-driven drainage, the porous medium domain is filled with liquid, thus starting from saturation state. The densities of liquid and gas are 0.28 and 0.0299 lattice units, respectively corresponding to a density ratio $\rho/\rho_c = 9.4$ at $T/T_c = 0.85$. To simulate as well as possible the experimental conditions, showing different contact angles of 102.5° for the covers and of 115° for the aggregates, also two contact angles of 100° ($w = 0.02$) for the walls, and 117° ($w = 0.04$) for the aggregates are imposed in LB simulations. Bounce back boundary conditions are imposed on all sides. The gravity acceleration is set to $2.12 \times 10^{-4}$ lattice units and applied as body force.

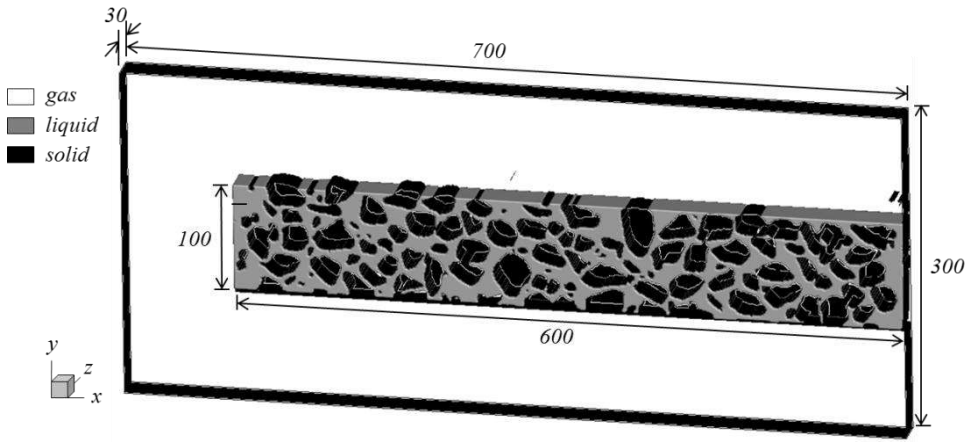

Fig.6.11. Schematic representation of the 3D computational domain for simulating gravity-driven drainage of the PA sample.

This 3D simulation is run with parallel computing based on Message Passing Interface (MPI) on the high performance computing cluster at Los Alamos National

Laboratory (LANL). The cluster aggregate performance is at 352 TFlop/s with 102.4 TB of memory for 38 400 cores. Each simulation is run on 700 processor cores (35 × 20 × 1) and requires 16 hrs to run 650 000 time steps.

### 6.4.2. Computational results

In this section, the 3D LB results are compared with the experimental data in terms of mass fraction and liquid distributions in the porous medium versus time. Fig.6.12 shows the global mass fraction in the porous medium versus time. By comparing LB results with experimental data, an overall good agreement is observed. After an initial faster decrease of mass fraction, the LB results follow the experimental drainage line for most of drainage time.

In the comparison of the experimental and LB drainage process, three different regimes can be distinguished. At non-dimensional time $t*$ less than one, the experimental mass fraction is higher than LB results, as highlighted by the first vertical lines in Fig.6.12 (a) and (b). In logarithmic scale, the difference becomes clearer. This underestimation of the mass fraction by LBM can be explained by the process used to fill the domain. The experimental domain is filled by pouring water at the top-right part as explained in section 6.2. Thus, void sections may still being filled, when water is already draining out. In contrast, in the LB simulation, the computational domain is fully saturated before start and the liquid moves out from the domain directly from start. Due to this difference initial filling, the LB results may show slightly lower mass fractions compared to the experimental data at initial stage. This regime can be thought as the 'initialization-controlled regime'.

When the non-dimensional time $t*$ is between 1 and 3, the LB results match the experimental data in Fig.6.12 very well. In this phase, gravity is the main factor controlling outflow and LB results and experimental data show a good agreement. In comparison with the 2D results, the 3D simulation result yields very similar changes of mass versus time. The regime is called the 'gravity-controlled regime'.

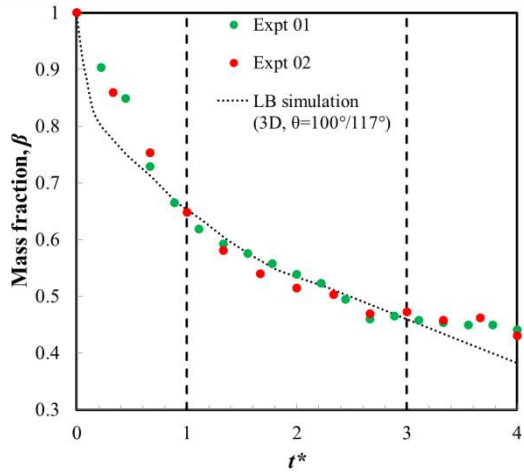

(a)

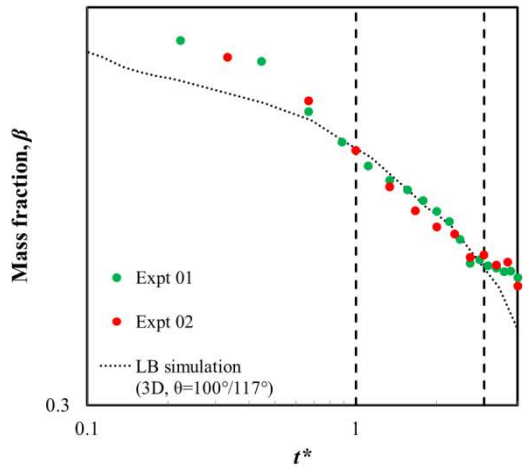

(b)

Fig.6.12. Evolution of mass fraction in the porous medium from LB simulations and gravity-driven drainage experiment versus (a) time and (b) logarithm of time.

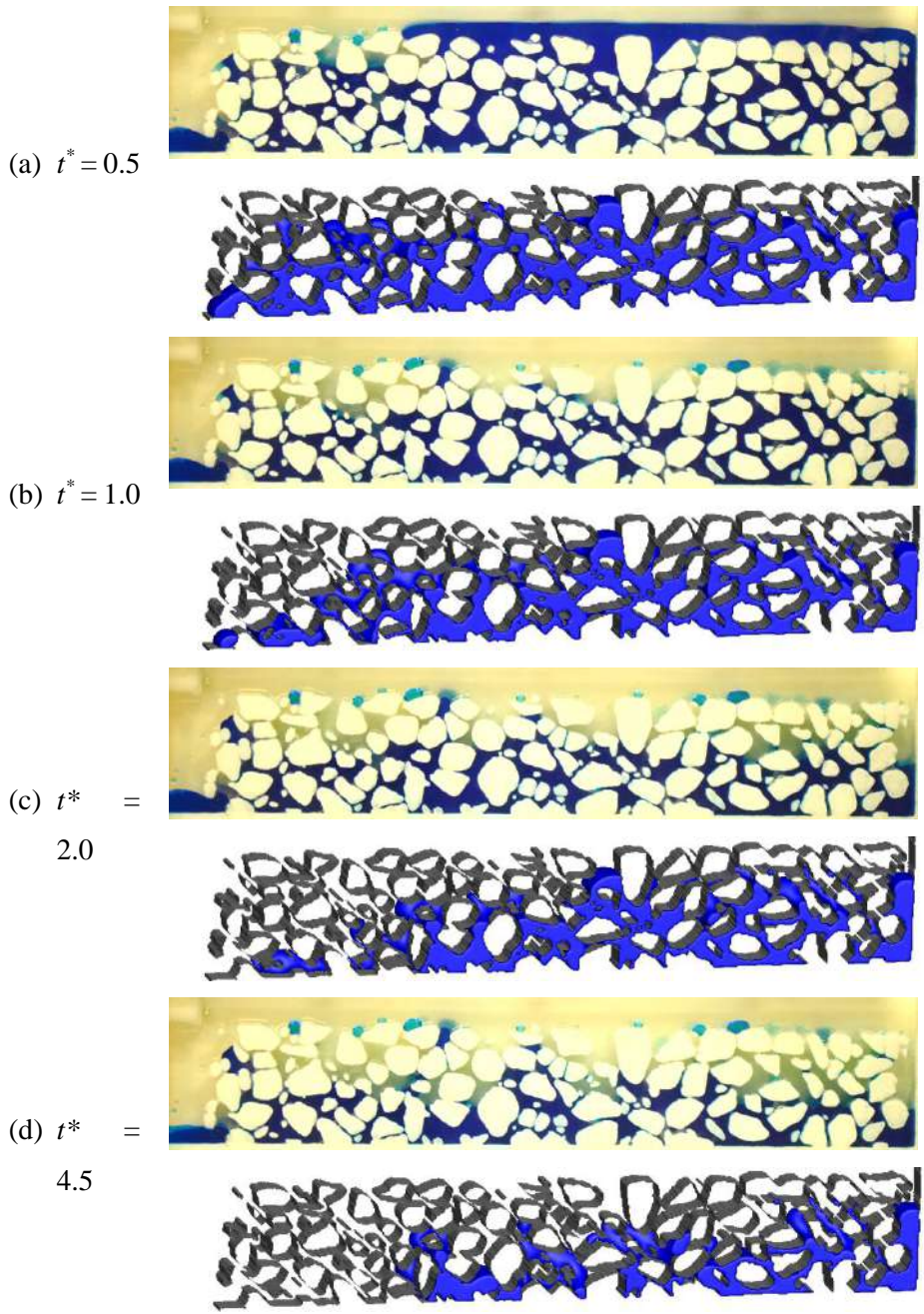

(a) $t^* = 0.5$

(b) $t^* = 1.0$

(c) $t^* = 2.0$

(d) $t^* = 4.5$

Fig.6.13. Comparison of water liquid configurations in PA of the first experiment and LB simulations versus time (a) $t^* = 0.5$; (b) $t^* = 1$; (c) $t^* = 2$; and (d) $t^* = 4.5$.

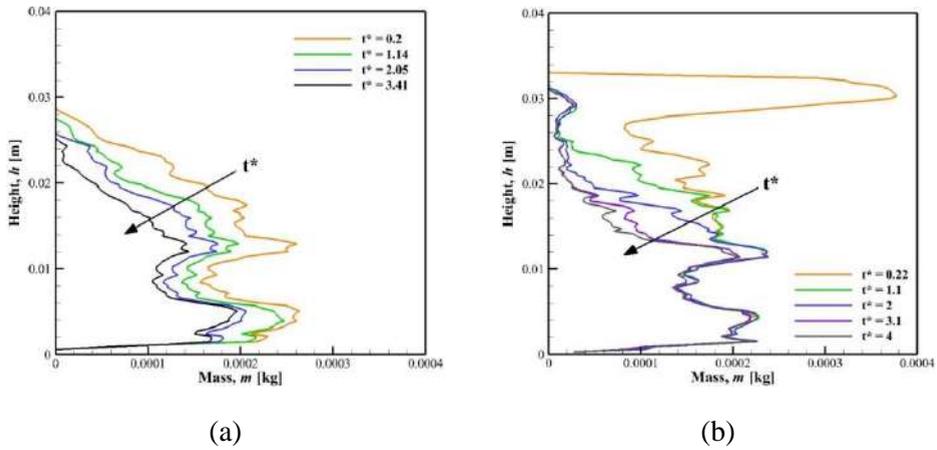

(a)                                    (b)

Fig.6.14. Temporal evolution of remaining water mass versus height from (a) 3D LB results and (b) the first experimental data.

At non-dimensional time $t*$ larger than 3, the mass fraction of experiment remains approximately at the same value of 0.4. In contrast, the mass fraction from the LB results continues to slightly decrease. The difference between experiment and simulation probably results from the modelling of the exit on left bottom side. The experimental domain seems to display a smaller area of exit, due to some fusing between PDMS and resin, compared to the exit area of the computational domain. This results in the liquid flow to become restricted or even restrained in the last part of the experiment leading to more liquid remaining inside the porous medium. This regime is called the 'geometry-controlled regime'.

Fig.6.13 presents the detailed liquid distributions inside PA at different non-dimensional times $t*$ of 0.5, 1.0, 2.0 and 4.5. In the LB results, a sloped liquid level is seen on the left part of the domain at $t*$ of 1.0 and 2.0. At $t* = 4.5$, the liquid on the left part of the domain is totally drained. At the right side, the liquid level becomes horizontal, due to the presence of smaller pores located in the middle of domain, which prevent the liquid to move through the smaller pores, liquid which, as result, remain inside the porous medium. In the experimental liquid configurations, the liquid level decreases horizontally although some liquid remains trapped as disconnected islands.

In Fig.6.14, the total mass distribution along height is plotted versus time. It is noted that the experimental result shows a high peak in the top part of the specimen at $t^* = 0.22$ due to over-pouring of water, which is not simulated in LBM. The LB results show triangular shaped graphs with higher mass at the bottom and low mass at the top due to gravity. Over time, the mass decreases gradually, while the profiles keep their triangular shape. Comparison with the experimental data indicates that the LBM provides qualitatively a similar evolution showing a decrease of mass in the top part. The larger mass at the bottom of the system is attributed to the movement of liquid to the bottom by gravity, replenishing any departed liquid. However, in LBM, the bottom part keeps on draining water, while the experiments show a rather constant mass in the lower part. As seen in Fig.6.13, in the LB results, the liquid drains out totally in the left part of the specimen at $t^*$ of 4.5 while being in the 'geometry controlled regime'.

Dividing horizontally the computational and experimental domains into six sections as presented in Fig.6.10 (a), the mass fraction of liquid in the different sections is plotted versus time in Fig.6.15. The LB results show wide divergence of the mass fraction curves for the different sections, while in the experiment the curves are more grouped. In the LB results, in the sections 01 and 02, the mass fractions decrease dramatically and almost drain out totally compared to what happens in the other sections 03, 04, 05 and 06. At the end of drainage $t^* = 4$, the mass fraction of section 01 approached zero while sections 03, 04, 05 and 06 still keeps the mass fraction with ranging from 0.6 to 0.8. In contrast, in the experimental data, the mass fraction of each section decreases following a similar evolution ranging between 0.4 and 0.6.

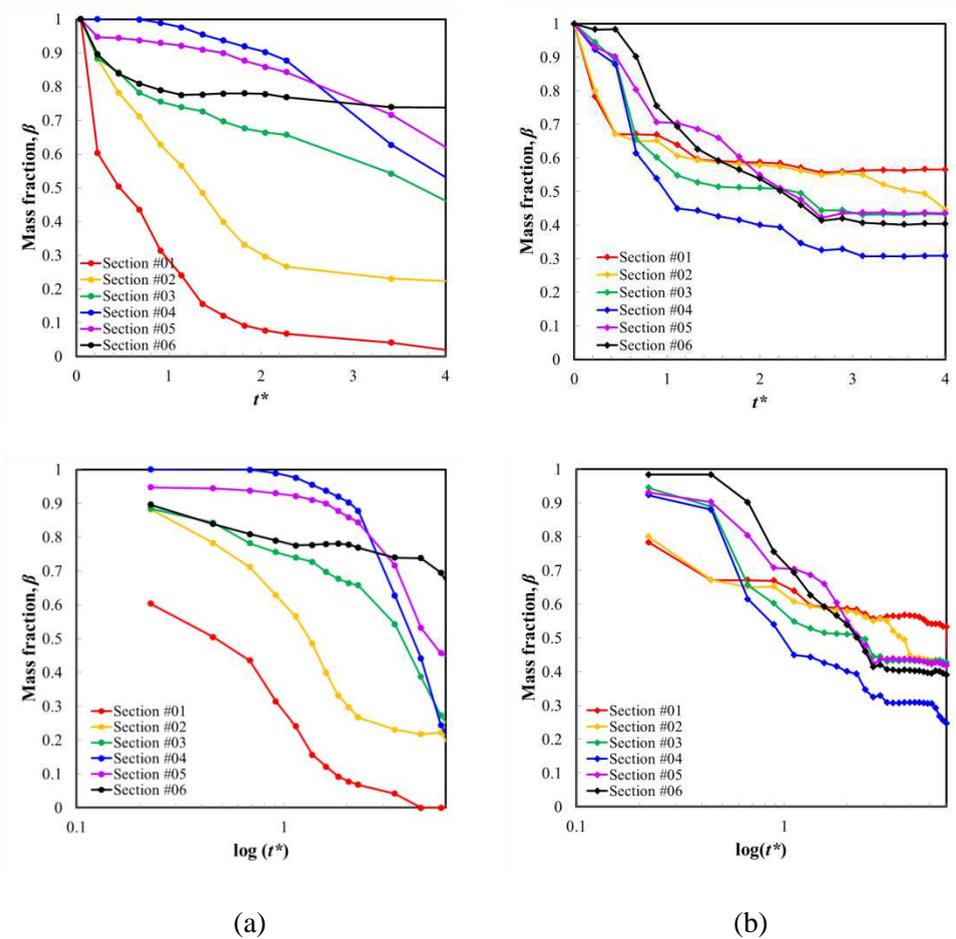

(a)                                                       (b)

Fig.6.15. Profiles of remaining mass fraction in each section as a function of time from (a) 3D LB results and (b) the first experimental data.

In particular, the discrepancies may be explained by the configuration of the aggregates at the bottom left of the systems, which show probably a smaller area of exit in the experiment due to some fusing between PDMS and resin during manufacturing of the microfluidic device. This probably leads to restricted or even restrained liquid outflow in the last part of the experiment. To validate and verify the LB results with experimental data, a more delicate manufacturing procedure of microfluidic device has to be required to avoid fusing. For LB simulation, higher resolution of computational domain may help to get better results.

## 6.5. Conclusion

Gravity-driven drainage in PA has been studied with 2D and 3D LBM for different contact angles. The 2D LB results do not agree well with the experimental data due to the assumption of a 2D domain, which oversimplifies the boundary conditions and cannot capture the effect of the different contact angles of the lateral plates compared to the aggregates. The 2D configuration of aggregates also reduces the possibility of outflow through gaps between the aggregates, leading to a restriction of the flow compared with experiments.

In contrast, the 3D LB simulations show an overall good agreement with experimental data. Three different regimes are observed: an initialization controlled regime at start, a gravity-controlled regime in the middle and a geometry controlled regime at the end of the drainage process. The 3D LBM showed an overall good prediction of the gravity-controlled regime. Some discrepancies between experiment and simulation were observed especially in the geometry controlled regime, indicating the importance of detailed geometric aspects, such as pore restrictions, on the remaining liquid distributions. It is suggested that a more complete set of pore geometries with different pore restrictions are tested and compared to LB simulation for further research.

In the next chapter, further explorations of diverse multiphase phenomena based on some experimental works are performed using 2D and 3D LBM.

# 7. FURTHER EXPLORATIONS
## 7.1. Introduction

In this chapter, further explorations of diverse multiphase phenomena are performed using 2D and 3D LBM. These explorations came along the development of this PhD as interesting related research questions. First, the question of a droplet finding its equilibrium state is studied on a flat surface covered with randomly organized patches of diverse contact angles. Second, the drop jumping over hydrophobic stepping stones is performed, as inspired by an experiment reported in literature. Finally, the question of the drying of a porous medium consisting of pillars between two parallel plates is studied, inspired again by experiments. This porous medium has either a regular or a hierarchical pillar nature and the time evolution of the liquid configuration within this porous medium is followed. The qualitative description of each simulation work highlights the potential of these LBM investigations to study complex interesting multiphase phenomena.

The 2D and 3D LB simulations are performed in similar ways as presented in the previous three chapters. The basics of LBM and boundary conditions are explained in chapter 3. The LB results in this chapter are illustrated with snapshots of liquid configurations overtime and discussed qualitatively. Further quantitative analysis and modeling is outside the scope of this thesis, but the results show the potentials of LBM to explore these research questions.

## 7.2. Droplet reaching equilibrium on a heterogeneous flat surface

This work is an explorative study as a next step to the study of the droplet deposited on a checkerboard heterogeneous surface as presented in section 5.4. Instead of two different surface properties, a range of contact angles is now applied on a heterogeneous surface and the droplet movement as the droplet finds its surface energy equilibrium is studied.

### 7.2.1. Simulation set-up and boundary conditions

For the study of a droplet on a randomly heterogeneous surface, the domain size is $200 \times 200 \times 200$ lattice$^3$. Periodic boundary conditions are imposed on all sides except top and bottom which are treated as bounce-back boundaries. The hemisphere droplet with a radius $R$ of 40 lattices is initially located in the middle of the bottom surface as shown in Fig.7.1. The densities of liquid and gas are 0.359 and $6.07 \times 10^{-3}$ lattice units respectively, corresponding to a density ratio $\rho/\rho_c = 59.1$ at $T/T_c = 0.7$. To generate a randomly heterogeneous surface, 800 circular patches with a radius ranging from 1 to 8 lattices are distributed randomly on the bottom surface. Each patch has a different contact angle varying from 63.1° to 99.5°, i.e. the solid-liquid interaction parameter $w$ is varied from -0.1 to 0.03.

One 3D simulation is performed by MPI parallel computing at the high performance computing cluster of Los Alamos National Laboratory (LANL). The simulation is run on 125 processor cores ($5 \times 5 \times 5$) and requires maximum 16 hrs to run 50 000 time steps.

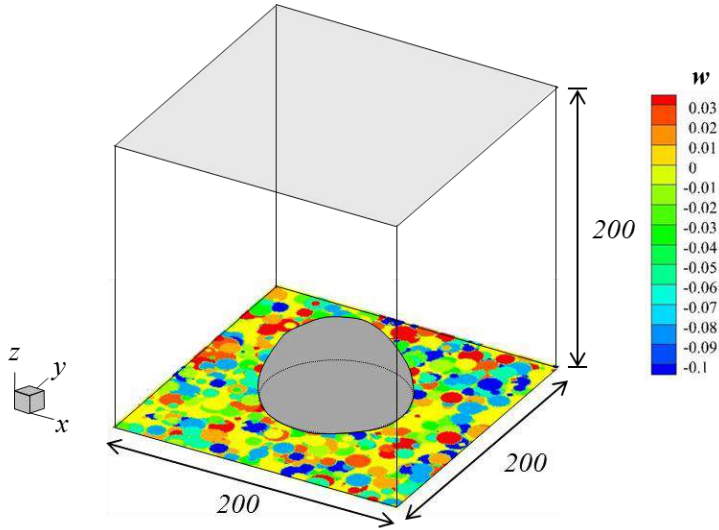

Fig.7.1. Initial condition of a single droplet on the randomly heterogeneous surface with a range of solid-liquid interaction parameter varying from -0.1 to 0.03.

### 7.2.2. Results

Fig.7.2 shows snapshots of the droplet slowly moving on the randomly heterogeneous surface at six iteration times (500, 1 000, 3 000, 5 000, 10 000 and 15 000) as it is reaching its equilibrium surface energy configuration. The droplet is placed initially at the center of the surface. By iteration time 1 000, the initial droplet spherical shape becomes distorted due to the interaction of the contact line with different hydrophilic or hydrophobic patches. Then, the droplet moves to the top-right direction from iteration 3 000 to 10 000. After iteration 10 000, the droplet remains at the same position since it has attained a minimum in surface energy.

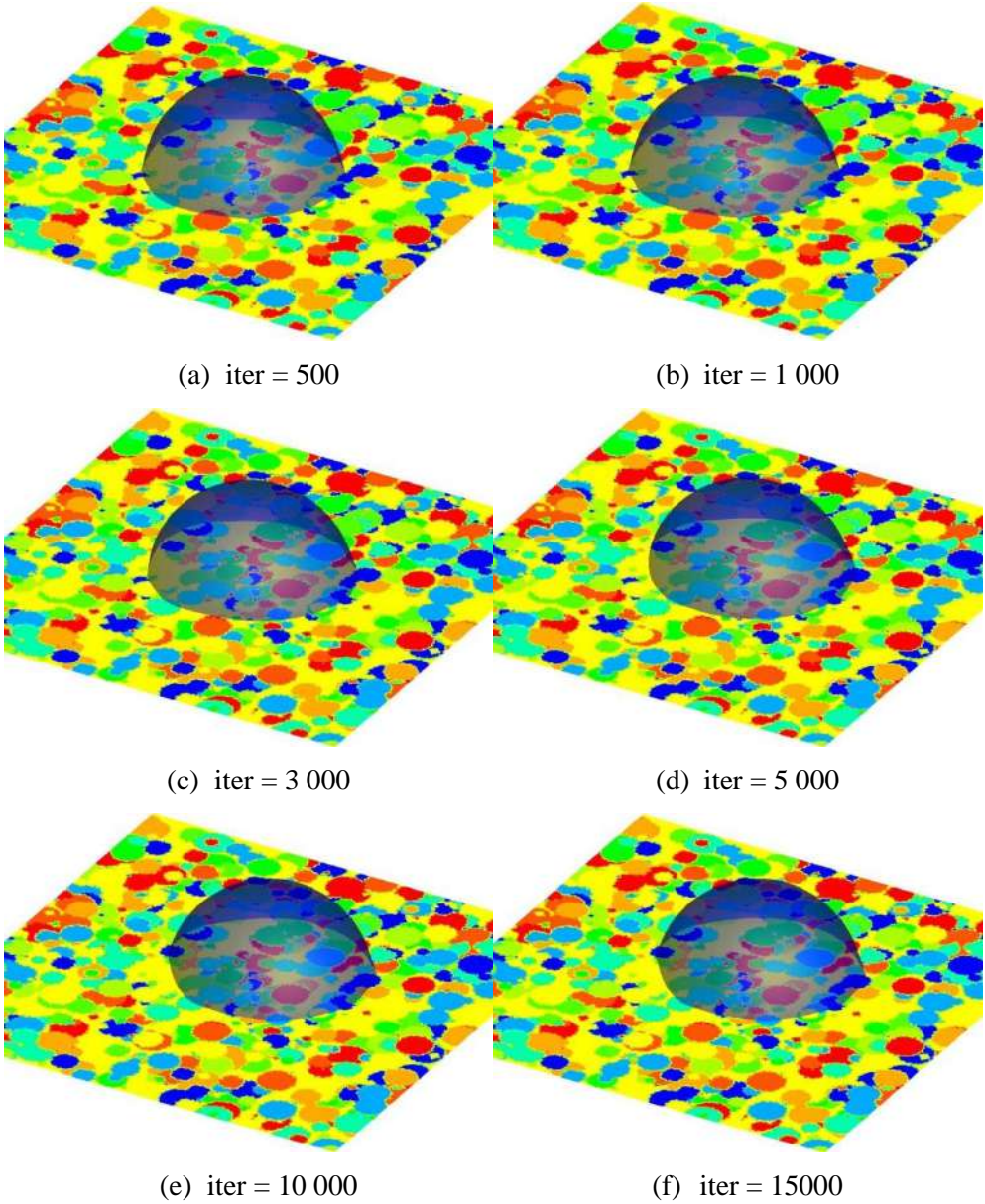

(a) iter = 500           (b) iter = 1 000

(c) iter = 3 000           (d) iter = 5 000

(e) iter = 10 000         (f) iter = 15000

Fig.7.2. Dynamic behavior of a droplet movement reaching its equilibrium configuration on a randomly heterogeneous surface at different iteration times.

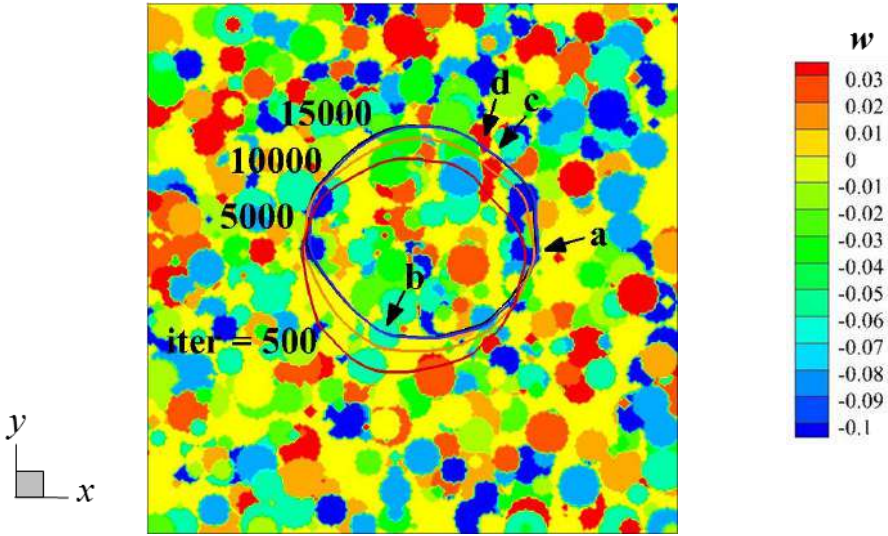

Fig.7.3. Contact line of the droplet on random heterogeneous surface at iteration times of 500 (red outline), 5 000 (orange), 10 000 (blue) and 15 000 (black).

In Fig.7.3, the movement of the droplet on the bottom surface is illustrated with its contact line position in function of time. The contact area of the droplet is found to remain constant over time. The droplet is initially (see red line in Fig.7.3) located mainly on hydrophobic patches (yellow, orange and red) with exception on the left and right sides, where the contact line contacts some hydrophilic patches (dark blue and blue, indicated by arrow **a**). The droplet then slides to top-right side where more hydrophilic patches are located finding a position with smaller surface energy. When the contact line is positioned on more hydrophilic patches (dark blue, light blue and green, indicated by arrow **b**), the droplet stops to slide further. At equilibrium, the contact line pins on a yellow patch (arrow **c**) since neighboring patches are more hydrophobic patches (red and orange, arrow **d**). This pinning can be explained by the fact that surface energy would increase again would the droplet slide further contacting more hydrophobic patches. From the LB results, it can be concluded that the final droplet position on a randomly heterogeneous surface is governed by a contact line sitting on more hydrophilic patches, where the neighboring patches show a more hydrophobic wettability.

The study could be extended to consider several droplets with different radii and initial droplet position. Droplets would then move and eventually coalescence. Also the design of surfaces with paths of increasing wettability in order to move droplets in a certain direction could be studied. LBM allows the optimization of such design, which can be used in several practical applications.

## 7.3. Droplet on stepping stones

Cira et al. (Cira, Benusiglio et al. 2015) developed a surface tension sorter for multiple droplets. In this experiment, multiple droplets, showing different surface tension, jump above channels filled with liquids with a different surface tension then theirs (see Fig.7.4). The liquids are a mixture of water and a color containing propylene glycol (PG). Six different concentrations of PG are used resulting in different surface tension values: 30% (red), 25% (orange), 20% (yellow), 15% (green), 10% (blue) and 5% (navy). The channels are made of hydrophobic glass (marked in black) making the droplet slide over the glass channels. The droplets move down by gravity. It was found that the droplets are finally trapped on the channel being filled with the liquid with same surface tension, from where stems the name surface tension sorter.

This surface tension sorter is used as inspiration for studying the movement of a single droplet over a set of hydrophobic channels until it becomes trapped in a liquid filled channel. This study can be seen as a further step with respect to the analysis of the displacement of a droplet on a surface with grooves (section 5.3).

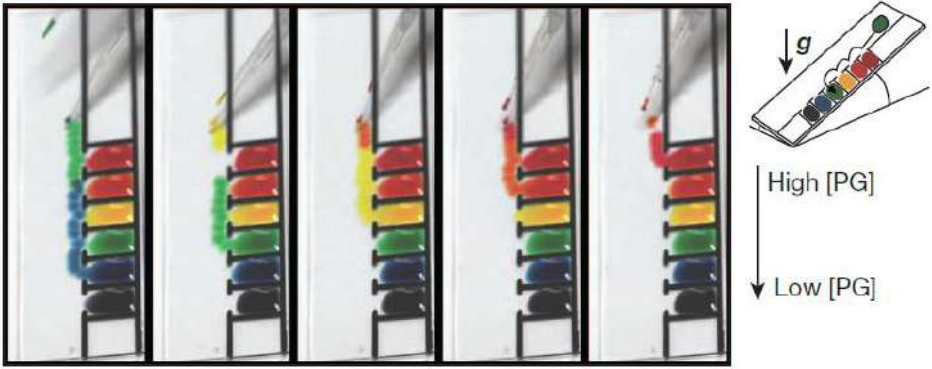

Fig.7.4. Schematic of the surface tension sorter of multiple droplets with various surface tension due to different concentrations of PG ranging from 30% (red), to 25% (orange), 20% (yellow), 15% (green), 10% (blue) to 5% (navy) (figure from Cira et al. (2015), permission pending).

### 7.3.1. Simulation set-up and boundary conditions

This experiment is simulated using 2D LBM. The 2D computational domain has a size of $500 \times 301$ lattice$^2$, as illustrated in Fig.7.5. The domain contains four channels with a width of 50 lattices and a height of 100 lattices. The channels are topped by flat surfaces of length of 25 lattices, called stepping stones which are at the same height as the 256 lattices long flat surface. The center of the semicircular droplet with a radius of 50 lattices is initially located at $x = 301$ lattices and $y = 101$ lattices on the surface as shown in Fig.7.5. The last channel is filled by liquid to a height of 90 lattices from the bottom. By applying a density ratio $\rho/\rho_c = 9.4$ at $T/T_c = 0.85$, the liquid and gas density are 0.28 and 0.0299 in lattice units, respectively. The surface inside the channels is hydrophobic with a contact angle of $123^\circ$ corresponding to a solid-fluid interaction parameter of 0.05. The horizontal top surface of the channel is more hydrophobic with a contact angle of $151^\circ$ corresponding to a solid-fluid interaction parameter of 0.08. The left and right sides of the domain are treated as periodic boundary conditions. A bounce back boundary condition is imposed on the top and bottom sides. The droplet moves from right to left side with an acceleration of $1 \times 10^{-5}$ lattice units. A body force $F_b$ is applied after 2 000 iterations, when the droplet initially having a semicircular, attains a shape according to the contact angle of 151° of the surface.

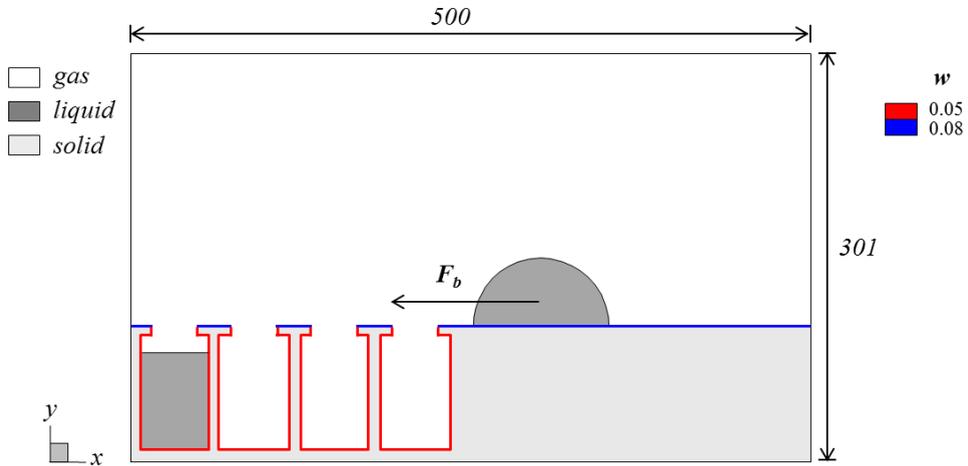

Fig.7.5. Schematic of the computational domain of the stepping stone domain with two different contact angles of 123o (red) and 151o (blue) corresponding to the solid-fluid interaction parameters $w$ of 0.05 and 0.08. The last channel is filled to 90% of its height with liquid.

### 7.3.2. Results

In Fig.7.6, snapshots of the droplet moving along the surface and the stepping stones are shown for different iteration times. Fig.7.7 shows the temporal evolution of the droplet interface.

In the beginning, before reaching the stepping stones, the droplet deforms due to the body force. The shape changes from a semicircle to a half-tear drop at iteration 5 500. The adhesive force between liquid and solid is weak on the hydrophobic surface, resulting in a small contact area between the droplet and the solid surface. As a result the droplet easily displaces along the surface. At iteration 7 000, the droplet arrives at the edge of the first channel and gets pinned. However, due to the inertia of the droplet, the bulk of the droplet is still displaced attaining an elongated form at iteration 7 700, while remaining pinned. At iteration 8 000, the droplet reaches the second channel edge making a bridge over the first channel. The droplet keeps on moving and repeats this procedure jumping over the stepping stones. When the droplet reaches the last channel partially filled with liquid at iteration 16 900, the droplet becomes trapped into the channel. As long as the droplet moves over the empty channels, the body force overrules the adhesive force on hydrophobic surface with contact angle of 151 º, as well as the adhesive force with the gas phase in the empty channel. However, when reaching the filled channel, the cohesive force between liquid of the droplet and of the liquid in the channel becomes dominant, leading to a merging of droplet and liquid in the channel. After merging, the droplet further elongates along the surface from iteration 18 000 to 20 000 due to the body force but it remains merged with the liquid in the channel. Between iteration 25 000 and 45 000, the droplet wiggles due to body, adhesion and cohesive forces, until coming at rest at iteration of 45 000 with half the amount of the droplet trapped in the channel and rest remaining on the surface.

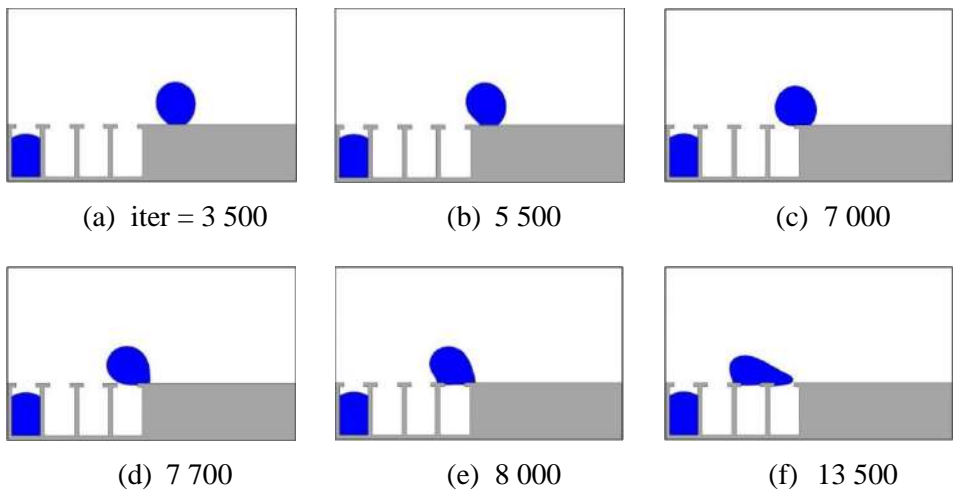

|          |          |          |
|----------|----------|----------|
| (a)  iter = 3 500 | (b)  5 500 | (c)  7 000 |
| (d)  7 700 | (e)  8 000 | (f)  13 500 |

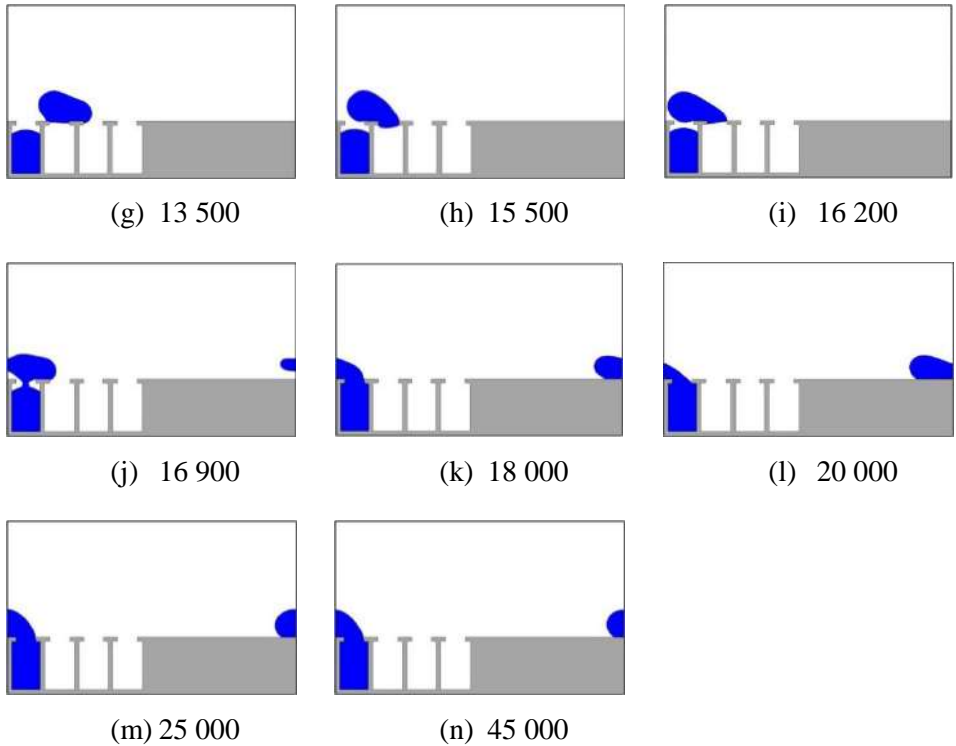

(g) 13 500        (h) 15 500        (i) 16 200

(j) 16 900        (k) 18 000        (l) 20 000

(m) 25 000        (n) 45 000

Fig.7.6. Dynamic behavior of a droplet on a stepping stone surface at different iteration times.

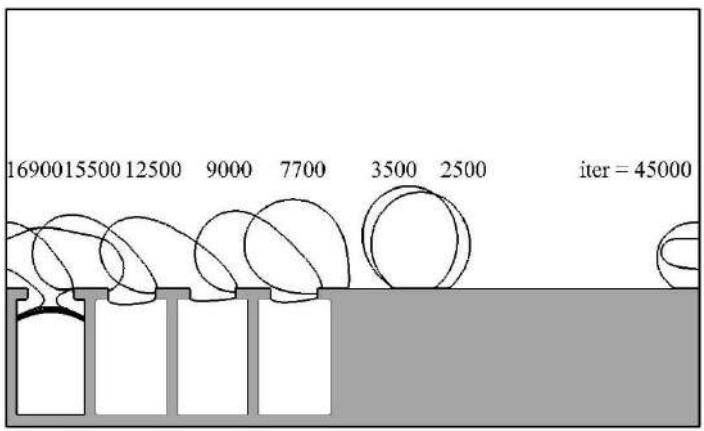

Fig.7.7. Shape and position of the droplet crossing the channels at different iteration times of 2 500, 3 500, 7 700, 9 000, 12 500, 15 500, 16 900 and 45 000.

In this section, the movement of a single droplet on hydrophobic stepping stones was presented inspired by the surface tension sorter experiment. The study could be extended by considering different liquid properties, such as viscosity and surface tension as was done in the surface tension sorter experiment.

## 7.4. Evaporative drying in regular and hierarchical porous structures

When gravity-driven drainage occurs in porous asphalt, some water may remain entrapped in the porous material. Thereafter, the entrapped water starts to evaporate and the PA dries out. In this section, the drying of water entrapped within regular and hierarchical porous structures is studied by LBM.

The work is inspired by experiments by Zurcher et al. (Zurcher, Chen et al. 2015, Zurcher, Yu et al. 2015), where the drying of liquid entrapped between two parallel plates filled with micropillars is visualized. The experiment aims to document drying in the gap between the chip and the substrate which is filled with Cu pillars of 70 μm or 90 μm. A liquid with a suspension of Cu nanoparticles fills the remaining space. A temperature of 60°C is imposed to induce evaporation. As the liquid with nanoparticles dries out, it forms capillary bridges rich of nanoparticles, which self-assembly forming interconnects between Cu pillars and Cu pad. Fig.7.8 shows the self-assembly patterns on the substrate after removing pillar chips. Fig. 7.9 shows fluorescent imaging of the liquid configuration between plates filled with a regular array of pillar for two different liquids. For the experiment, pillars are etched in a base plate and the top plate is glass allowing imaging.

This experimental work is an inspiration for analyzing the evaporation of a liquid from regular and hierarchical porous structures.

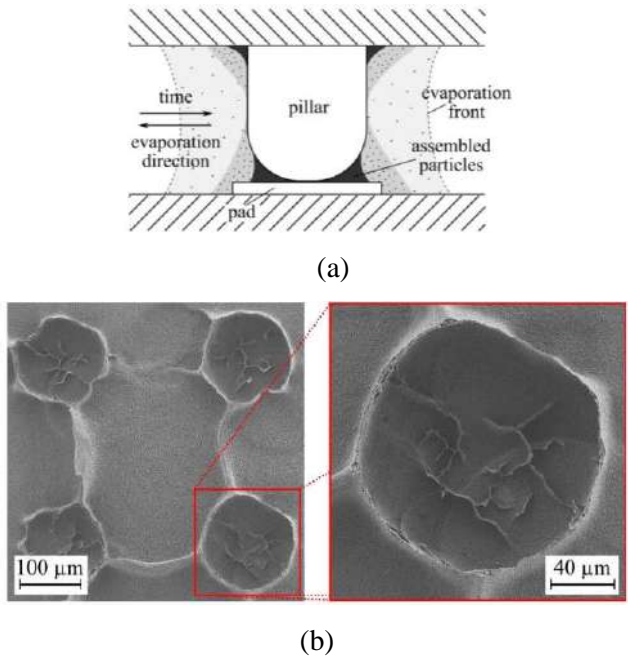

(a)

(b)

Fig.7.8. Schematic of (a) representation of the principle of the directed self-assembly of nanoparticles (NP) due to evaporation of Cu ink and (b) Cu nanoparticles self-assembly patterns on the substrate after removal of the pillar chips (Zurcher, Yu et al. 2015).

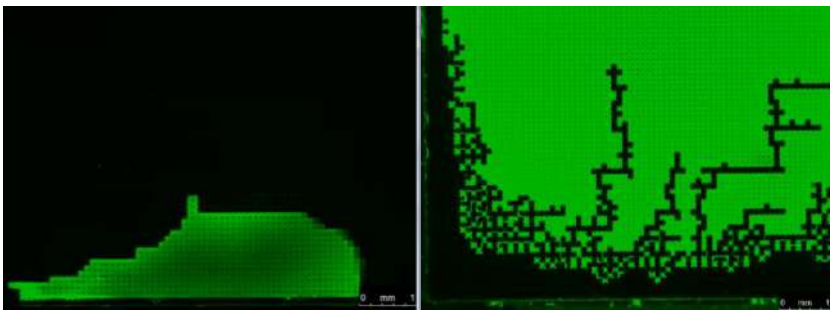

Fig. 7.9 Examples of images obtained during drying of liquid in a regular array of pillars (Zurcher 2015).

### 7.4.1. Simulation set-up and boundary conditions

The 2D computational domain has a size of 203 × 203 lattice$^2$ containing two different pillar arrangements: a regular (total of 784 pillars) and a hierarchical (900 pillars) organization as shown in Fig.7.10 with size of 1 lattice. The Zou-He's velocity boundary condition is imposed with a velocity of -0.01 lattices unit at the left and bottom sides, to generate diffusive drying. Bounce-back boundary conditions are imposed at the top and right sides. A circular droplet with a radius of 90 lattices is initially located at the center of the domain. The densities of liquid and gas are 0.359 and 6.07 × 10$^{-3}$ lattice units respectively, corresponding to a density ratio $\rho/\rho_c$ = 59.1 at $T/T_c$ = 0.7. The contact angle on all solid walls is equal to 90°, corresponding to a solid-fluid interaction parameter $w$ equals 0.

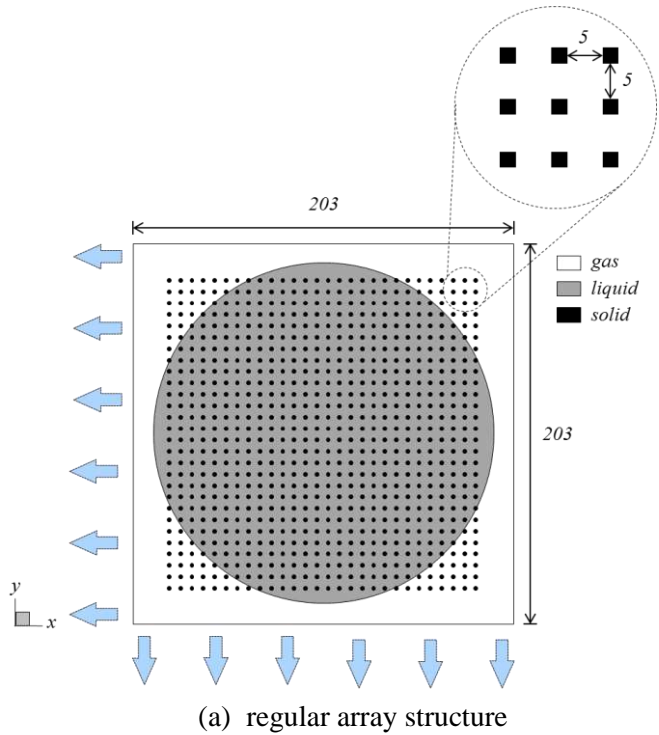

(a) regular array structure

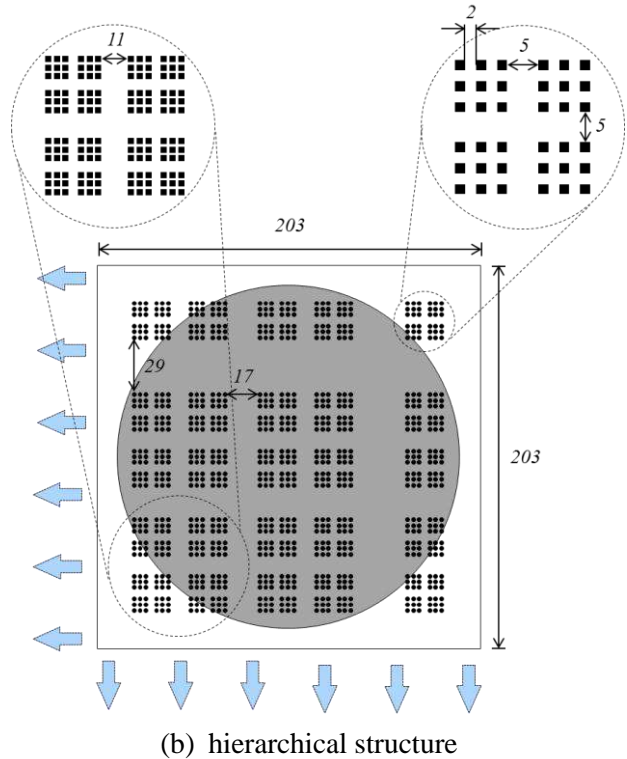

(b) hierarchical structure

Fig.7.10. Schematic of droplet located in in (a) regular and (b) hierarchical pillar structures.

## 7.4.2. Results

The liquid configuration within the regular pillar structure is shown in Fig.7.11 at iteration steps between 100 and 300 000. The initial circular droplet gets an octagonal shape at iteration of 10 000. From iteration step 10 000 to 160 000, the drying of the droplet represented by the movement of the liquid-vapor interface is faster close to the bottom and left sides, due to the outflow boundary conditions imposed on these sides, while bounce back boundary conditions are imposed at top and right sides. At 220 000, the liquid obtains a symmetric diamond shape and further evaporates keeping this shape until the liquid is totally evaporated.

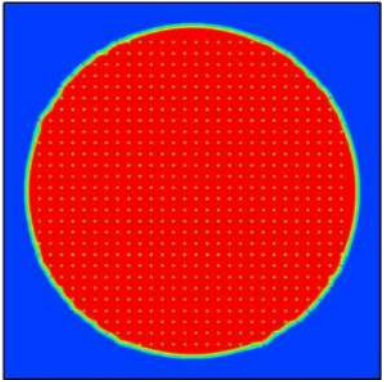

(a) iter = 100

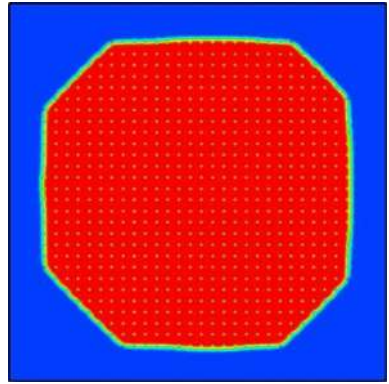

(b) 10 000

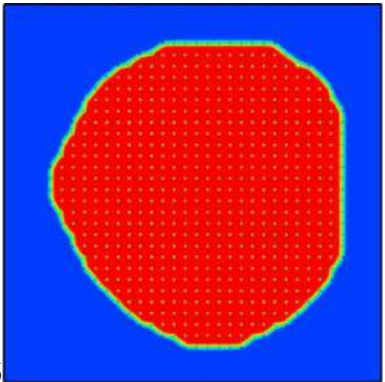

(c) 50 000

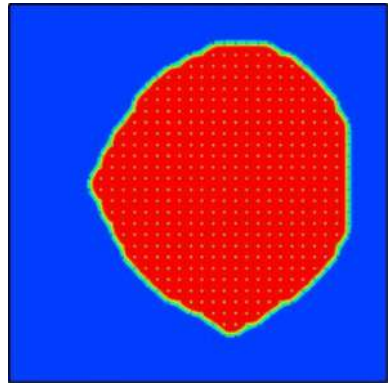

(d) 100 000

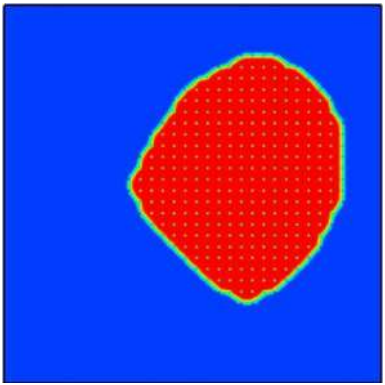

(e) 160 000

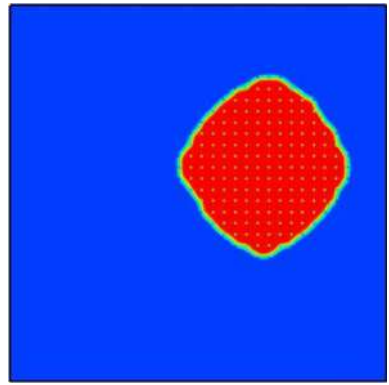

(f) 220 000

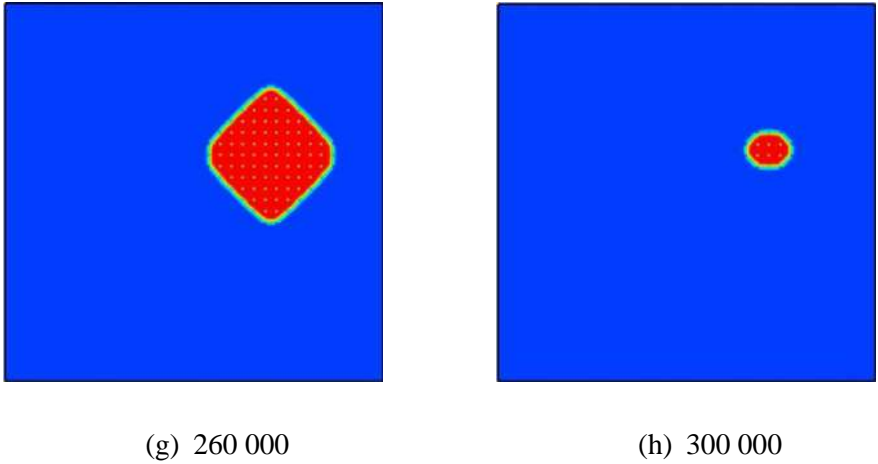

(g) 260 000          (h) 300 000

Fig.7.11. Liquid configurations in regular array of pillars versus time.

The liquid outline versus time as shown in Fig.7.12 displays clearly the faster evaporation and interface movement on the left and bottom sides. The liquid patterns are governed by the pinning of the interface between the pillars.

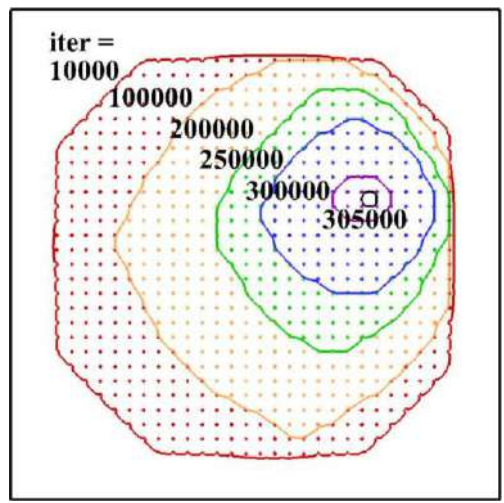

Fig.7.12. Shapes and positions of the liquid interface within the regular pillar array at iteration times of 10 000, 100 000, 200 000, 250 000, 300 000 and 305 000.

Fig.7.13 shows snapshots of the liquid configuration at different iteration times for the droplet evaporating from the hierarchical array of pillars. The evaporation occurs first mainly in the larger channels as shown at iteration times 50 000 and 100 000 (Fig.7.13 (c) and (d)).This is explained by the fact that the gas phase can more easily invade the larger than the narrow channels. As evaporation proceeds, the liquid front retracts forming distorted interfaces. Liquid remains preferentially trapped between the pillars that are more narrowly spaced.

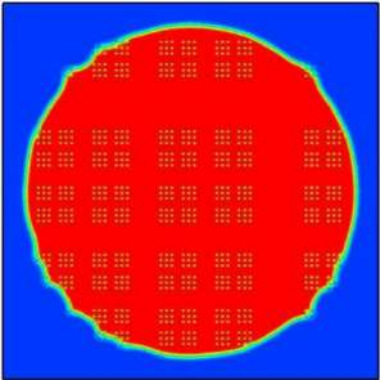

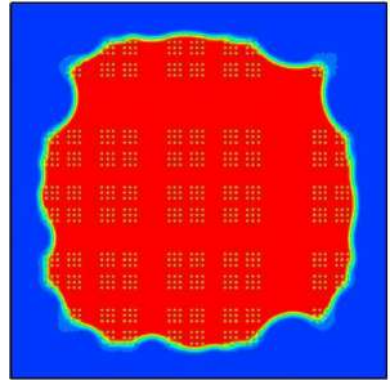

(a)  iter = 100                                  (b)  10 000

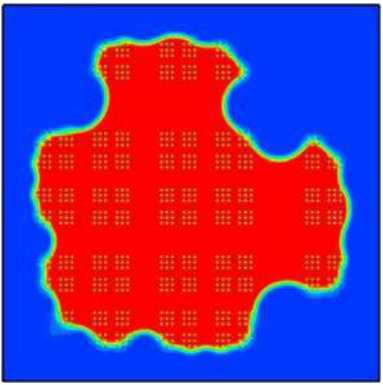

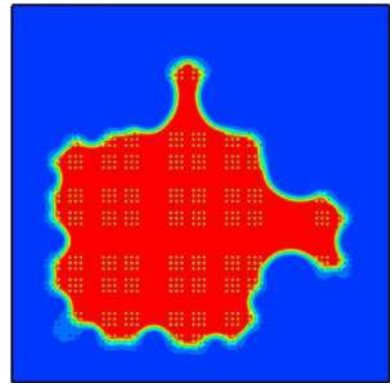

(c)  50 000                                       (d)  100 000

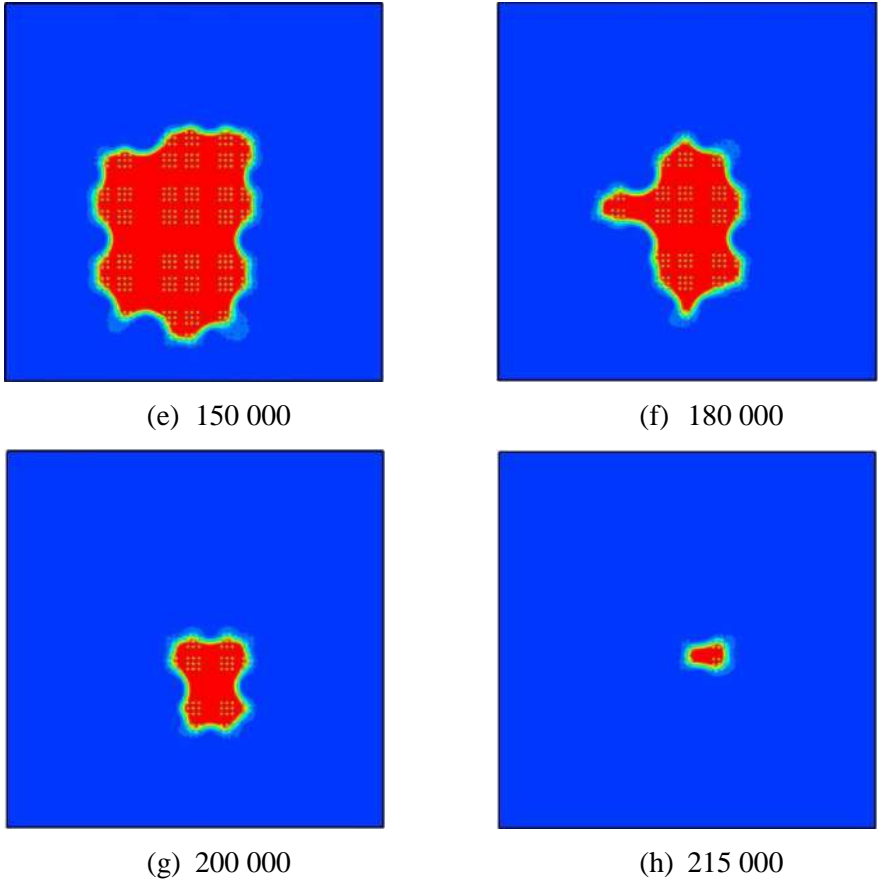

(e) 150 000                    (f) 180 000

(g) 200 000                    (h) 215 000

Fig.7.13. Liquid configurations during evaporation within hierarchical pillar array versus times.

Fig.7.14 shows the detailed liquid configurations and positions of the interface at different iteration times. The liquid remains for some time entrapped in the structure situated at the left-bottom corner, where the hierarchical structure does not contain the largest inter distance. The interface shows a distorted shape due to pinning at the pillars. The contact line moves more to the center of the domain after 150 000 iterations. It is noticed that, while the liquid in the regular structure remained entrapped preferentially into the right-top corner, the liquid in the hierarchical structure remains more entrapped at the bottom part. The evaporation of the liquid droplet goes also faster in the hierarchical structure compared to the regular structure. This can be explained by the fact that the gas phase can more easily penetrate through the larger channels, as such providing preferential pathways for drying of the liquid in the finer structure. In conclusion, these examples show the important influence of the hierarchical structure on the evaporation rate and pattern of the liquid remaining entrapped.

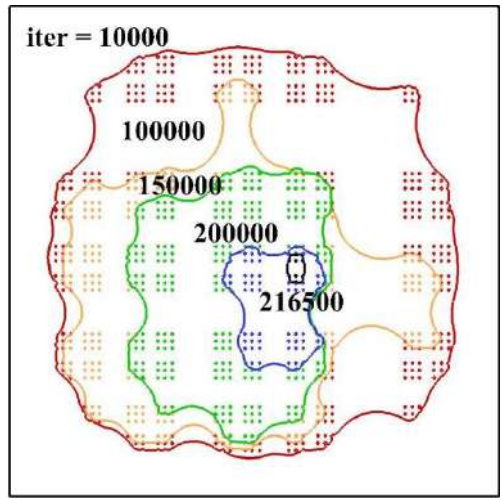

Fig.7.14. Shapes and positions of the liquid interface within hierarchical pillar array at the different iteration times of 10 000, 100 000, 150 000, 200 000 and 216 500.

## 7.5. Conclusion

In this chapter, diverse multiphase phenomena were explored, inspired in two cases by experimental studies. First, the movement of a droplet on a heterogeneous surface with randomly distributed patches with different radius and contact angle was studied. It was shown that the droplet moves until it finds its equilibrium position characterized by minimal surface energy. Second, the movement of a droplet over stepping stones was studied, inspired by the experiment of the surface tension sorter. It was shown that a droplet exposed to a body force will jump over hydrophobic unfilled channels due to inertia, but will be entrapped into a channel partially filled by liquid due to cohesive forces. Finally, evaporative drying in regular and hierarchical porous structures is studied inspired by experimental work on drying of fluids from micropillar arrays between two plates. The important influence of the hierarchical structure on the evaporation rate and pattern of the liquid remaining entrapped was illustrated.

LBM clearly showed in this chapter its potential to study in detail liquid configurations which are difficult to capture in experimental work. These explorative studies illustrated also the potential to analyze multiphase flow in complex materials like PA, including droplet wetting on the uneven surface of PA, droplet displacement inside PA covered by hydrophobic bitumen and evaporative drying of liquid island remaining in PA after drainage.

# 8. CONCLUSIONS AND OUTLOOK

## 8.1. Summary

The overall goal of the research of this thesis was to better understand the physics of multiphase flow at pore scale in porous asphalt (PA). The physical processes studied were capillary uptake, corner flow, droplet run-off and evaporation, drying and gravity-driven drainage. The Lattice Boltzmann method (LBM) was used for simulating these processes, and included also a comparison of LBM results with experimental data and analytical models. The specific objectives of this work were:

- To develop the three-dimensional Shan-Chen pseudopotential LBM with parallel computing, MPI, for the study of multiphase flow taking into account gravity.

- To develop a framework to validate and verify LBM by comparing with analytical solutions for Laplace pressure, contact angle and capillary uptake, and demonstrate the applicability of LBM to explore parametrically multiphase problems.

- To explore multiphase phenomena using LBM, through quantitative analysis and qualitative exploration. In particular:
  - To document capillary uptake phenomena, with special attention to the phenomenon of corner flow, in two- and three-dimensional geometries at pore scale;

o To model and understand the behavior of sessile droplets on different surfaces considering the influence of contact angle, surface geometry and surface structure;

o To explore and understand gravity-driven drainage for a complex porous medium.

The work performed and the main conclusions are summarized as follows:

*Capillary phenomena*

• Capillary rise in single pores was studied with two- and three-dimensional LBM. The uptake process, characterized by the capillary height versus time was compared with analytical solutions and an overall good agreement was obtained for pore configurations such as 2D parallel plates and 3D square tubes. In a parametric study, the rate of capillary uptake was found to increase with increasing size and decreasing contact angle. However, for capillary rise into a 3D circular pore geometry, the artificial roughness induced by the misalignment of the lattice with the physical boundary was found to lead to an over prediction of the capillary uptake rate.

• The LBM study of capillary rise in polygonal (square and triangular) tubes showed that the liquid configuration depends on a critical contact angle. A pore meniscus is formed at a contact angle larger than the critical contact angle, while a pore meniscus and corner arc menisci are formed at a contact angle smaller than critical contact angle. The relations of corner arc curvature versus saturation degree were found to agree very well with analytical relations. The lattice-induced roughness in the surfaces of triangular pores was found to lead to a differential wetting and corner menisci of different height in the corners, but the average menisci curvature and heights follow the expected trend as compared with the analytical solution.

• Corner flow was further explored for configurations including a straight path, a straight path with a U-bent and a staircase path. Since there is very little experimental investigation of corner flow related to more complex configurations, LBM was used as a first attempt to explore these geometries. It was found that the geometry of the path has not a significant effect on the temporal evolution of curvature and wetted area ratio at the corners, and that the corner flow is not hindered by bends in the corner path.

*Droplet-surface interactions*

- In a 2D LBM study of a droplet run-off on a grooved surface, different regimes have been identified depending on the groove size. The effects of groove geometry, wettability of the surface and of the groove, and of the tilt angle of the surface on the run-off and on the liquid remaining in the groove have been documented.

- The process of droplet evaporating on a set of pillars was studied with two-dimensional LBM while varying pillar and pitch widths. The pitch width was found to play a significant role in governing the droplet depinning mechanism during evaporation. LBM results provided detailed information in terms of internal liquid flow in the droplet and in the capillaries between the micropillars towards the triple point of the droplet. It was found that the internal flow tries to compensate for the high evaporative flux at the triple point. With increasing pitch width, the excess free energy and critical contact angle was found to decrease. The change of direction of the internal fluid flow in the space between the pillars was found to play an essential role in the depinning process of the droplet. The droplet becomes depinned when the internal flow cannot compensate anymore for the high evaporative flux at the contact point.

- The study of the process of droplet wetting of a heterogeneous surface, namely a checkerboard with regular arranged hydrophilic and hydrophobic patches, showed that the equilibrium droplet shape depends on the ratio of patch size versus droplet diameter. The local contact angle on hydrophilic (or hydrophobic) patch increases (or decreases) with decreasing patch/droplet ratio.

• The droplet is found to wet preferentially hydrophilic patches trying to accommodate its shape in such way to minimize the surface wetting energy. A further study of droplet wetting on randomly distributed heterogeneous patches showed that the droplet will move until the droplet finds its minimal surface energy at equilibrium.

*Drainage and drying from porous media*

• Gravity-driven drainage in a model porous asphalt specimen was studied with two- and three-dimensional LBM and the results were compared with experimental data in terms of liquid mass fraction versus time. The two-dimensional LB results do not reproduce accurately the drainage behavior as seen in the experiment due to the simplifications resulting from the 2D domain representation. On the contrary, three-dimensional LBM results show an overall good agreement with experimental data and three different regimes in the drainage process were identified.

• The drying of a liquid entrapped in 2D regular and hierarchical porous structures was studied with LBM, showing configurations of the liquid-vapor interface evolving with time during drying. The arrangement of the micropillar structure was found to highly determine the evolution of the interface patterns.

## 8.2. Contributions

This thesis contributed importantly to advances in LBM modeling which allowed for a better understanding of diverse multiphase flow processes. The modeling contributions (advances) in LBM can be summarized as:

• Implementation of body force as a forcing term, which is applied to calculate velocity part, in the two-dimensional Shan-Chen pseudopotential LBM by accounting for droplet run-off on vertical surface and gravity-driven drainage in porous media, especially PA.

- Extension of two-dimensional LBM code into three-dimensional LBM code to take into account multiphase phenomena which occur in only three-dimensional system including corner flow and liquid film that might have influence on liquid distribution and residence time and amount in the system. With some additional work to develop the code, implementation of a parallel computing program, Message Passing Interface (MPI), into three-dimensional LBM code required less computational cost for space and time to perform(investigate) complicated multiphase phenomena and geometry.

- Development of integrated methodology to perform parametric studies since LBM is bottom-up approach, meaning that no explicit relations for some parameters, and to verify and validate LBM works by comparison with analytical solutions before moving to solve complicate multiphase phenomena in porous media and geometry.

In terms of contributions to the scientific understanding of multiphase flow in general, this study includes:

- Investigation of capillary rise in a single capillary taking into account tube shape and size, surface wettability, topology and structure and comparison with analytical solutions to validate the LB results. The modeling could reproduce the formation of pore meniscus and the occurrence of corner arc menisci depending on liquid wettability. For a first time, with LBM, the temporal evolution of curvatures and saturation degrees at the corners depending on contact angle of surface could be observed, which not been done yet experimentally. The modeling shows that the initial wetting of the corners is extremely rapid, but almost meaningless in terms of mass. With time, the thin filament of liquid in the corner thickens to fill progressively the corner.

No significant effect of bends in the geometry, such in a staircase on corner flow was observed.

- Understanding the run-off droplet by gravity on flat and grooved vertical surfaces for different groove size, contact angle and surface tilt angle. On grooved surfaces, four regimes can be identified by studying the remaining liquid fraction and distribution inside the groove: 1) height controlled regime, 2) bottom surface controlled regime, 3) top-bottom surface controlled regime and 4) top surface controlled regime.

- Study of stick-slip behavior of an evaporating droplet deposited on a set of micropillars in terms of pillar and pitch widths. It is found that pitch width has a significant impact on the pinning/depinning process, the critical contact angle and excess Gibbs free energy. The internal liquid flow inside the droplet and capillaries has been documented showing that depinning occurs when the internal fluid flow cannot compensate anymore for the high evaporative flux at the triple point.

- Investigation of the equilibrium shape and local contact angle of a droplet deposited on a checkerboard heterogeneous surface to explore the domain of application of Cassie's equation. With increasing the ratio of patch size to droplet radius, the local contact angle on hydrophilic (or hydrophobic) patches decreases (or increases) and, inversely, the wetted area on hydrophilic (or hydrophobic) patches increases (or decreases). The detailed droplet shapes and the relation between contact angle and wetted area are discussed.

In terms of contributions of scientific understanding of multiphase flow in porous media, this study includes:

- Investigation of gravity-driven drainage in PA with two-and three-dimensional LBM, and with the experimental data. The LB results show

overall good agreement with experimental data, validating LBM. In the drainage process three regimes have been identified: 1) initialization-controlled regime, 2) gravity-controlled regime and 3) geometry-controlled regime. Gravity-controlled and geometry-controlled regimes have been found to depend highly on the detailed pore structures, especially narrow throats between the aggregates.

## 8.3.  Further perspectives

The conducted study has provided some answers to gaps in scientific knowledge as identified in the introduction. Nevertheless, a number of directions for further research can be formulated based on this study to provide better insight of multiphase flow with LBM investigations.

The capillary rise study could be pursued considering more realistic aspects as follows:

- Simulation of real porous asphalt geometries by three-dimensional LBM considering also the specific pore structure and complex connectivity composed by several pores and throats including also dead-end pores, the roughness of the pore surface, more complex phenomena like air entrapment and water islands formation. Such a study would provide better understanding of gravity driven drainage, the remaining water and the residence time of water in porous medium like PA.

- Gravity should to be taken into account by considering larger pore length scales than the capillary length to document properly the effect of gravity on capillary rise evolution over time.

The droplet study would provide more various insights in the following ways:

- The droplet study could be extended to consider several droplets simultaneously deposited on a heterogeneous surface including different

surface and droplet characteristics, such as surface tension, viscosity and density. The coalescence of several droplets will provide further insights to understand and explain complex wetting phenomena on different surfaces.

- The investigation of droplet impingement for different impact velocities or height (e.g. by using entropic LBM (Karlin, Ansumali et al. 2006)) on homogeneous and heterogeneous surfaces could provide answers on the dynamic wetting behavior by estimating the maximum spreading diameter, dynamic contact angle and the occurrence of droplet oscillations.

- The current LBM is isothermal. Enriching the current LBM to non-isothermal conditions, by including heat transfer would allow extending the study of evaporating taking into account latent heat and convective heat transfer processes.

The porous asphalt study would benefit from some improvements as follows:

- A detailed analysis of drainage and evaporation in porous asphalt including different pore networks, aggregate size and tilt angle, the understanding of liquid configurations and remaining liquid distributions inside PA would improve a lot, which could lead to the improvement of pavement design in future.

- LB results could be used for an upscaling to continuum approaches incorporating gravity-driven drainage, capillary uptake and drying in complex porous media.

Improvements of LBM could include:

- The mesh-alignment and zigzag boundary problem introducing an artificial roughness in the LB simulations could be solved by implementing the curved, half bounce-back or moving boundary condition approach (e.g. as per Mei, Luo et al. 1999, Mei, Shyy et al. 2000, Dorschner, Chikatamarla et al. 2015), which allow the tracking of the interface independently of the regular mesh. This LBM would allow to study complex domains, including curved and inclined shapes.

- To investigate complex multiphase phenomena, including evaporation and dissolution-precipitation processes, the current LBM model needs to be further enriched.

# Addendum

## A1 Contact angle of droplet

Temporal evolution of contact angle of evaporating droplet with pitch widths of (a) 4; (b) 7; and (c) 10 lattices.

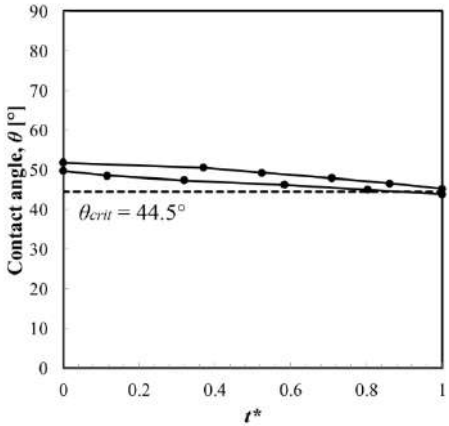

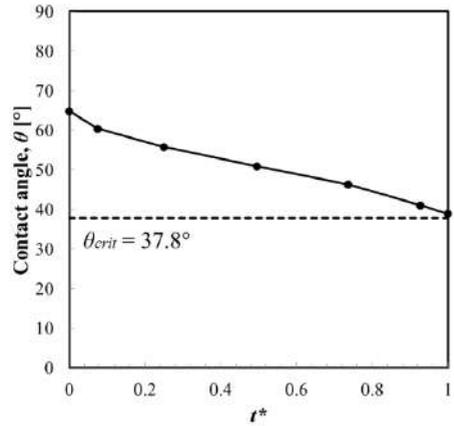

(a) b= 4 lattices                    (b) b = 7 lattices

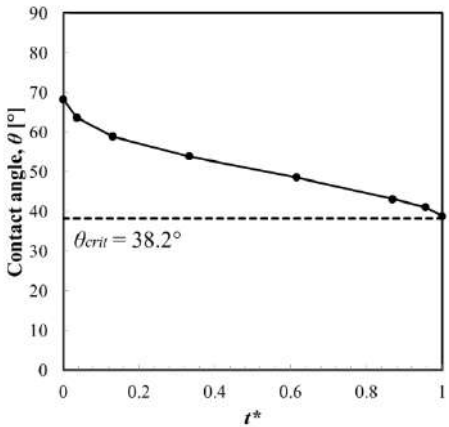

(c) b = 10 lattices

## **A2 Contact angle in the capillaries between the micropillars**

Temporal evolution of contact angle in the capillaries between the micropillars at edge, next edge and middle capillary with pitch widths of (a) 4; (b) 7; and (c) 10 lattices.

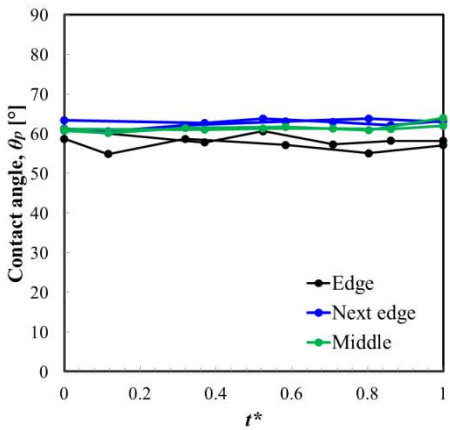

(a) b= 4 lattices

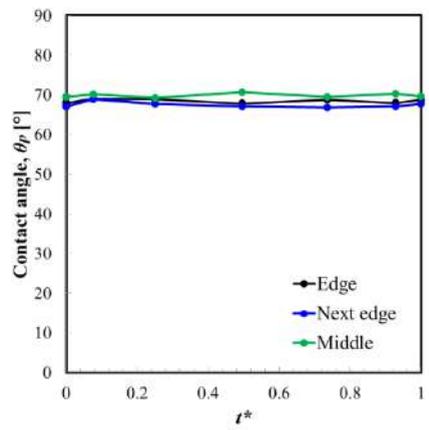

(b) b = 7 lattices

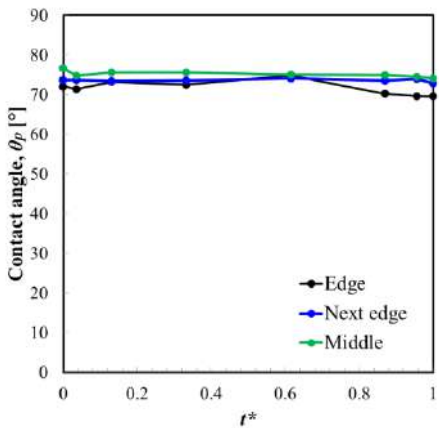

(c) b = 10 lattices

## <u>A3 Capillary pressures at meniscus inside capillaries</u>

Temporal evolution of capillary pressure inside the edge, next to edge and middle capillaries with pitch widths of (a) 4; (b) 7; and (c) 10 lattices. LB results in black are compared with Laplace equation (Eq. (5.11)) in red.

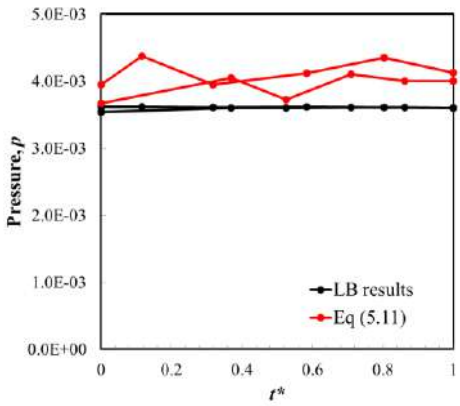

edge pillar

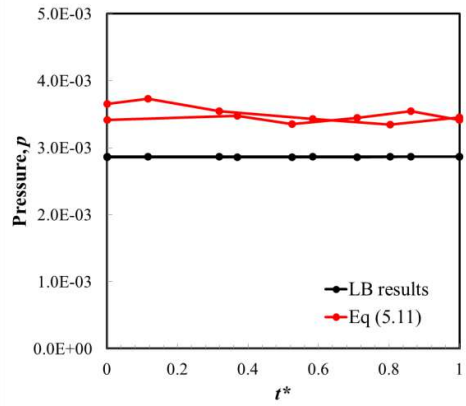

next edge pillar

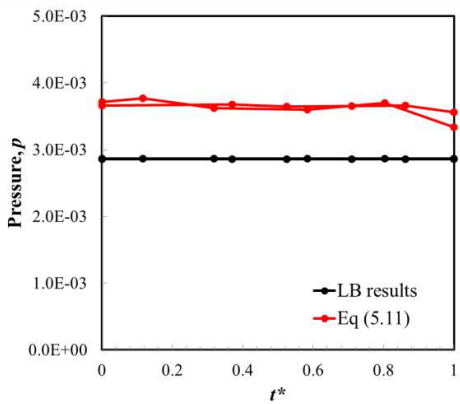

middle pillar

(a)  b = 4 lattices

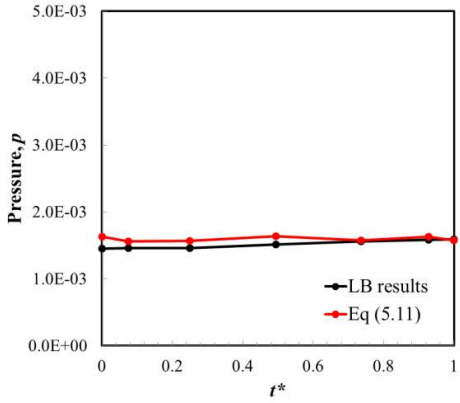

edge pillar

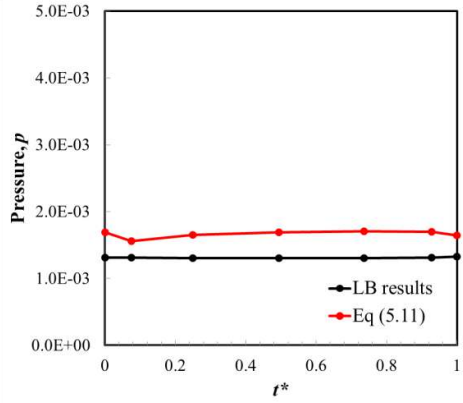

next edge pillar

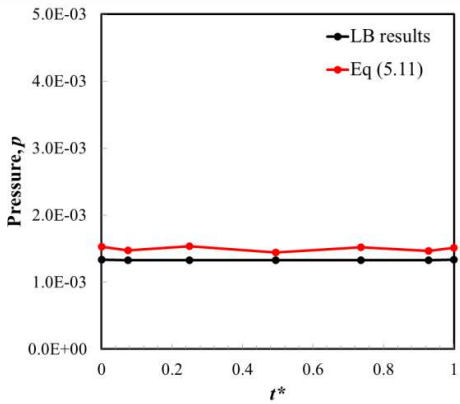

middle pillar

(b)  b = 7 lattices

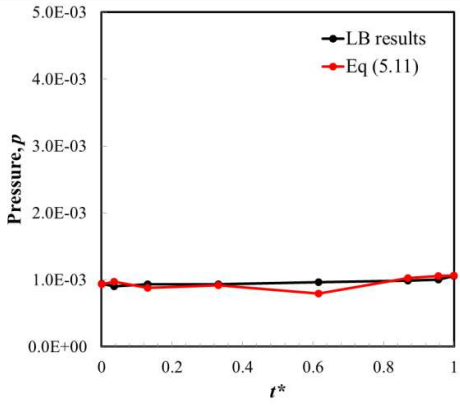

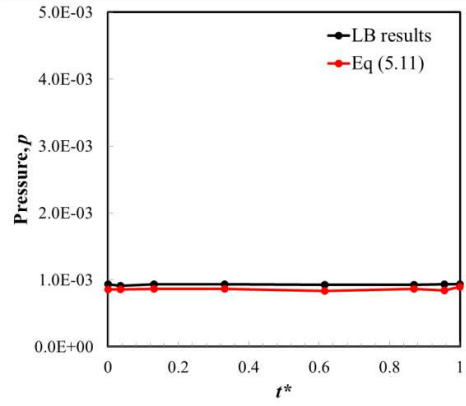

edge pillar                                    next edge pillar

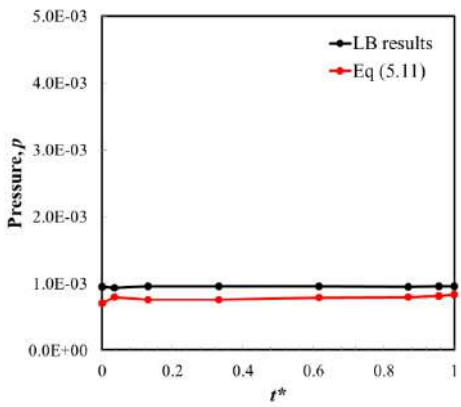

middle pillar

(c)  b = 10 lattices

# A4 Liquid pressures in the droplet

Temporal evolution of liquid pressure in bulk droplet (PL) and in three different capillaries at the edge (P1), the next edge (P2) and middle (P3) for pitch widths of 4, 7 and 10 lattices. Depending on locations inside micro pores, the second numbers refers to top or bottom: 1 for top and 2 for bottom.

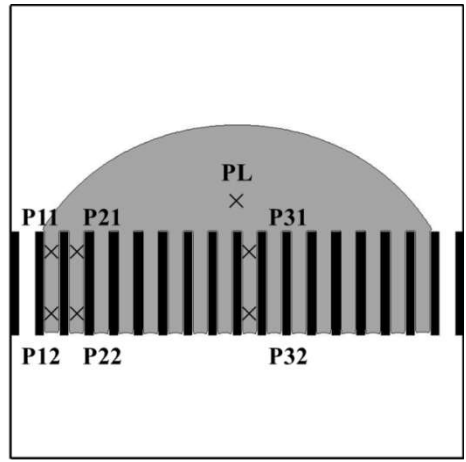

(a)  Schematic of locations inside droplet and pillars

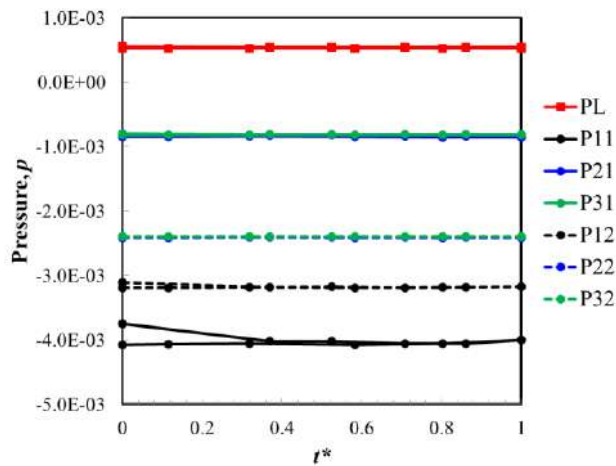

(b)  b = 4 lattices

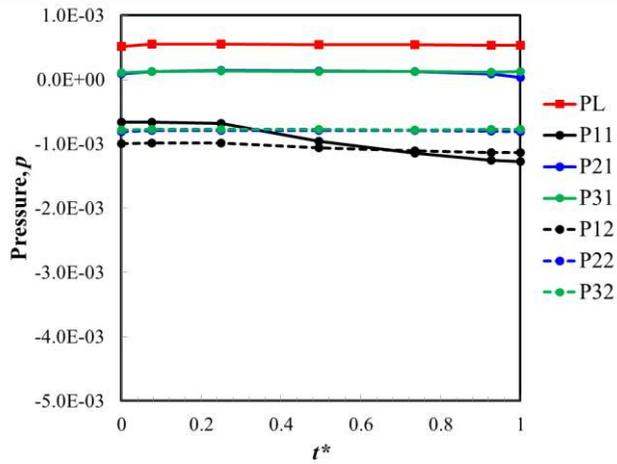

(c)  b = 7 lattices

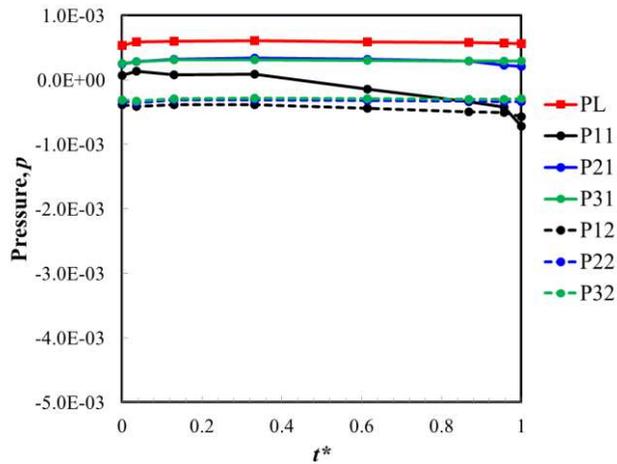

(d)  b = 10 lattices